%% file: flups_0_main.tex
\documentclass[a4paper,final,journal]{IEEEtran}

\newif\ifarxiv
\newif\ifclean

\arxivtrue
\cleantrue

\input{header}

\input{flups_author}

\definecolor{asparagus}{rgb}{0.53, 0.66, 0.42}
\definecolor{cerulean}{rgb}{0.0, 0.48, 0.65}
\definecolor{brickred}{rgb}{0.8, 0.25, 0.33}
\definecolor{ferngreen}{rgb}{0.31, 0.47, 0.26}
\hypersetup{
    colorlinks=true,
    linkcolor=cerulean,
    filecolor=magenta,      
    urlcolor=brickred,
    citecolor=ferngreen
    }

\ifarxiv
\makeatletter\@input{flups_0_sm_aux.tex}\makeatother
\nolinenumbers
\fi
\usepackage{xr}
\begin{document}
\maketitle

\textit{Note: published in \href{https://doi.org/10.1109/TPDS.2023.3254302}{IEEE Transactions on Parallel and Distributed Systems}.
This version includes updated results after the correction of a typo in \cite{Caprace:2021}, see \sect{sec2_biot_savart} and \sect{app:biosavart:conv}}.

\begin{abstract}
Massively parallel Fourier transforms are widely used in computational sciences, and specifically in computational fluid dynamics which involves unbounded Poisson problems.
In practice the latter is usually the most time-consuming operation due to its inescapable all-to-all communication pattern.
The original \flups library tackles that issue with an implementation of the distributed Fourier transform tailor-made for successive resolutions of unbounded Poisson problems.
However the proposed implementation lacks of flexibility as it only supports cell-centered data layout and features a plain communication strategy.
This work extends the library along two directions. First, \flups' implementation is generalized to support a node-centered data layout. Second,  three distinct approaches are provided to handle the communications: one all-to-all, and two non-blocking implementations relying on manual packing and \code{MPI\_Datatype} to communicate over the network.
The proposed software is validated against analytical solutions for unbounded, semi-unbounded, and periodic domains.
The performance of the approaches is then compared against \code{accFFT}, another distributed FFT implementation, using a periodic case.
Finally  the performance metrics of each implementation are analyzed and detailed on various top-tier European \revthree{facilities} up to $\mathbf{49,152}$ cores.
This work brings \flups up to a fully production-ready and performant distributed FFT library, featuring all the possible types of FFTs and with flexibility in the data-layout.
The code is available under a BSD-3 license at \url{github.com/vortexlab-uclouvain/flups}.
\end{abstract}


\input{flups_1_intro}

\input{flups_2_method}

\input{flups_3_implementation}

\input{flups_4_validation}

\input{flups_5_results}

\input{flups_6_conclusion}

\section*{Acknowledgments}
We would like to acknowledge the insightful discussions with Wim van Rees (MIT) and Matthieu Duponcheel (UCLouvain), as well as the help received from \typo{Ken Raffenetti (ANL), Hui Zhou (ANL)}, Barbara Krasovec (Vega), and Wahid Mainassara (MeluXina) to compile and run on different \revthree{supercomputers}.
Further we wish to acknowledge the financial support from the Wallonie-Bruxelles International (WBI) excellence fellowship (TG), the F.R.S.-FNRS postdoctoral fellowship (TG), and the Université catholique de Louvain (PB and PC).
Computational resources have been provided by the Consortium des Équipements de Calcul Intensif (CÉCI) funded by the Fonds de la Recherche Scientifique de Belgique (F.R.S.-FNRS) under Grant No. 2.5020.11 and by the Walloon Region. Additional resources include the Tier-1 supercomputer of the Fédération Wallonie-Bruxelles, infrastructure funded by the Walloon Region under the grant agreement n°1117545.
Additional computational resources have been provided by EuroHPC for the access to MeluXina and Vega (\textit{EHPC-BEN-2022B01}, \textit{EHPC-BEN-2022B06}), as well as LUMI-C (\textit{EHPC-DEV-2022D01}).

{\footnotesize
\bibliographystyle{IEEEtran}
\bibliography{IEEEabrv,AllPublications.bib}
}

\vspace*{-3\baselineskip}
\begin{IEEEbiography}[{\includegraphics[width=1in,height=1.25in,clip,keepaspectratio]{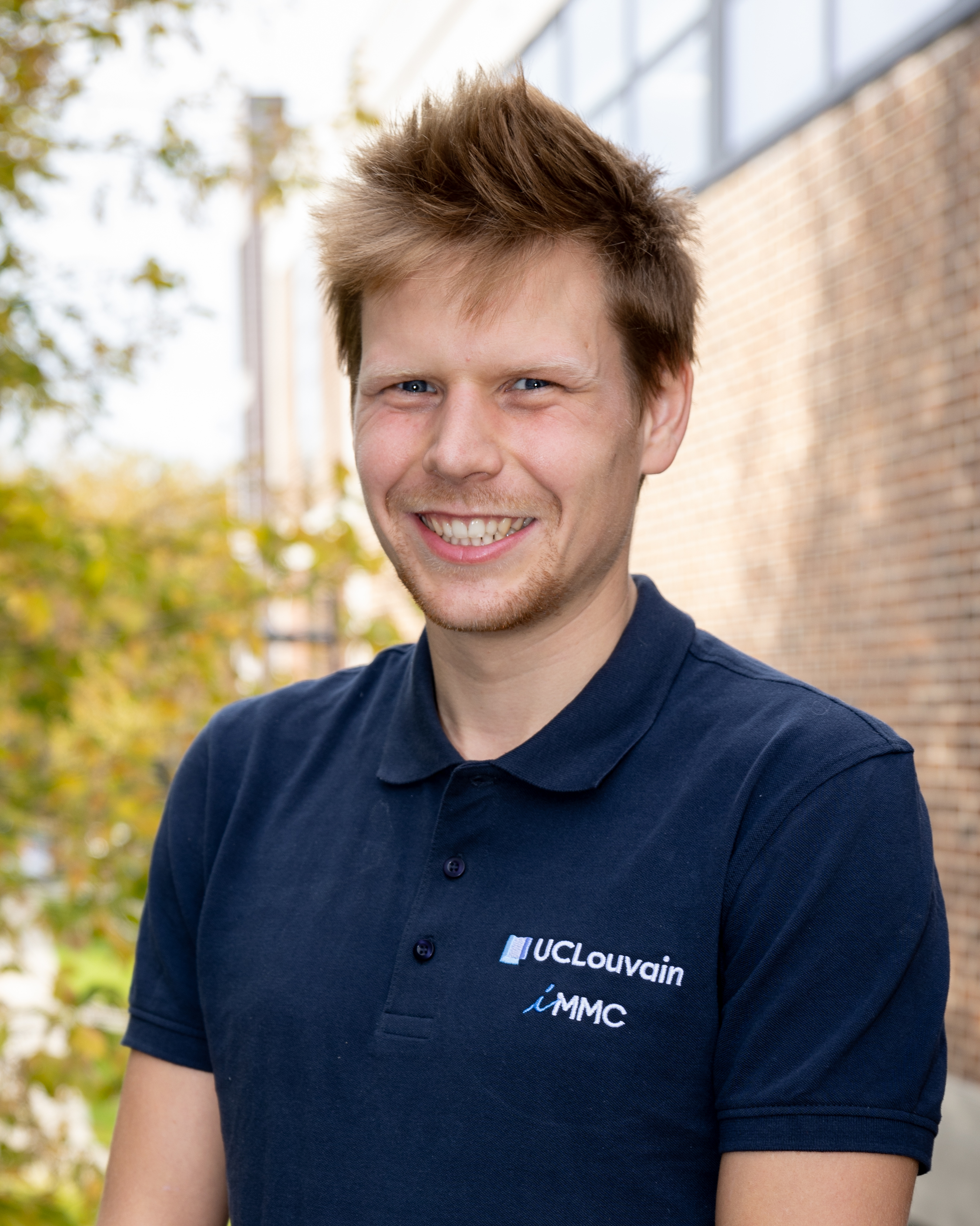}}]{Pierre Balty} obtained his Master degree from UCLouvain in 2019 and is currently a PhD student and teaching assistant in the institute for Mechanics, Material, and Civil engineering at UCLouvain. His research focuses on Lagrangian numerical methods, their deployment on distributed systems, and their applications to wind energy. 
\end{IEEEbiography}
\vspace*{-3\baselineskip}
\begin{IEEEbiography}[{\includegraphics[width=1in,height=1.25in,clip,keepaspectratio]{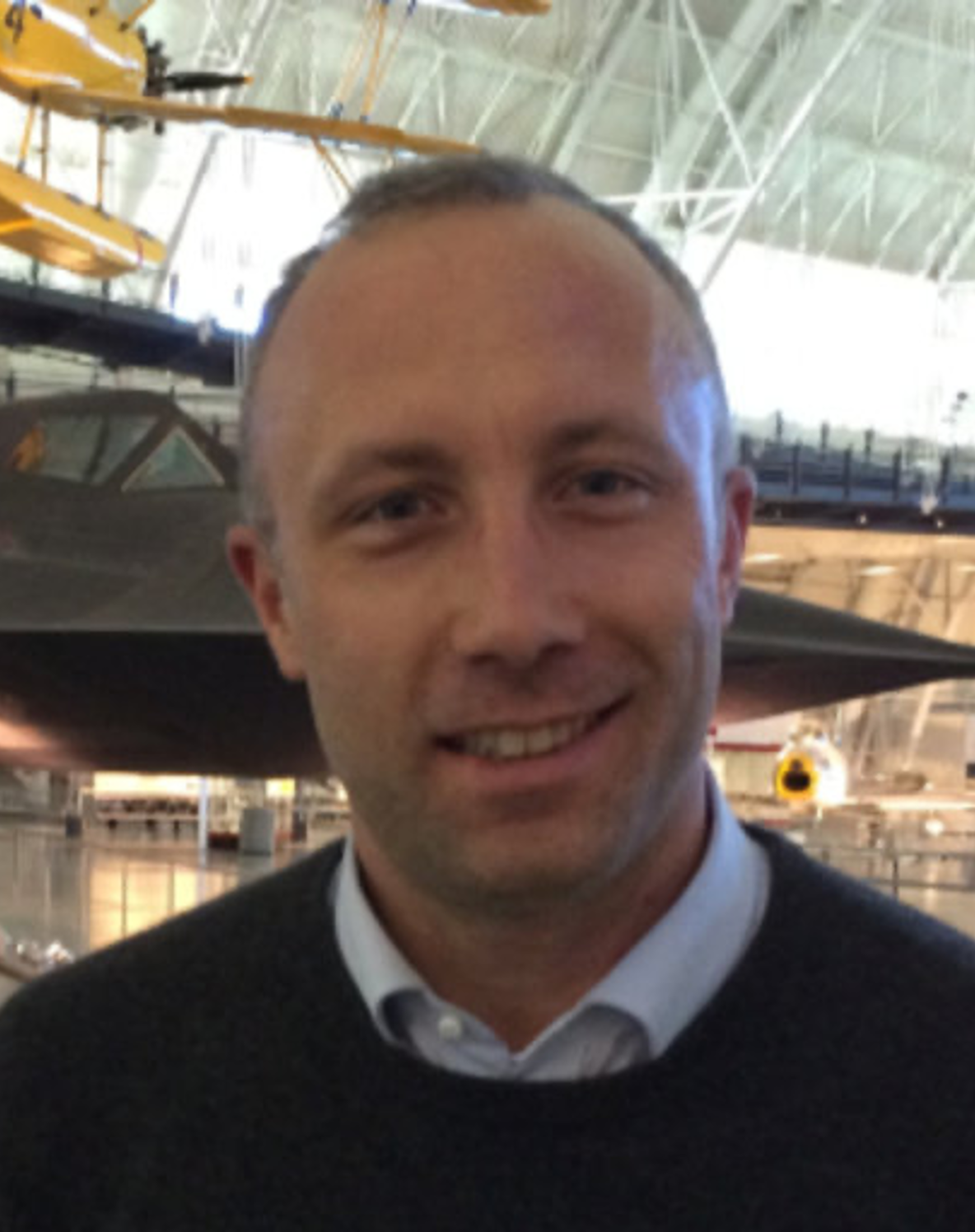}}]{Philippe Chatelain} obtained his Ph.D. from Caltech in 2005 and held a research associate position at ETH Zurich till 2009. He is currently full Professor at UCLouvain, leading the Vortex and Turbulence research group. His research interests cover fluid mechanics, Lagrangian numerical methods, their deployment in HPC environment, and their application to fundamental and applied problems in bio-propulsion, aeronautics, and wind energy.
\end{IEEEbiography}
\vspace*{-3\baselineskip}
\begin{IEEEbiography}[{\includegraphics[width=1in,height=1.25in,clip,keepaspectratio]{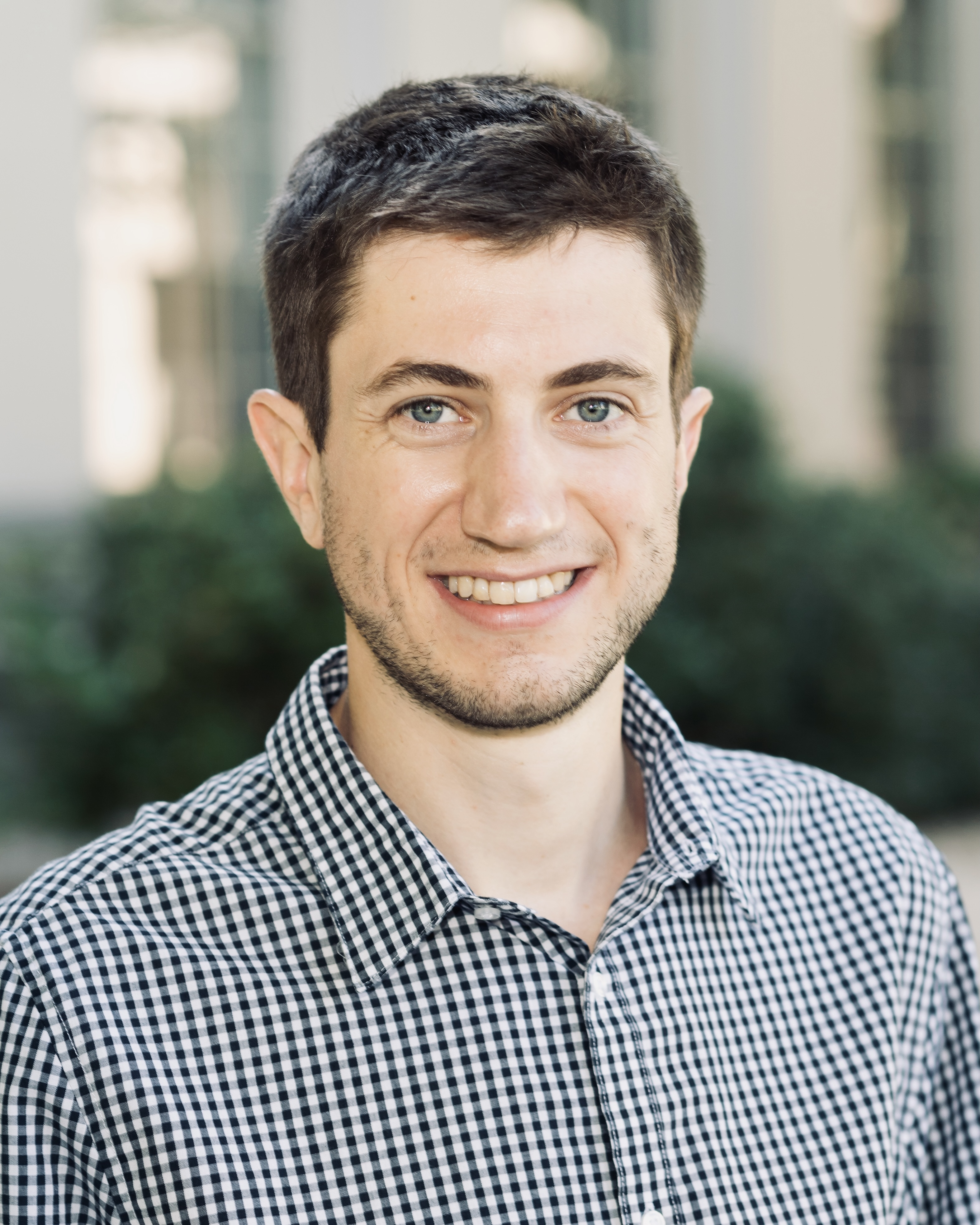}}]{Thomas Gillis} obtained his Ph.D. from UCLouvain in 2019. At the time of this work he was postdoctoral researcher at UCLouvain and visiting scientists at MIT. His research focuses on distributed systems, lossless compression for PDEs, and scalable and performant framework for computational fluid dynamics.
Thomas is now part of the Argonne National Lab working on distributed systems.
\end{IEEEbiography}

\ifarxiv
\clearpage
\input{flups_7_appendices}
\fi

\end{document}

%% file: header.tex
\usepackage[utf8]{inputenc}
\usepackage[english]{babel}
\usepackage{amsmath}
\usepackage{amsfonts}
\usepackage{amssymb}
\usepackage{amsopn}
\usepackage{bm}

\usepackage{fourier}
\usepackage[left=1.4cm,right=1.4cm,top=1.5cm,bottom=1.5cm]{geometry}

\usepackage{bm}
\usepackage{lineno}
\usepackage{hyperref}
\usepackage{cleveref}
\usepackage[boxruled,vlined,algo2e]{algorithm2e} 
\usepackage{setspace}
\usepackage{xcolor}
\usepackage{wrapfig}
\usepackage{graphicx}
\usepackage{pstool}
\usepackage{multirow}
\usepackage{setspace}
\usepackage{enumitem}
\usepackage{multirow}
\usepackage{subcaption}
\usepackage{comment}
\usepackage{xspace} %
\usepackage{tablefootnote} %
\usepackage{soul} %

\graphicspath{{./figures/}}

\usepackage{pgfplots}
\usepgfplotslibrary{external} 
\pgfplotsset{compat=newest}

\usepackage{tikz}%
\usetikzlibrary{shapes}
\usetikzlibrary{patterns}
\usetikzlibrary{shapes.misc}
\usetikzlibrary{arrows}
\usepgfplotslibrary{groupplots}
\usepgfplotslibrary{dateplot}
\usetikzlibrary{backgrounds}

\usepackage[numbers,sort&compress]{natbib}

\input{symbols_and_cmd}

%% file: symbols_and_cmd.tex
\SetCommentSty{mycommfont}

\SetKwFunction{FnStartAll}{MPI\_Startall}
\SetKwFunction{FnStart}{MPI\_Start}
\SetKwFunction{FnIsend}{MPI\_Isend}
\SetKwFunction{FnIrecv}{MPI\_Irecv}
\SetKwFunction{FnTestSome}{MPI\_Testsome}
\SetKwFunction{FnAlltoAll}{MPI\_Ialltoallv}
\SetKwFunction{FnMPIWait}{MPI\_Wait}
\SetKwFunction{FnResetData}{reset}
\SetKwFunction{FnShuffle}{shuffle}
\SetKwFunction{FnComputeOngoing}{ComputeOngoingSend}
\SetKwFunction{FnPack}{pack}
\SetKwFunction{FnUnpack}{unpack}
\SetKwFunction{FnSendBatch}{SendRqst}
\SetKwFunction{FnRecvBatch}{RecvRqst}

\newcommand{\block}{{b}}
\newcommand{\Block}{{B}}
\newcommand{\btosend}{{b_{\operatorname{send}}}}
\newcommand{\btorecv}{{b_{\operatorname{recv}}}}
\newcommand{\brecv}{{b_{\operatorname{recv}}}}

\newcommand{\btounpack}{{b_{\operatorname{unpack}}}}

\newcommand{\srequest}{{\operatorname{\textit{rqst}}_{\operatorname{send}}}}
\newcommand{\rrequest}{{\operatorname{\textit{rqst}}_{\operatorname{recv}}}}

\newcommand{\nsendbatch}{\operatorname{n}_{\operatorname{batch}}} %
\newcommand{\nsendpending}{\operatorname{n}_{\operatorname{max-pending}}} %

\newcommand{\ntosend}{\operatorname{n}_{\operatorname{to-send}}}
\newcommand{\nongoing}{\operatorname{n}_{\operatorname{ongoing}}}

\newcommand{\bufferdata}{\operatorname{data}}
\newcommand{\buffersend}{\operatorname{buf}_{\operatorname{send}}}
\newcommand{\bufferrecv}{\operatorname{buf}_{\operatorname{recv}}}
\newcommand{\blr}[1]{\left[ #1 \right]}  %

\newcommand{\code}[1]{\texttt{\small{#1}}}

\newcommand{\centercell}[1]{\multicolumn{1}{c}{#1}}

\newcommand{\ddsq}[2]{\frac{d^{2} #1}{d #2^{{2}}}}

\newcommand{\eg}{\textit{e.g.}}

\newcommand{\tf}{\tilde{f}}

\newcommand{\tG}{\tilde{G}}

\newcommand{\ic}{\mathsf{i} }

\newcommand{\ie}[1]{\textit{i.e.} #1}

\newcommand{\lapl}{\nabla^{2}}

\newcommand{\plr}[1]{\left( #1 \right)}  %

\newcommand{\x}{\mathbf{x}}

\newcommand{\dx}{\;\operatorname{d}\!{x}}

\newcommand{\ata}{\code{a2a}\xspace}
\newcommand{\isr}{\code{isr}\xspace}
\newcommand{\nb}{\code{nb}\xspace}

\newcommand{\fftw}{\code{fftw}\xspace}
\newcommand{\flups}{\code{flups}\xspace}
\newcommand{\accfft}{\code{accFFT}\xspace}

\SetKwFunction{FnGet}{MPI\_Get}
\SetKwFunction{FnPut}{MPI\_Put}
\SetKwFunction{FnPost}{MPI\_Win\_Post}
\SetKwFunction{FnComplete}{MPI\_Win\_Complete}
\SetKwFunction{FnWait}{MPI\_Win\_Wait}
\SetKwFunction{FnSyncStatus}{SyncStatus}

\SetKwFunction{FnCopy}{copy}
\SetKwFunction{FnSample}{sample}
\SetKwFunction{FnBounday}{BoundaryCondition}

\SetKwFunction{FnRefine}{WaveletRefine}
\SetKwFunction{FnCoarsen}{WaveletCoarsen}
\SetKwFunction{FnCriterion}{WaveletCriterion}
\SetKwFunction{FnOverwrite}{WaveletSubstitution}
\SetKwFunction{FnInterpolate}{WaveletInterpolate}
\SetKwFunction{FnSmooth}{WaveletUpdateAfterCoarsening}

\SetKwFunction{FnStatus}{StatusUpdate}
\SetKwFunction{FnPolicy}{EnforceStatusPolicy}

\SetKwFunction{FnTreeRefine}{p4estRefine}
\SetKwFunction{FnTreeCoarsen}{p4estCoarsen}
\SetKwFunction{FnTreeBalance}{p4estBalance}
\SetKwFunction{FnTreePartition}{p4estPartition}

\SetKwFunction{FnGhostPost}{GhostPull\_Post}
\SetKwFunction{FnGhostWait}{GhostPull\_Wait}
\SetKwFunction{FnGhost}{Ghost}

\SetKw{KwOr}{or}

\newcommand{\be}{\begin{equation}}
\newcommand{\eec}{\;\;,\end{equation}}
\newcommand{\ee}{ \end{equation}}
\newcommand{\eed}{\;\;.\end{equation}}
\newcommand{\bse}{\begin{subequations}}
\newcommand{\ese}{\end{subequations}}

\newcommand{\MBps}{M\!\!\:B/sec}

\newcommand{\Tref}[1]{\Cref{#1}}
\newcommand{\Fref}[1]{\Cref{#1}}
\newcommand{\Eqqref}[1]{\Cref{#1}}
\newcommand{\Algref}[1]{Algorithm~\ref{#1}}
\newcommand{\Frefs}[2]{\Cref{#1} and \Cref{#2}}
\newcommand{\Sect}[1]{\Cref{#1}}
\newcommand{\Appendix}[1]{Appendix~\ref{#1}}

\newcommand{\fref}[1]{\cref{#1}}

\newcommand{\algref}[1]{Algorithm~\ref{#1}}

\newcommand{\sect}[1]{\cref{#1}}

\definecolor{darkGreen}{RGB}{0,160,0}
\definecolor{darkBlue}{RGB}{51,51,204}
\definecolor{darkRed}{RGB}{160,0,0}
\definecolor{orange}{RGB}{255,120,0}
\definecolor{myGreen}{rgb}{0.0706, 0.5333, 0.1921}
\definecolor{myBrown}{rgb}{0.43, 0.21, 0.1}

\ifclean
\newcommand{\revone}[1]{{#1}}
\newcommand{\revtwo}[1]{{#1}}
\newcommand{\revthree}[1]{{#1}}

\newcommand{\typo}[1]{{#1}}
\newcommand{\delete}[1]{}
\newcommand{\us}[1]{{#1}}
\else
\newcommand{\revone}[1]{\textcolor{darkBlue}{#1}}
\newcommand{\revtwo}[1]{\textcolor{darkGreen}{#1}}
\newcommand{\revthree}[1]{\textcolor{orange}{#1}}

\newcommand{\typo}[1]{\textcolor{darkRed}{#1}}
\newcommand{\delete}[1]{\textcolor{darkRed}{\st{#1}}}
\newcommand{\us}[1]{\textcolor{myBrown}{{#1}}}
\fi

\definecolor{airforceblue}{rgb}{0.36, 0.54, 0.66}
\definecolor{carolinablue}{rgb}{0.6, 0.73, 0.89}
\definecolor{moonstoneblue}{rgb}{0.45, 0.66, 0.76}
\definecolor{amber}{rgb}{1.0, 0.75, 0.0}
\definecolor{cadmiumorange}{rgb}{0.93, 0.53, 0.18}
\definecolor{persianorange}{rgb}{0.85, 0.56, 0.35}
\definecolor{cadetgrey}{rgb}{0.57, 0.64, 0.69}
\definecolor{darkgray}{rgb}{0.66, 0.66, 0.66}
\definecolor{asparagus}{rgb}{0.53, 0.66, 0.42}
\definecolor{cambridgeblue}{rgb}{0.64, 0.76, 0.68}
\definecolor{olivine}{rgb}{0.6, 0.73, 0.45}
\definecolor{carnelian}{rgb}{0.7, 0.11, 0.11}
\definecolor{cardinal}{rgb}{0.77, 0.12, 0.23}
\definecolor{firebrick}{rgb}{0.7, 0.13, 0.13}

\def\figblue{moonstoneblue}
\def\figorange{persianorange}
\def\figgrey{darkgray}
\def\figgreen{olivine}

\definecolor{matlabBlue}{rgb}{0, 0.4470, 0.7410}
\definecolor{matlabOrange}{rgb}{0.8500, 0.3250, 0.0980}
\definecolor{matlabYellow}{rgb}{0.9290, 0.6940, 0.1250}
\definecolor{matlabPurple}{rgb}{0.4940, 0.1840, 0.5560}
\definecolor{matlabGreen}{rgb}{0.4660, 0.6740, 0.1880}
\definecolor{matlabRed}{rgb} {0.7765, 0.1569, 0.0902}
\definecolor{matlabBlack}{rgb}{0, 0, 0}

\definecolor{keyBlue}{rgb}{0.0392, 0.2274, 0.4235}
\definecolor{keyBlueTrans}{rgb}{0.4274, 0.4980, 0.5764}
\definecolor{keyGreen}{rgb}{0.05490, 0.38039, 0.003921}
\definecolor{keyGreenTrans}{rgb}{0.7019, 0.8313, 0.7176}
\definecolor{keyOrange}{rgb}{0.98039, 0.4980, 0.03529}
\definecolor{keyGray}{rgb}{0.50196, 0.50196, 0.50196}

\definecolor{pyBlue}{rgb}{0.11372,0.42352,0.67059}
\definecolor{pyBlueLight}{rgb}{0.6196,0.79215,0.88235}
\definecolor{pyOrange}{rgb}{1.0,0.4549,0.0627}
\definecolor{pyOrangeLight}{rgb}{0.9921568,0.643137,0.376470}
\definecolor{pyGreen}{rgb}{0.1529,0.5882,0.1529}
\definecolor{pyRed}{rgb}{0.81569,0.13725,0.14117}
\definecolor{pyPurple}{rgb}{0.53725,0.36078,0.7098}
\definecolor{myCacaDoie}{rgb}{0.580392156862745,0.43,0.27}
\definecolor{pyBrown}{rgb}{0.549019607843137,0.337254901960784,0.294117647058824}
\definecolor{pyRose}{rgb}{0.890196078431372,0.466666666666667,0.76078431372549}

\definecolor{pyOrange2}{rgb}{0.9921,0.5098,0.2078}
\definecolor{pyBlue2}{rgb}{0.3765,0.6431,0.8157}

\definecolor{pyBar1}{rgb}{0.1647,0.2745,0.2745}
\definecolor{pyBar2}{rgb}{0.0000,0.4588,0.4588}
\definecolor{pyBar3}{rgb}{0.4863,0.7804,0.9998}
\definecolor{pyBar4}{rgb}{0.4549,0.4549,0.4549}

\definecolor{keyGreen2}{rgb}{0.0627 , 0.4196 , 0.3882}%
\definecolor{keyOrange2}{rgb}{0.9843 , 0.2941 , 0.2431}%

\definecolor{bluegray}{rgb}{0.4, 0.6, 0.8}

\definecolor{coluni}{rgb}{0.58823,0.58823,0.58823}
\definecolor{col40}{rgb}{0.192156862745098,0.509803921568627,0.741176470588235}
\definecolor{col42}{rgb}{0.901960784313726,0.333333333333333,0.0509803921568627}
\definecolor{col20}{rgb}{0.388235294117647,0.474509803921569,0.223529411764706}
\definecolor{col22}{rgb}{0.549019607843137,0.427450980392157,0.192156862745098}
\definecolor{col60}{rgb}{0.458823529411765,0.419607843137255,0.694117647058824}
\definecolor{col62}{rgb}{0.482352941176471,0.254901960784314,0.450980392156863}

\definecolor{colchat2}{rgb}{0.192156862745098,0.509803921568627,0.741176470588235}
\definecolor{collgf2}{rgb}{0.619607843137255,0.792156862745098,0.882352941176471}
\definecolor{colhej2}{rgb}{0.901960784313726,0.333333333333333,0.0509803921568627}
\definecolor{colhej4}{rgb}{0.992156862745098,0.682352941176471,0.419607843137255}
\definecolor{colhej6}{rgb}{0.192156862745098,0.63921568627451,0.329411764705882}
\definecolor{colhej8}{rgb}{0.631372549019608,0.850980392156863,0.607843137254902}
\definecolor{colhej10}{rgb}{0.458823529411765,0.419607843137255,0.694117647058824}
\definecolor{colhej0}{rgb}{0.737254901960784,0.741176470588235,0.862745098039216}

\definecolor{cola2a}{rgb}{0.12156862745098,0.466666666666667,0.705882352941177}
\definecolor{colisr}{rgb}{0.388235294117647,0.474509803921569,0.223529411764706}
\definecolor{colnb}{rgb}{0.850980392156863,0.372549019607843,0.00784313725490196}

\definecolor{colaccfft}{RGB}{235,199,116}
\definecolor{colflups}{rgb}{0.388235294117647,0.474509803921569,0.223529411764706}

\definecolor{coldefault}{RGB}{31,119,180}
\definecolor{colfeature}{RGB}{255,127,14}

\definecolor{colweakstencil}{rgb}{0.419607843137255,0.682352941176471,0.83921568627451}
\definecolor{colweakadapt}{rgb}{0.992156862745098,0.552941176470588,0.235294117647059}
\definecolor{colweakghost}{rgb}{0.549019607843137,0.635294117647059,0.32156862745098}

\definecolor{colmlx}{RGB}{169,133,202}
\definecolor{colvega}{RGB}{136,190,222}
\definecolor{collumi}{RGB}{140,86,75}

\definecolor{brightmaroon}{rgb}{0.76, 0.13, 0.28}
\definecolor{camouflagegreen}{rgb}{0.47, 0.53, 0.42}
\definecolor{darkolivegreen}{rgb}{0.33, 0.42, 0.18}
\definecolor{airforceblue}{rgb}{0.36, 0.54, 0.66}
\definecolor{applegreen}{rgb}{0.55, 0.71, 0.0}
\definecolor{denim}{rgb}{0.08, 0.38, 0.74}
\definecolor{candyapplered}{rgb}{1.0, 0.03, 0.0}
\definecolor{chamoisee}{rgb}{0.63, 0.47, 0.35}
\definecolor{chromeyellow}{rgb}{1.0, 0.65, 0.0}
\definecolor{coralred}{rgb}{1.0, 0.25, 0.25}

\colorlet{coloddodd}{brightmaroon}  %
\colorlet{coloddeven}{darkolivegreen} %
\colorlet{colevenodd}{chromeyellow} %
\colorlet{coleveneven}{denim} %

\newcommand{\lineCircle}[3]{
        \tikz{
            \draw[-,#1,solid,line width = #3 pt](0,0) -- (13pt,0);
            \filldraw[draw=#1,fill=#2, line width=#3 pt] (6.5pt,0pt) circle (2pt);
        }
}
            
\newcommand{\ThickLineCircle}[2]{\lineCircle{#1}{#2}{1.0}}

\newcommand{\Circle}[3]{
        \tikz{
            \filldraw[draw=#1,fill=#2, line width=#3 pt] (2.5pt,0pt) circle (2pt);
        }
}
  
\newcommand{\ThickCircle}[2]{\Circle{#1}{#2}{1.0}}

\tikzset{cross/.style={cross out, draw, minimum size=2*(#1-\pgflinewidth), inner sep=0pt, outer sep=0pt}}

\newcommand{\BarLegend}[2]{
        \tikz{
            \filldraw[draw=#1,fill=#2, line width=1 pt] (0pt,-3pt) rectangle (2pt,3pt);
            \filldraw[draw=#1,fill=#2, line width=1 pt] (4pt,-3pt) rectangle (6pt,-1pt);
        }
}

\newcommand{\BufLegend}[1]{
        \tikz{
            \filldraw[draw=#1,fill=#1, line width=0.5 pt] (0pt,-3pt) rectangle (8pt,1pt);
        }
}

\newcommand{\oddeven}[2]{
	\tikz{
		\filldraw[draw=#1,fill=#2, line width=1.0 pt](-2pt,-2pt) --  (0pt,-6pt) -- (2pt, -2pt) -- (0pt,2pt) -- (-2pt,-2pt);
	}
}

\newcommand{\eveneven}[2]{
        \tikz{
                \filldraw[draw=#1,fill=#2, line width=1.0 pt](-2pt,-1pt) -- (2pt,-1pt) -- (0pt,3pt) -- (-2pt,-1pt);
        }
}

\newcommand{\evenodd}[2]{
        \tikz{
                \filldraw[draw=#1,fill=#2, line width=1 pt] (0pt,-1pt) rectangle (4pt,4pt);
        }
}

\newcommand{\CaptionUserBuf}{{\protect\BufLegend{\figblue}}}
\newcommand{\CaptionSendBuf}{{\protect\BufLegend{\figgreen}}}
\newcommand{\CaptionRecvBuf}{{\protect\BufLegend{\figorange}}}

%% file: flups_author.tex
\title{\textsc{FLUPS} - a flexible and performant massively parallel Fourier transform library}
\author{Pierre Balty, Philippe Chatelain, and Thomas Gillis}

%% file: flups_1_intro.tex
\section{Introduction}
\label{sec_intro}

The distributed implementation of the three-dimensional (3D) Fourier transform, or equivalently the successive application of three one-dimensional (1D) transforms, has been a computational challenge over the past decades.
Encouraged by its very broad impact many works have been proposed recently on the parallel implementation of the Fourier transform, with a particular emphasis on GPU support.
Among them, \code{accFFT}\cite{Gholami:2015} is the fastest on both CPU-only and heterogeneous architectures\cite{Ayala:2022}. Relying on a hybrid MPI and CUDA parallelism, it optimizes the overlap between communications and computations to hide the communication overhead. It also provides support for real-to-complex and complex-to-complex transforms.
Similarly \code{heFFTe}\cite{Ayala:2020} reaches good \revthree{performance} while removing some of the restrictions applied on the topology of the input data.
To hide the transpose of the local data in the communications, \textit{Dalcin et al.}\cite{Dalcin:2019} proposed an approach based on MPI datatypes and generalized all-to-all communications.
All of the mentioned libraries have been thoroughly benchmarked and compared to other software, such as P3DFFT\cite{Pekurovsky:2012} or SWFFT\cite{Pope:2017}, more details can be found in \cite{Ayala:2022}.

The applications of the Fourier transform are numerous, and in particular when solving PDEs in fields such as fluids dynamics or gravitational problems.
In the specific field of computational fluid dynamics, incompressibility is accounted for through a Poisson equation which is often solved using an FFT-based Poisson solver \cite{Caprace:2020a,Gillis:2019,Gabbard:2022}.
Despite its relevance, a software with enough flexibility to be used in practice while retaining its parallel \revthree{performance} at large scale is still missing, hence forcing the users to write their own, usually not optimized implementations.

Based on our computational fluid dynamics expertise, we identify three main requirements for a distributed implementation of the Fourier transform to be convenient:
\begin{enumerate}
\item the combination of various FFT types must be supported and efficient. We here refer to real-to-real, real-to-complex, and complex-to-complex FFTs, as many PDE simulations rely on real numbers;
\item the user-provided data should be given either in a cell-centered or a node-centered data layout, \revthree{as defined in \cref{sec2_point_convention}};
\item the required flexibility on the user side must not compromise the parallel \revthree{performance}, both measured \typo{by scalability metrics} and the time-to-solution.
\end{enumerate}

The original \code{flups} library\cite{Caprace:2021} focuses on the first and partially on the third requirement while omitting the second one.
\revtwo{
The authors proposed a software for the resolution of unbounded Poisson equations in 3D on uniform distributed grids with various boundary conditions (BCs) and all their combinations. 
They sort the different BCs into $4$ categories: (1) the even and odd BCs corresponding to DST/DCT, (2) semi-unbounded BCs using extra padding and a DCT/DST, (3) periodic BCs via the DFT, and (4) unbounded BCs through DFT and extra padding.
To reduce the computational cost, \flups reorders the transforms according to the cost entailed by the BC type, starting with the most affordable ones. It thus follows the order: (1), (2), (3), and finally (4).
}
However the implementation does not support the node-centered data layout and offers unsatisfying \revthree{time-to-solution}, especially with unconventional distribution of unknowns.

As most distributed FFT libraries, \flups uses \code{fftw}\cite{Frigo:2005}  as a 1D FFT transform library and implements the various memory transfers and communications between distributed resources\cite{Ayala:2021}.
The latter drives the overall performance and is considered as the algorithm bottleneck, especially \revthree{at large scale} \cite{Czechowski:2012}.
Two approaches stand out for the distribution of the 3D data among the resources \cite{Pekurovsky:2012}, which consequently impacts the application of the FFTs as well as the communication strategies: the slab or the pencil decomposition.
\revthree{Both strategies distribute a 3D computational grid, for simplicity assumed here of size $N$ in each dimension (hence with a total size of $N^3$), onto $P$ computational resources.}
The slab decomposition consists in dividing the grid into slices of data and performing 2D FFTs in the first two dimensions. It is followed by an all-to-all communication to reorder the data in the third direction and perform the remaining 1D FFT.
 \revthree{The scalability of this method breaks down when the number of processes $P$ exceeds the data size in the third direction $N$, as some of the processes do not own any data hence reducing the load balancing.}
On the other side, the pencil decomposition approach removes this partition-size limitation.
\revthree{For each of the three dimensions, it divides the data into 1D pencils, which leads to up to $N^{2}$ independent data chunks to be distributed among the processes.} %
\typo{With that strategy 1D FFTs are performed in the pencil direction,} and communications are needed to switch from one direction to another (and realign the data accordingly) between successive FFTs.
While the slab decomposition requires each MPI rank to communicate once with all the other ranks in the \revthree{communicator}, which makes it efficient on small partitions, the pencil decomposition requires each MPI rank to communicate twice with \typo{$\sqrt{P}$}, which reduces the communication cost on very large partitions.

To address the gap between the research efforts in computational science and the end-user applications, we propose a massively distributed FFT implementation that checks the three requirements we have identified: (1) mixed FFT types, (2) agnostic to the data layout, and (3) large scale scalability and performance.
This work builds upon the existing \flups library which has been used in computational fluid dynamics codes to solve unbounded Poisson problems.
Therefore our motivation as well as our presentation of the methodology and \typo{the obtained results is driven by this specific application}. However the performance results are not specific to this configuration and we expect the impact of our work to go beyond the application envisioned.

First we present in \cref{sec_methodo} the generalization of the implementation to be compatible with both node-centered and cell-centered applications.
Then in \cref{sec_implementation} we detail our communication strategies, as well as several optimizations to improve scalability and peformance.
The resulting codebase is validated against analytical solutions of the Poisson problem in \cref{sec_validation}.
In \cref{sec2_biot_savart}, we demonstrate the convergence of \flups and its ease of use by solving the Biot-Savart equation, a variation of the Poisson problems already considered.
Then, to ensure that the proposed changes and generalizations do not affect the performance, \typo{we compare our time-to-solution against \code{accFFT}} in \cref{sec:results}. Finally, we assess the parallel performance of each communication strategy and benchmark the code on three top-tier European clusters.
To conclude this work, we present a summary of the proposed innovations as well as our results in \cref{sec_conclusion}.

%% file: flups_2_method.tex
\section{Methodology}
\label{sec_methodo}

\revone{As an FFT-based approach, both the original and the proposed updated version of \flups,} solve the Poisson equation $\lapl u = f$ through a convolution between the right-hand side $f$ and a pre-computed Green's function $G$,
\be
\nabla^2 u = f \Rightarrow u = G * f
\label{eq:poisson}
\eed
The latter being performed as a point-wise multiplication in the spectral space: $\hat{u} = \hat{G} \cdot \hat{f}$.
The choice of the Green's function $G$ as well as its spectral representation $\hat{G}$ depends on the boundary conditions (BCs) and the chosen regularization at the origin. %
\revone{
The forward and backward multidimensional Fourier transform is performed through a succession of 1D FFTs, whose types are chosen to comply with the required boundary conditions (BC), as described in \Sect{sec_intro}.
We refer the reader to \cite{Chatelain:2010,Caprace:2021} for further details on the algorithm as well as the combination of unbounded and spectral directions.
}

As already mentioned \flups has been designed for cell-centered data layout exclusively.
To accommodate node-centered data layouts some parts of the algorithms have been generalized and/or rewritten.
In \cref{sec2_point_convention} we first detail the notation regarding the length of the domain in order to handle the two proposed data layouts.
Then in \sect{sec2_per_even_odd} we detail the choice of FFTs for the periodic, even, and odd boundary conditions, and in \sect{sec2_semi_unb} we finally detail the implementation of the \mbox{(semi-)unbounded} boundary condition.

\subsection{Point numbering conventions}
\label{sec2_point_convention}
The presented library relies on \fftw to take care of the FFT computations. For performance reasons, the authors of \fftw adapt the length of the provided data to avoid any trivially null computations. 
As \flups combines different types of FFTs and supports different boundary conditions, we have chosen to use the same convention for all the different cases (regardless of the type of FFTs):

\revthree{As commonly defined in PDE simulations, a multidimensional grid with a physical size of $L_i$ contains $N_i$ data in the $i$th dimension (\eg{} in 3D: $L_x \times L_y \times L_z$ with $N_x \times N_y \times N_z$ data).
The data might be organized either in a node-centered or cell-centered data layout, defined for the $i$th dimension as:}
\begin{itemize}
\item cell-centred layout: $f_j  \triangleq  f ( x_j) $ where $x_j = \left( j + 1/2 \right) h $ with $j \in \left[ 0 \;;\; N-1 \right]$ and $h = L/N$.
\item node-centred layout: $ f_j \triangleq f ( x_j) $ where $x_j = \left(\revtwo{ j }\right) h $ with $j \in \left[ 0 \;;\; N \right]$ and $h = L/N$.
\end{itemize}
We note that when considering the node-centered configuration the \us{data size} is $N+1$, instead of $N$ for the cell-centered layout.
Specifically to the node-centered layout, the last point \typo{$f_N$} is crucial in some configurations such as the even boundary condition. However the information is sometimes duplicated when considering periodic boundary conditions.
In the latter case, the boundary points ($0$ and $N$) must match the imposed boundary conditions and \revtwo{might be used (or not)} by \flups depending on the chosen FFT. 
Nevertheless for an improved usability we ensure that those duplicated points contain the correct information in the end result.
\revthree{Finally, as the rest of this section relates to 1D definitions, we simplify the notation by using $N$ instead of $N_i$ and $L$ for $L_i$ when appropriate.}

\subsection{Periodic, even, and odd boundary conditions}
\label{sec2_per_even_odd}
For both data layouts, the periodic boundary condition relies on the 1-D real-to-complex DFT defined as
\be
\tf_k = \sum_{j=0}^{N-1} f(x_{j}) \, \exp\blr{ -\ic \; \dfrac{2 \pi}{N} \;{j}\;{k}} =  \sum_{j=0}^{N-1} f(x_{j}) \, \exp\blr{ -\ic \; \omega_k \;{x_j} }
\label{sec2_dft_def}
\eec
with $\ic = \sqrt{-1}$, $k \in \blr{0 \;;\; N/2}$ and $\omega_k = k \frac{2\pi}{L}$ the frequency associated to the output $k$.
Note that we use $\tf$ instead of $\hat{f}$ to distinguish the partially spectral result obtained after a 1D DFT from the fully spectral output obtained through the 3D FFT.

The real-to-complex DFT produces $N/2 + 1$ complex modes where the mode $0$ and $\pi$ (constant and flip-flop modes respectively) are purely real.
As we use the standard complex storage of \fftw, we don't explicitly take advantage of the trivially null imaginary parts.
In the case of node-centered information, the last data provided by the user ($f_N$) is unused.

The choice of FFTs gets more diverse when considering real-to-real transforms used to impose even and odd boundary conditions.
The different combinations of the forward transform are summarized in \Tref{tab:fft-transforms} for both cell-centered and node-centered layouts.
For the cell-centered one we refer the reader to \cite{Caprace:2021}, and we detail hereunder the node-centered layout. %

\begin{table}[h]
  \centering
	\begin{tabular}{c|cc}
    & node-centered & cell-centered \\
	\hline         
	odd-odd   & type-I DST     & type-II DST \\
	odd-even  & type-III DST  & type-IV DST  \\
	even-odd  & type-III DCT  & type-IV DCT   \\
	even-even & type-I DCT   & type-II DCT \\
	\end{tabular}  
  \caption{Spectral boundary condition and the corresponding forward Fourier transforms}
  \label{tab:fft-transforms}
\end{table}

\subsubsection{Even-even boundary condition}
The even-even condition is imposed using a type-I DCT, which contains the real part of the DFT as given in \Eqqref{sec2_dft_def}.
This DCT is then defined as
\be
\tf_k =f(x_{0}) + \plr{-1}^{k} \,f(x_{N}) + 2 \; \sum_{j=1}^{N-1} f(x_{j}) \, \cos\blr{ \dfrac{2 \pi}{2 N} \;{j}\;{k}}
\eec
with $k \in \blr{0 \;;\; N}$. We note that this transform produces $N+1$ real spectral information corresponding to the frequencies $\omega_k = \frac{2 \pi}{2 L} k = \frac{2 \pi}{L } \frac{k}{2} =\frac{\pi}{L } {k}$.
The output of this transform is then equivalent to the frequencies obtained by using the DFT on a domain $2N$.
In order to improve consistency across the definitions and in the implementation, we will use the notation based on $\frac{2 \pi}{2L}$ throughout this section.
As illustrated in \Fref{fig:method-spectral} where we highlight the needed information from the user point of view, both the values in $f_0$ and $f_N$ are relevant in the even-even case.

\subsubsection{Odd-odd boundary condition}
To impose an odd-odd boundary condition, also illustrated in \Fref{fig:method-spectral}, we use the type-I DST which contains the imaginary part of the DFT as given in \Eqqref{sec2_dft_def}:
\be
\tf_k = 2 \, \sum_{j=1}^{N-1} f(x_{j}) \, \sin\blr{\dfrac{2\pi}{2N} \;{j}\;{k}}
\eec
with $k \in \blr{1 \; ; \; N-1}$, where $\tf_0 = 0$ and $\tf_{N} = 0$ since the input data are real.
To match this definition the first and last user-provided data are assumed to be zero and \fftw discards both information to reduce the memory footprint and the time-to-solution.
To adapt to this convention in \flups we consider $u_i$ for $i \in \blr{1 \;;\; N-1}$ only and the information located in $i=0$ and $i=N$ are overwritten to be zero.

\subsubsection{Odd-even boundary condition}
In order to impose mixed boundary conditions such as the odd-even one we use a type-III DST as defined by 
\be
\tf_k = \plr{-1}^{k}f(x_{N}) + 2 \, \sum_{j=1}^{N-1}f(x_{j}) \, \sin\blr{ \dfrac{2\pi}{2N}  \; {j} \; \plr{ k+\frac{1}{2} } }
\eec
with $k \in \blr{0 \; ; \; N-1}$.
Here, \flups overwrites the first user-provided information $f_0$ as represented in \Fref{fig:method-spectral}. 
Also, we note that the half modes produced have an associated frequency of $\omega_k = \frac{2\pi}{2L} \; \plr{ k+\frac{1}{2} } = \frac{2\pi}{4L} \; \plr{ 2k+1 }$.
Therefore they correspond to the odd frequencies that would have \revthree{been} obtained using a DFT on a domain of size $4N$ applying all the symmetries explicitly.

\subsubsection{Even-odd boundary condition}
\begin{figure}[h!tp]
\centering
\input{figures/tikz/method_spectral_case}
\caption{Examples of odd-odd ({\protect\ThickCircle{coloddodd}{coloddodd!5}}), odd-even ({\protect\oddeven{coloddeven}{coloddeven!5}}), even-odd  ({\protect\evenodd{colevenodd}{colevenodd!5}}), and even-even ({\protect\eveneven{coleveneven}{coleveneven!5}}) boundary conditions. The shaded area represents the symmetry imposed by the boundary conditions. Data located there are \revone{fictitious}. Filled shapes represent the points given to \fftw, while the empty shapes are the data assumed by the BCs.}
\label{fig:method-spectral}
\end{figure}
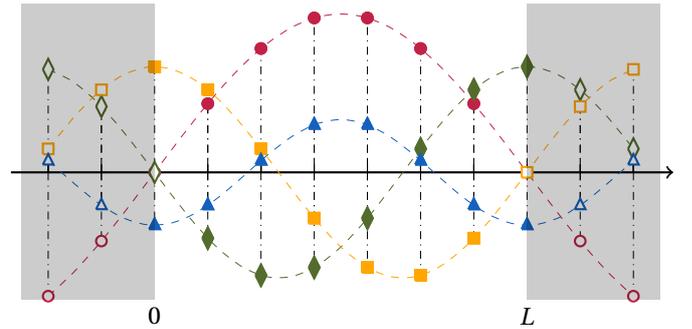

The final combination is \delete{the }the even-odd case which is obtained using the type-III DCT, defined as 
\be
\tf_k = f(x_{0}) + 2 \, \sum_{j=1}^{N-1}f(x_{j}) \, \cos\blr{ \dfrac{2\pi}{2N}  \; j \; \plr{k+\frac{1}{2}} }
\eec
with $k \in \blr{0 \; ; \; N-1}$ and where \flups overwrites the last user-provided information as illustrated in \Fref{fig:method-spectral}.
Similarly to the odd-even case, we note that the corresponding frequencies are the odd frequencies that would have been obtained using a DFT on a domain of size $4N$ applying explicitly all the boundary conditions.

\subsection{Semi-unbounded and unbounded boundary conditions}
\label{sec2_semi_unb}

\subsubsection{Unbounded directions} 
The unbounded boundary condition is imposed using the domain doubling technique of Hockney and Eastwood \cite{Hockney:1988}.
The algorithm extends the right-hand side to a domain of size $2L$ and fills the extension with zeros. 
The Green's function is also extended and symmetrized around $L$. The spectral representation of both extended fields $\tG_{ext}$ and $\tf_{ext}$ is then obtained with a DFT on the extended domain, and the periodic convolution is performed as a multiplication in the spectral space. We refer the reader to \cite{Caprace:2021} for more details on the unbounded boundary condition treatment and various expressions for the Green's function.

\begin{figure}[h!tp]
\centering
\begin{minipage}[b]{0.24\textwidth}
\centering
    \input{figures/tikz/method_unbounded}
    \subcaption{unbounded}
    \label{subfig:method-unbounded}
\end{minipage}
\hfill
\begin{minipage}[b]{0.24\textwidth}
    \input{figures/tikz/method_semi_unbounded}
    \subcaption{even/odd - unbounded}
    \label{subfig:method-semi-unbounded}
\end{minipage}
\caption{Extension of the right-hand side ({\protect\ThickCircle{brightmaroon}{brightmaroon}}) and of the Green function({\protect\ThickCircle{darkolivegreen}{darkolivegreen}}) for the unbounded and semi-unbounded boundary condition. The values of the extended fields are shaded. For the semi-unbounded case, the domain's extension can be done on both sides of the domain and is here represented on the right end of the domain.}
\label{fig:method-unbounded}
\end{figure}
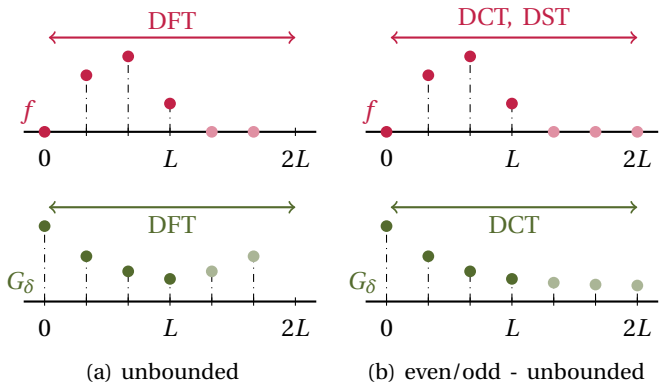

Compared to the \revthree{already} mentioned approach, the generalization to node-centered data layout only affects the padding sizes, and the expressions for the Green's function remain unchanged.
As illustrated in \Fref{fig:method-unbounded} the right-hand side is extended by $N-2$ points, all set to $0$, and we perform the DFT on a domain from $0$ to $2L$. 
On the Green's function side, the symmetry happens around $j=N$, and the domain is extended with $N-2$ information.

\subsubsection{Semi unbounded directions} \flups also supports semi-unbounded boundary conditions as a combination of an unbounded boundary condition at one end of the domain and a symmetry condition at the other end. As explained in \cite{Caprace:2021}, it relies on the domain doubling technique to impose the infinite boundary condition and on DSTs and DCTs to prescribe the correct symmetry while reducing memory usage. In this configuration, the right-hand side is first extended and padded on the unbounded side, unlike the fully unbounded case where we always extend it from $L$ to $2L$.
Then, a DST or DCT is executed on the extended domain to impose the proper symmetry condition.
The Green's function is evaluated on the extended domain, and a DCT transform imposes the proper symmetry conditions required by the domain doubling technique.

As highlighted in \Sect{sec2_per_even_odd} \fftw does not consider the same amount of data when computing a DST or a DCT.
Specifically when considering an even-unbounded boundary condition the right-hand side undergoes a DST while the Green's function a DCT.
Therefore the spectral information obtained as a result of \fftw has different wave-numbers and different sizes due to the use of \fftw's convention.
We solve this issue by allocating extra space for the DST and using a common indexing for both the right-hand side and the Green's function.

%% file: figures/tikz/method_spectral_case.tex
\begin{tikzpicture}[
    every path/.style = {},
  ]
  
\begin{scope}
    	\pgfmathsetmacro{\dx}{0.7}
    	\pgfmathsetmacro{\dy}{\dx}	      
    	\pgfmathsetmacro{\Lx}{7*\dx}
    	\pgfmathsetmacro{\Ly}{4	*\dy}   	
	\pgfmathsetmacro{\dashy}{0.25*\dy}   	
	\pgfmathsetmacro{\stepx}{\dx}   	
	
	  \tikzstyle{data_red} = [draw,minimum size=2em,thick,color=coloddodd, fill=coloddodd!0]
	  \tikzstyle{data_fred} = [draw,minimum size=2em,thick,color=coloddodd, fill=coloddodd]
	  
	  \tikzstyle{data_green} = [draw,minimum size=2em,thick,color=coloddeven, fill=coloddeven!0]
	  \tikzstyle{data_fgreen} = [draw,minimum size=2em,thick,color=coloddeven, fill=coloddeven]
	  
	  \tikzstyle{data_blue} = [draw,minimum size=2em,thick,color=coleveneven, fill=coleveneven!0]
	  \tikzstyle{data_fblue} = [draw,minimum size=2em,thick,color=coleveneven, fill=coleveneven]
	  
	  \tikzstyle{data_orange} = [draw,minimum size=2em,thick,color=colevenodd, fill=colevenodd!0]
	  \tikzstyle{data_forange} = [draw,minimum size=2em,thick,color=colevenodd, fill=colevenodd]

    	\draw[thick, ->] (-2.7*\dx,0) -- (\Lx + 2.75*\dx,0);

	\node[below] at (0,-0.6*\Ly){$0$};
	\node[below] at (\Lx,-0.6*\Ly){$L$};

    	\pgfmathsetmacro{\endx}{\Lx + 2*\dx};
	\pgfmathsetmacro{\sttx}{-2*\dx};
	\pgfmathsetmacro{\deltax}{-\dx};

	\draw[domain=\sttx:\endx, smooth, variable=\x, coloddodd, dashed, ultra thin ] plot ({\x}, {.75*\Ly *sin(deg(\x*pi/ \Lx)))});			%
	\draw[domain=\sttx:\endx, smooth, variable=\x, coloddeven, dashed, ultra thin ] plot ({\x}, {-0.5*\Ly *sin(deg(3*\x*pi/ \Lx /2)))});	%
	\draw[domain=\sttx:\endx, smooth, variable=\x, coleveneven, dashed, ultra thin ] plot ({\x}, {-0.25*\Ly*cos(deg(2*\x*pi/ \Lx )))});		%
	\draw[domain=\sttx:\endx, smooth, variable=\x, colevenodd, dashed, ultra thin ] plot ({\x}, {0.5*\Ly*cos(deg(1.5*\x*pi/ \Lx  )))});		%

	\foreach \x in {\sttx,\deltax,...,\endx} {
		\draw (\x, 0.0)--(\x,\dashy);
		
		\pgfmathsetmacro{\yoddodd}{.75*\Ly *sin(deg(\x*pi/ \Lx)))};
		\draw[dashdotted,ultra thin] (\x, 0.0)--(\x,\yoddodd);

		\pgfmathsetmacro{\yoddeven}{-0.5*\Ly*sin(deg(\x*3*pi/ \Lx /2)))};
		\draw[dashdotted, ultra thin] (\x, 0.0)--(\x,\yoddeven);
		
		\pgfmathsetmacro{\yevenodd}{0.5*\Ly*cos(deg(1.5*\x*pi/ \Lx  )))};
		\draw[dashdotted, ultra thin] (\x, 0.0)--(\x,\yevenodd);
		
		\pgfmathsetmacro{\yeveneven}{-0.25*\Ly*cos(deg(2*\x*pi/ \Lx )))};
		\draw[dashdotted,ultra thin] (\x, 0.0)--(\x,\yeveneven);		
		
		\filldraw[data_red](\x,\yoddodd) circle (0.1*\dx);
		\filldraw[data_green] (\x-0.1*\dx,\yoddeven) --  (\x,\yoddeven-0.2*\dx) -- (\x + 0.1*\dx,\yoddeven) -- (\x,\yoddeven+0.2*\dx) -- (\x-0.1*\dx,\yoddeven);	
		\filldraw[data_orange]  (\x-0.1*\dx,\yevenodd - 0.1*\dx) --  (\x+0.1*\dx,\yevenodd- 0.1*\dx) -- (\x + 0.1*\dx,\yevenodd + 0.1*\dx) -- (\x - 0.1*\dx,\yevenodd + 0.1*\dx) --(\x-0.1*\dx,\yevenodd - 0.1*\dx);
		\filldraw[data_blue](\x,\yeveneven)(\x - 0.1* \dx,\yeveneven-0.075*\dx) -- (\x + 0.1* \dx,\yeveneven-0.075*\dx) -- (\x,\yeveneven+0.125*\dx) -- (\x - 0.1* \dx,\yeveneven-0.075*\dx);

    	}

	\pgfmathsetmacro{\sttx}{\dx}
	\pgfmathsetmacro{\deltax}{2*\dx}
	\pgfmathsetmacro{\endx}{\Lx - \dx}
	\foreach \x in {\sttx,\deltax,...,\endx} {
		\pgfmathsetmacro{\yoddodd}{.75*\Ly *sin(deg(\x*pi/ \Lx)))};
		\filldraw[data_fred](\x,\yoddodd) circle (0.1*\dx);
    	}

	\pgfmathsetmacro{\sttx}{\dx}
	\pgfmathsetmacro{\deltax}{2*\dx}
	\pgfmathsetmacro{\endx}{\Lx}
	\foreach \x in {\sttx,\deltax,...,\endx} {
		\pgfmathsetmacro{\yoddeven}{-0.5*\Ly*sin(deg(\x*3*pi/ \Lx /2)))};
		\filldraw[data_fgreen](\x-0.1*\dx,\yoddeven) --  (\x,\yoddeven-0.2*\dx) -- (\x + 0.1*\dx,\yoddeven) -- (\x,\yoddeven+0.2*\dx) -- (\x-0.1*\dx,\yoddeven);
    	}

	\pgfmathsetmacro{\sttx}{0}
	\pgfmathsetmacro{\deltax}{\dx}
	\pgfmathsetmacro{\endx}{\Lx-\dx}
	\foreach \x in {\sttx,\deltax,...,\endx} {
		\pgfmathsetmacro{\yevenodd}{0.5*\Ly*cos(deg(1.5*\x*pi/ \Lx  )))};
		\filldraw[data_forange](\x,\yevenodd) (\x-0.1*\dx,\yevenodd - 0.1*\dx) --  (\x+0.1*\dx,\yevenodd- 0.1*\dx) -- (\x + 0.1*\dx,\yevenodd + 0.1*\dx) -- (\x - 0.1*\dx,\yevenodd + 0.1*\dx) --(\x-0.1*\dx,\yevenodd - 0.1*\dx);
	 }
	
	\pgfmathsetmacro{\sttx}{0}
	\pgfmathsetmacro{\deltax}{\dx}
	\pgfmathsetmacro{\endx}{\Lx}
	\foreach \x in {\sttx,\deltax,...,\endx} {
		\pgfmathsetmacro{\yeveneven}{-0.25*\Ly*cos(deg(2*\x*pi/ \Lx )))};
		\filldraw[data_fblue](\x,\yeveneven) (\x,\yeveneven)(\x - 0.1* \dx,\yeveneven-0.075*\dx) -- (\x + 0.1* \dx,\yeveneven-0.075*\dx) -- (\x,\yeveneven+0.125*\dx) -- (\x - 0.1* \dx,\yeveneven-0.075*\dx);
    	}
	
	\fill[opacity=0.2] (-2.5*\dx,-0.6*\Ly) rectangle (0,0.8*\Ly);
	\fill[opacity=0.2] (\Lx,-0.6*\Ly) rectangle (\Lx + 2.5*\dx,0.8*\Ly);

	\end{scope}

\end{tikzpicture}

%% file: figures/tikz/method_unbounded.tex
\begin{tikzpicture}[
    every path/.style = {},
  ]
\tikzstyle{dot} = [shape=circle,minimum size=0.5em,inner sep=0pt, outer sep=0pt]

  \begin{scope}

	\pgfmathsetmacro{\N}{4}
    	\pgfmathsetmacro{\dx}{0.55}
    	\pgfmathsetmacro{\dy}{0.5}	      
	\pgfmathsetmacro{\Domx}{\N*\dx -\dx}
    	\pgfmathsetmacro{\Lx}{7*\dx}
    	\pgfmathsetmacro{\Ly}{3*\dy}   	
	\pgfmathsetmacro{\dashy}{0.25*\dy}

	\def\Oy{0.0}
	\def\yRhs{{\Oy, \Oy+1.5*\dy,\Oy+ 2*\dy, \Oy+0.75*\dy, \Oy, \Oy}}
	\def\RhsColor{brightmaroon}
	\node[above left] at (0.0,\Oy) {\textcolor{\RhsColor}{$f$}};
	
	\draw[thick, -] (-0.5*\dx,\Oy) -- (\Lx - 0.5*\dx,\Oy);
	\node[below] at (0,\Oy-0.25*\dy){$0$};
	\node[below] at (\Domx,\Oy-0.25*\dy){$L$};
	\node[below] at (2*\Domx,\Oy-0.25*\dy){$2L$};

	\foreach \i in {0, 1, 2, 3, 4, 5, 6} {
	    	\pgfmathsetmacro{\x}{\i*\dx}
		\draw (\x, \Oy-0.1*\dx)--(\x,\Oy+0.1*\dx);
    	}

	\foreach \i in {0, 1, 2, 3} {
	    	\pgfmathsetmacro{\x}{\i*\dx}
		\pgfmathsetmacro{\y}{\yRhs[\i]}
		\draw[dashdotted] (\x, \Oy)--(\x,\y);
		\node[dot,fill=\RhsColor] (rhs_\i) at (\x,\y) {};
    	}
	
	\foreach \i in {4, 5} {
	    	\pgfmathsetmacro{\x}{\i*\dx}
		\pgfmathsetmacro{\y}{\yRhs[\i]}
		\node[dot,fill=\RhsColor!50] (rhs_\i) at (\x,\y) {};
    	}
	\draw[thick,<->,draw=\RhsColor](0.1*\dx,\Oy + \Ly - 0.5*\dy) -- node[above] {\textcolor{\RhsColor}{DFT}} (\Lx-\dx,\Oy + \Ly- 0.5*\dy);

	\def\Oy{-1.5*\Ly}	
	\def\Ox{0.0}	
	\def\yGreen{{\Oy+2*\dy, \Oy+1.2*\dy ,\Oy + 0.8*\dy, \Oy+0.6*\dy, \Oy+ 0.8*\dy,\Oy + 1.2*\dy}}   	
	\def\GreenColor{darkolivegreen}
	
	\node[above left] at (\Ox,\Oy) {\textcolor{\GreenColor}{$G_{\delta}$}};
	
	\draw[thick, -] (\Ox-0.5*\dx,\Oy) -- (\Ox+\Lx- 0.5*\dx,\Oy);
	\node[below] at (\Ox,\Oy-0.25*\dy){$0$};
	\node[below] at (\Ox+\Domx,\Oy-0.25*\dy){$L$};
	\node[below] at (\Ox+2*\Domx,\Oy-0.25*\dy){$2L$};
	\foreach \i in {0, 1, 2, 3, 4, 5, 6} {
	    	\pgfmathsetmacro{\x}{\Ox+\i*\dx}
		\draw (\x, \Oy-0.1*\dx)--(\x,\Oy+0.1*\dx);
    	}

	\foreach \i in {0,1,2,3} {
	    	\pgfmathsetmacro{\x}{\Ox+\i*\dx}
		\pgfmathsetmacro{\y}{\yGreen[\i]}
		\draw[dashdotted] (\x, \Oy)--(\x,\y);
		\node[dot,fill=\GreenColor] (green_\i) at (\x,\y) {};
    	}

	\foreach \i in {4, 5} {
	    	\pgfmathsetmacro{\x}{\Ox+\i*\dx}
		\pgfmathsetmacro{\y}{\yGreen[\i]}
		\draw[dashdotted] (\x, \Oy)--(\x,\y);
		\node[dot,fill=\GreenColor!50] (green_\i) at (\x,\y) {};
    	}

	\draw[thick,<->,draw=\GreenColor](\Ox+0.1*\dx,\Oy + \Ly - 0.5*\dy) -- node[below] {\textcolor{\GreenColor}{DFT}} (\Ox+\Lx-\dx,\Oy + \Ly- 0.5*\dy);
	\end{scope}
\end{tikzpicture}

%% file: figures/tikz/method_semi_unbounded.tex
\begin{tikzpicture}[
    every path/.style = {},
  ]
\tikzstyle{dot} = [shape=circle,minimum size=0.5em,inner sep=0pt, outer sep=0pt]

  \begin{scope}

	\pgfmathsetmacro{\N}{4}
    	\pgfmathsetmacro{\dx}{0.55}
    	\pgfmathsetmacro{\dy}{0.5}	      
	\pgfmathsetmacro{\Domx}{\N*\dx -\dx}
    	\pgfmathsetmacro{\Lx}{7*\dx}
    	\pgfmathsetmacro{\Ly}{3*\dy}   	
	\pgfmathsetmacro{\dashy}{0.25*\dy}

	\def\Oy{0.0}
	\def\yRhs{{\Oy, \Oy+1.5*\dy,\Oy+ 2*\dy, \Oy+0.75*\dy, \Oy, \Oy, \Oy}}
	\def\RhsColor{brightmaroon}
	\node[above left] at (0.0,\Oy) {\textcolor{\RhsColor}{$f$}};
	
	\draw[thick, -] (-0.5*\dx,\Oy) -- (\Lx - 0.5*\dx,\Oy);
	\node[below] at (0,\Oy-0.25*\dy){$0$};
	\node[below] at (\Domx,\Oy-0.25*\dy){$L$};
	\node[below] at (2*\Domx,\Oy-0.25*\dy){$2L$};

	\foreach \i in {0, 1, 2, 3, 4, 5, 6} {
	    	\pgfmathsetmacro{\x}{\i*\dx}
		\draw (\x, \Oy-0.1*\dx)--(\x,\Oy+0.1*\dx);
    	}

	\foreach \i in {0, 1, 2, 3} {
	    	\pgfmathsetmacro{\x}{\i*\dx}
		\pgfmathsetmacro{\y}{\yRhs[\i]}
		\draw[dashdotted] (\x, \Oy)--(\x,\y);
		\node[dot,fill=\RhsColor] (rhs_\i) at (\x,\y) {};
    	}
	
	\foreach \i in {4, 5,6} {
	    	\pgfmathsetmacro{\x}{\i*\dx}
		\pgfmathsetmacro{\y}{\yRhs[\i]}
		\node[dot,fill=\RhsColor!50] (rhs_\i) at (\x,\y) {};
    	}
	\draw[thick,<->,draw=\RhsColor](0.1*\dx,\Oy + \Ly - 0.5*\dy) -- node[above] {\textcolor{\RhsColor}{DCT, DST}} (\Lx-\dx,\Oy + \Ly- 0.5*\dy);

	\def\Oy{-1.5*\Ly}	
	\def\Ox{0.0}	
	\def\yGreen{{\Oy+2*\dy, \Oy+1.2*\dy ,\Oy + 0.8*\dy, \Oy+0.6*\dy, \Oy+ 0.5*\dy,\Oy + 0.45*\dy, \Oy + 0.425*\dy}}   	
	\def\GreenColor{darkolivegreen}
	
	\node[above left] at (\Ox,\Oy) {\textcolor{\GreenColor}{$G_{\delta}$}};
	
	\draw[thick, -] (\Ox-0.5*\dx,\Oy) -- (\Ox+\Lx- 0.5*\dx,\Oy);
	\node[below] at (\Ox,\Oy-0.25*\dy){$0$};
	\node[below] at (\Ox+\Domx,\Oy-0.25*\dy){$L$};
	\node[below] at (\Ox+2*\Domx,\Oy-0.25*\dy){$2L$};
	\foreach \i in {0, 1, 2, 3, 4, 5, 6} {
	    	\pgfmathsetmacro{\x}{\Ox+\i*\dx}
		\draw (\x, \Oy-0.1*\dx)--(\x,\Oy+0.1*\dx);
    	}

	\foreach \i in {0,1,2,3} {
	    	\pgfmathsetmacro{\x}{\Ox+\i*\dx}
		\pgfmathsetmacro{\y}{\yGreen[\i]}
		\draw[dashdotted] (\x, \Oy)--(\x,\y);
		\node[dot,fill=\GreenColor] (green_\i) at (\x,\y) {};
    	}

	\foreach \i in {4, 5,6} {
	    	\pgfmathsetmacro{\x}{\Ox+\i*\dx}
		\pgfmathsetmacro{\y}{\yGreen[\i]}
		\draw[dashdotted] (\x, \Oy)--(\x,\y);
		\node[dot,fill=\GreenColor!50] (green_\i) at (\x,\y) {};
    	}

	\draw[thick,<->,draw=\GreenColor](\Ox+0.1*\dx,\Oy + \Ly - 0.5*\dy) -- node[below] {\textcolor{\GreenColor}{DCT}} (\Ox+\Lx-\dx,\Oy + \Ly- 0.5*\dy);

	\end{scope}
\end{tikzpicture}

%% file: flups_3_implementation.tex
\section{Implementation}
\label{sec_implementation}

The implementation challenges of unbounded Poisson solvers as proposed in \flups are almost equivalent to the one of a parallel distributed FFT transform.
However a few key differences exist and are worth noting as they have motivated the authors of \flups toward their own implementation.
\begin{itemize}
\item ability to mix different types of FFTs together (DFTs, DCTs, and DSTs) to support the different boundary conditions;
\item the transform should happen in-place; 
\item the order of the transforms should be determined from the boundary conditions and not be imposed by the framework, to reduce the memory footprint and increase the solver \revthree{performance};
\item the unbounded boundary conditions should be supported, \ie{the FFT sizes might vary to accommodate zero-padding};
\end{itemize}

The original \flups library satisfies those requirements but suffers from a plain approach and is not generalizable to node-centered data layout.
The main problem is that the latter leads to an odd number of data to be distributed on a usually even number of processes, which results in a load in-balance very poorly managed by the original implementation.
To offer flexibility and performance to the user we have significantly improved the communication strategies for both the existing implementations (all-to-all and non-blocking), and we have introduced a third one based on \code{MPI\_Datatype}.

In this section, we detail the generalization and improvement of the communication strategies overarching \flups, equivalent to the distributed FFT algorithm.
For a pencil-based decomposition, the commonly used approach computes the forward or backward transform successively in the 3 dimensions.
For each dimension, we first decompose the domain into pencils and send the data over to their respective process. 
The data transfers are either performed from the domain decomposition chosen by the user to the first pencil decomposition or from one pencil decomposition to another. Then we re-order the data (this operation is also known as the \emph{shuffle}) and compute the FFT using \fftw on the resulting continuous array. Finally, we repeat the whole process for the next dimension.
In the rest of this section, we assume that the reader is familiar with the implementation of pencil-based distributed FFT and we refer to \cite{Caprace:2021} for further details.

\subsection{Communication strategies}
\label{sec:comm-strat}

The communications widely dominate the time-to-solution in a distributed FFT solver as already reported in several implementations~\cite{Caprace:2021,Ayala:2020,Gholami:2015}.
To achieve the highest level of performance, we propose three strategies based on different \code{MPI} functionalities: an implementation using the all-to-all function \code{MPI\_Ialltoallv} (noted \ata), an implementation relying on \emph{persistent} requests and manual (un)packing of the data (noted \nb), and an implementation based on non-blocking send and receives where \code{MPI\_Datatype} is used to avoid the manual packing/unpacking (noted \isr).
\revthree{Similar to the original implementation, all the \code{MPI} calls are performed in sub-communicators to reduce the memory footprint at large scale \cite{Guo:2017}.}

\subsubsection{Implementation using an all-to-all}

\begin{figure}[h!tp]
\centering
\input{figures/tikz/implementation_a2a}
\caption{Implementation of the \ata version. Data are packed from the user-provided buffer (\CaptionUserBuf) to the send buffer (\CaptionSendBuf). The communication is performed using \code{MPI\_Ialltoallv}. The data are then shuffled in the receive buffer (\CaptionRecvBuf) and copied back to the user buffer.}
\label{fig:implementation-alltoall}
\end{figure}
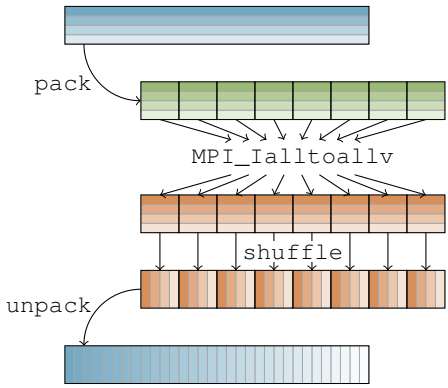

The most simple approach is to rely on \code{MPI\_Ialltoallv}  to perform the communications.
This approach is summarized in \Fref{fig:implementation-alltoall} and detailed in \algref{algo:a2a}. 
Each rank has a list of other ranks to send data to. The intersection of the data from the origin rank in the previous pencil decomposition with the destination rank in the new pencil decomposition leads to the definition of a \textit{block of memory} that has to be transferred, noted $\block$. For each rank we can pre-compute the list of blocks to send, $\btosend$, and the ones to receive, $\btorecv$. Finally each block $\block$ corresponds to a location in the communication buffers (send and receive) and the user-data, noted $\buffersend[\block]$, $\bufferrecv[\block]$, and $\bufferdata[\block]$ respectively.

With those definitions, the \ata{} approach consists in the following steps:
\begin{enumerate}
\item pack the non-regular data into the contiguous communication buffer in \FnPack{};
\item use \code{MPI\_Ialltoallv} to send the needed data to each of the corresponding ranks;
\item reset the data field to $0$ in \FnResetData{}, which is required to properly account for the unbounded boundary conditions and the use of in-place transforms;
\item wait for the communication to complete; 
\item shuffling the data to realign them in memory in \FnShuffle{} and unpacking of the communication buffer to the data in \FnUnpack{}.
\end{enumerate}

\begin{algorithm2e}[ht!]
\caption{All-to-all  implementation}
\label{algo:a2a}
\SetAlgoLined
\DontPrintSemicolon
\ForEach{$\block \in \btosend$}{
   		$\buffersend[\block]$ $\leftarrow$ \FnPack{$\bufferdata[\block]$} \;
}
\BlankLine \BlankLine
\FnAlltoAll{$\buffersend$, $\bufferrecv$} \;
\FnResetData{$\bufferdata$} \;
\FnMPIWait()\;
\BlankLine\BlankLine
\ForEach{$\block \in \btorecv$}{
	\FnShuffle{$\bufferrecv[\block]$} \;
	$\bufferdata[\block]$ $\leftarrow$ \FnUnpack{$\bufferrecv[\block]$} \;
}

\BlankLine
\end{algorithm2e}

Although this approach is straightforward to implement, it suffers from a main issue: it does not expose the parallelization structure of the algorithm to MPI.
It results in an implicit communicator-wide synchronization inherent to the \code{MPI\_Ialltoallv} call and there is only limited room for optimization such as overlapping the different tasks.
Therefore we expect the \revthree{performance} to be mitigated and driven by the MPI implementation.

\subsubsection{Implementation using non-blocking persistent requests}
To avoid the implicit barrier from the collective call and to expose more of the parallel structure of the algorithm to MPI, we have implemented a version based on persistent non-blocking \code{MPI\_Send} and \code{MPI\_Recv}, referred to as \nb{}.
Although similar to the approach proposed in \cite{Caprace:2021}, our implementation contains several improvements designed to increase the parallel \revthree{performance}.

\begin{figure}[h!tp]
\centering
\input{figures/tikz/implementation_nb}
\caption{Implementation of the \nb version. Data are copied from the user-provided buffer (\CaptionUserBuf) to the send buffer (\CaptionSendBuf). The communication is performed using \textit{persistent non-blocking} communications. The data are then shuffled in the receive buffer (\CaptionRecvBuf) and copied back to the user buffer.}
\label{fig:implementation-nb}
\end{figure}
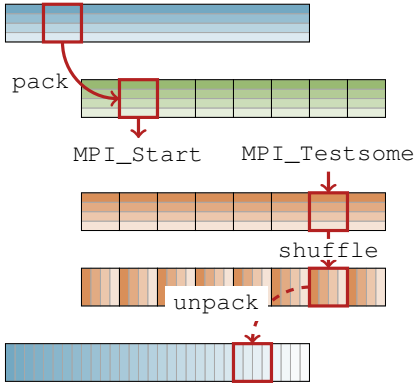

As for \ata{} we still have to perform the same \FnPack{}, \FnUnpack{}, and \FnShuffle{} tasks.
However we can overlap the computations, \ie{packing and unpacking, resetting the data field to $0$, and shuffling the data}, with the communication itself.
This implementation is detailed in \Algref{algo:nb_1} and summarized in \Fref{fig:implementation-nb}.
Here $\block$ is still used to refer to a block of memory, and $\btosend$ and $\btorecv$ corresponds to the list of blocks to be sent and to be received respectively. We use the notation $\btounpack$  to designate the list of blocks to unpack, \ie{copied from the receive buffer to the user-data}.

This approach relies on two phases: a first one pre-computes and stores different \code{MPI\_Requests}, which are used in the second one to communicate when needed.
During the initialization phase, each block to be sent or received is associated with one of these \code{MPI\_Requests}: $\srequest$ or $\rrequest$ respectively.
The communication phase is made of four distinct steps:
\begin{enumerate}
\item for the requests in $\srequest$, we manually pack the data into the continuous communication buffer and then start the corresponding request. This is done in the function $\FnSendBatch$ which takes as an argument $\Block$, an arbitrary list of blocks to be prepared and sent. For the requests in $\rrequest$, no particular operation is needed except activating the requests using \code{MPI\_Start};
\item reset the buffer if all the send requests have completed;
\item test the completion of some of the requests in $\rrequest$ and shuffle the one that have just completed;
\item unpack the shuffled requests if the buffer has been reset already.
\end{enumerate}

As described in \Algref{algo:nb_1}, the four tasks are organized as a \code{for} loop relying on two compile-time variables $\nsendbatch$ and $\nsendpending$ to control the granularity of the different steps.
The first one,  $\nsendbatch$, controls the number of requests gathered inside a batch and therefore treated one after another. 
The second one, $\nsendpending$, limits the total number of uncompleted send requests.
The function $\FnSendBatch$ will therefore activate the minimum number of requests to not overtake any of those two thresholds. 

In summary, the \nb{} approaches offer a control on the asynchronous granularity. First, it starts all the \emph{receive} requests, $\rrequest$, and a first batch of \emph{send} requests. Then, as long as there are ongoing send or receive requests, or block to unpack, the following steps are performed: 
\begin{itemize}
\item if some $\btosend$ have to be sent, we compute the number of blocks to send with respect to $\nsendbatch$ and $\nsendpending$ and we send them;
\item if all the $\btosend$ has been treated and if the user-provided data has not been reset yet, we reset the data field;
\item if some $\btorecv$ has not been received, we force progress by calling \code{MPI\_Testsome}, and we shuffle receive requests already completed;
\item if the data field has been reset and if some blocks have been shuffled and are ready to be unpacked, we unpack them.
\end{itemize}

In practice, we have observed the best \revthree{performance} with $\nsendbatch = 1$.
We attribute this behavior to the size of our communications which exceeds the eager / rendez-vous threshold. The messages can therefore not be sent right away and require a hand-shake to happen \revthree{beforehand}.
There is then no gain in grouping the send-requests together. On the other side, requesting progress more frequently on the receive requests via a call to \code{MPI\_Testsome} improves the \revthree{performance}.

\begin{algorithm2e}[ht!]
\caption{Non-blocking implementations}
\label{algo:nb_1}
\SetAlgoLined
\DontPrintSemicolon

$\nsendbatch$, $\nsendpending$  \tcp*{User defined variables}
\FnRecvBatch{$\btorecv$} \;
\FnSendBatch{$\btosend$} \;
\BlankLine

\While{ ($\btosend \neq \emptyset$  {\normalfont \textbf{or}}  $\btorecv \neq \emptyset$ {\normalfont \textbf{or}}  $\btounpack \neq \emptyset$) }{
	\If{($\btosend \neq \emptyset$) }{
	 	$\nongoing$ $\leftarrow$ \FnTestSome{$\srequest$}\;
		$\ntosend = \min( \: \nsendpending - \nongoing, \quad  \nsendbatch) $ \;
		 \FnSendBatch{$\btosend[\ntosend]$} \; 
	}
	\BlankLine\BlankLine	
	
	\If{($\btosend = \emptyset$ {\normalfont \textbf{and} data buffer is not reset}) }{
		\FnResetData{$\bufferdata$} \;
	}
	
	\BlankLine\BlankLine	
	
	\If{($\btorecv \neq \emptyset$ ) }{
		\FnTestSome{$\rrequest$}\; 
		\ForEach{$\block \in \brecv$}{
			\FnShuffle{$\bufferrecv[\block]$} \;
	 	}
	}
	
	\BlankLine\BlankLine	
	
	\If{( {\normalfont data buffer is reset \textbf{and} } $\btounpack \neq \emptyset$) }{
		\ForEach{$\block \in \btounpack$}{
   			$\bufferdata[\block]$ $\leftarrow$ \FnUnpack{$\bufferrecv[\block]$} \;
	 	}
	}
}

\BlankLine\BlankLine
\emph{---------------------------------------------------------------------------------}\;
\emph{----- non-blocking persistent requests}\;
  \SetKwProg{Fn}{Function}{:}{}
  \Fn{\FnSendBatch{$\Block$}}{
	\ForEach{$\block \in \Block$}{
   		$\buffersend[\block]$ $\leftarrow$ \FnPack{$\bufferdata[\block]$} \;
 	 	 \FnStart{$\srequest [\block] $}
	}
  }
  \BlankLine
  \Fn{\FnRecvBatch{$\Block$}}{
	\FnStartAll{$\rrequest[\block \in \Block]$} \;
  }
\BlankLine\BlankLine
\emph{---------------------------------------------------------------------------------}\;
\emph{----- non-blocking datatypes}\;
  \SetKwProg{Fn}{Function}{:}{}
  \Fn{\FnSendBatch{$\Block$}}{
  	\FnIsend{$\srequest [\block \in \Block] $}
  }
  \BlankLine
  \SetKwProg{Fn}{Function}{:}{}
  \Fn{\FnRecvBatch{$\Block$}}{
	\FnIrecv{$\srequest [\block \in \Block] $}
  }

\BlankLine
\BlankLine
\end{algorithm2e}

\subsubsection{Implementation using datatypes}
From the previous non-blocking implementation we have observed that a significant time is spent packing and unpacking the data.
Also we need to allocate the communication buffer to send the data which increases the memory footprint.
To tackle that issue we propose a third possible implementation where we take advantage of \code{MPI\_Datatype} to bypass the send-buffer allocation and packing on the send side.
However, we still expect the latter to lead to additional overhead from the MPI implementation.
On the receive side we still manually unpack the receive buffer to avoid waiting for the completion of all the send operations before starting the receive ones.
As the send requests are non-blocking, waiting for their completion could lead to an overflow on the network with too many send requests started and no receive requests ready.
This choice further allows us to overlap the shuffling of the received data with the overall communication scheme. 

\begin{figure}[h!tp]
\centering
\input{figures/tikz/implementation_isr}
\caption{Implementation of the \isr version. Data are sent and directly copied from the user-provided buffer (\CaptionUserBuf) to the receive buffer (\CaptionRecvBuf) thanks to \textit{non-blocking} communication and \code{MPI\_Datatype}. The data are then shuffled in the receive buffer and copied back to the user buffer.}
\label{fig:implementation-isr}
\end{figure}
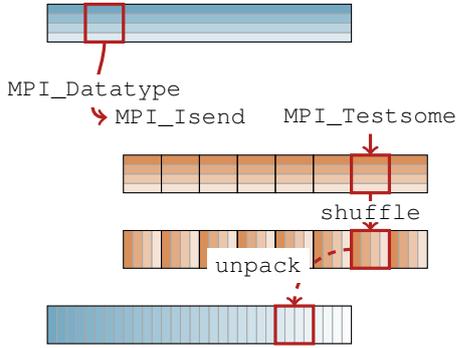

Compared to the \nb implementation, the approach remains unchanged as illustrated in \Fref{fig:implementation-isr}.  There are still four main steps: (1) the treatment of the send request, (2) the reset of the user-data, (3) the completion of the receive request and the shuffle of the associated buffer, and finally (4) the unpacking of the receive buffer in the user-data.  
The implementation differs in the treatment of the send request in the $\FnSendBatch$ function and with the use of non-blocking send  (\code{MPI\_Isend}) and receive (\code{MPI\_Irecv}) instead of persistent requests.
We also note that in $\FnSendBatch$ the packing is not needed anymore, as detailed in \Algref{algo:nb_1}.

%% file: figures/tikz/implementation_a2a.tex
\tikzstyle{dot} = [shape=circle,minimum size=0.2em,inner sep=0pt, outer sep=0pt]
\tikzstyle{usr} = [draw=black,fill=\figblue,very thin]
\tikzstyle{buf} = [draw=black,fill=\figorange,very thin]
\tikzstyle{sendbuf} = [draw=black,fill=\figgreen,very thin]

\begin{tikzpicture}
    \def\dx{0.125}
    \def\Dx{0.5}
    \def\Nx{8}
    \def\level{1.0}

    \def\posx{0.0}
    \def\posy{0.0}
    \pgfmathsetmacro{\Nstripe}{\Dx/\dx}
    \foreach \ii in {1,2,...,\Nstripe}{
        \pgfmathsetmacro{\ipx}{\ii/\Nstripe}
        \pgfmathsetmacro{\ippx}{(\ii-1)/\Nstripe}
        \pgfmathtruncatemacro{\icx}{100*(\ipx)} %
        \draw[usr,fill=\figblue!\icx,draw=\figgrey] (\posx,\posy+\ippx*\Dx) rectangle (\posx + \Nx*\Dx,\posy+\ipx*\Dx);
    };
    \draw[usr,fill=none,draw] (\posx,\posy) rectangle (\posx + \Nx*\Dx,\posy+\Dx);

    \draw[->] (\posx+0.5*\Dx,\posy) to[out=-90,in=180] node[left] {\tiny\code{pack}} (2.0*\Dx,-1.0*\level+0.5*\Dx);
    
    \def\posx{2.0*\Dx}
    \def\posy{-1.75*\level}
    \node[] at (\posx + \Nx*\Dx*0.5,\posy+\Dx*0.5) (a2a) {\tiny\code{MPI\_Ialltoallv}};

    \def\posx{2.0*\Dx}
    \def\posy{-1.0*\level}
    \foreach \ii in {1,2,...,\Nstripe}{
        \pgfmathsetmacro{\ipx}{\ii/\Nstripe}
        \pgfmathsetmacro{\ippx}{(\ii-1)/\Nstripe}
        \pgfmathtruncatemacro{\icx}{100*(\ipx)} %
        \draw[usr,fill=\figgreen!\icx,draw=\figgrey] (\posx,\posy+\ippx*\Dx) rectangle (\posx + \Nx*\Dx,\posy+\ipx*\Dx);
    };
    \foreach \ii in {1,2,...,\Nx}{
        \pgfmathsetmacro{\ipx}{\ii/\Nx}
        \pgfmathsetmacro{\ippx}{(\ii-1)/\Nx}
        \pgfmathsetmacro{\imx}{(\ii-0.5)/\Nx}
        \draw[buf,fill=none] (\posx+\ippx*\Dx*\Nx,\posy) rectangle (\posx +\ipx*\Dx*\Nx,\posy+\Dx);
        \draw[<-] (a2a) -- (\posx+\imx*\Dx*\Nx,\posy);
    };
    
    \def\posx{2.0*\Dx}
    \def\posy{-2.5*\level}
    \foreach \ii in {1,2,...,\Nstripe}{
        \pgfmathsetmacro{\ipx}{\ii/\Nstripe}
        \pgfmathsetmacro{\ippx}{(\ii-1)/\Nstripe}
        \pgfmathtruncatemacro{\icx}{100*(\ipx)} %
        \draw[usr,fill=\figorange!\icx,draw=\figgrey] (\posx,\posy+\ippx*\Dx) rectangle (\posx + \Nx*\Dx,\posy+\ipx*\Dx);
    };
    \foreach \ii in {1,2,...,\Nx}{
        \pgfmathsetmacro{\ipx}{\ii/\Nx}
        \pgfmathsetmacro{\ippx}{(\ii-1)/\Nx}
        \pgfmathsetmacro{\imx}{(\ii-0.5)/\Nx}
        \draw[buf,fill=none] (\posx+\ippx*\Dx*\Nx,\posy) rectangle (\posx +\ipx*\Dx*\Nx,\posy+\Dx);
        \draw[->] (a2a) -- (\posx+\imx*\Dx*\Nx,\posy+\Dx);
    };

    \def\posx{2.0*\Dx}
    \def\posy{-3.5*\level}
    \foreach \ii in {1,2,...,\Nx}{
        \foreach \jj in {1,2,...,\Nstripe}{
            \pgfmathsetmacro{\ipx}{((\ii-1)*\Nstripe + \jj)/(\Nx*\Nstripe)}
            \pgfmathsetmacro{\ippx}{((\ii-1)*\Nstripe + \jj - 1)/(\Nx*\Nstripe)}
            \pgfmathtruncatemacro{\icx}{100*(1-(\jj-1)/\Nstripe)} %
            \draw[usr,fill=\figorange!\icx,draw=\figgrey] (\posx+\ippx*\Nx*\Dx,\posy) rectangle (\posx+\ipx*\Nx*\Dx,\posy+\Dx);
        }
    }
    \foreach \ii in {1,2,...,\Nx}{
        \pgfmathsetmacro{\ipx}{\ii/\Nx}
        \pgfmathsetmacro{\ippx}{(\ii-1)/\Nx}
        \pgfmathsetmacro{\imx}{(\ii-0.5)/\Nx}
        \draw[buf,fill=none] (\posx+\ippx*\Dx*\Nx,\posy) rectangle (\posx +\ipx*\Dx*\Nx,\posy+\Dx);
        \draw[->] (\posx+\imx*\Dx*\Nx,-2.5*\level) -- (\posx+\imx*\Dx*\Nx,\posy+\Dx);
    };
    \node[fill=white,rounded corners=2pt,inner sep=1pt] at (\posx + \Nx*\Dx*0.5,\posy+\level*0.5+\Dx*0.5) {\tiny\code{shuffle}};

    \draw[->] (\posx,\posy+0.5*\Dx) to[out=180,in=90] node[left] {\tiny\code{unpack}} (0.5*\Dx,-4.5*\level+1.0*\Dx);

    \def\posx{0.0}
    \def\posy{-4.5*\level}
    \pgfmathsetmacro{\Nlist}{\Nx*\Nstripe}
    \foreach \ii in {1,2,...,\Nlist}{
        \pgfmathsetmacro{\ipx}{1.0-\ii/\Nlist}
        \pgfmathsetmacro{\ippx}{1.0-(\ii-1)/\Nlist}
        \pgfmathtruncatemacro{\icx}{100*(1.0-\ipx)} %
        \draw[usr,fill=\figblue!\icx,draw=\figgrey] (\posx+\ippx*\Nx*\Dx,\posy) rectangle (\posx+\ipx*\Nx*\Dx,\posy+\Dx);
    }
    \draw[usr,fill=none,draw] (\posx,\posy) rectangle (\posx + \Nx*\Dx,\posy+\Dx);

\end{tikzpicture}

%% file: figures/tikz/implementation_nb.tex
\tikzstyle{dot} = [shape=circle,minimum size=0.2em,inner sep=0pt, outer sep=0pt]
\tikzstyle{usr} = [draw=black,fill=\figblue,very thin]
\tikzstyle{buf} = [draw=black,fill=\figorange,very thin]
\def\highcol{firebrick}
\begin{tikzpicture}
    \def\dx{0.125}
    \def\Dx{0.5}
    \def\Nx{8}
    \def\level{1.0}
    \def\xred{1.0}

    \def\posx{0.0}
    \def\posy{0.0}
    \pgfmathsetmacro{\Nstripe}{\Dx/\dx}
    \foreach \ii in {1,2,...,\Nstripe}{
        \pgfmathsetmacro{\ipx}{\ii/\Nstripe}
        \pgfmathsetmacro{\ippx}{(\ii-1)/\Nstripe}
        \pgfmathtruncatemacro{\icx}{100*(\ipx)} %
        \draw[usr,fill=\figblue!\icx,draw=\figgrey] (\posx,\posy+\ippx*\Dx) rectangle (\posx + \Nx*\Dx,\posy+\ipx*\Dx);
    };
    \draw[usr,fill=none,draw] (\posx,\posy) rectangle (\posx + \Nx*\Dx,\posy+\Dx);

    \draw[very thick,draw=\highcol] (\posx+\xred*\Dx,\posy) rectangle (\posx+\xred*\Dx+\Dx,\posy+\Dx);
    
    \def\posx{2.0*\Dx}
    \def\posy{-1.75*\level}
    \node[] at (\posx + \xred*\Dx+0.5*\Dx,\posy+\Dx*0.5) (a2a) {\tiny\code{MPI\_Start}};

    \def\posx{2.0*\Dx}
    \def\posy{-1.0*\level}
    \foreach \ii in {1,2,...,\Nstripe}{
        \pgfmathsetmacro{\ipx}{\ii/\Nstripe}
        \pgfmathsetmacro{\ippx}{(\ii-1)/\Nstripe}
        \pgfmathtruncatemacro{\icx}{100*(\ipx)} %
        \draw[usr,fill=\figgreen!\icx,draw=\figgrey] (\posx,\posy+\ippx*\Dx) rectangle (\posx + \Nx*\Dx,\posy+\ipx*\Dx);
    };
    \foreach \ii in {1,2,...,\Nx}{
        \pgfmathsetmacro{\ipx}{\ii/\Nx}
        \pgfmathsetmacro{\ippx}{(\ii-1)/\Nx}
        \pgfmathsetmacro{\imx}{(\ii-0.5)/\Nx}
        \draw[buf,fill=none] (\posx+\ippx*\Dx*\Nx,\posy) rectangle (\posx +\ipx*\Dx*\Nx,\posy+\Dx);
    };

    \draw[very thick,draw=\highcol] (\posx+\xred*\Dx,\posy) rectangle (\posx+\xred*\Dx+\Dx,\posy+\Dx);
    \draw[->,very thick,draw=\highcol] (\xred*\Dx+0.5*\Dx,0.0) to[out=-90,in=180] node[left] {\tiny\code{pack}} (\posx+\xred*\Dx,\posy+0.5*\Dx);
    \pgfmathsetmacro{\imx}{(\xred+0.5)/\Nx}
    \draw[<-,very thick,draw=\highcol] (a2a) -- (\posx+\imx*\Dx*\Nx,\posy);
    
    \def\xred{6.0}
    \def\posx{2.0*\Dx}
    \def\posy{-1.75*\level}
    \node[] at (\posx + \xred*\Dx+0.5*\Dx,\posy+\Dx*0.5) (a2a) {\tiny\code{MPI\_Testsome}};
    \def\posx{2.0*\Dx}
    \def\posy{-2.5*\level}
    \foreach \ii in {1,2,...,\Nstripe}{
        \pgfmathsetmacro{\ipx}{\ii/\Nstripe}
        \pgfmathsetmacro{\ippx}{(\ii-1)/\Nstripe}
        \pgfmathtruncatemacro{\icx}{100*(\ipx)} %
        \draw[usr,fill=\figorange!\icx,draw=\figgrey] (\posx,\posy+\ippx*\Dx) rectangle (\posx + \Nx*\Dx,\posy+\ipx*\Dx);
    };
    \foreach \ii in {1,2,...,\Nx}{
        \pgfmathsetmacro{\ipx}{\ii/\Nx}
        \pgfmathsetmacro{\ippx}{(\ii-1)/\Nx}
        \pgfmathsetmacro{\imx}{(\ii-0.5)/\Nx}
        \draw[buf,fill=none] (\posx+\ippx*\Dx*\Nx,\posy) rectangle (\posx +\ipx*\Dx*\Nx,\posy+\Dx);
    };
    \pgfmathsetmacro{\imx}{(\xred+0.5)/\Nx}
    \draw[->,very thick,draw=\highcol] (a2a) -- (\posx+\imx*\Dx*\Nx,\posy+\Dx);
    \draw[very thick,draw=\highcol] (\posx+\xred*\Dx,\posy) rectangle (\posx+\xred*\Dx+\Dx,\posy+\Dx);

    \def\posx{2.0*\Dx}
    \def\posy{-3.5*\level}
    \foreach \ii in {1,2,...,\Nx}{
        \foreach \jj in {1,2,...,\Nstripe}{
            \pgfmathsetmacro{\ipx}{((\ii-1)*\Nstripe + \jj)/(\Nx*\Nstripe)}
            \pgfmathsetmacro{\ippx}{((\ii-1)*\Nstripe + \jj - 1)/(\Nx*\Nstripe)}
            \pgfmathtruncatemacro{\icx}{100*(1-(\jj-1)/\Nstripe)} %
            \draw[usr,fill=\figorange!\icx,draw=\figgrey] (\posx+\ippx*\Nx*\Dx,\posy) rectangle (\posx+\ipx*\Nx*\Dx,\posy+\Dx);
        }
    }
    \foreach \ii in {1,2,...,\Nx}{
        \pgfmathsetmacro{\ipx}{\ii/\Nx}
        \pgfmathsetmacro{\ippx}{(\ii-1)/\Nx}
        \pgfmathsetmacro{\imx}{(\ii-0.5)/\Nx}
        \draw[buf,fill=none] (\posx+\ippx*\Dx*\Nx,\posy) rectangle (\posx +\ipx*\Dx*\Nx,\posy+\Dx);
        
    };
    \pgfmathsetmacro{\imx}{(\xred+0.5)/\Nx}
    \draw[->,very thick,draw=\highcol] (\posx+\imx*\Dx*\Nx,-2.5*\level) -- (\posx+\imx*\Dx*\Nx,\posy+\Dx);
    \node[fill=white,rounded corners=2pt,inner sep=1pt] at (\posx+\xred*\Dx+0.5*\Dx,\posy+\level*0.5+\Dx*0.5) {\tiny\code{shuffle}};
    \draw[very thick,draw=\highcol] (\posx+\xred*\Dx,\posy) rectangle (\posx+\xred*\Dx+\Dx,\posy+\Dx);
    \draw[->,very thick,draw=\highcol,dashed] (\posx+\xred*\Dx,\posy+0.5*\Dx) to[out=180,in=90] node[left,fill=white] {\tiny\code{unpack}} (\xred*\Dx+0.5*\Dx,-4.5*\level+1.0*\Dx);

    \def\posx{0.0}
    \def\posy{-4.5*\level}
    \pgfmathsetmacro{\Nlist}{\Nx*\Nstripe}
    \foreach \ii in {1,2,...,\Nlist}{
        \pgfmathsetmacro{\ipx}{1.0-\ii/\Nlist}
        \pgfmathsetmacro{\ippx}{1.0-(\ii-1)/\Nlist}
        \pgfmathtruncatemacro{\icx}{100*(1.0-\ipx)} %
        \draw[usr,fill=\figblue!\icx,draw=\figgrey] (\posx+\ippx*\Nx*\Dx,\posy) rectangle (\posx+\ipx*\Nx*\Dx,\posy+\Dx);
    }
    \draw[usr,fill=none,draw] (\posx,\posy) rectangle (\posx + \Nx*\Dx,\posy+\Dx);
    \draw[very thick,draw=\highcol] (\posx+\xred*\Dx,\posy) rectangle (\posx+\xred*\Dx+\Dx,\posy+\Dx);

\end{tikzpicture}

%% file: figures/tikz/implementation_isr.tex
\tikzstyle{dot} = [shape=circle,minimum size=0.2em,inner sep=0pt, outer sep=0pt]
\tikzstyle{usr} = [draw=black,fill=\figblue,very thin]
\tikzstyle{buf} = [draw=black,fill=\figorange,very thin]
\def\highcol{firebrick}
\begin{tikzpicture}
    \def\dx{0.125}
    \def\Dx{0.5}
    \def\Nx{8}
    \def\level{1.0}
    \def\xred{1.0}

    \def\posx{0.0}
    \def\posy{0.0}
    \pgfmathsetmacro{\Nstripe}{\Dx/\dx}
    \foreach \ii in {1,2,...,\Nstripe}{
        \pgfmathsetmacro{\ipx}{\ii/\Nstripe}
        \pgfmathsetmacro{\ippx}{(\ii-1)/\Nstripe}
        \pgfmathtruncatemacro{\icx}{100*(\ipx)} %
        \draw[usr,fill=\figblue!\icx,draw=\figgrey] (\posx,\posy+\ippx*\Dx) rectangle (\posx + \Nx*\Dx,\posy+\ipx*\Dx);
    };
    \draw[usr,fill=none,draw] (\posx,\posy) rectangle (\posx + \Nx*\Dx,\posy+\Dx);

    \draw[very thick,draw=\highcol] (\posx+\xred*\Dx,\posy) rectangle (\posx+\xred*\Dx+\Dx,\posy+\Dx);
    
    \def\posx{2.0*\Dx}
    \def\posy{-1.25*\level}
    \node[] at (\posx + \xred*\Dx+0.5*\Dx,\posy+\Dx*0.5) (a2a) {\tiny\code{MPI\_Isend}};

    \def\posx{2.0*\Dx}
    \def\posy{-1.0*\level}

    \draw[->,very thick,draw=\highcol] (\xred*\Dx+0.5*\Dx,0.0) to[out=-90,in=180] node[fill=white] {\tiny\code{MPI\_Datatype}} (a2a);
    
    \def\xred{6.0}
    \def\posx{2.0*\Dx}
    \def\posy{-1.25*\level}
    \node[] at (\posx + \xred*\Dx+0.5*\Dx,\posy+\Dx*0.5) (a2a) {\tiny\code{MPI\_Testsome}};
    \def\posx{2.0*\Dx}
    \def\posy{-2.*\level}
    \foreach \ii in {1,2,...,\Nstripe}{
        \pgfmathsetmacro{\ipx}{\ii/\Nstripe}
        \pgfmathsetmacro{\ippx}{(\ii-1)/\Nstripe}
        \pgfmathtruncatemacro{\icx}{100*(\ipx)} %
        \draw[usr,fill=\figorange!\icx,draw=\figgrey] (\posx,\posy+\ippx*\Dx) rectangle (\posx + \Nx*\Dx,\posy+\ipx*\Dx);
    };
    \foreach \ii in {1,2,...,\Nx}{
        \pgfmathsetmacro{\ipx}{\ii/\Nx}
        \pgfmathsetmacro{\ippx}{(\ii-1)/\Nx}
        \pgfmathsetmacro{\imx}{(\ii-0.5)/\Nx}
        \draw[buf,fill=none] (\posx+\ippx*\Dx*\Nx,\posy) rectangle (\posx +\ipx*\Dx*\Nx,\posy+\Dx);
    };
    \pgfmathsetmacro{\imx}{(\xred+0.5)/\Nx}
    \draw[->,very thick,draw=\highcol] (a2a) -- (\posx+\imx*\Dx*\Nx,\posy+\Dx);
    \draw[very thick,draw=\highcol] (\posx+\xred*\Dx,\posy) rectangle (\posx+\xred*\Dx+\Dx,\posy+\Dx);

    \def\posx{2.0*\Dx}
    \def\posy{-3.*\level}
    \foreach \ii in {1,2,...,\Nx}{
        \foreach \jj in {1,2,...,\Nstripe}{
            \pgfmathsetmacro{\ipx}{((\ii-1)*\Nstripe + \jj)/(\Nx*\Nstripe)}
            \pgfmathsetmacro{\ippx}{((\ii-1)*\Nstripe + \jj - 1)/(\Nx*\Nstripe)}
            \pgfmathtruncatemacro{\icx}{100*(1-(\jj-1)/\Nstripe)} %
            \draw[usr,fill=\figorange!\icx,draw=\figgrey] (\posx+\ippx*\Nx*\Dx,\posy) rectangle (\posx+\ipx*\Nx*\Dx,\posy+\Dx);
        }
    }
    \foreach \ii in {1,2,...,\Nx}{
        \pgfmathsetmacro{\ipx}{\ii/\Nx}
        \pgfmathsetmacro{\ippx}{(\ii-1)/\Nx}
        \pgfmathsetmacro{\imx}{(\ii-0.5)/\Nx}
        \draw[buf,fill=none] (\posx+\ippx*\Dx*\Nx,\posy) rectangle (\posx +\ipx*\Dx*\Nx,\posy+\Dx);
        
    };
    \pgfmathsetmacro{\imx}{(\xred+0.5)/\Nx}
    \draw[->,very thick,draw=\highcol] (\posx+\imx*\Dx*\Nx,-2.0*\level) -- (\posx+\imx*\Dx*\Nx,\posy+\Dx);
    \node[fill=white,rounded corners=2pt,inner sep=1pt] at (\posx+\xred*\Dx+0.5*\Dx,\posy+\level*0.5+\Dx*0.5) {\tiny\code{shuffle}};
    \draw[very thick,draw=\highcol] (\posx+\xred*\Dx,\posy) rectangle (\posx+\xred*\Dx+\Dx,\posy+\Dx);
    \draw[->,very thick,draw=\highcol,dashed] (\posx+\xred*\Dx,\posy+0.5*\Dx) to[out=180,in=90] node[left,fill=white] {\tiny\code{unpack}} (\xred*\Dx+0.5*\Dx,-4.0*\level+1.0*\Dx);

    \def\posx{0.0}
    \def\posy{-4.*\level}
    \pgfmathsetmacro{\Nlist}{\Nx*\Nstripe}
    \foreach \ii in {1,2,...,\Nlist}{
        \pgfmathsetmacro{\ipx}{1.0-\ii/\Nlist}
        \pgfmathsetmacro{\ippx}{1.0-(\ii-1)/\Nlist}
        \pgfmathtruncatemacro{\icx}{100*(1.0-\ipx)} %
        \draw[usr,fill=\figblue!\icx,draw=\figgrey] (\posx+\ippx*\Nx*\Dx,\posy) rectangle (\posx+\ipx*\Nx*\Dx,\posy+\Dx);
    }
    \draw[usr,fill=none,draw] (\posx,\posy) rectangle (\posx + \Nx*\Dx,\posy+\Dx);
    \draw[very thick,draw=\highcol] (\posx+\xred*\Dx,\posy) rectangle (\posx+\xred*\Dx+\Dx,\posy+\Dx);

\end{tikzpicture}

%% file: flups_4_validation.tex
\section{Validation}
\label{sec_validation}
The parameter space used in \flups is very large due to its flexibility as it now supports $2$ data layouts, $1000$ boundary conditions, $8$ Green's functions, and $3$ communication strategies.
All the potential combinations (more than 48,000) have been validated thoroughly as part of our continuous integration framework.
Here for the sake of clarity, we present the validation of 3 different cases using the node-centered data layout.
They address the standard use of \flups when solving the Poisson equation with either fully spectral, fully unbounded, or semi-unbounded boundary conditions (BCs). All the possible Green's functions are tested. 
Since all the implementations produce the same results, we only present the order of convergence obtained with the \ata version of the framework.

As previously proposed in \cite{Caprace:2021} the test case is the Poisson~\Eqqref{eq:poisson} with various boundary conditions. The infinite norm of the error computed as 
\be
E_{\infty}  =  \sup_{x,y,z} \{ |\phi(x,y,z) - \phi_{ref}(x,y,z) |\}
\eec
where $\phi_{ref}$ is an analytical solution constructed as a product of 1-D functions:
\be
\phi_{ref}(x,y,z) = X(x) \; Y(y) \; Z(z)
\eed
The functions $X$, $Y$, $Z$ are chosen to match each set of boundary conditions. For example, sine and cosine with the proper wavelength are used with symmetric and periodic BCs while unbounded and semi-unbounded BCs are validated using Gaussian functions.
Given the analytical solution, we then compute the corresponding right-hand side as the Laplacian of the reference solution:
\be
f(x,y,z) = \ddsq{X}{x} \; Y(y) \; Z(z) + X(x) \; \ddsq{Y}{y} \; Z(z) + X(x) \; Y(y)\ddsq{Z}{z}
\eec
where the appropriate functions for each considered case is given in \Appendix{app:validation:anal}.

In the following sections, the Poisson equation is solved on a cubic domain of spatial extent $[0, L]$ in all directions.
We use the Green's function as defined in \cite{Caprace:2021}. More specifically, the one described in \cite{Chatelain:2010} hereafter referenced as \code{CHAT2} has a spectral accuracy for periodic or symmetric BCs and is of second order with unbounded conditions.
The regularized kernels \cite{Hejlesen:2013} are implemented from the order $m=2$ to the order $m=10$ together with the spectral-like regularization \cite{Hejlesen:2019}.
They are all named as a combination of the prefix \code{HEJ} followed by the respective order (spectral is labeled as order $0$).
Finally the Lattice Green's function \cite{Martinsson:2002a} has a convergence order of two and is labeled as \code{LGF2}.

\subsection{Domain with symmetric and periodic BCs}
\begin{figure}[h!tp]
\centering
\input{figures/python_fig/valid_spectral}
\caption{Convergence with symmetric and periodic BCs \code{CHAT2} ({\protect\ThickLineCircle{colchat2}{colchat2}}), \code{LGF2}({\protect\ThickLineCircle{collgf2}{collgf2}}), \code{HEJ2}({\protect\ThickLineCircle{colhej2}{colhej2}}), \code{HEJ4}({\protect\ThickLineCircle{colhej4}{colhej4}}), \code{HEJ6}({\protect\ThickLineCircle{colhej6}{colhej6}}), \code{HEJ8}({\protect\ThickLineCircle{colhej8}{colhej8}}), \code{HEJ10}({\protect\ThickLineCircle{colhej10}{colhej10}})}
\label{fig:validation-spectral}
\end{figure}
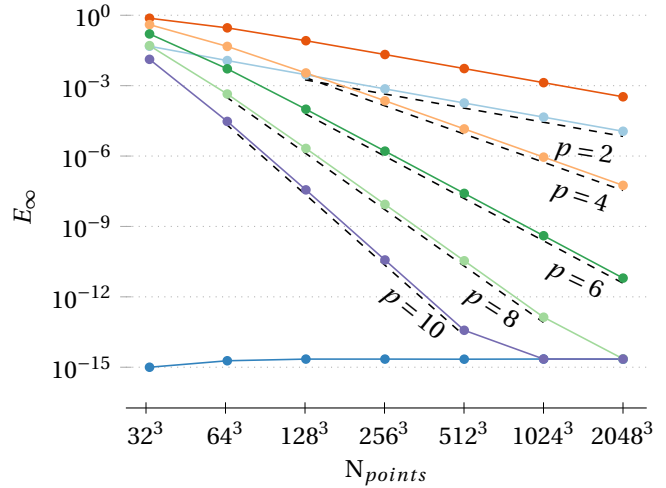

The chosen boundary conditions for this case are an even-even symmetry condition in the $x$-direction, an odd-even symmetry condition in the $y$-direction, and a periodic condition in the $z$-direction. The associated reference solution can be found in \Appendix{app:validation:spectral}.
The convergence results are shown in \Fref{fig:validation-spectral} and match the expected convergence order. 
As the present case does not involve unbounded conditions the singular Green's function \code{CHAT2} provides the exact solution regardless of the resolution.

\subsection{Fully unbounded boundary conditions}
\label{sec:valid-full-unb}
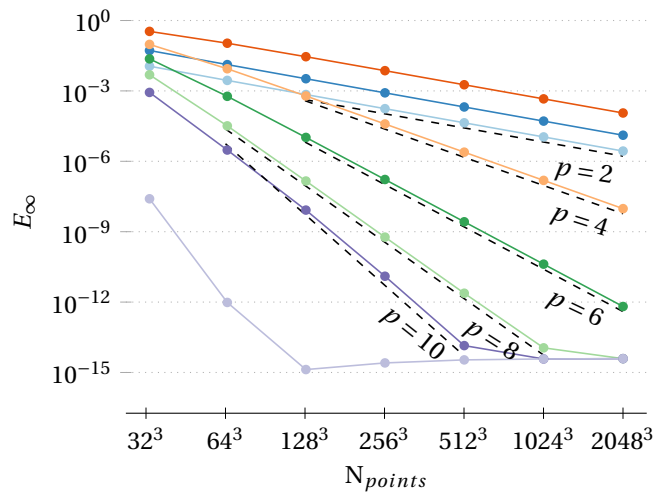
\begin{figure}[h!tp]
\centering
\input{figures/python_fig/valid_unbounded}
\caption{Convergence with fully unbounded BCs \code{CHAT2} ({\protect\ThickLineCircle{colchat2}{colchat2}}), \code{LGF2}({\protect\ThickLineCircle{collgf2}{collgf2}}), \code{HEJ2}({\protect\ThickLineCircle{colhej2}{colhej2}}), \code{HEJ4}({\protect\ThickLineCircle{colhej4}{colhej4}}), \code{HEJ6}({\protect\ThickLineCircle{colhej6}{colhej6}}), \code{HEJ8}({\protect\ThickLineCircle{colhej8}{colhej8}}), \code{HEJ10}({\protect\ThickLineCircle{colhej10}{colhej10}},  \code{HEJ0}({\protect\ThickLineCircle{colhej0}{colhej0}})}
\label{fig:validation-unbounded}
\end{figure}

The reference solution for the fully unbounded case is given in \Appendix{app:validation:unbounded}.
As depicted in \Fref{fig:validation-unbounded}, we observe a convergence corresponding to the theoretical order with all the Green's function.
We want to highlight that here we have added the spectrally truncated kernel \code{HEJ0} achieving a spectral-like convergence.

\subsection{Domain with two semi-infinite and one fully unbounded BCs}
\begin{figure}[h!tp]
\centering
\input{figures/python_fig/valid_mix_1}
\caption{Convergence with semi unbounded BCs \code{CHAT2} ({\protect\ThickLineCircle{colchat2}{colchat2}}), \code{LGF2}({\protect\ThickLineCircle{collgf2}{collgf2}}), \code{HEJ2}({\protect\ThickLineCircle{colhej2}{colhej2}}), \code{HEJ4}({\protect\ThickLineCircle{colhej4}{colhej4}}), \code{HEJ6}({\protect\ThickLineCircle{colhej6}{colhej6}}), \code{HEJ8}({\protect\ThickLineCircle{colhej8}{colhej8}}), \code{HEJ10}({\protect\ThickLineCircle{colhej10}{colhej10}}),  \code{HEJ0}({\protect\ThickLineCircle{colhej0}{colhej0}})}
\label{fig:validation-mix-1}
\end{figure}
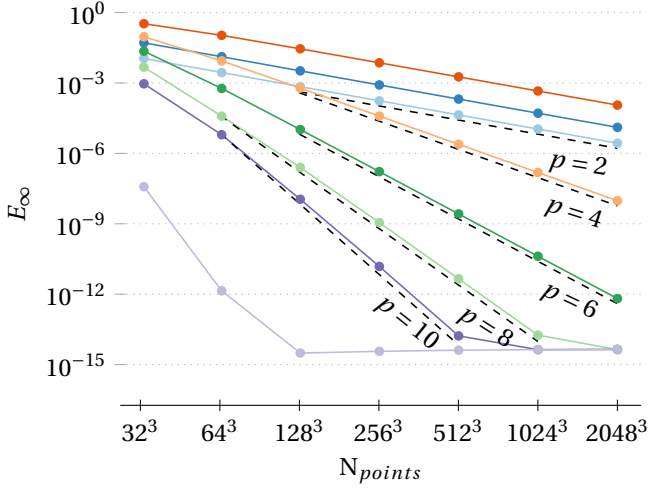

For the semi-unbounded BCs an even symmetry is imposed on the right side in the $x$-direction while an odd symmetry is applied on the left side in the $z$-direction. All the remaining boundaries are unbounded. 
The reference solution is then computed as indicated in \Appendix{app:validation:mix}.
As shown in \Fref{fig:validation-mix-1} all the Green's function reach the expected convergence orders.

%% file: figures/python_fig/valid_spectral.tex
\begin{tikzpicture}

\definecolor{darkgray}{RGB}{169,169,169}
\definecolor{darkgray176}{RGB}{176,176,176}
\definecolor{lightgreen161217155}{RGB}{161,217,155}
\definecolor{lightsteelblue158202225}{RGB}{158,202,225}
\definecolor{orangered2308513}{RGB}{230,85,13}
\definecolor{sandybrown253174107}{RGB}{253,174,107}
\definecolor{seagreen4916384}{RGB}{49,163,84}
\definecolor{slateblue117107177}{RGB}{117,107,177}
\definecolor{steelblue49130189}{RGB}{49,130,189}

\begin{axis}[
axis y line =left, axis x line =bottom, axis line style ={-}, axis x line shift=-5pt,,
log basis x={10},
log basis y={10},
minor xtick={},
minor ytick={},
tick align=center,
tick pos=left,
x grid style={darkgray176},
x grid style={draw=none},
xlabel near ticks,
xlabel={N\(\displaystyle _{points}\)},
xmajorgrids,
xmin=26.8449462631643, xmax=2518.79811332614,
xmode=log,
xtick style={color=black},
xtick={32,64,128,256,512,1024,2048},
xticklabels={$32^3$,$64^3$,$128^3$,$256^3$,$512^3$,$1024^3$,$2048^3$},
y axis line style={draw=none},
y grid style={darkgray176},
y grid style={dotted},
ylabel near ticks,
ylabel={\(\displaystyle E_{\infty}\)},
ymajorgrids,
ymin=5e-18, ymax=10,
ymode=log,
ytick style={color=darkgray},
ytick={1e-21,1e-18,1e-15,1e-12,1e-09,1e-06,0.001,1,1000,1000000}
]
\addplot [semithick, steelblue49130189, mark=*, mark size=1.5, mark options={solid}]
table {%
33 9.992007221626e-16
65 1.887379141863e-15
129 2.22044604925e-15
257 2.22044604925e-15
513 2.199629367539e-15
1025 2.234323837058e-15
2049 2.192690473635e-15
};
\addplot [semithick, lightsteelblue158202225, mark=*, mark size=1.5, mark options={solid}]
table {%
33 0.04787236290513
65 0.01173314141657
129 0.002918860814735
257 0.0007288176718652
513 0.0001821483845352
1025 4.553359501736e-05
2049 1.138317994998e-05
};
\addplot [semithick, black, dashed]
table {%
128 0.00176921388035339
128 0.00176921388035339
256 0.000442303470088348
512 0.000110575867522087
1024 2.76439668805217e-05
2048 6.91099172013043e-06
2048 6.91099172013043e-06
};
\addplot [semithick, orangered2308513, mark=*, mark size=1.5, mark options={solid}]
table {%
33 0.7467696399733
65 0.2906199948685
129 0.08225984194975
257 0.02123161252887
513 0.005350695053402
1025 0.00134036622852
2049 0.0003352601184711
};
\addplot [semithick, sandybrown253174107, mark=*, mark size=1.5, mark options={solid}]
table {%
33 0.3989689609666
65 0.04704449366043
129 0.003480126790429
257 0.0002270029736593
513 1.434056898142e-05
1025 8.98692429252e-07
2049 5.620595566747e-08
};
\addplot [semithick, black, dashed]
table {%
128 0.00223853067802378
128 0.00223853067802378
256 0.000139908167376486
512 8.74426046103037e-06
1024 5.46516278814398e-07
2048 3.41572674258999e-08
2048 3.41572674258999e-08
};
\addplot [semithick, seagreen4916384, mark=*, mark size=1.5, mark options={solid}]
table {%
33 0.1601245500817
65 0.005226973882875
129 9.886278768145e-05
257 1.620938330893e-06
513 2.56345409344e-08
1025 4.017504107168e-10
2049 6.282752096354e-12
};
\addplot [semithick, black, dashed]
table {%
128 6.41201626468781e-05
128 6.41201626468781e-05
256 1.00187754135747e-06
512 1.56543365837105e-08
1024 2.44599009120476e-10
2048 3.82185951750744e-12
2048 3.82185951750744e-12
};
\addplot [semithick, lightgreen161217155, mark=*, mark size=1.5, mark options={solid}]
table {%
33 0.0507771449839
65 0.0004407646741685
129 2.112448391212e-06
257 8.687086272552e-09
513 3.437394813233e-11
1025 1.353361867018e-13
2049 2.275957200482e-15
};
\addplot [semithick, black, dashed]
table {%
64 0.000349073400871193
64 0.000349073400871193
128 1.3635679721531e-06
256 5.32643739122304e-09
512 2.0806396059465e-11
1024 8.12749846072852e-14
1024 8.12749846072852e-14
};
\addplot [semithick, slateblue117107177, mark=*, mark size=1.5, mark options={solid}]
table {%
33 0.01323119103844
65 2.99117820427e-05
129 3.616239030446e-08
257 3.725930675103e-11
513 3.752553823233e-14
1025 2.234323837058e-15
2049 2.234323837058e-15
};
\addplot [semithick, black, dashed]
table {%
64 2.42776838406875e-05
64 2.42776838406875e-05
128 2.37086756256714e-08
256 2.31530035406948e-11
512 2.26103550202097e-14
512 2.26103550202097e-14
};
\draw (axis cs:1024,2.76439668805217e-05) ++(2pt,-2pt) node[
  scale=1.04166666666667,
  anchor=north west,
  text=black,
  rotate=350.9
]{$p = 2$};
\draw (axis cs:1024,5.46516278814398e-07) ++(2pt,-2pt) node[
  scale=1.04166666666667,
  anchor=north west,
  text=black,
  rotate=342.2
]{$p = 4$};
\draw (axis cs:1024,2.44599009120476e-10) ++(2pt,-2pt) node[
  scale=1.04166666666667,
  anchor=north west,
  text=black,
  rotate=334.3
]{$p = 6$};
\draw (axis cs:512,2.0806396059465e-11) ++(2pt,-2pt) node[
  scale=1.04166666666667,
  anchor=north west,
  text=black,
  rotate=327.3
]{$p = 8$};
\draw (axis cs:256,2.31530035406948e-11) ++(2pt,-2pt) node[
  scale=1.04166666666667,
  anchor=north west,
  text=black,
  rotate=321.2
]{$p = 10$};
\end{axis}

\end{tikzpicture}

%% file: figures/python_fig/valid_unbounded.tex
\begin{tikzpicture}

\definecolor{darkgray}{RGB}{169,169,169}
\definecolor{darkgray176}{RGB}{176,176,176}
\definecolor{lightgreen161217155}{RGB}{161,217,155}
\definecolor{lightsteelblue158202225}{RGB}{158,202,225}
\definecolor{lightsteelblue188189220}{RGB}{188,189,220}
\definecolor{orangered2308513}{RGB}{230,85,13}
\definecolor{sandybrown253174107}{RGB}{253,174,107}
\definecolor{seagreen4916384}{RGB}{49,163,84}
\definecolor{slateblue117107177}{RGB}{117,107,177}
\definecolor{steelblue49130189}{RGB}{49,130,189}

\begin{axis}[
axis y line =left, axis x line =bottom, axis line style ={-}, axis x line shift=-5pt,,
log basis x={10},
log basis y={10},
minor xtick={},
minor ytick={},
tick align=center,
tick pos=left,
x grid style={darkgray176},
x grid style={draw=none},
xlabel near ticks,
xlabel={N\(\displaystyle _{points}\)},
xmajorgrids,
xmin=26.8449462631643, xmax=2518.79811332614,
xmode=log,
xtick style={color=black},
xtick={32,64,128,256,512,1024,2048},
xticklabels={$32^3$,$64^3$,$128^3$,$256^3$,$512^3$,$1024^3$,$2048^3$},
y axis line style={draw=none},
y grid style={darkgray176},
y grid style={dotted},
ylabel near ticks,
ylabel={\(\displaystyle E_{\infty}\)},
ymajorgrids,
ymin=5e-18, ymax=10,
ymode=log,
ytick style={color=darkgray},
ytick={1e-21,1e-18,1e-15,1e-12,1e-09,1e-06,0.001,1,1000,1000000}
]
\addplot [semithick, steelblue49130189, mark=*, mark size=1.5, mark options={solid}]
table {%
33 0.05263276045406
65 0.01321175029864
129 0.003306282908244
257 0.0008267796288135
513 0.0002067079561279
1025 5.167780432969e-05
2049 1.291950202953e-05
};
\addplot [semithick, lightsteelblue158202225, mark=*, mark size=1.5, mark options={solid}]
table {%
33 0.01129626253568
65 0.002805084561202
129 0.0007000922324734
257 0.0001749494915277
513 4.373277676817e-05
1025 1.09329069633e-05
2049 2.73320878974e-06
};
\addplot [semithick, black, dashed]
table {%
128 0.000424804927090729
128 0.000424804927090729
256 0.000106201231772682
512 2.65503079431705e-05
1024 6.63757698579264e-06
2048 1.65939424644816e-06
2048 1.65939424644816e-06
};
\addplot [semithick, orangered2308513, mark=*, mark size=1.5, mark options={solid}]
table {%
33 0.3436999189742
65 0.1077740956147
129 0.02867815050632
257 0.007285048625973
513 0.001828598642503
1025 0.0004576100450299
2049 0.000114431314351
};
\addplot [semithick, sandybrown253174107, mark=*, mark size=1.5, mark options={solid}]
table {%
33 0.09598811809279
65 0.008791400516829
129 0.0006081716375965
257 3.900176915184e-05
513 2.4534006966e-06
1025 1.535854732282e-07
2049 9.602970418854e-09
};
\addplot [semithick, black, dashed]
table {%
128 0.00038246025047486
128 0.00038246025047486
256 2.39037656546788e-05
512 1.49398535341742e-06
1024 9.33740845885889e-08
2048 5.83588028678681e-09
2048 5.83588028678681e-09
};
\addplot [semithick, seagreen4916384, mark=*, mark size=1.5, mark options={solid}]
table {%
33 0.02320067531128
65 0.000588159687434
129 1.040679590414e-05
257 1.677698570068e-07
513 2.641999397035e-09
1025 4.136180287162e-11
2049 6.478151348688e-13
};
\addplot [semithick, black, dashed]
table {%
128 6.61143574915235e-06
128 6.61143574915235e-06
256 1.03303683580505e-07
512 1.6141200559454e-09
1024 2.52206258741468e-11
2048 3.94072279283544e-13
2048 3.94072279283544e-13
};
\addplot [semithick, lightgreen161217155, mark=*, mark size=1.5, mark options={solid}]
table {%
33 0.004859683170374
65 3.235093850429e-05
129 1.444360281377e-07
257 5.871219218266e-10
513 2.384865308708e-12
1025 1.121325254871e-14
2049 3.885780586188e-15
};
\addplot [semithick, black, dashed]
table {%
64 2.38674590868594e-05
64 2.38674590868594e-05
128 9.32322620580446e-08
256 3.64188523664237e-10
512 1.42261142056342e-12
1024 5.55707586157588e-15
1024 5.55707586157588e-15
};
\addplot [semithick, slateblue117107177, mark=*, mark size=1.5, mark options={solid}]
table {%
33 0.0008616949130498
65 3.020793967657e-06
129 8.32719010601e-09
257 1.264460633807e-11
513 1.397608936948e-14
1025 3.774758283726e-15
2049 3.885780586188e-15
};
\addplot [semithick, black, dashed]
table {%
64 5.59047361009427e-06
64 5.59047361009427e-06
128 5.45944688485768e-09
256 5.33149109849383e-12
512 5.20653427587288e-15
512 5.20653427587288e-15
};
\addplot [semithick, lightsteelblue188189220, mark=*, mark size=1.5, mark options={solid}]
table {%
33 2.524183947512e-08
65 9.654664220871e-13
129 1.33226762955e-15
257 2.553512956638e-15
513 3.441691376338e-15
1025 3.774758283726e-15
2049 3.885780586188e-15
};
\draw (axis cs:1024,6.63757698579264e-06) ++(2pt,-2pt) node[
  scale=1.04166666666667,
  anchor=north west,
  text=black,
  rotate=350.9
]{$p = 2$};
\draw (axis cs:1024,9.33740845885889e-08) ++(2pt,-2pt) node[
  scale=1.04166666666667,
  anchor=north west,
  text=black,
  rotate=342.2
]{$p = 4$};
\draw (axis cs:1024,2.52206258741468e-11) ++(2pt,-2pt) node[
  scale=1.04166666666667,
  anchor=north west,
  text=black,
  rotate=334.3
]{$p = 6$};
\draw (axis cs:512,1.42261142056342e-12) ++(2pt,-2pt) node[
  scale=1.04166666666667,
  anchor=north west,
  text=black,
  rotate=327.3
]{$p = 8$};
\draw (axis cs:256,5.33149109849383e-12) ++(2pt,-2pt) node[
  scale=1.04166666666667,
  anchor=north west,
  text=black,
  rotate=321.2
]{$p = 10$};
\end{axis}

\end{tikzpicture}

%% file: figures/python_fig/valid_mix_1.tex
\begin{tikzpicture}

\definecolor{darkgray}{RGB}{169,169,169}
\definecolor{darkgray176}{RGB}{176,176,176}
\definecolor{lightgreen161217155}{RGB}{161,217,155}
\definecolor{lightsteelblue158202225}{RGB}{158,202,225}
\definecolor{lightsteelblue188189220}{RGB}{188,189,220}
\definecolor{orangered2308513}{RGB}{230,85,13}
\definecolor{sandybrown253174107}{RGB}{253,174,107}
\definecolor{seagreen4916384}{RGB}{49,163,84}
\definecolor{slateblue117107177}{RGB}{117,107,177}
\definecolor{steelblue49130189}{RGB}{49,130,189}

\begin{axis}[
axis y line =left, axis x line =bottom, axis line style ={-}, axis x line shift=-5pt,,
log basis x={10},
log basis y={10},
minor xtick={},
minor ytick={},
tick align=center,
tick pos=left,
x grid style={darkgray176},
x grid style={draw=none},
xlabel near ticks,
xlabel={N\(\displaystyle _{points}\)},
xmajorgrids,
xmin=26.8449462631643, xmax=2518.79811332614,
xmode=log,
xtick style={color=black},
xtick={32,64,128,256,512,1024,2048},
xticklabels={$32^3$,$64^3$,$128^3$,$256^3$,$512^3$,$1024^3$,$2048^3$},
y axis line style={draw=none},
y grid style={darkgray176},
y grid style={dotted},
ylabel near ticks,
ylabel={\(\displaystyle E_{\infty}\)},
ymajorgrids,
ymin=5e-18, ymax=10,
ymode=log,
ytick style={color=darkgray},
ytick={1e-21,1e-18,1e-15,1e-12,1e-09,1e-06,0.001,1,1000,1000000}
]
\addplot [semithick, steelblue49130189, mark=*, mark size=1.5, mark options={solid}]
table {%
33 0.05167547510286
65 0.01319662247327
129 0.003302496348806
257 0.000826720421939
513 0.0002066931532828
1025 5.167757302627e-05
2049 1.291944420456e-05
};
\addplot [semithick, lightsteelblue158202225, mark=*, mark size=1.5, mark options={solid}]
table {%
33 0.01110282079313
65 0.002802752228912
129 0.0006995113037631
257 0.0001749816758663
513 4.374287033404e-05
1025 1.093575373734e-05
2049 2.733942973121e-06
};
\addplot [semithick, black, dashed]
table {%
128 0.00042491903645508
128 0.00042491903645508
256 0.00010622975911377
512 2.65574397784425e-05
1024 6.63935994461062e-06
2048 1.65983998615265e-06
2048 1.65983998615265e-06
};
\addplot [semithick, orangered2308513, mark=*, mark size=1.5, mark options={solid}]
table {%
33 0.3378443212264
65 0.1076527819304
129 0.02864544732221
257 0.007284527494031
513 0.001828467727471
1025 0.0004576079969525
2049 0.0001144308021774
};
\addplot [semithick, sandybrown253174107, mark=*, mark size=1.5, mark options={solid}]
table {%
33 0.09399061603496
65 0.008779417362971
129 0.0006073355960682
257 3.899840995214e-05
513 2.45318927905e-06
1025 1.535846472223e-07
2049 9.602919903706e-09
};
\addplot [semithick, black, dashed]
table {%
128 0.000382458238593609
128 0.000382458238593609
256 2.39036399121005e-05
512 1.49397749450628e-06
1024 9.33735934066427e-08
2048 5.83584958791517e-09
2048 5.83584958791517e-09
};
\addplot [semithick, seagreen4916384, mark=*, mark size=1.5, mark options={solid}]
table {%
33 0.02265987189851
65 0.000587287469793
129 1.03913536833e-05
257 1.677542872391e-07
513 2.641751484234e-09
1025 4.136280207234e-11
2049 6.48592290986e-13
};
\addplot [semithick, black, dashed]
table {%
128 6.61936720592046e-06
128 6.61936720592046e-06
256 1.03427612592507e-07
512 1.61605644675792e-09
1024 2.52508819805926e-11
2048 3.94545030946759e-13
2048 3.94545030946759e-13
};
\addplot [semithick, lightgreen161217155, mark=*, mark size=1.5, mark options={solid}]
table {%
33 0.004741994534202
65 3.871182650237e-05
129 2.524034890617e-07
257 1.124003257656e-09
513 4.533546281438e-12
1025 1.802290955366e-14
2049 4.329869796038e-15
};
\addplot [semithick, black, dashed]
table {%
64 4.17086375624883e-05
64 4.17086375624883e-05
128 1.6292436547847e-07
256 6.36423302650274e-10
512 2.48602852597763e-12
1024 9.71104892960012e-15
1024 9.71104892960012e-15
};
\addplot [semithick, slateblue117107177, mark=*, mark size=1.5, mark options={solid}]
table {%
33 0.0009236551104396
65 6.263975649929e-06
129 1.10804083556e-08
257 1.521904130497e-11
513 1.66949787328e-14
1025 4.329869796038e-15
2049 4.440892098501e-15
};
\addplot [semithick, black, dashed]
table {%
64 7.43885148681092e-06
64 7.43885148681092e-06
128 7.26450340508879e-09
256 7.09424160653202e-12
512 6.92797031887892e-15
512 6.92797031887892e-15
};
\addplot [semithick, lightsteelblue188189220, mark=*, mark size=1.5, mark options={solid}]
table {%
33 3.817465100572e-08
65 1.398325899515e-12
129 3.10862446895e-15
257 3.663735981263e-15
513 4.107825191113e-15
1025 4.329869796038e-15
2049 4.440892098501e-15
};
\draw (axis cs:1024,6.63935994461062e-06) ++(2pt,-2pt) node[
  scale=1.04166666666667,
  anchor=north west,
  text=black,
  rotate=350.9
]{$p = 2$};
\draw (axis cs:1024,9.33735934066427e-08) ++(2pt,-2pt) node[
  scale=1.04166666666667,
  anchor=north west,
  text=black,
  rotate=342.2
]{$p = 4$};
\draw (axis cs:1024,2.52508819805926e-11) ++(2pt,-2pt) node[
  scale=1.04166666666667,
  anchor=north west,
  text=black,
  rotate=334.3
]{$p = 6$};
\draw (axis cs:512,2.48602852597763e-12) ++(2pt,-2pt) node[
  scale=1.04166666666667,
  anchor=north west,
  text=black,
  rotate=327.3
]{$p = 8$};
\draw (axis cs:256,7.09424160653202e-12) ++(2pt,-2pt) node[
  scale=1.04166666666667,
  anchor=north west,
  text=black,
  rotate=321.2
]{$p = 10$};
\end{axis}

\end{tikzpicture}

%% file: flups_5_results.tex
\section{Application: the Biot-Savart solver}
\label{sec2_biot_savart}

To demonstrate the flexibility of \flups and its use in practice we now consider the Bio-Savart equation, a variation on the standard Poisson equation, given by
\be
\nabla^2 u = \nabla \times f
\label{eq:biotsavart}
\eed
This relation is particularly useful in application such as computational fluid dynamics, when one needs to recover the velocity from the vorticity field.
While the equation has its own Green's functions \cite{Hejlesen:2013}, one could take another approach that extends the work done previously and relies on the flexibility of \flups. %
First compute the forward FFT of the rhs $\hat{f}$, then compute the curl in the spectral space, compute the convolution with $\hat{G}$, and finally compute the FFT backward.

To evaluate the curl in spectral space we have to properly apply a spectral derivative on the result of the forward FFT.
With periodic or unbounded boundary conditions the result of the forward FFT is complex, and therefore the evaluation of a derivative becomes
\be
\dfrac{\partial}{\partial x} f \quad \rightarrow \quad   \plr{\ic \omega_k} \; \tf
\eed
where the differential operator $\plr{\ic \omega_k}$ is a purely imaginary number, and the derivation is then spectrally accurate.
Depending on the targeted application it can also be useful to consider a finite difference approximation instead of the actual derivative.
Then the following expressions are used for the order $2$:
\be
\dfrac{\partial}{\partial x} f + \mathcal{O}\left(h^2\right) \quad \rightarrow \quad   \dfrac{\ic}{h} \; \sin\left(\omega_k \; h \right) \; \tf
\eec
the order $4$:
\be
\dfrac{\partial}{\partial x} f + \mathcal{O}\left(h^4\right) \quad \rightarrow \quad   \dfrac{\ic}{h} \; \plr{ \dfrac{4}{3} \sin\left(\omega_k \; h \right) -\dfrac{1}{6} \sin\left(\omega_k \; 2h \right) } \; \tf
\eec
and the order $6$:
\begin{multline}
\dfrac{\partial}{\partial x} f + \mathcal{O}\left(h^6\right)   \\ 
\rightarrow \quad   \dfrac{\ic}{h} \; \plr{ \dfrac{3}{2} \sin\left(\omega_k \; h \right) -\dfrac{3}{10} \sin\left(\omega_k \; 2h \right)+ \dfrac{1}{30} \sin\left(\omega_k \; 3h \right) } \; \tf ~.
\end{multline}

For the even or odd BCs, applying a derivative will inverse the original condition: an even condition will become odd, and an odd one will become even. 
Therefore, the type of the backward FFT used must be adapted in order to reflect that change.
However, changing a DST for a DCT (and the opposite) should be done with care, as already discussed earlier.
If the input of a DCT is $f$ then the output $\tf$ corresponds to the complex number $\plr{\tf + 0\ic }$. On the other hand, if the input of a DST is $f$, then its output $\tf$ corresponds to the complex number $\plr{0 - i \tf}$. 
Consequently, if a DST is used as the forward transform, taking the derivative of the output leads to $\plr{ \ic \; \omega_k}  \; \plr{ 0 - \ic \tf} =  \tf  \omega_k$ which can be used directly as the input of the backward DCT.
On the other hand, if a DCT is used as the forward transform, the output is $\plr{\tf + 0 \ic }$ and the first derivative is $\plr{ \ic \; \omega_k}  \; \tf =  \ic  \omega_k \tf$ whose sign must be changed to be used as the input of the DST leading to $- \tf  \omega_k$.
The same approach applies to the derivatives computed using the finite differences, where $\omega_k$ is replaced by the appropriate formula.

The obtained Biot-Savart solver is validated with the case of a vortex tube aligned in the $z$-direction. The tube is centered within a cubic domain of size $[0,L]^3$ and fully unbounded boundary conditions are imposed in the $x$- and $y$-direction while symmetry conditions are used in the $z$-axis: $\omega_{x}$ and $\omega_{y}$ undergo odd symmetry conditions while $\omega_{z}$ satisfies an even symmetry on the domain boundaries.

The tube is compact and has the following expression: 

\be
\mathbf{\omega}(x,y,z) = \Bigl\{0, 0, -\omega_z(r)\Bigl\}
\eec
where $r$ is defined as $r = \sqrt{(x-0.5\,L)^2 + (y-0.5\,L)^2} $ and $\omega_z(r)$ is computed as
\be
\omega_z(r)=
    \begin{cases}
        \frac{1}{2\pi} \frac{2}{R^2} \frac{1}{\text{E}_2(1)} \exp\plr{-\frac{1}{\plr{1-\plr{\frac{r}{R}}^2}}} & \text{if}\ r <= R \\
        0 & \text{otherwise,}
    \end{cases}
\ee
using $R$ as the radius of the vortex tube, and E$_2$ is the generalized exponential integral function.

The corresponding analytical velocity can be retrieved through \cite{Winckelmans:2004a}:
\be
\mathbf{u}(x,y,z) = \Bigl\{ -\sin(\theta) u_\theta(r), \; \cos(\theta) u_\theta(r),\; 0\Bigl \}
\eec
where $u_\theta (r)$ is given by
\be
u_\theta(r)=
    \begin{cases}
        \frac{1}{2\pi r} \blr{1 - \frac{1}{\text{E}_2(1)} \plr{1-\plr{\frac{r}{R}}^2 } \text{E}_2 \plr{\frac{1}{1-\plr{\frac{r}{R}}^2} }} & \text{if}\ r <= R \\
        \us{\frac{1}{2\pi r}} & \text{otherwise.}
    \end{cases}
\ee

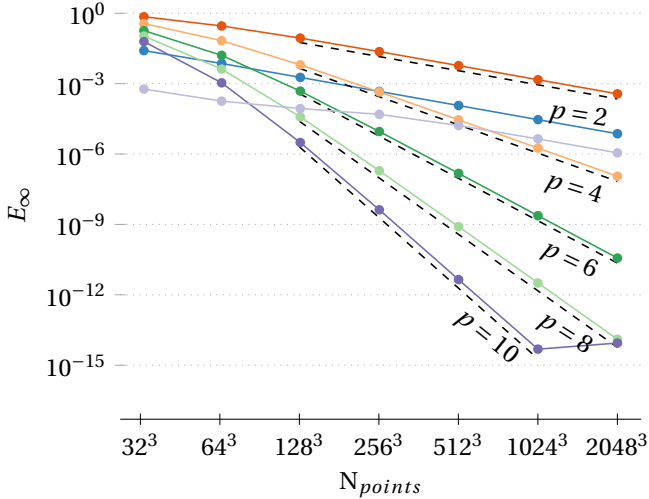
\begin{figure}[h!tp]
\centering
\input{figures/erratum_biot_savart/valid_tube_spectral}
\caption{Convergence of the Biot-Savart solver using spectral differentiation \code{CHAT2} ({\protect\ThickLineCircle{colchat2}{colchat2}}),  \code{HEJ2}({\protect\ThickLineCircle{colhej2}{colhej2}}), \code{HEJ4}({\protect\ThickLineCircle{colhej4}{colhej4}}), \code{HEJ6}({\protect\ThickLineCircle{colhej6}{colhej6}}), \code{HEJ8}({\protect\ThickLineCircle{colhej8}{colhej8}}), \code{HEJ10}({\protect\ThickLineCircle{colhej10}{colhej10}}),  \code{HEJ0}({\protect\ThickLineCircle{colhej0}{colhej0}})}
\label{fig:validation-bs-tube-spectral}
\end{figure}

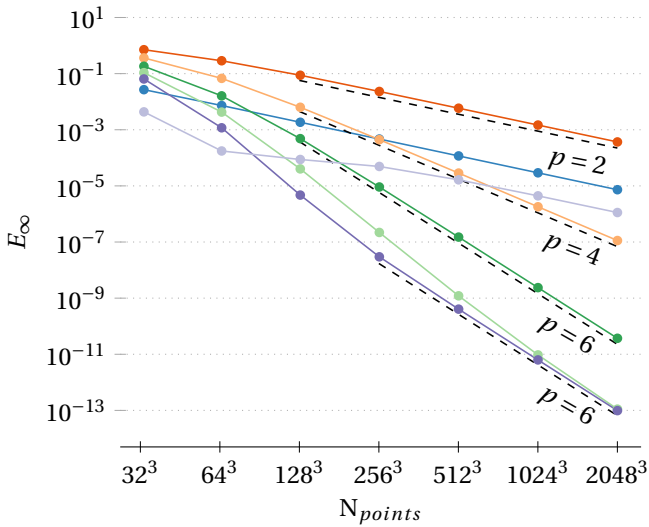
\begin{figure}[h!tp]
\centering
\input{figures/erratum_biot_savart/valid_tube_order_6}
\caption{Convergence of the Biot-Savart solver using 6th order differentiation   \code{CHAT2} ({\protect\ThickLineCircle{colchat2}{colchat2}}),  \code{HEJ2}({\protect\ThickLineCircle{colhej2}{colhej2}}), \code{HEJ4}({\protect\ThickLineCircle{colhej4}{colhej4}}), \code{HEJ6}({\protect\ThickLineCircle{colhej6}{colhej6}}), \code{HEJ8}({\protect\ThickLineCircle{colhej8}{colhej8}}), \code{HEJ10}({\protect\ThickLineCircle{colhej10}{colhej10}}),  \code{HEJ0}({\protect\ThickLineCircle{colhej0}{colhej0}})}
\label{fig:validation-bs-tube-FD6}
\end{figure}

\Fref{fig:validation-bs-tube-spectral} shows the convergence of the error computed as the infinite norm of the velocity field when using the spectral differentiation of the curl\footnote{We have corrected a typo from \cite[B.10]{Caprace:2021}, which should be $\frac{1}{8\pi}  \left(\pi - 6 + 2\log\left(\frac{\pi}{2} r_{eq}^2\right)\right)$ when $k_z = 0$.}.
One can also replace the differentiation by an approximation using finite differences of order $6$, whose convergence is given in \Fref{fig:validation-bs-tube-FD6}. For further details, the convergence of the finite difference of order $2$ and $4$ are given in \Sect{app:biosavart:conv} (\Fref{fig:validation-bs-tube-FD2} and \Fref{fig:validation-bs-tube-FD4} respectively).
 As expected the measured convergence orders correspond to the minimum between the differentiation order and the kernel order.

We also note that \code{HEJ0} kernel is the only one not achieving the expected convergence.
We attribute this behavior to the truncated infinite sum taking place when computing this non-singular Green's function~\cite{Hejlesen:2019} in the case of only two unbounded directions.
The truncation entails an approximation of the kernel and affects its accuracy.
In this case, the kernel is consequently bounded to the second order.

\section{Parallel performance analysis}
\label{sec:results}
This section presents the results of performance tests carried out on different massively parallel architectures. 
To obtain the results presented in this section we have set $\nsendbatch$ to $1$ and $\nsendpending$ to \code{INT\_MAX} for both the \isr and the \nb implementations. 
From a technical perspective, \flups has been compiled with the \code{-DNDEBUG -O3} flags.
We have used \code{mpich} 4.1a1 compiled with \code{--enable-fast=03,ndebug,alwaysinline} as the MPI implementation.
On Infiniband (IB)-based networks the communication library is \code{ucx} 1.13.1, while \code{libfabric} 15.0.0 from the vendor has been used for the slingshot-based infrastructure.
Still for IB networks the simulations were run with the \code{DC} transport layer.

First, we compare our \nb and \isr implementation against the \accfft library.
Then we study our scalability for the three proposed implementations.
Finally, we provide a comparison of the performance on three different \revthree{systems}.

\subsection{Comparison with \accfft}
\label{sec_5_accfft}
They are many other implementations of the distributed FFT algorithm, but only a handful of them provide the \textit{real-to-complex} FFT computation. 
Among them, \accfft \cite{Gholami:2015}, an implementation for both CPU and GPU partitions, is usually considered as one of the fastest~\cite{Ayala:2022}.

To fit in the framework proposed by \accfft we had to use a very specific test-case, here again highlighting the flexibility of \flups.
We use a cell-centered data-layout and perform a forward 3D FFT followed by a backward 3D FFT.
The first topology is aligned in the direction of the first FFT so that only two topology switches are performed and three 1D FFTs. In our case we then start with pencils in the $X$ direction while \accfft starts with pencils in the $Z$ direction by default.
The number of unknowns per rank is fixed to $256^3$, \revtwo{and the process distribution over the domain is given in \Cref{sec_detail_testcase_accfft}.}
To have the fairest comparison possible, a single executable calling both libraries is created, with both libraries compiled using the same binaries of \code{FFTW} and \code{MPI}. Both libraries call \code{FFTW} with the \code{FFTW\_MEASURE} flag so that the differences in timings are only due to differences in the implementation.
The tests have been done with exclusive node allocations on MeluXina, detailed in \Tref{tab:table-clusters}.

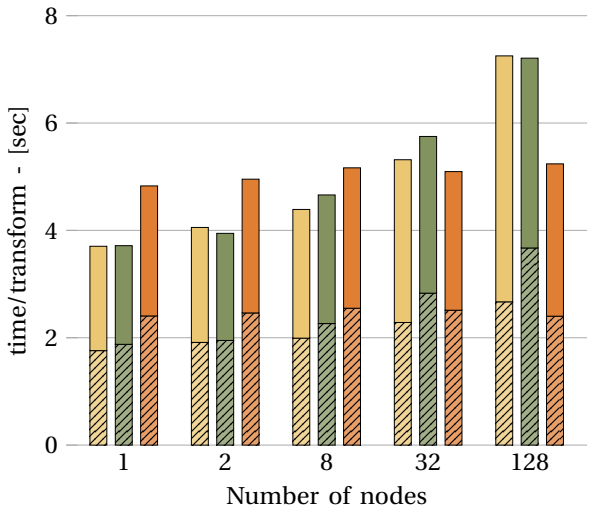
\begin{figure}[h!tp]
	\centering
	\input{figures/python_fig/mlx/accfft}
	\caption{Comparison with \code{accFFT}: weak scaling time-to-solution for \accfft ({\protect\BarLegend{colaccfft}{colaccfft}}), the \isr ({\protect\BarLegend{colisr}{colisr}}), and the \nb ({\protect\BarLegend{colnb}{colnb}}) versions of \flups. A forward and a backward 3D FFTs are performed with $256^{3}$ per rank on MeluXina. The hashed part of the bar corresponds to the forward transform, the plain part represents the backward transform.}
	\label{fig:comp-accfft}
\end{figure}

The obtained time-to-solution and weak efficiency for \accfft and the \isr and \nb implementation is presented in \Fref{fig:comp-accfft} using up to $16,384$ cores ($128$ nodes).
For reference, we also convert those times-to-solution into throughput per rank [\MBps], provided in \Tref{tab:perf-accfft}.
This metric is particularly useful to compare time-to-solution results across \revthree{infrastructure} and testcases. 
In \Fref{fig:comp-accfft} we first note that the \isr implementation and the \accfft one are very similar for a small number of nodes. For a larger count, the \isr implementation is slightly slower yet remains very competitive compared to \accfft.
The \nb implementation follows another path where the penalty of manual packing and unpacking vanishes as the node count increases.
From \Tref{tab:perf-accfft}, at $128$ nodes the \nb implementation runs $27.7\%$ faster than \accfft. We attribute this difference to the overhead coming from the use of \code{MPI\_Datatype}, which can also be observed in the scalability of the \nb and \isr approaches presented in \Sect{sec:weak-scaling}.
We conclude from this comparison that \flups offers more flexibility compared to \accfft, but it does not come with significant performance degradations.

\begin{table}[htp]
\begin{center}
\input{figures/python_tab/mlx/comp_acc_time_unknowns_rank}
\end{center}
\caption{Comparison with \code{accFFT}: throughput per rank [\MBps] for a forward and backward FFT.}
\label{tab:perf-accfft}
\end{table}

\subsection{General comments on weak and strong scalability}
\typo{
Ahead of our weak and strong scalability analysis, we would like to refer the reader to \Cref{sec_app_perfs} for details on the performance metrics used in this section.
In particular we will use the sequential percentage of a program, $\beta$, as a measure of the quality of our implementation, together with the speedup $s_P$ (strong scalability) and the efficiency $\eta_P$ (weak scalability).}

In this section, we present both weak and strong scalability tests.
The tests were performed on the CPU nodes of MeluXina (LuxConnect’s Data Center DC2, Luxembourg).
The CPU partition is made of 573 nodes, each of them composed of 2 AMD EPYC 7H12 CPUs with 64 cores per CPU, which make a total of 128 cores per node. The nodes are connected with InfiniBand HDR200 Gb/s and organized in a Dragonfly+ topology.
In both cases, we considered a 3D gaussian function in a fully unbounded domain (see \Sect{sec:valid-full-unb} for details on the test cases).
All the \typo{times-to-solution} are presented as the average execution times over the ranks.

\subsection{Weak Scalability}
\label{sec:weak-scaling}

\revtwo{We start our weak scalability test on a single node and run it up to $384$ nodes, \ie{$49,152$ cores}, where we have used $96^3$ points per core in the user domain with the process distribution given in \Cref{sec_detail_testcase_weak}.
}

\begin{figure}[h!tp]
\centering
\input{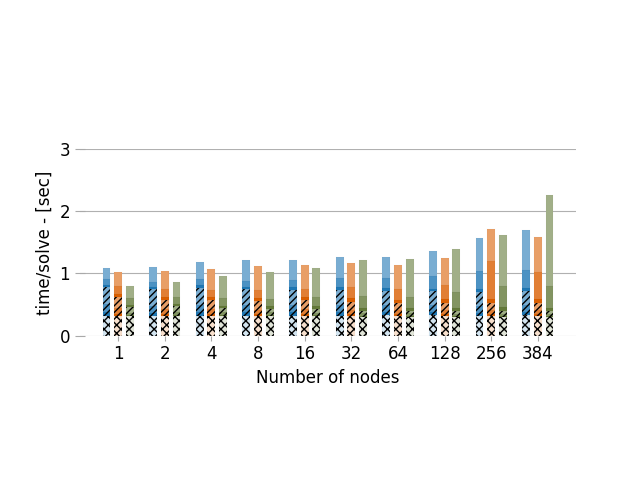}
\caption{Weak scaling: time-to-solution for \code{a2a}({\protect\BarLegend{cola2a}{cola2a}}), \code{isr}({\protect\BarLegend{colisr}{colisr}}), and \code{nb}({\protect\BarLegend{colnb}{colnb}}) versions. Tests on MeluXina with $96^{3}$ unknowns per rank and a fully unbounded testcase.}
\label{fig:results-weak-comp-mlx}
\end{figure}

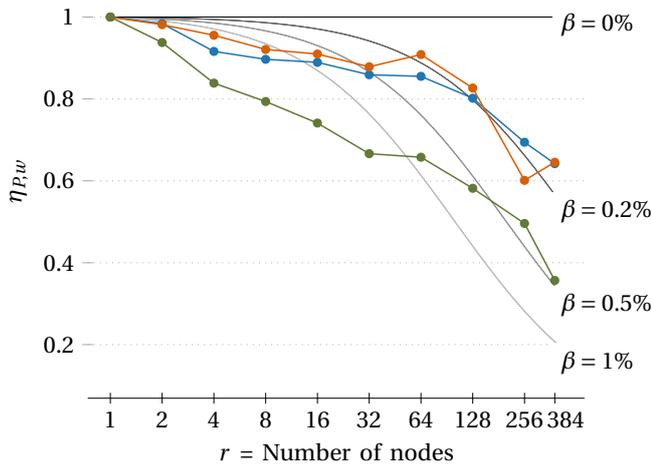
\begin{figure}
\centering
\resizebox{0.45\textwidth}{!}{%
	\begin{tikzpicture}
		\node[anchor=south west] (image1) at (0,0) {
			\input{figures/python_fig/mlx/unbounded/weak-eff}
                 };
                 \useasboundingbox;    
		\node[anchor=south west,fill=white] at (7.7,6) (note) {$\beta=0\%$};
		\node[anchor=south west,fill=white] at (7.7,3.4) (note) {$\beta=0.2\%$};
		\node[anchor=south west,fill=white] at (7.7,2.1) (note) {$\beta=0.5\%$};
		\node[anchor=south west,fill=white] at (7.7,1.3) (note) {$\beta=1\%$};
	\end{tikzpicture}
}
	\caption{Weak scaling: efficiency $\eta_{P,w}$ for \code{a2a}({\protect\ThickLineCircle{cola2a}{cola2a}}), \code{isr}({\protect\ThickLineCircle{colisr}{colisr}}), and \code{nb}({\protect\ThickLineCircle{colnb}{colnb}}) versions. Tests on MeluXina with $96^{3}$ unknowns per rank and a fully unbounded testcase.}
	\label{fig:results-weak-eff-mlx}
\end{figure}

\paragraph{Time-to-solution}
\Fref{fig:results-weak-comp-mlx} presents the evolution of the time-to-solution needed for a solve for each communication strategy, while \Fref{fig:results-weak-eff-mlx} shows the associated weak efficiency.
On a single node, the \nb strategy is close to the \ata one, yet \typo{slightly} faster.
The timing difference between both implementations remains constant when increasing the number of nodes, achieving therefore a similar weak efficiency.  
On the other side, the \isr method is the fastest one up to 16 nodes. Afterwards, the communication timings reach those of the \ata approach.
The difference in scalability between the implementation relying on the \code{MPI\_Datatype} (\isr) and those using manual packing and unpacking (\ata and \nb) is significant. Even if the time-to-solution is lower at small count, the gain of the \code{MPI\_Datatype} appears to vanish when increasing the partition count. This behavior varies with the different MPI versions as the treatment of \code{MPI\_Datatypes} is specific to each implementation.
Finally, we attribute the peaks happening on large partitions to the congestion on the network at the time of the testing (Dragonfly topology).

To have a meaningful comparison with a periodic case such as the one used in the previous section, the throughput per rank obtained for a fully unbounded domain in \Tref{tab:perf-mlx} must be normalized by a factor of $14/3$.
This factor comes from the doubling technique in which the first transform is performed on $2N$ data, the second on $4N$, and the last on $8N$, which makes an average of $14/3 N$ data per transform for the whole 3D FFT.
Compared to the results presented in \Tref{tab:perf-accfft} we here apply three topology switches instead of $2$.

\paragraph{Step-by-step analysis}
For further details we distinguish the different steps from within a call to \flups inside the bar plots, see \fref{fig:results-weak-comp-mlx}.
The crossed section represents the time spent performing computations only: 1D FFTs, spectral multiplication, and copy of the data provided by the user to the work buffer.
The lined section shows the time spent in computations overlapping communication: copying back and forth data from the work buffer to the communication buffers, shuffling the data, and resetting the work buffer to $0$.
The remaining non-hatched regions correspond to the time spent only communicating.
The different colors shades further differentiate the topology switches, the darker color corresponds to the first topology switch, and the lighter color to the last one.

As expected in the case of a weak scalability the computation-only part of the code scales perfectly and represents a fixed cost regardless of the communication strategy. The \isr timings for the computation (hatched colored region, overlapping with communication in the case of \isr and \nb) are the fastest, as the \isr implementation removes the manual packing.
However, the benefits are lessened as the time spent in the communication-only part of the algorithm increases.
As explained in \Sect{sec_implementation}, the \ata strategy only resets the user buffer to zero while waiting for the all-to-all communication to complete and the other operations are done sequentially.
This translates into timings associated with the computation-communication overlap section of the code slightly behind the timings of the other strategies.
Finally, in a fully unbounded case, the domain is expanded between the topology switches. It increases the number of points to be exchanged between the \code{MPI} processes and raises the communication cost of the topology switches.
Also, due to rank distribution among the nodes, the first topology switch is mostly happening intra-node, while the second and third ones are inter-node mostly. Those two factors together explain the increasing time from one topology switch to another.

\paragraph{Weak efficiency}
The associated weak efficiency of the software is presented in  \Fref{fig:results-weak-eff-mlx}.
In our cases, increasing the number of resources leads to an increase of the number of communications and congestion on the network. As expected from the previous results, the \isr approach shows the poorest scalability, and we can estimate its theoretical sequential percentage to $\beta = 0.5\%$.
We also note that the \nb and the \ata version have very similar weak scaling efficiency. Their serial percentage is estimated at $\beta_{nb} \approx 0.2 \%$. 
At the light of those results we anticipate the \nb version, having both smaller timings and higher efficiency, to be the version the most suited for very large-scale simulations. 

\begin{table}[htp]
\begin{center}
\input{figures/python_tab/mlx/unbounded/weak_time_unknowns_rank}
\end{center}
\caption{Weak scaling:  throughput per rank [\MBps] for a solve with the three different code versions. To account for the domain-doubling technique for unbounded BCs a normalization factor of $14/3$ has been applied.}
\label{tab:perf-mlx} 
\end{table}

\subsection{Strong scalability}
\label{sec:strong-scaling}
Similarly to the weak scalability testing, the strong analysis covers a range \typo{from one to $384$ nodes}, \revtwo{where the total problem size is fixed to $1280^{3}$ unknowns and the process distribution is given in  \Cref{sec_detail_testcase_strong}.}

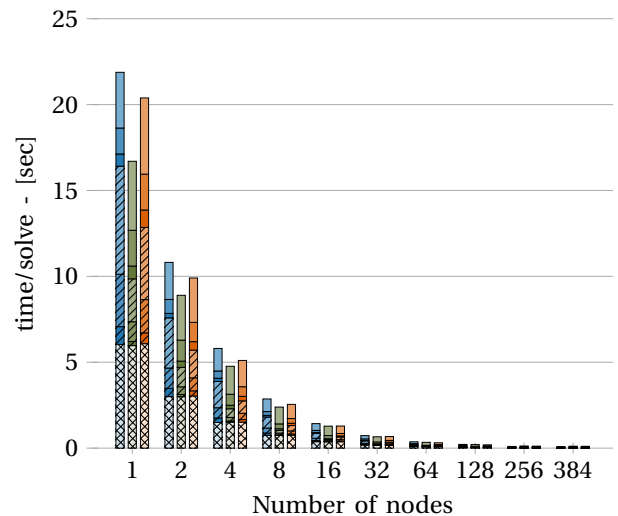
\begin{figure}[h!tp]
\centering
\input{figures/python_fig/mlx/unbounded/strong-comp}
\caption{Strong scaling: time-to-solution for \code{a2a}(\protect\BarLegend{cola2a}{cola2a}), \code{isr}(\protect\BarLegend{colisr}{colisr}), and \code{nb}(\protect\BarLegend{colnb}{colnb}) versions. Tests on MeluXina with $1280^{3}$ unknowns in total on a fully unbounded testcase.}
\label{fig:results-strong-timing-mlx}
\end{figure}

\Fref{fig:results-strong-timing-mlx} shows the strong scalability time-to-solution. As in the weak scalability, the \isr approach has the shortest resolution timings for a small number of \code{MPI} ranks. It is followed by the \nb and finally the \ata version.
The timings gap between the \isr and the other methods steadily decreased when increasing the nodes number. 
In agreement with those results, the computation-only part of the code shows a linear speed-up: the computation time is inversely proportional to the number of resources.

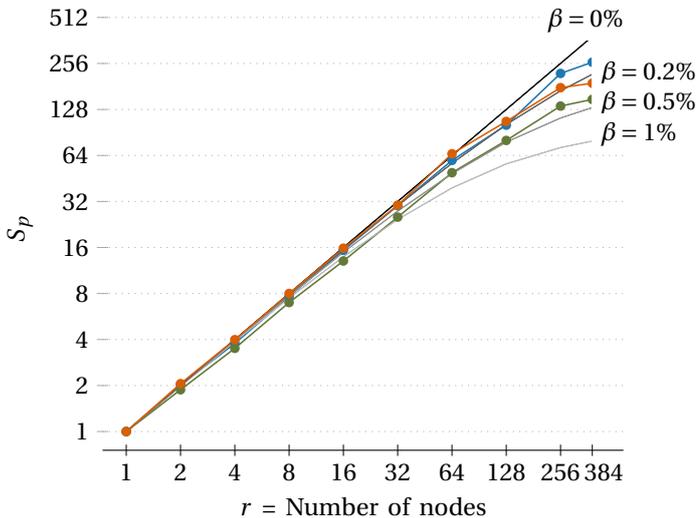
\begin{figure}
\centering
	\begin{tikzpicture}
		\node[anchor=south west] (image1) at (0,0) {
			\input{figures/python_fig/mlx/unbounded/strong-eff}
                 };
                 \useasboundingbox;                 
		\node[anchor=south west,fill=white] at (7.2,6.5) (note) {$\beta=0\%$};
		\node[anchor=south west,fill=white] at (7.9, 5.8) (note) {$\beta=0.2\%$};
		\node[anchor=south west,fill=white] at (7.9, 5.4) (note) {$\beta=0.5\%$};
		\node[anchor=south west,fill=white] at (7.9,5.0) (note) {$\beta=1\%$};
	\end{tikzpicture}
	\caption{Strong scaling: speedup $s_P$ for \code{a2a}({\protect\ThickLineCircle{cola2a}{cola2a}}), \code{isr}({\protect\ThickLineCircle{colisr}{colisr}}), and \code{nb}({\protect\ThickLineCircle{colnb}{colnb}}) versions. Tests on MeluXina with $1280^{3}$ unknowns in total on a fully unbounded testcase.}
	\label{fig:results-strong-eff-mlx}
\end{figure}

In \Fref{fig:results-strong-eff-mlx} we estimate the percentage of the software running in parallel, $\beta$, now based on the effective speed-up $s_P$.
The values found using the speedup are similar to those of the weak scaling tests, stating that approximately $99.5-99.8\%$ of our implementation is parallelized, while the remaining is sequential.
\revone{
We also observe that the scalability gap between the implementations illustrates the associated software latency: the \nb{} approach is expected to involve more software operations than the \ata{}, and the \isr has a higher latency than the \nb{} due to the allocation of extra buffers in the MPI implementation to pack and unpack the data.
}

\subsection{Comparison of main European \revthree{systems}}
\label{sec:eu_comparison}
We present in this section the results of the same weak scalability test on three main European \revthree{systems}: Lumi, Vega, and MeluXina, summarized in \Tref{tab:table-clusters}.
Vega (\revthree{Slovenia}) is equipped with similar nodes as MeluXina (AMD 7H12, 128 cores/node) but with slower interconnect: IB-HDR 100Gb/s instead of 200Gb/s for MeluXina . Lumi (Finland) and specifically on the Lumi-C partition each node has two AMD EPYC 7763 CPUs with a total of 128 cores per node, which are connected with a 200 Gb/s slingshot-11 network. 
Vega has then the slowest bandwidth and Lumi CPUs have a slightly faster CPUs clock speed.

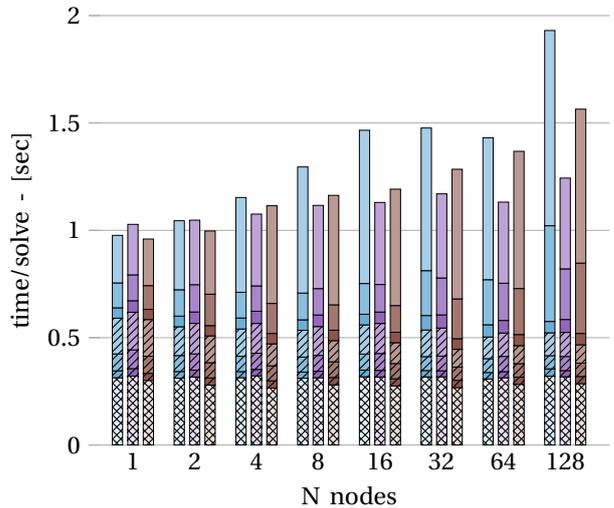
\begin{figure}[h!tp]
	\centering
	\input{figures/python_fig/weak-cluster-comp-nb}
	\caption{Implementation \nb: time-to-solution on Vega ({\protect\BarLegend{colvega}{colvega}}), MeluXina ({\protect\BarLegend{colmlx}{colmlx}}), and Lumi ({\protect\BarLegend{collumi}{collumi}}). Weak scalability tests performed with $96^{3}$ unknowns per rank on a fully unbounded test case.}
	\label{fig:benchmark-nb}
\end{figure}

\begin{figure}[h!tp]
	\centering		
	\begin{tikzpicture}
		\node[anchor=south west] (image1) at (0,0) {
			\input{figures/python_fig/weak-cluster-eff-nb}
                 };
                 \useasboundingbox;                     
		\node[anchor=south west,fill=white] at (7.6,6) (note) {$\beta=0\%$};
		\node[anchor=south west,fill=white] at (7.6,4.7) (note) {$\beta=0.2\%$};
		\node[anchor=south west,fill=white] at (7.6,3.5) (note) {$\beta=0.5\%$};
		\node[anchor=south west,fill=white] at (7.6,2.5) (note) {$\beta=1\%$};
	\end{tikzpicture}
	\caption{Implementation \nb: Weak efficiency $\eta_{P,w}$ on Vega ({\protect\ThickLineCircle{colvega}{colvega}}), MeluXina ({\protect\ThickLineCircle{colmlx}{colmlx}}), and Lumi ({\protect\ThickLineCircle{collumi}{collumi}}). Weak scalability tests performed with $96^{3}$ unknowns per rank on a fully unbounded test case.}
	\label{fig:benchmark-nb-weak-eff}
\end{figure}
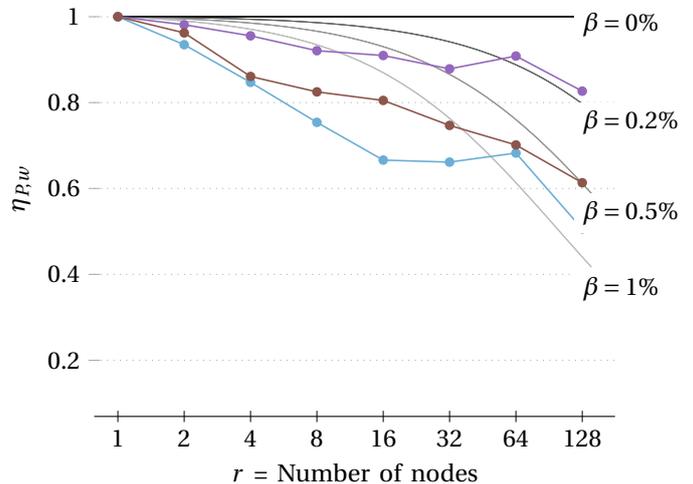

\Fref{fig:benchmark-nb} and \Fref{fig:benchmark-nb-weak-eff} show the results of the weak scalability tests for the \nb approach. For the interested reader, the throughput per rank as well as the results for the \isr and \ata versions are presented in \Sect{app:eu-comp}.
On a few nodes, the time-to-solutions are almost similar for all the architectures. However, the timings diverge with the increasing number of nodes. 
The Vega timings increase steeply than for Lumi and MeluXina. Lumi \revthree{results} are close to Meluxina ones, and the latter shows the best weak efficiency with a time increase of only $20\%$ when multiplying the resources by 128. 
As discussed in \Sect{sec:weak-scaling}, a weak scalability test raises the number of resources together with the number of unknowns hence increasing the number of communications and the network congestion. The Vega result is therefore explained by a lower bandwidth which saturates with the growing resources. On the other hand, we attribute the smaller times of MeluXina compared to Lumi to the hardware differences, the first being equipped with IB interconnect while the second uses slingshot technology.
As in \Sect{sec:weak-scaling}, we note that the computation time remains constant regardless of the number of nodes. Lumi spends slightly less time on computations while MeluXina and Vega have precisely the same computations timings. We attribute those differences to the CPUs properties as Vega and MeluXina have identical CPUs whereas Lumi has a slightly higher clock speed.

%% file: figures/erratum_biot_savart/valid_tube_spectral.tex
\begin{tikzpicture}

\definecolor{darkgray}{RGB}{169,169,169}
\definecolor{darkgray176}{RGB}{176,176,176}
\definecolor{lightgreen161217155}{RGB}{161,217,155}
\definecolor{lightsteelblue188189220}{RGB}{188,189,220}
\definecolor{orangered2308513}{RGB}{230,85,13}
\definecolor{sandybrown253174107}{RGB}{253,174,107}
\definecolor{seagreen4916384}{RGB}{49,163,84}
\definecolor{slateblue117107177}{RGB}{117,107,177}
\definecolor{steelblue49130189}{RGB}{49,130,189}

\begin{axis}[
axis y line =left, axis x line =bottom, axis line style ={-},
log basis x={10},
log basis y={10},
minor xtick={},
minor ytick={},
tick align=center,
tick pos=left,
x grid style={darkgray176},
x grid style={draw=none},
xlabel near ticks,
xlabel={N\(\displaystyle _{points}\)},
xmajorgrids,
xmin=26.8449462631643, xmax=2518.79811332614,
xmode=log,
xtick style={color=black},
xtick={32,64,128,256,512,1024,2048},
xticklabels={$32^3$,$64^3$,$128^3$,$256^3$,$512^3$,$1024^3$,$2048^3$},
y axis line style={draw=none},
y grid style={darkgray176},
y grid style={dotted},
ylabel near ticks,
ylabel={\(\displaystyle E_{\infty}\)},
ymajorgrids,
ymin=5e-18, ymax=10,
ymode=log,
ytick style={color=darkgray},
ytick={1e-21,1e-18,1e-15,1e-12,1e-09,1e-06,0.001,1,1000,1000000}
]
\addplot [semithick, steelblue49130189, mark=*, mark size=1.5, mark options={solid}]
table {%
33 0.02560729532124
65 0.007229962469887
129 0.001849313121153
257 0.0004667156874738
513 0.0001168337466244
1025 2.922471852718e-05
2049 7.306792328965e-06
};
\addplot [semithick, orangered2308513, mark=*, mark size=1.5, mark options={solid}]
table {%
33 0.7114351040324
65 0.2874551623362
129 0.08745968417294
257 0.02314215004994
513 0.005863870673902
1025 0.001471256136321
2049 0.0003681304775889
};
\addplot [semithick, black, dashed]
table {%
128 0.0572161341201101
128 0.0572161341201101
256 0.0143040335300275
512 0.00357600838250688
1024 0.00089400209562672
2048 0.00022350052390668
2048 0.00022350052390668
};
\addplot [semithick, sandybrown253174107, mark=*, mark size=1.5, mark options={solid}]
table {%
33 0.3624430263832
65 0.06772511897716
129 0.006326010549039
257 0.0004444731602955
513 2.863490331184e-05
1025 1.803629254393e-06
2049 1.129540427858e-07
};
\addplot [semithick, black, dashed]
table {%
128 0.00449865298045568
128 0.00449865298045568
256 0.00028116581127848
512 1.7572863204905e-05
1024 1.09830395030656e-06
2048 6.86439968941601e-08
2048 6.86439968941601e-08
};
\addplot [semithick, seagreen4916384, mark=*, mark size=1.5, mark options={solid}]
table {%
33 0.182082010645
65 0.01606201925554
129 0.000480598903275
257 9.114547344424e-06
513 1.49398029281e-07
1025 2.363337525679e-09
2049 3.704114792669e-11
};
\addplot [semithick, black, dashed]
table {%
128 0.00037803249169496
128 0.00037803249169496
256 5.90675768273374e-06
512 9.22930887927147e-08
1024 1.44207951238617e-09
2048 2.25324923810339e-11
2048 2.25324923810339e-11
};
\addplot [semithick, lightgreen161217155, mark=*, mark size=1.5, mark options={solid}]
table {%
33 0.1064721771105
65 0.004174996648993
129 3.867980647354e-05
257 1.935880293846e-07
513 8.072001955739e-10
1025 3.207434318142e-12
2049 1.276756478319e-14
};
\addplot [semithick, black, dashed]
table {%
128 2.49674952987319e-05
128 2.49674952987319e-05
256 9.75292785106714e-08
512 3.8097374418231e-10
1024 1.48817868821215e-12
2048 5.81319800082871e-15
2048 5.81319800082871e-15
};
\addplot [semithick, slateblue117107177, mark=*, mark size=1.5, mark options={solid}]
table {%
33 0.06229027016928
65 0.001082705360188
129 3.10069366638e-06
257 4.206367365533e-09
513 4.462319402876e-12
1025 4.843347944927e-15
2049 8.801986917106e-15
};
\addplot [semithick, black, dashed]
table {%
128 2.03286728924306e-06
128 2.03286728924306e-06
256 1.98522196215143e-09
512 1.9386933224135e-12
1024 1.89325519766944e-15
1024 1.89325519766944e-15
};
\addplot [semithick, lightsteelblue188189220, mark=*, mark size=1.5, mark options={solid}]
table {%
33 0.0005857467416752
65 0.0001800059163661
129 8.65464308748e-05
257 4.921072503716e-05
513 1.6426012583e-05
1025 4.41484711966e-06
2049 1.12377782302e-06
};
\draw (axis cs:1024,0.00089400209562672) ++(2pt,-2pt) node[
  scale=1.04166666666667,
  anchor=north west,
  text=black,
  rotate=350.9
]{$p = 2$};
\draw (axis cs:1024,1.09830395030656e-06) ++(2pt,-2pt) node[
  scale=1.04166666666667,
  anchor=north west,
  text=black,
  rotate=342.2
]{$p = 4$};
\draw (axis cs:1024,1.44207951238617e-09) ++(2pt,-2pt) node[
  scale=1.04166666666667,
  anchor=north west,
  text=black,
  rotate=334.3
]{$p = 6$};
\draw (axis cs:1024,1.48817868821215e-12) ++(2pt,-2pt) node[
  scale=1.04166666666667,
  anchor=north west,
  text=black,
  rotate=327.3
]{$p = 8$};
\draw (axis cs:512,1.9386933224135e-12) ++(2pt,-2pt) node[
  scale=1.04166666666667,
  anchor=north west,
  text=black,
  rotate=321.2
]{$p = 10$};
\end{axis}

\end{tikzpicture}

%% file: figures/erratum_biot_savart/valid_tube_order_6.tex
\begin{tikzpicture}

\definecolor{darkgray}{RGB}{169,169,169}
\definecolor{darkgray176}{RGB}{176,176,176}
\definecolor{lightgreen161217155}{RGB}{161,217,155}
\definecolor{lightsteelblue188189220}{RGB}{188,189,220}
\definecolor{orangered2308513}{RGB}{230,85,13}
\definecolor{sandybrown253174107}{RGB}{253,174,107}
\definecolor{seagreen4916384}{RGB}{49,163,84}
\definecolor{slateblue117107177}{RGB}{117,107,177}
\definecolor{steelblue49130189}{RGB}{49,130,189}

\begin{axis}[
axis y line =left, axis x line =bottom, axis line style ={-},
log basis x={10},
log basis y={10},
minor xtick={},
minor ytick={},
tick align=center,
tick pos=left,
x grid style={darkgray176},
x grid style={draw=none},
xlabel near ticks,
xlabel={N\(\displaystyle _{points}\)},
xmajorgrids,
xmin=26.8449462631643, xmax=2518.79811332614,
xmode=log,
xtick style={color=black},
xtick={32,64,128,256,512,1024,2048},
xticklabels={$32^3$,$64^3$,$128^3$,$256^3$,$512^3$,$1024^3$,$2048^3$},
y axis line style={draw=none},
y grid style={darkgray176},
y grid style={dotted},
ylabel near ticks,
ylabel={\(\displaystyle E_{\infty}\)},
ymajorgrids,
ymin=5e-15, ymax=10,
ymode=log,
ytick style={color=darkgray},
ytick={1e-17,1e-15,1e-13,1e-11,1e-09,1e-07,1e-05,0.001,0.1,10,1000}
]
\addplot [semithick, steelblue49130189, mark=*, mark size=1.5, mark options={solid}]
table {%
33 0.02683127172044
65 0.007297421588565
129 0.001850446481994
257 0.0004667303350199
513 0.0001168339762869
1025 2.922472233591e-05
2049 7.306792388473e-06
};
\addplot [semithick, orangered2308513, mark=*, mark size=1.5, mark options={solid}]
table {%
33 0.7116303066968
65 0.2874829334485
129 0.08746029578123
257 0.02314216403258
513 0.005863870900943
1025 0.001471256140118
2049 0.0003681304776486
};
\addplot [semithick, black, dashed]
table {%
128 0.0572161341293889
128 0.0572161341293889
256 0.0143040335323472
512 0.00357600838308681
1024 0.000894002095771702
2048 0.000223500523942925
2048 0.000223500523942925
};
\addplot [semithick, sandybrown253174107, mark=*, mark size=1.5, mark options={solid}]
table {%
33 0.3629938860375
65 0.06777625130409
129 0.006327414204496
257 0.000444496404149
513 2.863526931929e-05
1025 1.803634853914e-06
2049 1.129541311595e-07
};
\addplot [semithick, black, dashed]
table {%
128 0.00449865650014031
128 0.00449865650014031
256 0.000281166031258769
512 1.75728769536731e-05
1024 1.09830480960457e-06
2048 6.86440506002854e-08
2048 6.86440506002854e-08
};
\addplot [semithick, seagreen4916384, mark=*, mark size=1.5, mark options={solid}]
table {%
33 0.1830243110501
65 0.01614041683023
129 0.0004821870475998
257 9.139419847037e-06
513 1.497959519758e-07
1025 2.369513585343e-09
2049 3.713807039674e-11
};
\addplot [semithick, black, dashed]
table {%
128 0.000379021657660517
128 0.000379021657660517
256 5.92221340094557e-06
512 9.25345843897746e-08
1024 1.44585288109023e-09
2048 2.25914512670348e-11
2048 2.25914512670348e-11
};
\addplot [semithick, lightgreen161217155, mark=*, mark size=1.5, mark options={solid}]
table {%
33 0.1081642340915
65 0.004262595006432
129 4.027858359712e-05
257 2.191448287281e-07
513 1.208547706355e-09
1025 9.495959574224e-12
2049 1.11133324765e-13
};
\addplot [semithick, slateblue117107177, mark=*, mark size=1.5, mark options={solid}]
table {%
33 0.06455378686296
65 0.001174269878135
129 4.700877559438e-06
257 2.967835688317e-08
513 4.058763325432e-10
1025 6.292744103575e-12
2049 9.825473767933e-14
};
\addplot [semithick, black, dashed]
table {%
256 1.71725854716934e-08
256 1.71725854716934e-08
512 2.68321647995209e-10
1024 4.19252574992514e-12
2048 6.55082148425804e-14
2048 6.55082148425804e-14
};
\addplot [semithick, lightsteelblue188189220, mark=*, mark size=1.5, mark options={solid}]
table {%
33 0.004335250143036
65 0.0001752386515468
129 8.722638207903e-05
257 4.920351231408e-05
513 1.642578285943e-05
1025 4.414843527867e-06
2049 1.123777765288e-06
};
\draw (axis cs:1024,0.000894002095771702) ++(2pt,-2pt) node[
  scale=1.04166666666667,
  anchor=north west,
  text=black,
  rotate=349.1
]{$p = 2$};
\draw (axis cs:1024,1.09830480960457e-06) ++(2pt,-2pt) node[
  scale=1.04166666666667,
  anchor=north west,
  text=black,
  rotate=339.0
]{$p = 4$};
\draw (axis cs:1024,1.44585288109023e-09) ++(2pt,-2pt) node[
  scale=1.04166666666667,
  anchor=north west,
  text=black,
  rotate=330.0
]{$p = 6$};
\draw (axis cs:1024,4.19252574992514e-12) ++(2pt,-2pt) node[
  scale=1.04166666666667,
  anchor=north west,
  text=black,
  rotate=330.0
]{$p = 6$};
\end{axis}

\end{tikzpicture}

%% file: figures/python_fig/mlx/accfft.tex
\begin{tikzpicture}

\definecolor{burlywood235199116}{RGB}{235,199,116}
\definecolor{darkgray}{RGB}{169,169,169}
\definecolor{darkgray176}{RGB}{176,176,176}
\definecolor{darkseagreen161174136}{RGB}{161,174,136}
\definecolor{gray13014796}{RGB}{130,147,96}
\definecolor{khaki240213151}{RGB}{240,213,151}
\definecolor{peru22412652}{RGB}{224,126,52}
\definecolor{sandybrown232159103}{RGB}{232,159,103}

\begin{axis}[
axis line style={draw=none},
tick pos=left,
x grid style={darkgray176},
x grid style={draw=none},
xlabel near ticks,
xlabel={Number of nodes},
xmajorgrids,
xmin=-0.566666666666667, xmax=4.56666666666667,
xtick style={color=darkgray},
xtick={0,1,2,3,4},
xticklabels={$1$,$2$,$8$,$32$,$128$},
xticklabels={1,2,8,32,128},
y grid style={darkgray176},
ylabel near ticks,
ylabel={time/transform - [sec]},
ymajorgrids,
ymin=0, ymax=8,
ytick style={color=darkgray}
]
\draw[draw=black,fill=khaki240213151,line width=0.004pt,postaction={pattern=north east lines}] (axis cs:-0.333333333333333,0) rectangle (axis cs:-0.166666666666667,1.7596862495);
\draw[draw=black,fill=burlywood235199116,line width=0.004pt] (axis cs:-0.333333333333333,1.7596862495) rectangle (axis cs:-0.166666666666667,3.704094056);
\draw[draw=black,fill=darkseagreen161174136,line width=0.004pt,postaction={pattern=north east lines}] (axis cs:-0.0833333333333333,0) rectangle (axis cs:0.0833333333333333,1.8771377075);
\draw[draw=black,fill=gray13014796,line width=0.004pt] (axis cs:-0.0833333333333333,1.8771377075) rectangle (axis cs:0.0833333333333333,3.715047617);
\draw[draw=black,fill=sandybrown232159103,line width=0.004pt,postaction={pattern=north east lines}] (axis cs:0.166666666666667,0) rectangle (axis cs:0.333333333333333,2.405952024);
\draw[draw=black,fill=peru22412652,line width=0.004pt] (axis cs:0.166666666666667,2.405952024) rectangle (axis cs:0.333333333333333,4.8282417095);
\draw[draw=black,fill=khaki240213151,line width=0.004pt,postaction={pattern=north east lines}] (axis cs:0.666666666666667,0) rectangle (axis cs:0.833333333333333,1.912995006);
\draw[draw=black,fill=burlywood235199116,line width=0.004pt] (axis cs:0.666666666666667,1.912995006) rectangle (axis cs:0.833333333333333,4.0547422025);
\draw[draw=black,fill=darkseagreen161174136,line width=0.004pt,postaction={pattern=north east lines}] (axis cs:0.916666666666667,0) rectangle (axis cs:1.08333333333333,1.949456017);
\draw[draw=black,fill=gray13014796,line width=0.004pt] (axis cs:0.916666666666667,1.949456017) rectangle (axis cs:1.08333333333333,3.9448116055);
\draw[draw=black,fill=sandybrown232159103,line width=0.004pt,postaction={pattern=north east lines}] (axis cs:1.16666666666667,0) rectangle (axis cs:1.33333333333333,2.46205997);
\draw[draw=black,fill=peru22412652,line width=0.004pt] (axis cs:1.16666666666667,2.46205997) rectangle (axis cs:1.33333333333333,4.9539379595);
\draw[draw=black,fill=khaki240213151,line width=0.004pt,postaction={pattern=north east lines}] (axis cs:1.66666666666667,0) rectangle (axis cs:1.83333333333333,1.99216905);
\draw[draw=black,fill=burlywood235199116,line width=0.004pt] (axis cs:1.66666666666667,1.99216905) rectangle (axis cs:1.83333333333333,4.3901135135);
\draw[draw=black,fill=darkseagreen161174136,line width=0.004pt,postaction={pattern=north east lines}] (axis cs:1.91666666666667,0) rectangle (axis cs:2.08333333333333,2.2639710375);
\draw[draw=black,fill=gray13014796,line width=0.004pt] (axis cs:1.91666666666667,2.2639710375) rectangle (axis cs:2.08333333333333,4.66012);
\draw[draw=black,fill=sandybrown232159103,line width=0.004pt,postaction={pattern=north east lines}] (axis cs:2.16666666666667,0) rectangle (axis cs:2.33333333333333,2.5494606505);
\draw[draw=black,fill=peru22412652,line width=0.004pt] (axis cs:2.16666666666667,2.5494606505) rectangle (axis cs:2.33333333333333,5.165354895);
\draw[draw=black,fill=khaki240213151,line width=0.004pt,postaction={pattern=north east lines}] (axis cs:2.66666666666667,0) rectangle (axis cs:2.83333333333333,2.284220073);
\draw[draw=black,fill=burlywood235199116,line width=0.004pt] (axis cs:2.66666666666667,2.284220073) rectangle (axis cs:2.83333333333333,5.316124533);
\draw[draw=black,fill=darkseagreen161174136,line width=0.004pt,postaction={pattern=north east lines}] (axis cs:2.91666666666667,0) rectangle (axis cs:3.08333333333333,2.830092265);
\draw[draw=black,fill=gray13014796,line width=0.004pt] (axis cs:2.91666666666667,2.830092265) rectangle (axis cs:3.08333333333333,5.749824747);
\draw[draw=black,fill=sandybrown232159103,line width=0.004pt,postaction={pattern=north east lines}] (axis cs:3.16666666666667,0) rectangle (axis cs:3.33333333333333,2.512451938);
\draw[draw=black,fill=peru22412652,line width=0.004pt] (axis cs:3.16666666666667,2.512451938) rectangle (axis cs:3.33333333333333,5.095220028);
\draw[draw=black,fill=khaki240213151,line width=0.004pt,postaction={pattern=north east lines}] (axis cs:3.66666666666667,0) rectangle (axis cs:3.83333333333333,2.6686458905);
\draw[draw=black,fill=burlywood235199116,line width=0.004pt] (axis cs:3.66666666666667,2.6686458905) rectangle (axis cs:3.83333333333333,7.2507786325);
\draw[draw=black,fill=darkseagreen161174136,line width=0.004pt,postaction={pattern=north east lines}] (axis cs:3.91666666666667,0) rectangle (axis cs:4.08333333333333,3.672276844);
\draw[draw=black,fill=gray13014796,line width=0.004pt] (axis cs:3.91666666666667,3.672276844) rectangle (axis cs:4.08333333333333,7.208488785);
\draw[draw=black,fill=sandybrown232159103,line width=0.004pt,postaction={pattern=north east lines}] (axis cs:4.16666666666667,0) rectangle (axis cs:4.33333333333333,2.4015123305);
\draw[draw=black,fill=peru22412652,line width=0.004pt] (axis cs:4.16666666666667,2.4015123305) rectangle (axis cs:4.33333333333333,5.2393198445);
\end{axis}

\end{tikzpicture}

%% file: figures/python_tab/mlx/comp_acc_time_unknowns_rank.tex
\begin{tabular}{lrrrrr}
\hline
  N nodes      &   \centercell{$1$ } &   \centercell{$2$ } &   \centercell{$8$ } &   \centercell{$32$ } &   \centercell{$128$ } \\
\hline
 \code{accFFT} &               36.23 &               33.10 &               30.57 &                25.25 &                 18.51 \\
 \isr          &               36.13 &               34.02 &               28.80 &                23.34 &                 18.62 \\
 \nb           &               27.80 &               27.09 &               25.98 &                26.34 &                 25.62 \\
\hline
\end{tabular}

%% file: figures/python_fig/mlx/unbounded/weak-comp.tex
\begin{tikzpicture}

\definecolor{bisque247223204}{RGB}{247,223,204}
\definecolor{chocolate217952}{RGB}{217,95,2}
\definecolor{cornflowerblue120173210}{RGB}{120,173,210}
\definecolor{darkgray}{RGB}{169,169,169}
\definecolor{darkgray176}{RGB}{176,176,176}
\definecolor{darkolivegreen9912157}{RGB}{99,121,57}
\definecolor{darkseagreen161174136}{RGB}{161,174,136}
\definecolor{gainsboro223228215}{RGB}{223,228,215}
\definecolor{gray13014796}{RGB}{130,147,96}
\definecolor{lavender210227240}{RGB}{210,227,240}
\definecolor{peru22412652}{RGB}{224,126,52}
\definecolor{sandybrown232159103}{RGB}{232,159,103}
\definecolor{steelblue31119180}{RGB}{31,119,180}
\definecolor{steelblue75146195}{RGB}{75,146,195}

\begin{axis}[
axis line style={draw=none},
tick pos=left,
x grid style={darkgray176},
x grid style={draw=none},
xlabel near ticks,
xlabel={Number of nodes},
xmajorgrids,
xmin=-0.816666666666667, xmax=9.81666666666667,
xtick style={color=darkgray},
xtick={0,1,2,3,4,5,6,7,8,9},
xticklabels={$1$,$2$,$4$,$8$,$16$,$32$,$64$,$128$,$256$,$384$},
xticklabels={1,2,4,8,16,32,64,128,256,384},
y grid style={darkgray176},
ylabel near ticks,
ylabel={time/solve - [sec]},
ymajorgrids,
ymin=0, ymax=3,
ytick style={color=darkgray}
]
\draw[draw=black,fill=lavender210227240,line width=0.004pt,postaction={pattern=crosshatch}] (axis cs:-0.333333333333333,0) rectangle (axis cs:-0.166666666666667,0.318895546);
\draw[draw=black,fill=steelblue31119180,line width=0.004pt,postaction={pattern=north east lines}] (axis cs:-0.333333333333333,0.318895546) rectangle (axis cs:-0.166666666666667,0.3745067295);
\draw[draw=black,fill=steelblue75146195,line width=0.004pt,postaction={pattern=north east lines}] (axis cs:-0.333333333333333,0.3745067295) rectangle (axis cs:-0.166666666666667,0.5043525595);
\draw[draw=black,fill=cornflowerblue120173210,line width=0.004pt,postaction={pattern=north east lines}] (axis cs:-0.333333333333333,0.5043525595) rectangle (axis cs:-0.166666666666667,0.778857539);
\draw[draw=black,fill=steelblue31119180,line width=0.004pt] (axis cs:-0.333333333333333,0.778857539) rectangle (axis cs:-0.166666666666667,0.820389006);
\draw[draw=black,fill=steelblue75146195,line width=0.004pt] (axis cs:-0.333333333333333,0.820389006) rectangle (axis cs:-0.166666666666667,0.91084508);
\draw[draw=black,fill=cornflowerblue120173210,line width=0.004pt] (axis cs:-0.333333333333333,0.91084508) rectangle (axis cs:-0.166666666666667,1.088675068);
\draw[draw=black,fill=lavender210227240,line width=0.004pt,postaction={pattern=crosshatch}] (axis cs:0.666666666666667,0) rectangle (axis cs:0.833333333333333,0.3186631645);
\draw[draw=black,fill=steelblue31119180,line width=0.004pt,postaction={pattern=north east lines}] (axis cs:0.666666666666667,0.3186631645) rectangle (axis cs:0.833333333333333,0.368269101);
\draw[draw=black,fill=steelblue75146195,line width=0.004pt,postaction={pattern=north east lines}] (axis cs:0.666666666666667,0.368269101) rectangle (axis cs:0.833333333333333,0.481165614);
\draw[draw=black,fill=cornflowerblue120173210,line width=0.004pt,postaction={pattern=north east lines}] (axis cs:0.666666666666667,0.481165614) rectangle (axis cs:0.833333333333333,0.744513847);
\draw[draw=black,fill=steelblue31119180,line width=0.004pt] (axis cs:0.666666666666667,0.744513847) rectangle (axis cs:0.833333333333333,0.7758217555);
\draw[draw=black,fill=steelblue75146195,line width=0.004pt] (axis cs:0.666666666666667,0.7758217555) rectangle (axis cs:0.833333333333333,0.867516589);
\draw[draw=black,fill=cornflowerblue120173210,line width=0.004pt] (axis cs:0.666666666666667,0.867516589) rectangle (axis cs:0.833333333333333,1.1063788225);
\draw[draw=black,fill=lavender210227240,line width=0.004pt,postaction={pattern=crosshatch}] (axis cs:1.66666666666667,0) rectangle (axis cs:1.83333333333333,0.3203934215);
\draw[draw=black,fill=steelblue31119180,line width=0.004pt,postaction={pattern=north east lines}] (axis cs:1.66666666666667,0.3203934215) rectangle (axis cs:1.83333333333333,0.374975716);
\draw[draw=black,fill=steelblue75146195,line width=0.004pt,postaction={pattern=north east lines}] (axis cs:1.66666666666667,0.374975716) rectangle (axis cs:1.83333333333333,0.496181119);
\draw[draw=black,fill=cornflowerblue120173210,line width=0.004pt,postaction={pattern=north east lines}] (axis cs:1.66666666666667,0.496181119) rectangle (axis cs:1.83333333333333,0.766342383);
\draw[draw=black,fill=steelblue31119180,line width=0.004pt] (axis cs:1.66666666666667,0.766342383) rectangle (axis cs:1.83333333333333,0.809961216);
\draw[draw=black,fill=steelblue75146195,line width=0.004pt] (axis cs:1.66666666666667,0.809961216) rectangle (axis cs:1.83333333333333,0.9051660255);
\draw[draw=black,fill=cornflowerblue120173210,line width=0.004pt] (axis cs:1.66666666666667,0.9051660255) rectangle (axis cs:1.83333333333333,1.188427556);
\draw[draw=black,fill=lavender210227240,line width=0.004pt,postaction={pattern=crosshatch}] (axis cs:2.66666666666667,0) rectangle (axis cs:2.83333333333333,0.3216185865);
\draw[draw=black,fill=steelblue31119180,line width=0.004pt,postaction={pattern=north east lines}] (axis cs:2.66666666666667,0.3216185865) rectangle (axis cs:2.83333333333333,0.376055395);
\draw[draw=black,fill=steelblue75146195,line width=0.004pt,postaction={pattern=north east lines}] (axis cs:2.66666666666667,0.376055395) rectangle (axis cs:2.83333333333333,0.4966798365);
\draw[draw=black,fill=cornflowerblue120173210,line width=0.004pt,postaction={pattern=north east lines}] (axis cs:2.66666666666667,0.4966798365) rectangle (axis cs:2.83333333333333,0.743724191);
\draw[draw=black,fill=steelblue31119180,line width=0.004pt] (axis cs:2.66666666666667,0.743724191) rectangle (axis cs:2.83333333333333,0.786988862);
\draw[draw=black,fill=steelblue75146195,line width=0.004pt] (axis cs:2.66666666666667,0.786988862) rectangle (axis cs:2.83333333333333,0.882010453);
\draw[draw=black,fill=cornflowerblue120173210,line width=0.004pt] (axis cs:2.66666666666667,0.882010453) rectangle (axis cs:2.83333333333333,1.214015122);
\draw[draw=black,fill=lavender210227240,line width=0.004pt,postaction={pattern=crosshatch}] (axis cs:3.66666666666667,0) rectangle (axis cs:3.83333333333333,0.3223934345);
\draw[draw=black,fill=steelblue31119180,line width=0.004pt,postaction={pattern=north east lines}] (axis cs:3.66666666666667,0.3223934345) rectangle (axis cs:3.83333333333333,0.377215677);
\draw[draw=black,fill=steelblue75146195,line width=0.004pt,postaction={pattern=north east lines}] (axis cs:3.66666666666667,0.377215677) rectangle (axis cs:3.83333333333333,0.4969262105);
\draw[draw=black,fill=cornflowerblue120173210,line width=0.004pt,postaction={pattern=north east lines}] (axis cs:3.66666666666667,0.4969262105) rectangle (axis cs:3.83333333333333,0.7407181345);
\draw[draw=black,fill=steelblue31119180,line width=0.004pt] (axis cs:3.66666666666667,0.7407181345) rectangle (axis cs:3.83333333333333,0.7843835285);
\draw[draw=black,fill=steelblue75146195,line width=0.004pt] (axis cs:3.66666666666667,0.7843835285) rectangle (axis cs:3.83333333333333,0.887952079);
\draw[draw=black,fill=cornflowerblue120173210,line width=0.004pt] (axis cs:3.66666666666667,0.887952079) rectangle (axis cs:3.83333333333333,1.224180271);
\draw[draw=black,fill=lavender210227240,line width=0.004pt,postaction={pattern=crosshatch}] (axis cs:4.66666666666667,0) rectangle (axis cs:4.83333333333333,0.3226837415);
\draw[draw=black,fill=steelblue31119180,line width=0.004pt,postaction={pattern=north east lines}] (axis cs:4.66666666666667,0.3226837415) rectangle (axis cs:4.83333333333333,0.376428799);
\draw[draw=black,fill=steelblue75146195,line width=0.004pt,postaction={pattern=north east lines}] (axis cs:4.66666666666667,0.376428799) rectangle (axis cs:4.83333333333333,0.4947181175);
\draw[draw=black,fill=cornflowerblue120173210,line width=0.004pt,postaction={pattern=north east lines}] (axis cs:4.66666666666667,0.4947181175) rectangle (axis cs:4.83333333333333,0.7367447625);
\draw[draw=black,fill=steelblue31119180,line width=0.004pt] (axis cs:4.66666666666667,0.7367447625) rectangle (axis cs:4.83333333333333,0.7826634395);
\draw[draw=black,fill=steelblue75146195,line width=0.004pt] (axis cs:4.66666666666667,0.7826634395) rectangle (axis cs:4.83333333333333,0.9322976675);
\draw[draw=black,fill=cornflowerblue120173210,line width=0.004pt] (axis cs:4.66666666666667,0.9322976675) rectangle (axis cs:4.83333333333333,1.267331512);
\draw[draw=black,fill=lavender210227240,line width=0.004pt,postaction={pattern=crosshatch}] (axis cs:5.66666666666667,0) rectangle (axis cs:5.83333333333333,0.325898754);
\draw[draw=black,fill=steelblue31119180,line width=0.004pt,postaction={pattern=north east lines}] (axis cs:5.66666666666667,0.325898754) rectangle (axis cs:5.83333333333333,0.379624845);
\draw[draw=black,fill=steelblue75146195,line width=0.004pt,postaction={pattern=north east lines}] (axis cs:5.66666666666667,0.379624845) rectangle (axis cs:5.83333333333333,0.495069975);
\draw[draw=black,fill=cornflowerblue120173210,line width=0.004pt,postaction={pattern=north east lines}] (axis cs:5.66666666666667,0.495069975) rectangle (axis cs:5.83333333333333,0.7177937175);
\draw[draw=black,fill=steelblue31119180,line width=0.004pt] (axis cs:5.66666666666667,0.7177937175) rectangle (axis cs:5.83333333333333,0.763620278);
\draw[draw=black,fill=steelblue75146195,line width=0.004pt] (axis cs:5.66666666666667,0.763620278) rectangle (axis cs:5.83333333333333,0.920273322);
\draw[draw=black,fill=cornflowerblue120173210,line width=0.004pt] (axis cs:5.66666666666667,0.920273322) rectangle (axis cs:5.83333333333333,1.273096441);
\draw[draw=black,fill=lavender210227240,line width=0.004pt,postaction={pattern=crosshatch}] (axis cs:6.66666666666667,0) rectangle (axis cs:6.83333333333333,0.3272236035);
\draw[draw=black,fill=steelblue31119180,line width=0.004pt,postaction={pattern=north east lines}] (axis cs:6.66666666666667,0.3272236035) rectangle (axis cs:6.83333333333333,0.3811352675);
\draw[draw=black,fill=steelblue75146195,line width=0.004pt,postaction={pattern=north east lines}] (axis cs:6.66666666666667,0.3811352675) rectangle (axis cs:6.83333333333333,0.4948544945);
\draw[draw=black,fill=cornflowerblue120173210,line width=0.004pt,postaction={pattern=north east lines}] (axis cs:6.66666666666667,0.4948544945) rectangle (axis cs:6.83333333333333,0.711501018);
\draw[draw=black,fill=steelblue31119180,line width=0.004pt] (axis cs:6.66666666666667,0.711501018) rectangle (axis cs:6.83333333333333,0.7575044165);
\draw[draw=black,fill=steelblue75146195,line width=0.004pt] (axis cs:6.66666666666667,0.7575044165) rectangle (axis cs:6.83333333333333,0.9607041545);
\draw[draw=black,fill=cornflowerblue120173210,line width=0.004pt] (axis cs:6.66666666666667,0.9607041545) rectangle (axis cs:6.83333333333333,1.357799428);
\draw[draw=black,fill=lavender210227240,line width=0.004pt,postaction={pattern=crosshatch}] (axis cs:7.66666666666667,0) rectangle (axis cs:7.83333333333333,0.3231361715);
\draw[draw=black,fill=steelblue31119180,line width=0.004pt,postaction={pattern=north east lines}] (axis cs:7.66666666666667,0.3231361715) rectangle (axis cs:7.83333333333333,0.3722550105);
\draw[draw=black,fill=steelblue75146195,line width=0.004pt,postaction={pattern=north east lines}] (axis cs:7.66666666666667,0.3722550105) rectangle (axis cs:7.83333333333333,0.4838369435);
\draw[draw=black,fill=cornflowerblue120173210,line width=0.004pt,postaction={pattern=north east lines}] (axis cs:7.66666666666667,0.4838369435) rectangle (axis cs:7.83333333333333,0.701142781);
\draw[draw=black,fill=steelblue31119180,line width=0.004pt] (axis cs:7.66666666666667,0.701142781) rectangle (axis cs:7.83333333333333,0.7486098425);
\draw[draw=black,fill=steelblue75146195,line width=0.004pt] (axis cs:7.66666666666667,0.7486098425) rectangle (axis cs:7.83333333333333,1.0415767725);
\draw[draw=black,fill=cornflowerblue120173210,line width=0.004pt] (axis cs:7.66666666666667,1.0415767725) rectangle (axis cs:7.83333333333333,1.568197574);
\draw[draw=black,fill=lavender210227240,line width=0.004pt,postaction={pattern=crosshatch}] (axis cs:8.66666666666667,0) rectangle (axis cs:8.83333333333333,0.326201897);
\draw[draw=black,fill=steelblue31119180,line width=0.004pt,postaction={pattern=north east lines}] (axis cs:8.66666666666667,0.326201897) rectangle (axis cs:8.83333333333333,0.3752720355);
\draw[draw=black,fill=steelblue75146195,line width=0.004pt,postaction={pattern=north east lines}] (axis cs:8.66666666666667,0.3752720355) rectangle (axis cs:8.83333333333333,0.4890422665);
\draw[draw=black,fill=cornflowerblue120173210,line width=0.004pt,postaction={pattern=north east lines}] (axis cs:8.66666666666667,0.4890422665) rectangle (axis cs:8.83333333333333,0.711941412);
\draw[draw=black,fill=steelblue31119180,line width=0.004pt] (axis cs:8.66666666666667,0.711941412) rectangle (axis cs:8.83333333333333,0.75933272);
\draw[draw=black,fill=steelblue75146195,line width=0.004pt] (axis cs:8.66666666666667,0.75933272) rectangle (axis cs:8.83333333333333,1.0540549805);
\draw[draw=black,fill=cornflowerblue120173210,line width=0.004pt] (axis cs:8.66666666666667,1.0540549805) rectangle (axis cs:8.83333333333333,1.695710964);
\draw[draw=black,fill=bisque247223204,line width=0.004pt,postaction={pattern=crosshatch}] (axis cs:-0.0833333333333333,0) rectangle (axis cs:0.0833333333333333,0.3210666065);
\draw[draw=black,fill=chocolate217952,line width=0.004pt,postaction={pattern=north east lines}] (axis cs:-0.0833333333333333,0.3210666065) rectangle (axis cs:0.0833333333333333,0.3559028375);
\draw[draw=black,fill=peru22412652,line width=0.004pt,postaction={pattern=north east lines}] (axis cs:-0.0833333333333333,0.3559028375) rectangle (axis cs:0.0833333333333333,0.443289068);
\draw[draw=black,fill=sandybrown232159103,line width=0.004pt,postaction={pattern=north east lines}] (axis cs:-0.0833333333333333,0.443289068) rectangle (axis cs:0.0833333333333333,0.6186279805);
\draw[draw=black,fill=chocolate217952,line width=0.004pt] (axis cs:-0.0833333333333333,0.6186279805) rectangle (axis cs:0.0833333333333333,0.6719930955);
\draw[draw=black,fill=peru22412652,line width=0.004pt] (axis cs:-0.0833333333333333,0.6719930955) rectangle (axis cs:0.0833333333333333,0.792448798);
\draw[draw=black,fill=sandybrown232159103,line width=0.004pt] (axis cs:-0.0833333333333333,0.792448798) rectangle (axis cs:0.0833333333333333,1.0277880955);
\draw[draw=black,fill=bisque247223204,line width=0.004pt,postaction={pattern=crosshatch}] (axis cs:0.916666666666667,0) rectangle (axis cs:1.08333333333333,0.316878601);
\draw[draw=black,fill=chocolate217952,line width=0.004pt,postaction={pattern=north east lines}] (axis cs:0.916666666666667,0.316878601) rectangle (axis cs:1.08333333333333,0.34870027);
\draw[draw=black,fill=peru22412652,line width=0.004pt,postaction={pattern=north east lines}] (axis cs:0.916666666666667,0.34870027) rectangle (axis cs:1.08333333333333,0.4246208985);
\draw[draw=black,fill=sandybrown232159103,line width=0.004pt,postaction={pattern=north east lines}] (axis cs:0.916666666666667,0.4246208985) rectangle (axis cs:1.08333333333333,0.5665687855);
\draw[draw=black,fill=chocolate217952,line width=0.004pt] (axis cs:0.916666666666667,0.5665687855) rectangle (axis cs:1.08333333333333,0.619709677);
\draw[draw=black,fill=peru22412652,line width=0.004pt] (axis cs:0.916666666666667,0.619709677) rectangle (axis cs:1.08333333333333,0.746839338);
\draw[draw=black,fill=sandybrown232159103,line width=0.004pt] (axis cs:0.916666666666667,0.746839338) rectangle (axis cs:1.08333333333333,1.047342582);
\draw[draw=black,fill=bisque247223204,line width=0.004pt,postaction={pattern=crosshatch}] (axis cs:1.91666666666667,0) rectangle (axis cs:2.08333333333333,0.322396057);
\draw[draw=black,fill=chocolate217952,line width=0.004pt,postaction={pattern=north east lines}] (axis cs:1.91666666666667,0.322396057) rectangle (axis cs:2.08333333333333,0.35207707);
\draw[draw=black,fill=peru22412652,line width=0.004pt,postaction={pattern=north east lines}] (axis cs:1.91666666666667,0.35207707) rectangle (axis cs:2.08333333333333,0.4273433365);
\draw[draw=black,fill=sandybrown232159103,line width=0.004pt,postaction={pattern=north east lines}] (axis cs:1.91666666666667,0.4273433365) rectangle (axis cs:2.08333333333333,0.5662799185);
\draw[draw=black,fill=chocolate217952,line width=0.004pt] (axis cs:1.91666666666667,0.5662799185) rectangle (axis cs:2.08333333333333,0.62292775);
\draw[draw=black,fill=peru22412652,line width=0.004pt] (axis cs:1.91666666666667,0.62292775) rectangle (axis cs:2.08333333333333,0.7411521745);
\draw[draw=black,fill=sandybrown232159103,line width=0.004pt] (axis cs:1.91666666666667,0.7411521745) rectangle (axis cs:2.08333333333333,1.0756606445);
\draw[draw=black,fill=bisque247223204,line width=0.004pt,postaction={pattern=crosshatch}] (axis cs:2.91666666666667,0) rectangle (axis cs:3.08333333333333,0.3135838485);
\draw[draw=black,fill=chocolate217952,line width=0.004pt,postaction={pattern=north east lines}] (axis cs:2.91666666666667,0.3135838485) rectangle (axis cs:3.08333333333333,0.343568668);
\draw[draw=black,fill=peru22412652,line width=0.004pt,postaction={pattern=north east lines}] (axis cs:2.91666666666667,0.343568668) rectangle (axis cs:3.08333333333333,0.417248557);
\draw[draw=black,fill=sandybrown232159103,line width=0.004pt,postaction={pattern=north east lines}] (axis cs:2.91666666666667,0.417248557) rectangle (axis cs:3.08333333333333,0.551845153);
\draw[draw=black,fill=chocolate217952,line width=0.004pt] (axis cs:2.91666666666667,0.551845153) rectangle (axis cs:3.08333333333333,0.605658658);
\draw[draw=black,fill=peru22412652,line width=0.004pt] (axis cs:2.91666666666667,0.605658658) rectangle (axis cs:3.08333333333333,0.728477756);
\draw[draw=black,fill=sandybrown232159103,line width=0.004pt] (axis cs:2.91666666666667,0.728477756) rectangle (axis cs:3.08333333333333,1.116099435);
\draw[draw=black,fill=bisque247223204,line width=0.004pt,postaction={pattern=crosshatch}] (axis cs:3.91666666666667,0) rectangle (axis cs:4.08333333333333,0.3167856195);
\draw[draw=black,fill=chocolate217952,line width=0.004pt,postaction={pattern=north east lines}] (axis cs:3.91666666666667,0.3167856195) rectangle (axis cs:4.08333333333333,0.347585438);
\draw[draw=black,fill=peru22412652,line width=0.004pt,postaction={pattern=north east lines}] (axis cs:3.91666666666667,0.347585438) rectangle (axis cs:4.08333333333333,0.4262232585);
\draw[draw=black,fill=sandybrown232159103,line width=0.004pt,postaction={pattern=north east lines}] (axis cs:3.91666666666667,0.4262232585) rectangle (axis cs:4.08333333333333,0.566199851);
\draw[draw=black,fill=chocolate217952,line width=0.004pt] (axis cs:3.91666666666667,0.566199851) rectangle (axis cs:4.08333333333333,0.6194036225);
\draw[draw=black,fill=peru22412652,line width=0.004pt] (axis cs:3.91666666666667,0.6194036225) rectangle (axis cs:4.08333333333333,0.7476323125);
\draw[draw=black,fill=sandybrown232159103,line width=0.004pt] (axis cs:3.91666666666667,0.7476323125) rectangle (axis cs:4.08333333333333,1.129873575);
\draw[draw=black,fill=bisque247223204,line width=0.004pt,postaction={pattern=crosshatch}] (axis cs:4.91666666666667,0) rectangle (axis cs:5.08333333333333,0.317855948);
\draw[draw=black,fill=chocolate217952,line width=0.004pt,postaction={pattern=north east lines}] (axis cs:4.91666666666667,0.317855948) rectangle (axis cs:5.08333333333333,0.3459035055);
\draw[draw=black,fill=peru22412652,line width=0.004pt,postaction={pattern=north east lines}] (axis cs:4.91666666666667,0.3459035055) rectangle (axis cs:5.08333333333333,0.414425989);
\draw[draw=black,fill=sandybrown232159103,line width=0.004pt,postaction={pattern=north east lines}] (axis cs:4.91666666666667,0.414425989) rectangle (axis cs:5.08333333333333,0.5448022405);
\draw[draw=black,fill=chocolate217952,line width=0.004pt] (axis cs:4.91666666666667,0.5448022405) rectangle (axis cs:5.08333333333333,0.606055196);
\draw[draw=black,fill=peru22412652,line width=0.004pt] (axis cs:4.91666666666667,0.606055196) rectangle (axis cs:5.08333333333333,0.778055971);
\draw[draw=black,fill=sandybrown232159103,line width=0.004pt] (axis cs:4.91666666666667,0.778055971) rectangle (axis cs:5.08333333333333,1.1698969815);
\draw[draw=black,fill=bisque247223204,line width=0.004pt,postaction={pattern=crosshatch}] (axis cs:5.91666666666667,0) rectangle (axis cs:6.08333333333333,0.313023139);
\draw[draw=black,fill=chocolate217952,line width=0.004pt,postaction={pattern=north east lines}] (axis cs:5.91666666666667,0.313023139) rectangle (axis cs:6.08333333333333,0.341868088);
\draw[draw=black,fill=peru22412652,line width=0.004pt,postaction={pattern=north east lines}] (axis cs:5.91666666666667,0.341868088) rectangle (axis cs:6.08333333333333,0.412765279);
\draw[draw=black,fill=sandybrown232159103,line width=0.004pt,postaction={pattern=north east lines}] (axis cs:5.91666666666667,0.412765279) rectangle (axis cs:6.08333333333333,0.5216306885);
\draw[draw=black,fill=chocolate217952,line width=0.004pt] (axis cs:5.91666666666667,0.5216306885) rectangle (axis cs:6.08333333333333,0.5801603815);
\draw[draw=black,fill=peru22412652,line width=0.004pt] (axis cs:5.91666666666667,0.5801603815) rectangle (axis cs:6.08333333333333,0.7535133);
\draw[draw=black,fill=sandybrown232159103,line width=0.004pt] (axis cs:5.91666666666667,0.7535133) rectangle (axis cs:6.08333333333333,1.131439556);
\draw[draw=black,fill=bisque247223204,line width=0.004pt,postaction={pattern=crosshatch}] (axis cs:6.91666666666667,0) rectangle (axis cs:7.08333333333333,0.3179651735);
\draw[draw=black,fill=chocolate217952,line width=0.004pt,postaction={pattern=north east lines}] (axis cs:6.91666666666667,0.3179651735) rectangle (axis cs:7.08333333333333,0.346761818);
\draw[draw=black,fill=peru22412652,line width=0.004pt,postaction={pattern=north east lines}] (axis cs:6.91666666666667,0.346761818) rectangle (axis cs:7.08333333333333,0.412317423);
\draw[draw=black,fill=sandybrown232159103,line width=0.004pt,postaction={pattern=north east lines}] (axis cs:6.91666666666667,0.412317423) rectangle (axis cs:7.08333333333333,0.5245089645);
\draw[draw=black,fill=chocolate217952,line width=0.004pt] (axis cs:6.91666666666667,0.5245089645) rectangle (axis cs:7.08333333333333,0.584591277);
\draw[draw=black,fill=peru22412652,line width=0.004pt] (axis cs:6.91666666666667,0.584591277) rectangle (axis cs:7.08333333333333,0.820368144);
\draw[draw=black,fill=sandybrown232159103,line width=0.004pt] (axis cs:6.91666666666667,0.820368144) rectangle (axis cs:7.08333333333333,1.243305256);
\draw[draw=black,fill=bisque247223204,line width=0.004pt,postaction={pattern=crosshatch}] (axis cs:7.91666666666667,0) rectangle (axis cs:8.08333333333333,0.319245784);
\draw[draw=black,fill=chocolate217952,line width=0.004pt,postaction={pattern=north east lines}] (axis cs:7.91666666666667,0.319245784) rectangle (axis cs:8.08333333333333,0.3515903455);
\draw[draw=black,fill=peru22412652,line width=0.004pt,postaction={pattern=north east lines}] (axis cs:7.91666666666667,0.3515903455) rectangle (axis cs:8.08333333333333,0.4148839455);
\draw[draw=black,fill=sandybrown232159103,line width=0.004pt,postaction={pattern=north east lines}] (axis cs:7.91666666666667,0.4148839455) rectangle (axis cs:8.08333333333333,0.5329934215);
\draw[draw=black,fill=chocolate217952,line width=0.004pt] (axis cs:7.91666666666667,0.5329934215) rectangle (axis cs:8.08333333333333,0.5901769935);
\draw[draw=black,fill=peru22412652,line width=0.004pt] (axis cs:7.91666666666667,0.5901769935) rectangle (axis cs:8.08333333333333,1.201762328);
\draw[draw=black,fill=sandybrown232159103,line width=0.004pt] (axis cs:7.91666666666667,1.201762328) rectangle (axis cs:8.08333333333333,1.7097539275);
\draw[draw=black,fill=bisque247223204,line width=0.004pt,postaction={pattern=crosshatch}] (axis cs:8.91666666666667,0) rectangle (axis cs:9.08333333333333,0.317627248);
\draw[draw=black,fill=chocolate217952,line width=0.004pt,postaction={pattern=north east lines}] (axis cs:8.91666666666667,0.317627248) rectangle (axis cs:9.08333333333333,0.349247222);
\draw[draw=black,fill=peru22412652,line width=0.004pt,postaction={pattern=north east lines}] (axis cs:8.91666666666667,0.349247222) rectangle (axis cs:9.08333333333333,0.4120896735);
\draw[draw=black,fill=sandybrown232159103,line width=0.004pt,postaction={pattern=north east lines}] (axis cs:8.91666666666667,0.4120896735) rectangle (axis cs:9.08333333333333,0.5258890075);
\draw[draw=black,fill=chocolate217952,line width=0.004pt] (axis cs:8.91666666666667,0.5258890075) rectangle (axis cs:9.08333333333333,0.582852337);
\draw[draw=black,fill=peru22412652,line width=0.004pt] (axis cs:8.91666666666667,0.582852337) rectangle (axis cs:9.08333333333333,1.0242729225);
\draw[draw=black,fill=sandybrown232159103,line width=0.004pt] (axis cs:8.91666666666667,1.0242729225) rectangle (axis cs:9.08333333333333,1.5925455475);
\draw[draw=black,fill=gainsboro223228215,line width=0.004pt,postaction={pattern=crosshatch}] (axis cs:0.166666666666667,0) rectangle (axis cs:0.333333333333333,0.3131803425);
\draw[draw=black,fill=darkolivegreen9912157,line width=0.004pt,postaction={pattern=north east lines}] (axis cs:0.166666666666667,0.3131803425) rectangle (axis cs:0.333333333333333,0.3262096605);
\draw[draw=black,fill=gray13014796,line width=0.004pt,postaction={pattern=north east lines}] (axis cs:0.166666666666667,0.3262096605) rectangle (axis cs:0.333333333333333,0.3695229205);
\draw[draw=black,fill=darkseagreen161174136,line width=0.004pt,postaction={pattern=north east lines}] (axis cs:0.166666666666667,0.3695229205) rectangle (axis cs:0.333333333333333,0.454220782);
\draw[draw=black,fill=darkolivegreen9912157,line width=0.004pt] (axis cs:0.166666666666667,0.454220782) rectangle (axis cs:0.333333333333333,0.494498673);
\draw[draw=black,fill=gray13014796,line width=0.004pt] (axis cs:0.166666666666667,0.494498673) rectangle (axis cs:0.333333333333333,0.602061583);
\draw[draw=black,fill=darkseagreen161174136,line width=0.004pt] (axis cs:0.166666666666667,0.602061583) rectangle (axis cs:0.333333333333333,0.8068698685);
\draw[draw=black,fill=gainsboro223228215,line width=0.004pt,postaction={pattern=crosshatch}] (axis cs:1.16666666666667,0) rectangle (axis cs:1.33333333333333,0.3219412555);
\draw[draw=black,fill=darkolivegreen9912157,line width=0.004pt,postaction={pattern=north east lines}] (axis cs:1.16666666666667,0.3219412555) rectangle (axis cs:1.33333333333333,0.335544663);
\draw[draw=black,fill=gray13014796,line width=0.004pt,postaction={pattern=north east lines}] (axis cs:1.16666666666667,0.335544663) rectangle (axis cs:1.33333333333333,0.3803152365);
\draw[draw=black,fill=darkseagreen161174136,line width=0.004pt,postaction={pattern=north east lines}] (axis cs:1.16666666666667,0.3803152365) rectangle (axis cs:1.33333333333333,0.4717677835);
\draw[draw=black,fill=darkolivegreen9912157,line width=0.004pt] (axis cs:1.16666666666667,0.4717677835) rectangle (axis cs:1.33333333333333,0.511654312);
\draw[draw=black,fill=gray13014796,line width=0.004pt] (axis cs:1.16666666666667,0.511654312) rectangle (axis cs:1.33333333333333,0.6273247245);
\draw[draw=black,fill=darkseagreen161174136,line width=0.004pt] (axis cs:1.16666666666667,0.6273247245) rectangle (axis cs:1.33333333333333,0.8604125095);
\draw[draw=black,fill=gainsboro223228215,line width=0.004pt,postaction={pattern=crosshatch}] (axis cs:2.16666666666667,0) rectangle (axis cs:2.33333333333333,0.3167791635);
\draw[draw=black,fill=darkolivegreen9912157,line width=0.004pt,postaction={pattern=north east lines}] (axis cs:2.16666666666667,0.3167791635) rectangle (axis cs:2.33333333333333,0.3307854995);
\draw[draw=black,fill=gray13014796,line width=0.004pt,postaction={pattern=north east lines}] (axis cs:2.16666666666667,0.3307854995) rectangle (axis cs:2.33333333333333,0.373768301);
\draw[draw=black,fill=darkseagreen161174136,line width=0.004pt,postaction={pattern=north east lines}] (axis cs:2.16666666666667,0.373768301) rectangle (axis cs:2.33333333333333,0.4373708245);
\draw[draw=black,fill=darkolivegreen9912157,line width=0.004pt] (axis cs:2.16666666666667,0.4373708245) rectangle (axis cs:2.33333333333333,0.481189448);
\draw[draw=black,fill=gray13014796,line width=0.004pt] (axis cs:2.16666666666667,0.481189448) rectangle (axis cs:2.33333333333333,0.609244789);
\draw[draw=black,fill=darkseagreen161174136,line width=0.004pt] (axis cs:2.16666666666667,0.609244789) rectangle (axis cs:2.33333333333333,0.9621874315);
\draw[draw=black,fill=gainsboro223228215,line width=0.004pt,postaction={pattern=crosshatch}] (axis cs:3.16666666666667,0) rectangle (axis cs:3.33333333333333,0.3169454135);
\draw[draw=black,fill=darkolivegreen9912157,line width=0.004pt,postaction={pattern=north east lines}] (axis cs:3.16666666666667,0.3169454135) rectangle (axis cs:3.33333333333333,0.330893929);
\draw[draw=black,fill=gray13014796,line width=0.004pt,postaction={pattern=north east lines}] (axis cs:3.16666666666667,0.330893929) rectangle (axis cs:3.33333333333333,0.3734772125);
\draw[draw=black,fill=darkseagreen161174136,line width=0.004pt,postaction={pattern=north east lines}] (axis cs:3.16666666666667,0.3734772125) rectangle (axis cs:3.33333333333333,0.4306532825);
\draw[draw=black,fill=darkolivegreen9912157,line width=0.004pt] (axis cs:3.16666666666667,0.4306532825) rectangle (axis cs:3.33333333333333,0.4750181235);
\draw[draw=black,fill=gray13014796,line width=0.004pt] (axis cs:3.16666666666667,0.4750181235) rectangle (axis cs:3.33333333333333,0.5952132435);
\draw[draw=black,fill=darkseagreen161174136,line width=0.004pt] (axis cs:3.16666666666667,0.5952132435) rectangle (axis cs:3.33333333333333,1.016622728);
\draw[draw=black,fill=gainsboro223228215,line width=0.004pt,postaction={pattern=crosshatch}] (axis cs:4.16666666666667,0) rectangle (axis cs:4.33333333333333,0.3156866215);
\draw[draw=black,fill=darkolivegreen9912157,line width=0.004pt,postaction={pattern=north east lines}] (axis cs:4.16666666666667,0.3156866215) rectangle (axis cs:4.33333333333333,0.329673557);
\draw[draw=black,fill=gray13014796,line width=0.004pt,postaction={pattern=north east lines}] (axis cs:4.16666666666667,0.329673557) rectangle (axis cs:4.33333333333333,0.37084059);
\draw[draw=black,fill=darkseagreen161174136,line width=0.004pt,postaction={pattern=north east lines}] (axis cs:4.16666666666667,0.37084059) rectangle (axis cs:4.33333333333333,0.427272454);
\draw[draw=black,fill=darkolivegreen9912157,line width=0.004pt] (axis cs:4.16666666666667,0.427272454) rectangle (axis cs:4.33333333333333,0.4721085735);
\draw[draw=black,fill=gray13014796,line width=0.004pt] (axis cs:4.16666666666667,0.4721085735) rectangle (axis cs:4.33333333333333,0.6209276545);
\draw[draw=black,fill=darkseagreen161174136,line width=0.004pt] (axis cs:4.16666666666667,0.6209276545) rectangle (axis cs:4.33333333333333,1.0887241035);
\draw[draw=black,fill=gainsboro223228215,line width=0.004pt,postaction={pattern=crosshatch}] (axis cs:5.16666666666667,0) rectangle (axis cs:5.33333333333333,0.3068474565);
\draw[draw=black,fill=darkolivegreen9912157,line width=0.004pt,postaction={pattern=north east lines}] (axis cs:5.16666666666667,0.3068474565) rectangle (axis cs:5.33333333333333,0.321241185);
\draw[draw=black,fill=gray13014796,line width=0.004pt,postaction={pattern=north east lines}] (axis cs:5.16666666666667,0.321241185) rectangle (axis cs:5.33333333333333,0.357739982);
\draw[draw=black,fill=darkseagreen161174136,line width=0.004pt,postaction={pattern=north east lines}] (axis cs:5.16666666666667,0.357739982) rectangle (axis cs:5.33333333333333,0.4013377585);
\draw[draw=black,fill=darkolivegreen9912157,line width=0.004pt] (axis cs:5.16666666666667,0.4013377585) rectangle (axis cs:5.33333333333333,0.449570246);
\draw[draw=black,fill=gray13014796,line width=0.004pt] (axis cs:5.16666666666667,0.449570246) rectangle (axis cs:5.33333333333333,0.6390834765);
\draw[draw=black,fill=darkseagreen161174136,line width=0.004pt] (axis cs:5.16666666666667,0.6390834765) rectangle (axis cs:5.33333333333333,1.210944598);
\draw[draw=black,fill=gainsboro223228215,line width=0.004pt,postaction={pattern=crosshatch}] (axis cs:6.16666666666667,0) rectangle (axis cs:6.33333333333333,0.306652815);
\draw[draw=black,fill=darkolivegreen9912157,line width=0.004pt,postaction={pattern=north east lines}] (axis cs:6.16666666666667,0.306652815) rectangle (axis cs:6.33333333333333,0.321189027);
\draw[draw=black,fill=gray13014796,line width=0.004pt,postaction={pattern=north east lines}] (axis cs:6.16666666666667,0.321189027) rectangle (axis cs:6.33333333333333,0.35803039);
\draw[draw=black,fill=darkseagreen161174136,line width=0.004pt,postaction={pattern=north east lines}] (axis cs:6.16666666666667,0.35803039) rectangle (axis cs:6.33333333333333,0.395563924);
\draw[draw=black,fill=darkolivegreen9912157,line width=0.004pt] (axis cs:6.16666666666667,0.395563924) rectangle (axis cs:6.33333333333333,0.443450597);
\draw[draw=black,fill=gray13014796,line width=0.004pt] (axis cs:6.16666666666667,0.443450597) rectangle (axis cs:6.33333333333333,0.622660325);
\draw[draw=black,fill=darkseagreen161174136,line width=0.004pt] (axis cs:6.16666666666667,0.622660325) rectangle (axis cs:6.33333333333333,1.2268986195);
\draw[draw=black,fill=gainsboro223228215,line width=0.004pt,postaction={pattern=crosshatch}] (axis cs:7.16666666666667,0) rectangle (axis cs:7.33333333333333,0.3021572655);
\draw[draw=black,fill=darkolivegreen9912157,line width=0.004pt,postaction={pattern=north east lines}] (axis cs:7.16666666666667,0.3021572655) rectangle (axis cs:7.33333333333333,0.3165073005);
\draw[draw=black,fill=gray13014796,line width=0.004pt,postaction={pattern=north east lines}] (axis cs:7.16666666666667,0.3165073005) rectangle (axis cs:7.33333333333333,0.349636095);
\draw[draw=black,fill=darkseagreen161174136,line width=0.004pt,postaction={pattern=north east lines}] (axis cs:7.16666666666667,0.349636095) rectangle (axis cs:7.33333333333333,0.3892873675);
\draw[draw=black,fill=darkolivegreen9912157,line width=0.004pt] (axis cs:7.16666666666667,0.3892873675) rectangle (axis cs:7.33333333333333,0.4403530925);
\draw[draw=black,fill=gray13014796,line width=0.004pt] (axis cs:7.16666666666667,0.4403530925) rectangle (axis cs:7.33333333333333,0.698434321);
\draw[draw=black,fill=darkseagreen161174136,line width=0.004pt] (axis cs:7.16666666666667,0.698434321) rectangle (axis cs:7.33333333333333,1.3868527435);
\draw[draw=black,fill=gainsboro223228215,line width=0.004pt,postaction={pattern=crosshatch}] (axis cs:8.16666666666667,0) rectangle (axis cs:8.33333333333333,0.3040396025);
\draw[draw=black,fill=darkolivegreen9912157,line width=0.004pt,postaction={pattern=north east lines}] (axis cs:8.16666666666667,0.3040396025) rectangle (axis cs:8.33333333333333,0.318685368);
\draw[draw=black,fill=gray13014796,line width=0.004pt,postaction={pattern=north east lines}] (axis cs:8.16666666666667,0.318685368) rectangle (axis cs:8.33333333333333,0.354919132);
\draw[draw=black,fill=darkseagreen161174136,line width=0.004pt,postaction={pattern=north east lines}] (axis cs:8.16666666666667,0.354919132) rectangle (axis cs:8.33333333333333,0.3951335355);
\draw[draw=black,fill=darkolivegreen9912157,line width=0.004pt] (axis cs:8.16666666666667,0.3951335355) rectangle (axis cs:8.33333333333333,0.456592291);
\draw[draw=black,fill=gray13014796,line width=0.004pt] (axis cs:8.16666666666667,0.456592291) rectangle (axis cs:8.33333333333333,0.790952947);
\draw[draw=black,fill=darkseagreen161174136,line width=0.004pt] (axis cs:8.16666666666667,0.790952947) rectangle (axis cs:8.33333333333333,1.6270061045);
\draw[draw=black,fill=gainsboro223228215,line width=0.004pt,postaction={pattern=crosshatch}] (axis cs:9.16666666666667,0) rectangle (axis cs:9.33333333333333,0.299422131);
\draw[draw=black,fill=darkolivegreen9912157,line width=0.004pt,postaction={pattern=north east lines}] (axis cs:9.16666666666667,0.299422131) rectangle (axis cs:9.33333333333333,0.314213246);
\draw[draw=black,fill=gray13014796,line width=0.004pt,postaction={pattern=north east lines}] (axis cs:9.16666666666667,0.314213246) rectangle (axis cs:9.33333333333333,0.350004032);
\draw[draw=black,fill=darkseagreen161174136,line width=0.004pt,postaction={pattern=north east lines}] (axis cs:9.16666666666667,0.350004032) rectangle (axis cs:9.33333333333333,0.39031369);
\draw[draw=black,fill=darkolivegreen9912157,line width=0.004pt] (axis cs:9.16666666666667,0.39031369) rectangle (axis cs:9.33333333333333,0.4513137525);
\draw[draw=black,fill=gray13014796,line width=0.004pt] (axis cs:9.16666666666667,0.4513137525) rectangle (axis cs:9.33333333333333,0.7996290895);
\draw[draw=black,fill=darkseagreen161174136,line width=0.004pt] (axis cs:9.16666666666667,0.7996290895) rectangle (axis cs:9.33333333333333,2.2612386325);
\end{axis}

\end{tikzpicture}

%% file: figures/python_fig/mlx/unbounded/weak-eff.tex
\begin{tikzpicture}

\definecolor{chocolate217952}{RGB}{217,95,2}
\definecolor{darkgray}{RGB}{169,169,169}
\definecolor{darkgray150}{RGB}{150,150,150}
\definecolor{darkgray176}{RGB}{176,176,176}
\definecolor{darkolivegreen9912157}{RGB}{99,121,57}
\definecolor{dimgray99}{RGB}{99,99,99}
\definecolor{silver189}{RGB}{189,189,189}
\definecolor{steelblue31119180}{RGB}{31,119,180}

\begin{axis}[
axis y line =left, axis x line =bottom, axis x line shift=5pt, axis line style ={-},
log basis x={10},
minor xtick={},
minor ytick={},
tick align=center,
tick pos=left,
x grid style={darkgray176},
x grid style={draw=none},
xlabel near ticks,
xlabel={\(\displaystyle r\) = Number of nodes},
xmajorgrids,
xmin=0.739118043551982, xmax=571.491933778359,
xmode=log,
xtick style={color=black},
xtick={1,2,4,8,16,32,64,128,256,384},
xticklabels={$1$,$2$,$4$,$8$,$16$,$32$,$64$,$128$,$256$,\quad $384$},
y axis line style={draw=none},
y grid style={darkgray176},
y grid style={dotted},
ylabel near ticks,
ylabel={\(\displaystyle \eta_{P,w}\)},
ymajorgrids,
ymin=0.1, ymax=1.1,
ytick style={color=darkgray},
ytick={0,0.2,0.4,0.6,0.8,1,1.2}
]
\addplot [very thin, silver189]
table {%
1 1
1.1313219208544 0.998688503074411
1.27988928860569 0.997208919050557
1.44796680846636 0.995540309846982
1.63811659108758 0.993659295178581
1.85323720841266 0.99153981337605
2.09660787842026 0.989152871679542
2.37193845229287 0.986466289653339
2.68342596599638 0.983444440920398
3.0358186183216 0.980048000340578
3.43448815064515 0.976233706102335
3.88551173173954 0.971954149003378
4.39576459585389 0.967157604480928
4.97302484620518 0.961787926704239
5.62609202146551 0.955784528198603
6.36492123262797 0.949082472896428
7.20077491498363 0.941612714973619
8.14639450845946 0.933302519965847
9.21619468334809 0.924076107948635
10.4264830721335 0.913855560304177
11.7957088569219 0.9025620308918
13.3447440018522 0.890117298218699
15.0972014174856 0.876445686288978
17.0797949071556 0.861476367011014
19.3227463821625 0.845146035378667
21.8602465532505 0.827401919587683
24.7309761209741 0.808205052081931
27.9786954097847 0.787533685688617
31.6529114339978 0.765386694429058
35.8096325641446 0.741786755871604
40.5122222975584 0.716783077160903
45.8323651477542 0.690453407275249
51.8511593762575 0.662905080832537
58.6603532240753 0.634274869712328
66.3637434874584 0.604727480710331
75.0787577573201 0.574452628731463
84.9382444413736 0.543660728652182
96.0924978554153 0.51257737278094
108.711549253486 0.481436878976636
122.987758720511 0.450475290062739
139.138747437266 0.419923263543422
157.410715016003 0.389999302461908
178.082192474969 0.360903741618233
201.468288060745 0.332813824198923
227.92549064013 0.305880094587292
257.857103882673 0.280224209948409
291.719394070498 0.25593815284725
330.028545250318 0.233084723865294
373.368527749373 0.21169911652763
422.4 0.191791331031837
};
\addplot [very thin, darkgray150]
table {%
1 1
1.1313219208544 0.999343821248998
1.27988928860569 0.998602509270402
1.44796680846636 0.997765171602392
1.63811659108758 0.996819564487898
1.85323720841266 0.995751936985077
2.09660787842026 0.994546860387206
2.37193845229287 0.993187042530169
2.68342596599638 0.991653126884704
3.0358186183216 0.989923476776325
3.43448815064515 0.987973945680474
3.88551173173954 0.985777635341675
4.39576459585389 0.983304644506235
4.97302484620518 0.980521812385727
5.62609202146551 0.977392462633845
6.36492123262797 0.973876155672415
7.20077491498363 0.969928459689154
8.14639450845946 0.965500753583391
9.21619468334809 0.960540078566688
10.4264830721335 0.954989059006034
11.7957088569219 0.948785917344032
13.3447440018522 0.941864612378893
15.0972014174856 0.93415513456434
17.0797949071556 0.925583995884183
19.3227463821625 0.916074954690752
21.8602465532505 0.905550016905278
24.7309761209741 0.893930753208967
27.9786954097847 0.881139966193402
31.6529114339978 0.86710373069464
35.8096325641446 0.851753813572211
40.5122222975584 0.835030455153681
45.8323651477542 0.816885463158852
51.8511593762575 0.797285531776296
58.6603532240753 0.776215655600182
66.3637434874584 0.753682463819525
75.0787577573201 0.729717259507896
84.9382444413736 0.704378518622896
96.0924978554153 0.677753590665638
108.711549253486 0.649959354743765
122.987758720511 0.621141625988343
139.138747437266 0.591473179325906
157.410715016003 0.561150357084575
178.082192474969 0.53038834501122
201.468288060745 0.499415324415558
227.92549064013 0.46846581987906
257.857103882673 0.437773645851771
291.719394070498 0.407564898425973
330.028545250318 0.378051433699793
373.368527749373 0.349425222219025
422.4 0.321853878339231
};
\addplot [very thin, dimgray99]
table {%
1 1
1.1313219208544 0.999737425121966
1.27988928860569 0.999440534599535
1.44796680846636 0.999104868361594
1.63811659108758 0.998725393512918
1.85323720841266 0.998296432677228
2.09660787842026 0.997811583911807
2.37193845229287 0.997263631354144
2.68342596599638 0.99664444572241
3.0358186183216 0.995944873766329
3.43448815064515 0.995154615759746
3.88551173173954 0.994262090148903
4.39576459585389 0.993254284531814
4.97302484620518 0.992116592257259
5.62609202146551 0.99083263411344
6.36492123262797 0.989384064846562
7.20077491498363 0.98775036463343
8.14639450845946 0.985908616159273
9.21619468334809 0.983833268657511
10.4264830721335 0.981495891192608
11.7957088569219 0.978864918655874
13.3447440018522 0.975905395446377
15.0972014174856 0.972578723675957
17.0797949071556 0.968842425016757
19.3227463821625 0.964649928041836
21.8602465532505 0.959950396116251
24.7309761209741 0.954688614569375
27.9786954097847 0.948804959963693
31.6529114339978 0.942235478646175
35.8096325641446 0.934912106206372
40.5122222975584 0.926763063625706
45.8323651477542 0.917713469287756
51.8511593762575 0.90768620795164
58.6603532240753 0.896603097403793
66.3637434874584 0.884386389752795
75.0787577573201 0.870960636051551
84.9382444413736 0.856254928940867
96.0924978554153 0.840205517296705
108.711549253486 0.822758758845705
122.987758720511 0.803874341560267
139.138747437266 0.783528663645601
157.410715016003 0.761718217820083
178.082192474969 0.738462782741822
201.468288060745 0.713808188782185
227.92549064013 0.687828403925831
257.857103882673 0.660626685585691
291.719394070498 0.632335571568673
330.028545250318 0.60311554151496
373.368527749373 0.573152267757692
422.4 0.54265248534838
};
\addplot [very thin, black]
table {%
1 1
1.1313219208544 1
1.27988928860569 1
1.44796680846636 1
1.63811659108758 1
1.85323720841266 1
2.09660787842026 1
2.37193845229287 1
2.68342596599638 1
3.0358186183216 1
3.43448815064515 1
3.88551173173954 1
4.39576459585389 1
4.97302484620518 1
5.62609202146551 1
6.36492123262797 1
7.20077491498363 1
8.14639450845946 1
9.21619468334809 1
10.4264830721335 1
11.7957088569219 1
13.3447440018522 1
15.0972014174856 1
17.0797949071556 1
19.3227463821625 1
21.8602465532505 1
24.7309761209741 1
27.9786954097847 1
31.6529114339978 1
35.8096325641446 1
40.5122222975584 1
45.8323651477542 1
51.8511593762575 1
58.6603532240753 1
66.3637434874584 1
75.0787577573201 1
84.9382444413736 1
96.0924978554153 1
108.711549253486 1
122.987758720511 1
139.138747437266 1
157.410715016003 1
178.082192474969 1
201.468288060745 1
227.92549064013 1
257.857103882673 1
291.719394070498 1
330.028545250318 1
373.368527749373 1
422.4 1
};
\addplot [very thin, silver189]
table {%
1 1
1.1313219208544 0.998688503074411
1.27988928860569 0.997208919050557
1.44796680846636 0.995540309846982
1.63811659108758 0.993659295178581
1.85323720841266 0.99153981337605
2.09660787842026 0.989152871679542
2.37193845229287 0.986466289653339
2.68342596599638 0.983444440920398
3.0358186183216 0.980048000340578
3.43448815064515 0.976233706102335
3.88551173173954 0.971954149003378
4.39576459585389 0.967157604480928
4.97302484620518 0.961787926704239
5.62609202146551 0.955784528198603
6.36492123262797 0.949082472896428
7.20077491498363 0.941612714973619
8.14639450845946 0.933302519965847
9.21619468334809 0.924076107948635
10.4264830721335 0.913855560304177
11.7957088569219 0.9025620308918
13.3447440018522 0.890117298218699
15.0972014174856 0.876445686288978
17.0797949071556 0.861476367011014
19.3227463821625 0.845146035378667
21.8602465532505 0.827401919587683
24.7309761209741 0.808205052081931
27.9786954097847 0.787533685688617
31.6529114339978 0.765386694429058
35.8096325641446 0.741786755871604
40.5122222975584 0.716783077160903
45.8323651477542 0.690453407275249
51.8511593762575 0.662905080832537
58.6603532240753 0.634274869712328
66.3637434874584 0.604727480710331
75.0787577573201 0.574452628731463
84.9382444413736 0.543660728652182
96.0924978554153 0.51257737278094
108.711549253486 0.481436878976636
122.987758720511 0.450475290062739
139.138747437266 0.419923263543422
157.410715016003 0.389999302461908
178.082192474969 0.360903741618233
201.468288060745 0.332813824198923
227.92549064013 0.305880094587292
257.857103882673 0.280224209948409
291.719394070498 0.25593815284725
330.028545250318 0.233084723865294
373.368527749373 0.21169911652763
422.4 0.191791331031837
};
\addplot [very thin, darkgray150]
table {%
1 1
1.1313219208544 0.999343821248998
1.27988928860569 0.998602509270402
1.44796680846636 0.997765171602392
1.63811659108758 0.996819564487898
1.85323720841266 0.995751936985077
2.09660787842026 0.994546860387206
2.37193845229287 0.993187042530169
2.68342596599638 0.991653126884704
3.0358186183216 0.989923476776325
3.43448815064515 0.987973945680474
3.88551173173954 0.985777635341675
4.39576459585389 0.983304644506235
4.97302484620518 0.980521812385727
5.62609202146551 0.977392462633845
6.36492123262797 0.973876155672415
7.20077491498363 0.969928459689154
8.14639450845946 0.965500753583391
9.21619468334809 0.960540078566688
10.4264830721335 0.954989059006034
11.7957088569219 0.948785917344032
13.3447440018522 0.941864612378893
15.0972014174856 0.93415513456434
17.0797949071556 0.925583995884183
19.3227463821625 0.916074954690752
21.8602465532505 0.905550016905278
24.7309761209741 0.893930753208967
27.9786954097847 0.881139966193402
31.6529114339978 0.86710373069464
35.8096325641446 0.851753813572211
40.5122222975584 0.835030455153681
45.8323651477542 0.816885463158852
51.8511593762575 0.797285531776296
58.6603532240753 0.776215655600182
66.3637434874584 0.753682463819525
75.0787577573201 0.729717259507896
84.9382444413736 0.704378518622896
96.0924978554153 0.677753590665638
108.711549253486 0.649959354743765
122.987758720511 0.621141625988343
139.138747437266 0.591473179325906
157.410715016003 0.561150357084575
178.082192474969 0.53038834501122
201.468288060745 0.499415324415558
227.92549064013 0.46846581987906
257.857103882673 0.437773645851771
291.719394070498 0.407564898425973
330.028545250318 0.378051433699793
373.368527749373 0.349425222219025
422.4 0.321853878339231
};
\addplot [very thin, dimgray99]
table {%
1 1
1.1313219208544 0.999737425121966
1.27988928860569 0.999440534599535
1.44796680846636 0.999104868361594
1.63811659108758 0.998725393512918
1.85323720841266 0.998296432677228
2.09660787842026 0.997811583911807
2.37193845229287 0.997263631354144
2.68342596599638 0.99664444572241
3.0358186183216 0.995944873766329
3.43448815064515 0.995154615759746
3.88551173173954 0.994262090148903
4.39576459585389 0.993254284531814
4.97302484620518 0.992116592257259
5.62609202146551 0.99083263411344
6.36492123262797 0.989384064846562
7.20077491498363 0.98775036463343
8.14639450845946 0.985908616159273
9.21619468334809 0.983833268657511
10.4264830721335 0.981495891192608
11.7957088569219 0.978864918655874
13.3447440018522 0.975905395446377
15.0972014174856 0.972578723675957
17.0797949071556 0.968842425016757
19.3227463821625 0.964649928041836
21.8602465532505 0.959950396116251
24.7309761209741 0.954688614569375
27.9786954097847 0.948804959963693
31.6529114339978 0.942235478646175
35.8096325641446 0.934912106206372
40.5122222975584 0.926763063625706
45.8323651477542 0.917713469287756
51.8511593762575 0.90768620795164
58.6603532240753 0.896603097403793
66.3637434874584 0.884386389752795
75.0787577573201 0.870960636051551
84.9382444413736 0.856254928940867
96.0924978554153 0.840205517296705
108.711549253486 0.822758758845705
122.987758720511 0.803874341560267
139.138747437266 0.783528663645601
157.410715016003 0.761718217820083
178.082192474969 0.738462782741822
201.468288060745 0.713808188782185
227.92549064013 0.687828403925831
257.857103882673 0.660626685585691
291.719394070498 0.632335571568673
330.028545250318 0.60311554151496
373.368527749373 0.573152267757692
422.4 0.54265248534838
};
\addplot [very thin, black]
table {%
1 1
1.1313219208544 1
1.27988928860569 1
1.44796680846636 1
1.63811659108758 1
1.85323720841266 1
2.09660787842026 1
2.37193845229287 1
2.68342596599638 1
3.0358186183216 1
3.43448815064515 1
3.88551173173954 1
4.39576459585389 1
4.97302484620518 1
5.62609202146551 1
6.36492123262797 1
7.20077491498363 1
8.14639450845946 1
9.21619468334809 1
10.4264830721335 1
11.7957088569219 1
13.3447440018522 1
15.0972014174856 1
17.0797949071556 1
19.3227463821625 1
21.8602465532505 1
24.7309761209741 1
27.9786954097847 1
31.6529114339978 1
35.8096325641446 1
40.5122222975584 1
45.8323651477542 1
51.8511593762575 1
58.6603532240753 1
66.3637434874584 1
75.0787577573201 1
84.9382444413736 1
96.0924978554153 1
108.711549253486 1
122.987758720511 1
139.138747437266 1
157.410715016003 1
178.082192474969 1
201.468288060745 1
227.92549064013 1
257.857103882673 1
291.719394070498 1
330.028545250318 1
373.368527749373 1
422.4 1
};
\addplot [very thin, silver189]
table {%
1 1
1.1313219208544 0.998688503074411
1.27988928860569 0.997208919050557
1.44796680846636 0.995540309846982
1.63811659108758 0.993659295178581
1.85323720841266 0.99153981337605
2.09660787842026 0.989152871679542
2.37193845229287 0.986466289653339
2.68342596599638 0.983444440920398
3.0358186183216 0.980048000340578
3.43448815064515 0.976233706102335
3.88551173173954 0.971954149003378
4.39576459585389 0.967157604480928
4.97302484620518 0.961787926704239
5.62609202146551 0.955784528198603
6.36492123262797 0.949082472896428
7.20077491498363 0.941612714973619
8.14639450845946 0.933302519965847
9.21619468334809 0.924076107948635
10.4264830721335 0.913855560304177
11.7957088569219 0.9025620308918
13.3447440018522 0.890117298218699
15.0972014174856 0.876445686288978
17.0797949071556 0.861476367011014
19.3227463821625 0.845146035378667
21.8602465532505 0.827401919587683
24.7309761209741 0.808205052081931
27.9786954097847 0.787533685688617
31.6529114339978 0.765386694429058
35.8096325641446 0.741786755871604
40.5122222975584 0.716783077160903
45.8323651477542 0.690453407275249
51.8511593762575 0.662905080832537
58.6603532240753 0.634274869712328
66.3637434874584 0.604727480710331
75.0787577573201 0.574452628731463
84.9382444413736 0.543660728652182
96.0924978554153 0.51257737278094
108.711549253486 0.481436878976636
122.987758720511 0.450475290062739
139.138747437266 0.419923263543422
157.410715016003 0.389999302461908
178.082192474969 0.360903741618233
201.468288060745 0.332813824198923
227.92549064013 0.305880094587292
257.857103882673 0.280224209948409
291.719394070498 0.25593815284725
330.028545250318 0.233084723865294
373.368527749373 0.21169911652763
422.4 0.191791331031837
};
\addplot [very thin, darkgray150]
table {%
1 1
1.1313219208544 0.999343821248998
1.27988928860569 0.998602509270402
1.44796680846636 0.997765171602392
1.63811659108758 0.996819564487898
1.85323720841266 0.995751936985077
2.09660787842026 0.994546860387206
2.37193845229287 0.993187042530169
2.68342596599638 0.991653126884704
3.0358186183216 0.989923476776325
3.43448815064515 0.987973945680474
3.88551173173954 0.985777635341675
4.39576459585389 0.983304644506235
4.97302484620518 0.980521812385727
5.62609202146551 0.977392462633845
6.36492123262797 0.973876155672415
7.20077491498363 0.969928459689154
8.14639450845946 0.965500753583391
9.21619468334809 0.960540078566688
10.4264830721335 0.954989059006034
11.7957088569219 0.948785917344032
13.3447440018522 0.941864612378893
15.0972014174856 0.93415513456434
17.0797949071556 0.925583995884183
19.3227463821625 0.916074954690752
21.8602465532505 0.905550016905278
24.7309761209741 0.893930753208967
27.9786954097847 0.881139966193402
31.6529114339978 0.86710373069464
35.8096325641446 0.851753813572211
40.5122222975584 0.835030455153681
45.8323651477542 0.816885463158852
51.8511593762575 0.797285531776296
58.6603532240753 0.776215655600182
66.3637434874584 0.753682463819525
75.0787577573201 0.729717259507896
84.9382444413736 0.704378518622896
96.0924978554153 0.677753590665638
108.711549253486 0.649959354743765
122.987758720511 0.621141625988343
139.138747437266 0.591473179325906
157.410715016003 0.561150357084575
178.082192474969 0.53038834501122
201.468288060745 0.499415324415558
227.92549064013 0.46846581987906
257.857103882673 0.437773645851771
291.719394070498 0.407564898425973
330.028545250318 0.378051433699793
373.368527749373 0.349425222219025
422.4 0.321853878339231
};
\addplot [very thin, dimgray99]
table {%
1 1
1.1313219208544 0.999737425121966
1.27988928860569 0.999440534599535
1.44796680846636 0.999104868361594
1.63811659108758 0.998725393512918
1.85323720841266 0.998296432677228
2.09660787842026 0.997811583911807
2.37193845229287 0.997263631354144
2.68342596599638 0.99664444572241
3.0358186183216 0.995944873766329
3.43448815064515 0.995154615759746
3.88551173173954 0.994262090148903
4.39576459585389 0.993254284531814
4.97302484620518 0.992116592257259
5.62609202146551 0.99083263411344
6.36492123262797 0.989384064846562
7.20077491498363 0.98775036463343
8.14639450845946 0.985908616159273
9.21619468334809 0.983833268657511
10.4264830721335 0.981495891192608
11.7957088569219 0.978864918655874
13.3447440018522 0.975905395446377
15.0972014174856 0.972578723675957
17.0797949071556 0.968842425016757
19.3227463821625 0.964649928041836
21.8602465532505 0.959950396116251
24.7309761209741 0.954688614569375
27.9786954097847 0.948804959963693
31.6529114339978 0.942235478646175
35.8096325641446 0.934912106206372
40.5122222975584 0.926763063625706
45.8323651477542 0.917713469287756
51.8511593762575 0.90768620795164
58.6603532240753 0.896603097403793
66.3637434874584 0.884386389752795
75.0787577573201 0.870960636051551
84.9382444413736 0.856254928940867
96.0924978554153 0.840205517296705
108.711549253486 0.822758758845705
122.987758720511 0.803874341560267
139.138747437266 0.783528663645601
157.410715016003 0.761718217820083
178.082192474969 0.738462782741822
201.468288060745 0.713808188782185
227.92549064013 0.687828403925831
257.857103882673 0.660626685585691
291.719394070498 0.632335571568673
330.028545250318 0.60311554151496
373.368527749373 0.573152267757692
422.4 0.54265248534838
};
\addplot [very thin, black]
table {%
1 1
1.1313219208544 1
1.27988928860569 1
1.44796680846636 1
1.63811659108758 1
1.85323720841266 1
2.09660787842026 1
2.37193845229287 1
2.68342596599638 1
3.0358186183216 1
3.43448815064515 1
3.88551173173954 1
4.39576459585389 1
4.97302484620518 1
5.62609202146551 1
6.36492123262797 1
7.20077491498363 1
8.14639450845946 1
9.21619468334809 1
10.4264830721335 1
11.7957088569219 1
13.3447440018522 1
15.0972014174856 1
17.0797949071556 1
19.3227463821625 1
21.8602465532505 1
24.7309761209741 1
27.9786954097847 1
31.6529114339978 1
35.8096325641446 1
40.5122222975584 1
45.8323651477542 1
51.8511593762575 1
58.6603532240753 1
66.3637434874584 1
75.0787577573201 1
84.9382444413736 1
96.0924978554153 1
108.711549253486 1
122.987758720511 1
139.138747437266 1
157.410715016003 1
178.082192474969 1
201.468288060745 1
227.92549064013 1
257.857103882673 1
291.719394070498 1
330.028545250318 1
373.368527749373 1
422.4 1
};
\addplot [semithick, steelblue31119180, mark=*, mark size=1.5, mark options={solid}]
table {%
1 1
2 0.983998469475404
4 0.916063467649853
8 0.896755772042187
16 0.889309437335312
32 0.859029431282697
64 0.855139510990118
128 0.801793730023578
256 0.694220604628993
384 0.642016883249922
};
\addplot [semithick, chocolate217952, mark=*, mark size=1.5, mark options={solid}]
table {%
1 1
2 0.981329426649818
4 0.95549474711678
8 0.920875025351124
16 0.909648759154315
32 0.878528718128845
64 0.908389750075169
128 0.826657886741854
256 0.601132174033272
384 0.645374380100736
};
\addplot [semithick, darkolivegreen9912157, mark=*, mark size=1.5, mark options={solid}]
table {%
1 1
2 0.937770964033154
4 0.838578682369746
8 0.793676795016528
16 0.741115096015691
32 0.666314437367844
64 0.657649992978902
128 0.581799237360776
256 0.495923073839948
384 0.356826500707682
};
\end{axis}

\end{tikzpicture}

%% file: figures/python_tab/mlx/unbounded/weak_time_unknowns_rank.tex
\begin{tabular}{lrrrrr}
\hline
  N nodes   &   \centercell{$1$ } &   \centercell{$2$ } &   \centercell{$8$ } &   \centercell{$64$ } &   \centercell{$128$ } \\
\hline
 \ata       &               30.34 &               29.85 &               27.21 &                25.94 &                 24.33 \\
 \nb        &               32.14 &               31.54 &               29.59 &                29.19 &                 26.57 \\
 \isr       &               40.94 &               38.39 &               32.49 &                26.92 &                 23.82 \\
\hline
\end{tabular}

%% file: figures/python_fig/mlx/unbounded/strong-comp.tex
\begin{tikzpicture}

\definecolor{bisque247223204}{RGB}{247,223,204}
\definecolor{chocolate217952}{RGB}{217,95,2}
\definecolor{cornflowerblue120173210}{RGB}{120,173,210}
\definecolor{darkgray}{RGB}{169,169,169}
\definecolor{darkgray176}{RGB}{176,176,176}
\definecolor{darkolivegreen9912157}{RGB}{99,121,57}
\definecolor{darkseagreen161174136}{RGB}{161,174,136}
\definecolor{gainsboro223228215}{RGB}{223,228,215}
\definecolor{gray13014796}{RGB}{130,147,96}
\definecolor{lavender210227240}{RGB}{210,227,240}
\definecolor{peru22412652}{RGB}{224,126,52}
\definecolor{sandybrown232159103}{RGB}{232,159,103}
\definecolor{steelblue31119180}{RGB}{31,119,180}
\definecolor{steelblue75146195}{RGB}{75,146,195}

\begin{axis}[
axis line style={draw=none},
axis y line =left, axis x line =bottom, axis line style ={-},
tick pos=left,
x grid style={darkgray176},
x grid style={draw=none},
xlabel near ticks,
xlabel={Number of nodes},
xmajorgrids,
xmin=-0.816666666666667, xmax=9.81666666666667,
xtick style={color=darkgray},
xtick={0,1,2,3,4,5,6,7,8,9},
xticklabels={$1$,$2$,$4$,$8$,$16$,$32$,$64$,$128$,$256$,$384$},
xticklabels={1,2,4,8,16,32,64,128,256,384},
y axis line style={draw=none},
y grid style={darkgray176},
ylabel near ticks,
ylabel={time/solve - [sec]},
ymajorgrids,
ymin=0, ymax=25,
ytick style={color=darkgray}
]
\draw[draw=black,fill=lavender210227240,line width=0.004pt,postaction={pattern=crosshatch}] (axis cs:-0.333333333333333,1e-16) rectangle (axis cs:-0.166666666666667,6.0409708725);
\draw[draw=black,fill=steelblue31119180,line width=0.004pt,postaction={pattern=north east lines}] (axis cs:-0.333333333333333,6.0409708725) rectangle (axis cs:-0.166666666666667,7.072429006);
\draw[draw=black,fill=steelblue75146195,line width=0.004pt,postaction={pattern=north east lines}] (axis cs:-0.333333333333333,7.072429006) rectangle (axis cs:-0.166666666666667,10.1252579545);
\draw[draw=black,fill=cornflowerblue120173210,line width=0.004pt,postaction={pattern=north east lines}] (axis cs:-0.333333333333333,10.1252579545) rectangle (axis cs:-0.166666666666667,16.413732444);
\draw[draw=black,fill=steelblue31119180,line width=0.004pt] (axis cs:-0.333333333333333,16.413732444) rectangle (axis cs:-0.166666666666667,17.1233320915);
\draw[draw=black,fill=steelblue75146195,line width=0.004pt] (axis cs:-0.333333333333333,17.1233320915) rectangle (axis cs:-0.166666666666667,18.6349273405);
\draw[draw=black,fill=cornflowerblue120173210,line width=0.004pt] (axis cs:-0.333333333333333,18.6349273405) rectangle (axis cs:-0.166666666666667,21.880232988);
\draw[draw=black,fill=lavender210227240,line width=0.004pt,postaction={pattern=crosshatch}] (axis cs:0.666666666666667,1e-16) rectangle (axis cs:0.833333333333333,3.001882495);
\draw[draw=black,fill=steelblue31119180,line width=0.004pt,postaction={pattern=north east lines}] (axis cs:0.666666666666667,3.001882495) rectangle (axis cs:0.833333333333333,3.4810170995);
\draw[draw=black,fill=steelblue75146195,line width=0.004pt,postaction={pattern=north east lines}] (axis cs:0.666666666666667,3.4810170995) rectangle (axis cs:0.833333333333333,4.664137867);
\draw[draw=black,fill=cornflowerblue120173210,line width=0.004pt,postaction={pattern=north east lines}] (axis cs:0.666666666666667,4.664137867) rectangle (axis cs:0.833333333333333,7.581900125);
\draw[draw=black,fill=steelblue31119180,line width=0.004pt] (axis cs:0.666666666666667,7.581900125) rectangle (axis cs:0.833333333333333,7.844831557);
\draw[draw=black,fill=steelblue75146195,line width=0.004pt] (axis cs:0.666666666666667,7.844831557) rectangle (axis cs:0.833333333333333,8.655820175);
\draw[draw=black,fill=cornflowerblue120173210,line width=0.004pt] (axis cs:0.666666666666667,8.655820175) rectangle (axis cs:0.833333333333333,10.813991909);
\draw[draw=black,fill=lavender210227240,line width=0.004pt,postaction={pattern=crosshatch}] (axis cs:1.66666666666667,1e-16) rectangle (axis cs:1.83333333333333,1.492307842);
\draw[draw=black,fill=steelblue31119180,line width=0.004pt,postaction={pattern=north east lines}] (axis cs:1.66666666666667,1.492307842) rectangle (axis cs:1.83333333333333,1.754679162);
\draw[draw=black,fill=steelblue75146195,line width=0.004pt,postaction={pattern=north east lines}] (axis cs:1.66666666666667,1.754679162) rectangle (axis cs:1.83333333333333,2.350314721);
\draw[draw=black,fill=cornflowerblue120173210,line width=0.004pt,postaction={pattern=north east lines}] (axis cs:1.66666666666667,2.350314721) rectangle (axis cs:1.83333333333333,3.882714663);
\draw[draw=black,fill=steelblue31119180,line width=0.004pt] (axis cs:1.66666666666667,3.882714663) rectangle (axis cs:1.83333333333333,4.0699330995);
\draw[draw=black,fill=steelblue75146195,line width=0.004pt] (axis cs:1.66666666666667,4.0699330995) rectangle (axis cs:1.83333333333333,4.4908144945);
\draw[draw=black,fill=cornflowerblue120173210,line width=0.004pt] (axis cs:1.66666666666667,4.4908144945) rectangle (axis cs:1.83333333333333,5.80342419);
\draw[draw=black,fill=lavender210227240,line width=0.004pt,postaction={pattern=crosshatch}] (axis cs:2.66666666666667,1e-16) rectangle (axis cs:2.83333333333333,0.748641443);
\draw[draw=black,fill=steelblue31119180,line width=0.004pt,postaction={pattern=north east lines}] (axis cs:2.66666666666667,0.748641443) rectangle (axis cs:2.83333333333333,0.878428151);
\draw[draw=black,fill=steelblue75146195,line width=0.004pt,postaction={pattern=north east lines}] (axis cs:2.66666666666667,0.878428151) rectangle (axis cs:2.83333333333333,1.1708924);
\draw[draw=black,fill=cornflowerblue120173210,line width=0.004pt,postaction={pattern=north east lines}] (axis cs:2.66666666666667,1.1708924) rectangle (axis cs:2.83333333333333,1.8088900225);
\draw[draw=black,fill=steelblue31119180,line width=0.004pt] (axis cs:2.66666666666667,1.8088900225) rectangle (axis cs:2.83333333333333,1.9085964585);
\draw[draw=black,fill=steelblue75146195,line width=0.004pt] (axis cs:2.66666666666667,1.9085964585) rectangle (axis cs:2.83333333333333,2.127708781);
\draw[draw=black,fill=cornflowerblue120173210,line width=0.004pt] (axis cs:2.66666666666667,2.127708781) rectangle (axis cs:2.83333333333333,2.863012497);
\draw[draw=black,fill=lavender210227240,line width=0.004pt,postaction={pattern=crosshatch}] (axis cs:3.66666666666667,1e-16) rectangle (axis cs:3.83333333333333,0.3740410855);
\draw[draw=black,fill=steelblue31119180,line width=0.004pt,postaction={pattern=north east lines}] (axis cs:3.66666666666667,0.3740410855) rectangle (axis cs:3.83333333333333,0.4377723115);
\draw[draw=black,fill=steelblue75146195,line width=0.004pt,postaction={pattern=north east lines}] (axis cs:3.66666666666667,0.4377723115) rectangle (axis cs:3.83333333333333,0.5777579715);
\draw[draw=black,fill=cornflowerblue120173210,line width=0.004pt,postaction={pattern=north east lines}] (axis cs:3.66666666666667,0.5777579715) rectangle (axis cs:3.83333333333333,0.8620991375);
\draw[draw=black,fill=steelblue31119180,line width=0.004pt] (axis cs:3.66666666666667,0.8620991375) rectangle (axis cs:3.83333333333333,0.9125253145);
\draw[draw=black,fill=steelblue75146195,line width=0.004pt] (axis cs:3.66666666666667,0.9125253145) rectangle (axis cs:3.83333333333333,1.031958538);
\draw[draw=black,fill=cornflowerblue120173210,line width=0.004pt] (axis cs:3.66666666666667,1.031958538) rectangle (axis cs:3.83333333333333,1.426380573);
\draw[draw=black,fill=lavender210227240,line width=0.004pt,postaction={pattern=crosshatch}] (axis cs:4.66666666666667,1e-16) rectangle (axis cs:4.83333333333333,0.1853601435);
\draw[draw=black,fill=steelblue31119180,line width=0.004pt,postaction={pattern=north east lines}] (axis cs:4.66666666666667,0.1853601435) rectangle (axis cs:4.83333333333333,0.214705524);
\draw[draw=black,fill=steelblue75146195,line width=0.004pt,postaction={pattern=north east lines}] (axis cs:4.66666666666667,0.214705524) rectangle (axis cs:4.83333333333333,0.281794723);
\draw[draw=black,fill=cornflowerblue120173210,line width=0.004pt,postaction={pattern=north east lines}] (axis cs:4.66666666666667,0.281794723) rectangle (axis cs:4.83333333333333,0.41921316);
\draw[draw=black,fill=steelblue31119180,line width=0.004pt] (axis cs:4.66666666666667,0.41921316) rectangle (axis cs:4.83333333333333,0.4462214525);
\draw[draw=black,fill=steelblue75146195,line width=0.004pt] (axis cs:4.66666666666667,0.4462214525) rectangle (axis cs:4.83333333333333,0.537727267);
\draw[draw=black,fill=cornflowerblue120173210,line width=0.004pt] (axis cs:4.66666666666667,0.537727267) rectangle (axis cs:4.83333333333333,0.728101596);
\draw[draw=black,fill=lavender210227240,line width=0.004pt,postaction={pattern=crosshatch}] (axis cs:5.66666666666667,1e-16) rectangle (axis cs:5.83333333333333,0.0892128465000001);
\draw[draw=black,fill=steelblue31119180,line width=0.004pt,postaction={pattern=north east lines}] (axis cs:5.66666666666667,0.0892128465000001) rectangle (axis cs:5.83333333333333,0.0985508950000001);
\draw[draw=black,fill=steelblue75146195,line width=0.004pt,postaction={pattern=north east lines}] (axis cs:5.66666666666667,0.0985508950000001) rectangle (axis cs:5.83333333333333,0.1267044825);
\draw[draw=black,fill=cornflowerblue120173210,line width=0.004pt,postaction={pattern=north east lines}] (axis cs:5.66666666666667,0.1267044825) rectangle (axis cs:5.83333333333333,0.1869367305);
\draw[draw=black,fill=steelblue31119180,line width=0.004pt] (axis cs:5.66666666666667,0.1869367305) rectangle (axis cs:5.83333333333333,0.2011060225);
\draw[draw=black,fill=steelblue75146195,line width=0.004pt] (axis cs:5.66666666666667,0.2011060225) rectangle (axis cs:5.83333333333333,0.2611023445);
\draw[draw=black,fill=cornflowerblue120173210,line width=0.004pt] (axis cs:5.66666666666667,0.2611023445) rectangle (axis cs:5.83333333333333,0.368065898);
\draw[draw=black,fill=lavender210227240,line width=0.004pt,postaction={pattern=crosshatch}] (axis cs:6.66666666666667,1e-16) rectangle (axis cs:6.83333333333333,0.0412305935000001);
\draw[draw=black,fill=steelblue31119180,line width=0.004pt,postaction={pattern=north east lines}] (axis cs:6.66666666666667,0.0412305935000001) rectangle (axis cs:6.83333333333333,0.0441698115000001);
\draw[draw=black,fill=steelblue75146195,line width=0.004pt,postaction={pattern=north east lines}] (axis cs:6.66666666666667,0.0441698115000001) rectangle (axis cs:6.83333333333333,0.0531962150000001);
\draw[draw=black,fill=cornflowerblue120173210,line width=0.004pt,postaction={pattern=north east lines}] (axis cs:6.66666666666667,0.0531962150000001) rectangle (axis cs:6.83333333333333,0.0789337160000001);
\draw[draw=black,fill=steelblue31119180,line width=0.004pt] (axis cs:6.66666666666667,0.0789337160000001) rectangle (axis cs:6.83333333333333,0.0896365765000001);
\draw[draw=black,fill=steelblue75146195,line width=0.004pt] (axis cs:6.66666666666667,0.0896365765000001) rectangle (axis cs:6.83333333333333,0.157714074);
\draw[draw=black,fill=cornflowerblue120173210,line width=0.004pt] (axis cs:6.66666666666667,0.157714074) rectangle (axis cs:6.83333333333333,0.216520988);
\draw[draw=black,fill=lavender210227240,line width=0.004pt,postaction={pattern=crosshatch}] (axis cs:7.66666666666667,1e-16) rectangle (axis cs:7.83333333333333,0.0127078505000001);
\draw[draw=black,fill=steelblue31119180,line width=0.004pt,postaction={pattern=north east lines}] (axis cs:7.66666666666667,0.0127078505000001) rectangle (axis cs:7.83333333333333,0.0134287725000001);
\draw[draw=black,fill=steelblue75146195,line width=0.004pt,postaction={pattern=north east lines}] (axis cs:7.66666666666667,0.0134287725000001) rectangle (axis cs:7.83333333333333,0.0166784530000001);
\draw[draw=black,fill=cornflowerblue120173210,line width=0.004pt,postaction={pattern=north east lines}] (axis cs:7.66666666666667,0.0166784530000001) rectangle (axis cs:7.83333333333333,0.0208570050000001);
\draw[draw=black,fill=steelblue31119180,line width=0.004pt] (axis cs:7.66666666666667,0.0208570050000001) rectangle (axis cs:7.83333333333333,0.0269110565000001);
\draw[draw=black,fill=steelblue75146195,line width=0.004pt] (axis cs:7.66666666666667,0.0269110565000001) rectangle (axis cs:7.83333333333333,0.0664566610000001);
\draw[draw=black,fill=cornflowerblue120173210,line width=0.004pt] (axis cs:7.66666666666667,0.0664566610000001) rectangle (axis cs:7.83333333333333,0.0993471540000001);
\draw[draw=black,fill=lavender210227240,line width=0.004pt,postaction={pattern=crosshatch}] (axis cs:8.66666666666667,1e-16) rectangle (axis cs:8.83333333333333,0.00761424400000011);
\draw[draw=black,fill=steelblue31119180,line width=0.004pt,postaction={pattern=north east lines}] (axis cs:8.66666666666667,0.00761424400000011) rectangle (axis cs:8.83333333333333,0.00791782100000011);
\draw[draw=black,fill=steelblue75146195,line width=0.004pt,postaction={pattern=north east lines}] (axis cs:8.66666666666667,0.00791782100000011) rectangle (axis cs:8.83333333333333,0.0098149215000001);
\draw[draw=black,fill=cornflowerblue120173210,line width=0.004pt,postaction={pattern=north east lines}] (axis cs:8.66666666666667,0.0098149215000001) rectangle (axis cs:8.83333333333333,0.0117911510000001);
\draw[draw=black,fill=steelblue31119180,line width=0.004pt] (axis cs:8.66666666666667,0.0117911510000001) rectangle (axis cs:8.83333333333333,0.0162263730000001);
\draw[draw=black,fill=steelblue75146195,line width=0.004pt] (axis cs:8.66666666666667,0.0162263730000001) rectangle (axis cs:8.83333333333333,0.0456256375000001);
\draw[draw=black,fill=cornflowerblue120173210,line width=0.004pt] (axis cs:8.66666666666667,0.0456256375000001) rectangle (axis cs:8.83333333333333,0.0839018975000001);
\draw[draw=black,fill=gainsboro223228215,line width=0.004pt,postaction={pattern=crosshatch}] (axis cs:-0.0833333333333333,1e-16) rectangle (axis cs:0.0833333333333333,5.9690528665);
\draw[draw=black,fill=darkolivegreen9912157,line width=0.004pt,postaction={pattern=north east lines}] (axis cs:-0.0833333333333333,5.9690528665) rectangle (axis cs:0.0833333333333333,6.209761352);
\draw[draw=black,fill=gray13014796,line width=0.004pt,postaction={pattern=north east lines}] (axis cs:-0.0833333333333333,6.209761352) rectangle (axis cs:0.0833333333333333,7.3571658515);
\draw[draw=black,fill=darkseagreen161174136,line width=0.004pt,postaction={pattern=north east lines}] (axis cs:-0.0833333333333333,7.3571658515) rectangle (axis cs:0.0833333333333333,9.854755044);
\draw[draw=black,fill=darkolivegreen9912157,line width=0.004pt] (axis cs:-0.0833333333333333,9.854755044) rectangle (axis cs:0.0833333333333333,10.5987119065);
\draw[draw=black,fill=gray13014796,line width=0.004pt] (axis cs:-0.0833333333333333,10.5987119065) rectangle (axis cs:0.0833333333333333,12.682746721);
\draw[draw=black,fill=darkseagreen161174136,line width=0.004pt] (axis cs:-0.0833333333333333,12.682746721) rectangle (axis cs:0.0833333333333333,16.7001163725);
\draw[draw=black,fill=gainsboro223228215,line width=0.004pt,postaction={pattern=crosshatch}] (axis cs:0.916666666666667,1e-16) rectangle (axis cs:1.08333333333333,2.9961217385);
\draw[draw=black,fill=darkolivegreen9912157,line width=0.004pt,postaction={pattern=north east lines}] (axis cs:0.916666666666667,2.9961217385) rectangle (axis cs:1.08333333333333,3.1251621475);
\draw[draw=black,fill=gray13014796,line width=0.004pt,postaction={pattern=north east lines}] (axis cs:0.916666666666667,3.1251621475) rectangle (axis cs:1.08333333333333,3.5710042275);
\draw[draw=black,fill=darkseagreen161174136,line width=0.004pt,postaction={pattern=north east lines}] (axis cs:0.916666666666667,3.5710042275) rectangle (axis cs:1.08333333333333,4.70317304);
\draw[draw=black,fill=darkolivegreen9912157,line width=0.004pt] (axis cs:0.916666666666667,4.70317304) rectangle (axis cs:1.08333333333333,5.065567321);
\draw[draw=black,fill=gray13014796,line width=0.004pt] (axis cs:0.916666666666667,5.065567321) rectangle (axis cs:1.08333333333333,6.2870256145);
\draw[draw=black,fill=darkseagreen161174136,line width=0.004pt] (axis cs:0.916666666666667,6.2870256145) rectangle (axis cs:1.08333333333333,8.896939861);
\draw[draw=black,fill=gainsboro223228215,line width=0.004pt,postaction={pattern=crosshatch}] (axis cs:1.91666666666667,1e-16) rectangle (axis cs:2.08333333333333,1.51236762);
\draw[draw=black,fill=darkolivegreen9912157,line width=0.004pt,postaction={pattern=north east lines}] (axis cs:1.91666666666667,1.51236762) rectangle (axis cs:2.08333333333333,1.5785207);
\draw[draw=black,fill=gray13014796,line width=0.004pt,postaction={pattern=north east lines}] (axis cs:1.91666666666667,1.5785207) rectangle (axis cs:2.08333333333333,1.789205846);
\draw[draw=black,fill=darkseagreen161174136,line width=0.004pt,postaction={pattern=north east lines}] (axis cs:1.91666666666667,1.789205846) rectangle (axis cs:2.08333333333333,2.29058932);
\draw[draw=black,fill=darkolivegreen9912157,line width=0.004pt] (axis cs:1.91666666666667,2.29058932) rectangle (axis cs:2.08333333333333,2.4950383165);
\draw[draw=black,fill=gray13014796,line width=0.004pt] (axis cs:1.91666666666667,2.4950383165) rectangle (axis cs:2.08333333333333,3.1356633935);
\draw[draw=black,fill=darkseagreen161174136,line width=0.004pt] (axis cs:1.91666666666667,3.1356633935) rectangle (axis cs:2.08333333333333,4.766389787);
\draw[draw=black,fill=gainsboro223228215,line width=0.004pt,postaction={pattern=crosshatch}] (axis cs:2.91666666666667,1e-16) rectangle (axis cs:3.08333333333333,0.7435030495);
\draw[draw=black,fill=darkolivegreen9912157,line width=0.004pt,postaction={pattern=north east lines}] (axis cs:2.91666666666667,0.7435030495) rectangle (axis cs:3.08333333333333,0.776156979);
\draw[draw=black,fill=gray13014796,line width=0.004pt,postaction={pattern=north east lines}] (axis cs:2.91666666666667,0.776156979) rectangle (axis cs:3.08333333333333,0.8789627965);
\draw[draw=black,fill=darkseagreen161174136,line width=0.004pt,postaction={pattern=north east lines}] (axis cs:2.91666666666667,0.8789627965) rectangle (axis cs:3.08333333333333,1.026516546);
\draw[draw=black,fill=darkolivegreen9912157,line width=0.004pt] (axis cs:2.91666666666667,1.026516546) rectangle (axis cs:3.08333333333333,1.1288557695);
\draw[draw=black,fill=gray13014796,line width=0.004pt] (axis cs:2.91666666666667,1.1288557695) rectangle (axis cs:3.08333333333333,1.410585457);
\draw[draw=black,fill=darkseagreen161174136,line width=0.004pt] (axis cs:2.91666666666667,1.410585457) rectangle (axis cs:3.08333333333333,2.390334494);
\draw[draw=black,fill=gainsboro223228215,line width=0.004pt,postaction={pattern=crosshatch}] (axis cs:3.91666666666667,1e-16) rectangle (axis cs:4.08333333333333,0.3678282645);
\draw[draw=black,fill=darkolivegreen9912157,line width=0.004pt,postaction={pattern=north east lines}] (axis cs:3.91666666666667,0.3678282645) rectangle (axis cs:4.08333333333333,0.3840345415);
\draw[draw=black,fill=gray13014796,line width=0.004pt,postaction={pattern=north east lines}] (axis cs:3.91666666666667,0.3840345415) rectangle (axis cs:4.08333333333333,0.431247587);
\draw[draw=black,fill=darkseagreen161174136,line width=0.004pt,postaction={pattern=north east lines}] (axis cs:3.91666666666667,0.431247587) rectangle (axis cs:4.08333333333333,0.4976388655);
\draw[draw=black,fill=darkolivegreen9912157,line width=0.004pt] (axis cs:3.91666666666667,0.4976388655) rectangle (axis cs:4.08333333333333,0.5479691645);
\draw[draw=black,fill=gray13014796,line width=0.004pt] (axis cs:3.91666666666667,0.5479691645) rectangle (axis cs:4.08333333333333,0.732306017);
\draw[draw=black,fill=darkseagreen161174136,line width=0.004pt] (axis cs:3.91666666666667,0.732306017) rectangle (axis cs:4.08333333333333,1.27919451);
\draw[draw=black,fill=gainsboro223228215,line width=0.004pt,postaction={pattern=crosshatch}] (axis cs:4.91666666666667,1e-16) rectangle (axis cs:5.08333333333333,0.178280355);
\draw[draw=black,fill=darkolivegreen9912157,line width=0.004pt,postaction={pattern=north east lines}] (axis cs:4.91666666666667,0.178280355) rectangle (axis cs:5.08333333333333,0.1852179625);
\draw[draw=black,fill=gray13014796,line width=0.004pt,postaction={pattern=north east lines}] (axis cs:4.91666666666667,0.1852179625) rectangle (axis cs:5.08333333333333,0.2037443065);
\draw[draw=black,fill=darkseagreen161174136,line width=0.004pt,postaction={pattern=north east lines}] (axis cs:4.91666666666667,0.2037443065) rectangle (axis cs:5.08333333333333,0.2281671515);
\draw[draw=black,fill=darkolivegreen9912157,line width=0.004pt] (axis cs:4.91666666666667,0.2281671515) rectangle (axis cs:5.08333333333333,0.2545601285);
\draw[draw=black,fill=gray13014796,line width=0.004pt] (axis cs:4.91666666666667,0.2545601285) rectangle (axis cs:5.08333333333333,0.360111171);
\draw[draw=black,fill=darkseagreen161174136,line width=0.004pt] (axis cs:4.91666666666667,0.360111171) rectangle (axis cs:5.08333333333333,0.6603822205);
\draw[draw=black,fill=gainsboro223228215,line width=0.004pt,postaction={pattern=crosshatch}] (axis cs:5.91666666666667,1e-16) rectangle (axis cs:6.08333333333333,0.0818457455000001);
\draw[draw=black,fill=darkolivegreen9912157,line width=0.004pt,postaction={pattern=north east lines}] (axis cs:5.91666666666667,0.0818457455000001) rectangle (axis cs:6.08333333333333,0.0842201430000001);
\draw[draw=black,fill=gray13014796,line width=0.004pt,postaction={pattern=north east lines}] (axis cs:5.91666666666667,0.0842201430000001) rectangle (axis cs:6.08333333333333,0.0927952720000001);
\draw[draw=black,fill=darkseagreen161174136,line width=0.004pt,postaction={pattern=north east lines}] (axis cs:5.91666666666667,0.0927952720000001) rectangle (axis cs:6.08333333333333,0.1022303405);
\draw[draw=black,fill=darkolivegreen9912157,line width=0.004pt] (axis cs:5.91666666666667,0.1022303405) rectangle (axis cs:6.08333333333333,0.112997564);
\draw[draw=black,fill=gray13014796,line width=0.004pt] (axis cs:5.91666666666667,0.112997564) rectangle (axis cs:6.08333333333333,0.166074061);
\draw[draw=black,fill=darkseagreen161174136,line width=0.004pt] (axis cs:5.91666666666667,0.166074061) rectangle (axis cs:6.08333333333333,0.338279419);
\draw[draw=black,fill=gainsboro223228215,line width=0.004pt,postaction={pattern=crosshatch}] (axis cs:6.91666666666667,1e-16) rectangle (axis cs:7.08333333333333,0.0352255280000001);
\draw[draw=black,fill=darkolivegreen9912157,line width=0.004pt,postaction={pattern=north east lines}] (axis cs:6.91666666666667,0.0352255280000001) rectangle (axis cs:7.08333333333333,0.0354898105000001);
\draw[draw=black,fill=gray13014796,line width=0.004pt,postaction={pattern=north east lines}] (axis cs:6.91666666666667,0.0354898105000001) rectangle (axis cs:7.08333333333333,0.0385290470000001);
\draw[draw=black,fill=darkseagreen161174136,line width=0.004pt,postaction={pattern=north east lines}] (axis cs:6.91666666666667,0.0385290470000001) rectangle (axis cs:7.08333333333333,0.0414690295000001);
\draw[draw=black,fill=darkolivegreen9912157,line width=0.004pt] (axis cs:6.91666666666667,0.0414690295000001) rectangle (axis cs:7.08333333333333,0.0489434795000001);
\draw[draw=black,fill=gray13014796,line width=0.004pt] (axis cs:6.91666666666667,0.0489434795000001) rectangle (axis cs:7.08333333333333,0.1095095325);
\draw[draw=black,fill=darkseagreen161174136,line width=0.004pt] (axis cs:6.91666666666667,0.1095095325) rectangle (axis cs:7.08333333333333,0.2080653695);
\draw[draw=black,fill=gainsboro223228215,line width=0.004pt,postaction={pattern=crosshatch}] (axis cs:7.91666666666667,1e-16) rectangle (axis cs:8.08333333333333,0.0115961765000001);
\draw[draw=black,fill=darkolivegreen9912157,line width=0.004pt,postaction={pattern=north east lines}] (axis cs:7.91666666666667,0.0115961765000001) rectangle (axis cs:8.08333333333333,0.0117031425000001);
\draw[draw=black,fill=gray13014796,line width=0.004pt,postaction={pattern=north east lines}] (axis cs:7.91666666666667,0.0117031425000001) rectangle (axis cs:8.08333333333333,0.0132999135000001);
\draw[draw=black,fill=darkseagreen161174136,line width=0.004pt,postaction={pattern=north east lines}] (axis cs:7.91666666666667,0.0132999135000001) rectangle (axis cs:8.08333333333333,0.0141274690000001);
\draw[draw=black,fill=darkolivegreen9912157,line width=0.004pt] (axis cs:7.91666666666667,0.0141274690000001) rectangle (axis cs:8.08333333333333,0.0188138645000001);
\draw[draw=black,fill=gray13014796,line width=0.004pt] (axis cs:7.91666666666667,0.0188138645000001) rectangle (axis cs:8.08333333333333,0.0812500730000001);
\draw[draw=black,fill=darkseagreen161174136,line width=0.004pt] (axis cs:7.91666666666667,0.0812500730000001) rectangle (axis cs:8.08333333333333,0.1239894065);
\draw[draw=black,fill=gainsboro223228215,line width=0.004pt,postaction={pattern=crosshatch}] (axis cs:8.91666666666667,1e-16) rectangle (axis cs:9.08333333333333,0.00698967350000008);
\draw[draw=black,fill=darkolivegreen9912157,line width=0.004pt,postaction={pattern=north east lines}] (axis cs:8.91666666666667,0.00698967350000008) rectangle (axis cs:9.08333333333333,0.00705454700000008);
\draw[draw=black,fill=gray13014796,line width=0.004pt,postaction={pattern=north east lines}] (axis cs:8.91666666666667,0.00705454700000008) rectangle (axis cs:9.08333333333333,0.00822678650000008);
\draw[draw=black,fill=darkseagreen161174136,line width=0.004pt,postaction={pattern=north east lines}] (axis cs:8.91666666666667,0.00822678650000008) rectangle (axis cs:9.08333333333333,0.00866953850000008);
\draw[draw=black,fill=darkolivegreen9912157,line width=0.004pt] (axis cs:8.91666666666667,0.00866953850000008) rectangle (axis cs:9.08333333333333,0.0117902270000001);
\draw[draw=black,fill=gray13014796,line width=0.004pt] (axis cs:8.91666666666667,0.0117902270000001) rectangle (axis cs:9.08333333333333,0.0691779390000001);
\draw[draw=black,fill=darkseagreen161174136,line width=0.004pt] (axis cs:8.91666666666667,0.0691779390000001) rectangle (axis cs:9.08333333333333,0.1121216115);
\draw[draw=black,fill=bisque247223204,line width=0.004pt,postaction={pattern=crosshatch}] (axis cs:0.166666666666667,1e-16) rectangle (axis cs:0.333333333333333,6.0810810935);
\draw[draw=black,fill=chocolate217952,line width=0.004pt,postaction={pattern=north east lines}] (axis cs:0.166666666666667,6.0810810935) rectangle (axis cs:0.333333333333333,6.710027381);
\draw[draw=black,fill=peru22412652,line width=0.004pt,postaction={pattern=north east lines}] (axis cs:0.166666666666667,6.710027381) rectangle (axis cs:0.333333333333333,8.6466023945);
\draw[draw=black,fill=sandybrown232159103,line width=0.004pt,postaction={pattern=north east lines}] (axis cs:0.166666666666667,8.6466023945) rectangle (axis cs:0.333333333333333,12.8476190205);
\draw[draw=black,fill=chocolate217952,line width=0.004pt] (axis cs:0.166666666666667,12.8476190205) rectangle (axis cs:0.333333333333333,13.8636372325);
\draw[draw=black,fill=peru22412652,line width=0.004pt] (axis cs:0.166666666666667,13.8636372325) rectangle (axis cs:0.333333333333333,15.9496005585);
\draw[draw=black,fill=sandybrown232159103,line width=0.004pt] (axis cs:0.166666666666667,15.9496005585) rectangle (axis cs:0.333333333333333,20.3871096535);
\draw[draw=black,fill=bisque247223204,line width=0.004pt,postaction={pattern=crosshatch}] (axis cs:1.16666666666667,1e-16) rectangle (axis cs:1.33333333333333,3.0374713405);
\draw[draw=black,fill=chocolate217952,line width=0.004pt,postaction={pattern=north east lines}] (axis cs:1.16666666666667,3.0374713405) rectangle (axis cs:1.33333333333333,3.3369468755);
\draw[draw=black,fill=peru22412652,line width=0.004pt,postaction={pattern=north east lines}] (axis cs:1.16666666666667,3.3369468755) rectangle (axis cs:1.33333333333333,4.0922097275);
\draw[draw=black,fill=sandybrown232159103,line width=0.004pt,postaction={pattern=north east lines}] (axis cs:1.16666666666667,4.0922097275) rectangle (axis cs:1.33333333333333,5.7079680555);
\draw[draw=black,fill=chocolate217952,line width=0.004pt] (axis cs:1.16666666666667,5.7079680555) rectangle (axis cs:1.33333333333333,6.1977963845);
\draw[draw=black,fill=peru22412652,line width=0.004pt] (axis cs:1.16666666666667,6.1977963845) rectangle (axis cs:1.33333333333333,7.322402572);
\draw[draw=black,fill=sandybrown232159103,line width=0.004pt] (axis cs:1.16666666666667,7.322402572) rectangle (axis cs:1.33333333333333,9.90559118);
\draw[draw=black,fill=bisque247223204,line width=0.004pt,postaction={pattern=crosshatch}] (axis cs:2.16666666666667,1e-16) rectangle (axis cs:2.33333333333333,1.513989342);
\draw[draw=black,fill=chocolate217952,line width=0.004pt,postaction={pattern=north east lines}] (axis cs:2.16666666666667,1.513989342) rectangle (axis cs:2.33333333333333,1.663215041);
\draw[draw=black,fill=peru22412652,line width=0.004pt,postaction={pattern=north east lines}] (axis cs:2.16666666666667,1.663215041) rectangle (axis cs:2.33333333333333,2.030145667);
\draw[draw=black,fill=sandybrown232159103,line width=0.004pt,postaction={pattern=north east lines}] (axis cs:2.16666666666667,2.030145667) rectangle (axis cs:2.33333333333333,2.7510111885);
\draw[draw=black,fill=chocolate217952,line width=0.004pt] (axis cs:2.16666666666667,2.7510111885) rectangle (axis cs:2.33333333333333,3.0154451025);
\draw[draw=black,fill=peru22412652,line width=0.004pt] (axis cs:2.16666666666667,3.0154451025) rectangle (axis cs:2.33333333333333,3.576110293);
\draw[draw=black,fill=sandybrown232159103,line width=0.004pt] (axis cs:2.16666666666667,3.576110293) rectangle (axis cs:2.33333333333333,5.1052968705);
\draw[draw=black,fill=bisque247223204,line width=0.004pt,postaction={pattern=crosshatch}] (axis cs:3.16666666666667,1e-16) rectangle (axis cs:3.33333333333333,0.751727895);
\draw[draw=black,fill=chocolate217952,line width=0.004pt,postaction={pattern=north east lines}] (axis cs:3.16666666666667,0.751727895) rectangle (axis cs:3.33333333333333,0.824213351);
\draw[draw=black,fill=peru22412652,line width=0.004pt,postaction={pattern=north east lines}] (axis cs:3.16666666666667,0.824213351) rectangle (axis cs:3.33333333333333,1.004551926);
\draw[draw=black,fill=sandybrown232159103,line width=0.004pt,postaction={pattern=north east lines}] (axis cs:3.16666666666667,1.004551926) rectangle (axis cs:3.33333333333333,1.315362827);
\draw[draw=black,fill=chocolate217952,line width=0.004pt] (axis cs:3.16666666666667,1.315362827) rectangle (axis cs:3.33333333333333,1.4489065335);
\draw[draw=black,fill=peru22412652,line width=0.004pt] (axis cs:3.16666666666667,1.4489065335) rectangle (axis cs:3.33333333333333,1.722233609);
\draw[draw=black,fill=sandybrown232159103,line width=0.004pt] (axis cs:3.16666666666667,1.722233609) rectangle (axis cs:3.33333333333333,2.546875466);
\draw[draw=black,fill=bisque247223204,line width=0.004pt,postaction={pattern=crosshatch}] (axis cs:4.16666666666667,1e-16) rectangle (axis cs:4.33333333333333,0.3703436255);
\draw[draw=black,fill=chocolate217952,line width=0.004pt,postaction={pattern=north east lines}] (axis cs:4.16666666666667,0.3703436255) rectangle (axis cs:4.33333333333333,0.4053181905);
\draw[draw=black,fill=peru22412652,line width=0.004pt,postaction={pattern=north east lines}] (axis cs:4.16666666666667,0.4053181905) rectangle (axis cs:4.33333333333333,0.4924839595);
\draw[draw=black,fill=sandybrown232159103,line width=0.004pt,postaction={pattern=north east lines}] (axis cs:4.16666666666667,0.4924839595) rectangle (axis cs:4.33333333333333,0.64925635);
\draw[draw=black,fill=chocolate217952,line width=0.004pt] (axis cs:4.16666666666667,0.64925635) rectangle (axis cs:4.33333333333333,0.7121878575);
\draw[draw=black,fill=peru22412652,line width=0.004pt] (axis cs:4.16666666666667,0.7121878575) rectangle (axis cs:4.33333333333333,0.853357638);
\draw[draw=black,fill=sandybrown232159103,line width=0.004pt] (axis cs:4.16666666666667,0.853357638) rectangle (axis cs:4.33333333333333,1.287791312);
\draw[draw=black,fill=bisque247223204,line width=0.004pt,postaction={pattern=crosshatch}] (axis cs:5.16666666666667,1e-16) rectangle (axis cs:5.33333333333333,0.180881929);
\draw[draw=black,fill=chocolate217952,line width=0.004pt,postaction={pattern=north east lines}] (axis cs:5.16666666666667,0.180881929) rectangle (axis cs:5.33333333333333,0.1947692465);
\draw[draw=black,fill=peru22412652,line width=0.004pt,postaction={pattern=north east lines}] (axis cs:5.16666666666667,0.1947692465) rectangle (axis cs:5.33333333333333,0.2318932095);
\draw[draw=black,fill=sandybrown232159103,line width=0.004pt,postaction={pattern=north east lines}] (axis cs:5.16666666666667,0.2318932095) rectangle (axis cs:5.33333333333333,0.3097899505);
\draw[draw=black,fill=chocolate217952,line width=0.004pt] (axis cs:5.16666666666667,0.3097899505) rectangle (axis cs:5.33333333333333,0.3452060275);
\draw[draw=black,fill=peru22412652,line width=0.004pt] (axis cs:5.16666666666667,0.3452060275) rectangle (axis cs:5.33333333333333,0.44858612);
\draw[draw=black,fill=sandybrown232159103,line width=0.004pt] (axis cs:5.16666666666667,0.44858612) rectangle (axis cs:5.33333333333333,0.6738959055);
\draw[draw=black,fill=bisque247223204,line width=0.004pt,postaction={pattern=crosshatch}] (axis cs:6.16666666666667,1e-16) rectangle (axis cs:6.33333333333333,0.0850633510000001);
\draw[draw=black,fill=chocolate217952,line width=0.004pt,postaction={pattern=north east lines}] (axis cs:6.16666666666667,0.0850633510000001) rectangle (axis cs:6.33333333333333,0.0908847115000001);
\draw[draw=black,fill=peru22412652,line width=0.004pt,postaction={pattern=north east lines}] (axis cs:6.16666666666667,0.0908847115000001) rectangle (axis cs:6.33333333333333,0.1066026815);
\draw[draw=black,fill=sandybrown232159103,line width=0.004pt,postaction={pattern=north east lines}] (axis cs:6.16666666666667,0.1066026815) rectangle (axis cs:6.33333333333333,0.136917829);
\draw[draw=black,fill=chocolate217952,line width=0.004pt] (axis cs:6.16666666666667,0.136917829) rectangle (axis cs:6.33333333333333,0.15235187);
\draw[draw=black,fill=peru22412652,line width=0.004pt] (axis cs:6.16666666666667,0.15235187) rectangle (axis cs:6.33333333333333,0.209611773);
\draw[draw=black,fill=sandybrown232159103,line width=0.004pt] (axis cs:6.16666666666667,0.209611773) rectangle (axis cs:6.33333333333333,0.310492818);
\draw[draw=black,fill=bisque247223204,line width=0.004pt,postaction={pattern=crosshatch}] (axis cs:7.16666666666667,1e-16) rectangle (axis cs:7.33333333333333,0.0392894775000001);
\draw[draw=black,fill=chocolate217952,line width=0.004pt,postaction={pattern=north east lines}] (axis cs:7.16666666666667,0.0392894775000001) rectangle (axis cs:7.33333333333333,0.0410394935000001);
\draw[draw=black,fill=peru22412652,line width=0.004pt,postaction={pattern=north east lines}] (axis cs:7.16666666666667,0.0410394935000001) rectangle (axis cs:7.33333333333333,0.0467122960000001);
\draw[draw=black,fill=sandybrown232159103,line width=0.004pt,postaction={pattern=north east lines}] (axis cs:7.16666666666667,0.0467122960000001) rectangle (axis cs:7.33333333333333,0.0599331675000001);
\draw[draw=black,fill=chocolate217952,line width=0.004pt] (axis cs:7.16666666666667,0.0599331675000001) rectangle (axis cs:7.33333333333333,0.0706164420000001);
\draw[draw=black,fill=peru22412652,line width=0.004pt] (axis cs:7.16666666666667,0.0706164420000001) rectangle (axis cs:7.33333333333333,0.12565855);
\draw[draw=black,fill=sandybrown232159103,line width=0.004pt] (axis cs:7.16666666666667,0.12565855) rectangle (axis cs:7.33333333333333,0.1907607085);
\draw[draw=black,fill=bisque247223204,line width=0.004pt,postaction={pattern=crosshatch}] (axis cs:8.16666666666667,1e-16) rectangle (axis cs:8.33333333333333,0.0124351110000001);
\draw[draw=black,fill=chocolate217952,line width=0.004pt,postaction={pattern=north east lines}] (axis cs:8.16666666666667,0.0124351110000001) rectangle (axis cs:8.33333333333333,0.0129829290000001);
\draw[draw=black,fill=peru22412652,line width=0.004pt,postaction={pattern=north east lines}] (axis cs:8.16666666666667,0.0129829290000001) rectangle (axis cs:8.33333333333333,0.0160062895000001);
\draw[draw=black,fill=sandybrown232159103,line width=0.004pt,postaction={pattern=north east lines}] (axis cs:8.16666666666667,0.0160062895000001) rectangle (axis cs:8.33333333333333,0.0188966990000001);
\draw[draw=black,fill=chocolate217952,line width=0.004pt] (axis cs:8.16666666666667,0.0188966990000001) rectangle (axis cs:8.33333333333333,0.0243579305000001);
\draw[draw=black,fill=peru22412652,line width=0.004pt] (axis cs:8.16666666666667,0.0243579305000001) rectangle (axis cs:8.33333333333333,0.0888870425000001);
\draw[draw=black,fill=sandybrown232159103,line width=0.004pt] (axis cs:8.16666666666667,0.0888870425000001) rectangle (axis cs:8.33333333333333,0.114663191);
\draw[draw=black,fill=bisque247223204,line width=0.004pt,postaction={pattern=crosshatch}] (axis cs:9.16666666666667,1e-16) rectangle (axis cs:9.33333333333333,0.0073311515000001);
\draw[draw=black,fill=chocolate217952,line width=0.004pt,postaction={pattern=north east lines}] (axis cs:9.16666666666667,0.0073311515000001) rectangle (axis cs:9.33333333333333,0.0075810660000001);
\draw[draw=black,fill=peru22412652,line width=0.004pt,postaction={pattern=north east lines}] (axis cs:9.16666666666667,0.0075810660000001) rectangle (axis cs:9.33333333333333,0.0096996060000001);
\draw[draw=black,fill=sandybrown232159103,line width=0.004pt,postaction={pattern=north east lines}] (axis cs:9.16666666666667,0.0096996060000001) rectangle (axis cs:9.33333333333333,0.0109075310000001);
\draw[draw=black,fill=chocolate217952,line width=0.004pt] (axis cs:9.16666666666667,0.0109075310000001) rectangle (axis cs:9.33333333333333,0.0148838510000001);
\draw[draw=black,fill=peru22412652,line width=0.004pt] (axis cs:9.16666666666667,0.0148838510000001) rectangle (axis cs:9.33333333333333,0.0706647790000001);
\draw[draw=black,fill=sandybrown232159103,line width=0.004pt] (axis cs:9.16666666666667,0.0706647790000001) rectangle (axis cs:9.33333333333333,0.107268202);
\end{axis}

\end{tikzpicture}

%% file: figures/python_fig/mlx/unbounded/strong-eff.tex
\begin{tikzpicture}

\definecolor{chocolate217952}{RGB}{217,95,2}
\definecolor{darkgray}{RGB}{169,169,169}
\definecolor{darkgray150}{RGB}{150,150,150}
\definecolor{darkgray176}{RGB}{176,176,176}
\definecolor{darkolivegreen9912157}{RGB}{99,121,57}
\definecolor{dimgray99}{RGB}{99,99,99}
\definecolor{silver189}{RGB}{189,189,189}
\definecolor{steelblue31119180}{RGB}{31,119,180}

\begin{axis}[
axis y line =left, axis x line =bottom, axis x line shift=2.5pt, axis line style ={-},
log basis x={10},
log basis y={10},
minor xtick={},
minor ytick={},
tick pos=left,
x grid style={darkgray176},
x grid style={draw=none},
xlabel near ticks,
xlabel={\(\displaystyle r\) = Number of nodes},
xmajorgrids,
xmin=0.739118043551982, xmax=571.491933778359,
xmode=log,
xtick style={color=black},
xtick={1,2,4,8,16,32,64,128,256,384},
xticklabels={$1$,$2$,$4$,$8$,$16$,$32$,$64$,$128$,$256$,\quad $384$},
y axis line style={draw=none},
y grid style={darkgray176},
y grid style={dotted},
ylabel near ticks,
ylabel={\(\displaystyle S_p\)},
ymajorgrids,
ymin=0.833333333333333, ymax=537.6,
ymode=log,
ytick style={color=darkgray},
ytick={1,2,4,8,16,32,64,128,256,512},
yticklabels={$1$,$2$,$4$,$8$,$16$,$32$,$64$,$128$,$256$,$512$}
]
\addplot [very thin, black]
table {%
1 1
2 2
4 4
8 8
16 16
32 32
64 64
128 128
256 256
384 384
422.4 422.4
};
\addplot [very thin, silver189]
table {%
1 1
2 1.98019801980198
4 3.88349514563107
8 7.47663551401869
16 13.9130434782609
32 24.4274809160305
64 39.2638036809816
128 56.3876651982379
256 72.112676056338
384 79.503105590062
422.4 81.012658227848
};
\addplot [very thin, darkgray150]
table {%
1 1
2 1.99004975124378
4 3.94088669950739
8 7.72946859903382
16 14.8837209302326
32 27.7056277056277
64 48.6692015209125
128 78.2874617737003
256 112.527472527472
384 131.732418524871
422.4 135.951078210492
};
\addplot [very thin, dimgray99]
table {%
1 1
2 1.99600798403194
4 3.97614314115308
8 7.88954635108481
16 15.5339805825243
32 30.1318267419962
64 56.8383658969805
128 102.07336523126
256 169.53642384106
384 217.440543601359
422.4 229.216409811157
};
\addplot [semithick, steelblue31119180, mark=*, mark size=1.5, mark options={solid}]
table {%
1 1
2 2.02332618445831
4 3.77022810528003
8 7.642381236871
16 15.3396880202742
32 30.0510713178
64 59.4465097334282
128 101.05363544711
256 220.240158948086
384 260.783529812303
};
\addplot [very thin, black]
table {%
1 1
2 2
4 4
8 8
16 16
32 32
64 64
128 128
256 256
384 384
422.4 422.4
};
\addplot [very thin, silver189]
table {%
1 1
2 1.98019801980198
4 3.88349514563107
8 7.47663551401869
16 13.9130434782609
32 24.4274809160305
64 39.2638036809816
128 56.3876651982379
256 72.112676056338
384 79.503105590062
422.4 81.012658227848
};
\addplot [very thin, darkgray150]
table {%
1 1
2 1.99004975124378
4 3.94088669950739
8 7.72946859903382
16 14.8837209302326
32 27.7056277056277
64 48.6692015209125
128 78.2874617737003
256 112.527472527472
384 131.732418524871
422.4 135.951078210492
};
\addplot [very thin, dimgray99]
table {%
1 1
2 1.99600798403194
4 3.97614314115308
8 7.88954635108481
16 15.5339805825243
32 30.1318267419962
64 56.8383658969805
128 102.07336523126
256 169.53642384106
384 217.440543601359
422.4 229.216409811157
};
\addplot [semithick, darkolivegreen9912157, mark=*, mark size=1.5, mark options={solid}]
table {%
1 1
2 1.87706297147241
4 3.50372443689948
8 6.98651858742745
16 13.0551813988789
32 25.2885614634745
64 49.3678167648148
128 80.2637960013812
256 134.689864593392
384 148.946453311546
};
\addplot [very thin, black]
table {%
1 1
2 2
4 4
8 8
16 16
32 32
64 64
128 128
256 256
384 384
422.4 422.4
};
\addplot [very thin, silver189]
table {%
1 1
2 1.98019801980198
4 3.88349514563107
8 7.47663551401869
16 13.9130434782609
32 24.4274809160305
64 39.2638036809816
128 56.3876651982379
256 72.112676056338
384 79.503105590062
422.4 81.012658227848
};
\addplot [very thin, darkgray150]
table {%
1 1
2 1.99004975124378
4 3.94088669950739
8 7.72946859903382
16 14.8837209302326
32 27.7056277056277
64 48.6692015209125
128 78.2874617737003
256 112.527472527472
384 131.732418524871
422.4 135.951078210492
};
\addplot [very thin, dimgray99]
table {%
1 1
2 1.99600798403194
4 3.97614314115308
8 7.88954635108481
16 15.5339805825243
32 30.1318267419962
64 56.8383658969805
128 102.07336523126
256 169.53642384106
384 217.440543601359
422.4 229.216409811157
};
\addplot [semithick, chocolate217952, mark=*, mark size=1.5, mark options={solid}]
table {%
1 1
2 2.05814163769072
4 3.99332500550616
8 8.00475324595239
16 15.8310663098339
32 30.252609471449
64 65.6604870438581
128 106.872687849657
256 177.799950234247
384 190.057344799161
};
\end{axis}

\end{tikzpicture}

%% file: figures/python_fig/weak-cluster-comp-nb.tex
\begin{tikzpicture}

\definecolor{cornflowerblue107174214}{RGB}{107,174,214}
\definecolor{darkgray}{RGB}{169,169,169}
\definecolor{darkgray176}{RGB}{176,176,176}
\definecolor{gainsboro232221219}{RGB}{232,221,219}
\definecolor{gray163119111}{RGB}{163,119,111}
\definecolor{lavender225238246}{RGB}{225,238,246}
\definecolor{lavender233224241}{RGB}{233,224,241}
\definecolor{lightblue166206230}{RGB}{166,206,230}
\definecolor{mediumpurple148103189}{RGB}{148,103,189}
\definecolor{mediumpurple169133202}{RGB}{169,133,202}
\definecolor{plum190163215}{RGB}{190,163,215}
\definecolor{rosybrown186153147}{RGB}{186,153,147}
\definecolor{sienna1408675}{RGB}{140,86,75}
\definecolor{skyblue136190222}{RGB}{136,190,222}

\begin{axis}[
axis line style={draw=none},
tick pos=left,
x grid style={darkgray176},
x grid style={draw=none},
xlabel near ticks,
xlabel={N nodes},
xmajorgrids,
xmin=-0.716666666666667, xmax=7.71666666666667,
xtick style={color=darkgray},
xtick={0,1,2,3,4,5,6,7},
xticklabels={$1$,$2$,$4$,$8$,$16$,$32$,$64$,$128$},
xticklabels={1,2,4,8,16,32,64,128},
y grid style={darkgray176},
ylabel near ticks,
ylabel={time/solve - [sec]},
ymajorgrids,
ymin=0, ymax=2,
ytick style={color=darkgray}
]
\draw[draw=black,fill=lavender225238246,line width=0.004pt,postaction={pattern=crosshatch}] (axis cs:-0.333333333333333,0) rectangle (axis cs:-0.166666666666667,0.313366726);
\draw[draw=black,fill=cornflowerblue107174214,line width=0.004pt,postaction={pattern=north east lines}] (axis cs:-0.333333333333333,0.313366726) rectangle (axis cs:-0.166666666666667,0.345685233);
\draw[draw=black,fill=skyblue136190222,line width=0.004pt,postaction={pattern=north east lines}] (axis cs:-0.333333333333333,0.345685233) rectangle (axis cs:-0.166666666666667,0.4247058205);
\draw[draw=black,fill=lightblue166206230,line width=0.004pt,postaction={pattern=north east lines}] (axis cs:-0.333333333333333,0.4247058205) rectangle (axis cs:-0.166666666666667,0.5906478175);
\draw[draw=black,fill=cornflowerblue107174214,line width=0.004pt] (axis cs:-0.333333333333333,0.5906478175) rectangle (axis cs:-0.166666666666667,0.639074498);
\draw[draw=black,fill=skyblue136190222,line width=0.004pt] (axis cs:-0.333333333333333,0.639074498) rectangle (axis cs:-0.166666666666667,0.754894611);
\draw[draw=black,fill=lightblue166206230,line width=0.004pt] (axis cs:-0.333333333333333,0.754894611) rectangle (axis cs:-0.166666666666667,0.976532825);
\draw[draw=black,fill=lavender225238246,line width=0.004pt,postaction={pattern=crosshatch}] (axis cs:0.666666666666667,0) rectangle (axis cs:0.833333333333333,0.311158109);
\draw[draw=black,fill=cornflowerblue107174214,line width=0.004pt,postaction={pattern=north east lines}] (axis cs:0.666666666666667,0.311158109) rectangle (axis cs:0.833333333333333,0.3424704035);
\draw[draw=black,fill=skyblue136190222,line width=0.004pt,postaction={pattern=north east lines}] (axis cs:0.666666666666667,0.3424704035) rectangle (axis cs:0.833333333333333,0.416736666);
\draw[draw=black,fill=lightblue166206230,line width=0.004pt,postaction={pattern=north east lines}] (axis cs:0.666666666666667,0.416736666) rectangle (axis cs:0.833333333333333,0.550905671);
\draw[draw=black,fill=cornflowerblue107174214,line width=0.004pt] (axis cs:0.666666666666667,0.550905671) rectangle (axis cs:0.833333333333333,0.6002511605);
\draw[draw=black,fill=skyblue136190222,line width=0.004pt] (axis cs:0.666666666666667,0.6002511605) rectangle (axis cs:0.833333333333333,0.723115027);
\draw[draw=black,fill=lightblue166206230,line width=0.004pt] (axis cs:0.666666666666667,0.723115027) rectangle (axis cs:0.833333333333333,1.0446452285);
\draw[draw=black,fill=lavender225238246,line width=0.004pt,postaction={pattern=crosshatch}] (axis cs:1.66666666666667,0) rectangle (axis cs:1.83333333333333,0.313730212);
\draw[draw=black,fill=cornflowerblue107174214,line width=0.004pt,postaction={pattern=north east lines}] (axis cs:1.66666666666667,0.313730212) rectangle (axis cs:1.83333333333333,0.342080978);
\draw[draw=black,fill=skyblue136190222,line width=0.004pt,postaction={pattern=north east lines}] (axis cs:1.66666666666667,0.342080978) rectangle (axis cs:1.83333333333333,0.413208959);
\draw[draw=black,fill=lightblue166206230,line width=0.004pt,postaction={pattern=north east lines}] (axis cs:1.66666666666667,0.413208959) rectangle (axis cs:1.83333333333333,0.540241504);
\draw[draw=black,fill=cornflowerblue107174214,line width=0.004pt] (axis cs:1.66666666666667,0.540241504) rectangle (axis cs:1.83333333333333,0.591624588);
\draw[draw=black,fill=skyblue136190222,line width=0.004pt] (axis cs:1.66666666666667,0.591624588) rectangle (axis cs:1.83333333333333,0.711410056);
\draw[draw=black,fill=lightblue166206230,line width=0.004pt] (axis cs:1.66666666666667,0.711410056) rectangle (axis cs:1.83333333333333,1.1528297265);
\draw[draw=black,fill=lavender225238246,line width=0.004pt,postaction={pattern=crosshatch}] (axis cs:2.66666666666667,0) rectangle (axis cs:2.83333333333333,0.3110470145);
\draw[draw=black,fill=cornflowerblue107174214,line width=0.004pt,postaction={pattern=north east lines}] (axis cs:2.66666666666667,0.3110470145) rectangle (axis cs:2.83333333333333,0.3399978295);
\draw[draw=black,fill=skyblue136190222,line width=0.004pt,postaction={pattern=north east lines}] (axis cs:2.66666666666667,0.3399978295) rectangle (axis cs:2.83333333333333,0.4098068095);
\draw[draw=black,fill=lightblue166206230,line width=0.004pt,postaction={pattern=north east lines}] (axis cs:2.66666666666667,0.4098068095) rectangle (axis cs:2.83333333333333,0.534339482);
\draw[draw=black,fill=cornflowerblue107174214,line width=0.004pt] (axis cs:2.66666666666667,0.534339482) rectangle (axis cs:2.83333333333333,0.58318717);
\draw[draw=black,fill=skyblue136190222,line width=0.004pt] (axis cs:2.66666666666667,0.58318717) rectangle (axis cs:2.83333333333333,0.707586291);
\draw[draw=black,fill=lightblue166206230,line width=0.004pt] (axis cs:2.66666666666667,0.707586291) rectangle (axis cs:2.83333333333333,1.295198587);
\draw[draw=black,fill=lavender225238246,line width=0.004pt,postaction={pattern=crosshatch}] (axis cs:3.66666666666667,0) rectangle (axis cs:3.83333333333333,0.318010739);
\draw[draw=black,fill=cornflowerblue107174214,line width=0.004pt,postaction={pattern=north east lines}] (axis cs:3.66666666666667,0.318010739) rectangle (axis cs:3.83333333333333,0.348896087);
\draw[draw=black,fill=skyblue136190222,line width=0.004pt,postaction={pattern=north east lines}] (axis cs:3.66666666666667,0.348896087) rectangle (axis cs:3.83333333333333,0.4237620485);
\draw[draw=black,fill=lightblue166206230,line width=0.004pt,postaction={pattern=north east lines}] (axis cs:3.66666666666667,0.4237620485) rectangle (axis cs:3.83333333333333,0.5595357455);
\draw[draw=black,fill=cornflowerblue107174214,line width=0.004pt] (axis cs:3.66666666666667,0.5595357455) rectangle (axis cs:3.83333333333333,0.6089868645);
\draw[draw=black,fill=skyblue136190222,line width=0.004pt] (axis cs:3.66666666666667,0.6089868645) rectangle (axis cs:3.83333333333333,0.7521749065);
\draw[draw=black,fill=lightblue166206230,line width=0.004pt] (axis cs:3.66666666666667,0.7521749065) rectangle (axis cs:3.83333333333333,1.466088458);
\draw[draw=black,fill=lavender225238246,line width=0.004pt,postaction={pattern=crosshatch}] (axis cs:4.66666666666667,0) rectangle (axis cs:4.83333333333333,0.3175183115);
\draw[draw=black,fill=cornflowerblue107174214,line width=0.004pt,postaction={pattern=north east lines}] (axis cs:4.66666666666667,0.3175183115) rectangle (axis cs:4.83333333333333,0.347019639);
\draw[draw=black,fill=skyblue136190222,line width=0.004pt,postaction={pattern=north east lines}] (axis cs:4.66666666666667,0.347019639) rectangle (axis cs:4.83333333333333,0.4118761245);
\draw[draw=black,fill=lightblue166206230,line width=0.004pt,postaction={pattern=north east lines}] (axis cs:4.66666666666667,0.4118761245) rectangle (axis cs:4.83333333333333,0.535799317);
\draw[draw=black,fill=cornflowerblue107174214,line width=0.004pt] (axis cs:4.66666666666667,0.535799317) rectangle (axis cs:4.83333333333333,0.6031437285);
\draw[draw=black,fill=skyblue136190222,line width=0.004pt] (axis cs:4.66666666666667,0.6031437285) rectangle (axis cs:4.83333333333333,0.812060418);
\draw[draw=black,fill=lightblue166206230,line width=0.004pt] (axis cs:4.66666666666667,0.812060418) rectangle (axis cs:4.83333333333333,1.476504445);
\draw[draw=black,fill=lavender225238246,line width=0.004pt,postaction={pattern=crosshatch}] (axis cs:5.66666666666667,0) rectangle (axis cs:5.83333333333333,0.3077922085);
\draw[draw=black,fill=cornflowerblue107174214,line width=0.004pt,postaction={pattern=north east lines}] (axis cs:5.66666666666667,0.3077922085) rectangle (axis cs:5.83333333333333,0.337867371);
\draw[draw=black,fill=skyblue136190222,line width=0.004pt,postaction={pattern=north east lines}] (axis cs:5.66666666666667,0.337867371) rectangle (axis cs:5.83333333333333,0.402401777);
\draw[draw=black,fill=lightblue166206230,line width=0.004pt,postaction={pattern=north east lines}] (axis cs:5.66666666666667,0.402401777) rectangle (axis cs:5.83333333333333,0.503204826);
\draw[draw=black,fill=cornflowerblue107174214,line width=0.004pt] (axis cs:5.66666666666667,0.503204826) rectangle (axis cs:5.83333333333333,0.5601186695);
\draw[draw=black,fill=skyblue136190222,line width=0.004pt] (axis cs:5.66666666666667,0.5601186695) rectangle (axis cs:5.83333333333333,0.769929625);
\draw[draw=black,fill=lightblue166206230,line width=0.004pt] (axis cs:5.66666666666667,0.769929625) rectangle (axis cs:5.83333333333333,1.4306674795);
\draw[draw=black,fill=lavender225238246,line width=0.004pt,postaction={pattern=crosshatch}] (axis cs:6.66666666666667,0) rectangle (axis cs:6.83333333333333,0.320560654);
\draw[draw=black,fill=cornflowerblue107174214,line width=0.004pt,postaction={pattern=north east lines}] (axis cs:6.66666666666667,0.320560654) rectangle (axis cs:6.83333333333333,0.3549529575);
\draw[draw=black,fill=skyblue136190222,line width=0.004pt,postaction={pattern=north east lines}] (axis cs:6.66666666666667,0.3549529575) rectangle (axis cs:6.83333333333333,0.4163114385);
\draw[draw=black,fill=lightblue166206230,line width=0.004pt,postaction={pattern=north east lines}] (axis cs:6.66666666666667,0.4163114385) rectangle (axis cs:6.83333333333333,0.52262434);
\draw[draw=black,fill=cornflowerblue107174214,line width=0.004pt] (axis cs:6.66666666666667,0.52262434) rectangle (axis cs:6.83333333333333,0.5755550475);
\draw[draw=black,fill=skyblue136190222,line width=0.004pt] (axis cs:6.66666666666667,0.5755550475) rectangle (axis cs:6.83333333333333,1.0215415725);
\draw[draw=black,fill=lightblue166206230,line width=0.004pt] (axis cs:6.66666666666667,1.0215415725) rectangle (axis cs:6.83333333333333,1.930609155);
\draw[draw=black,fill=lavender233224241,line width=0.004pt,postaction={pattern=crosshatch}] (axis cs:-0.0833333333333333,0) rectangle (axis cs:0.0833333333333333,0.3210666065);
\draw[draw=black,fill=mediumpurple148103189,line width=0.004pt,postaction={pattern=north east lines}] (axis cs:-0.0833333333333333,0.3210666065) rectangle (axis cs:0.0833333333333333,0.3559028375);
\draw[draw=black,fill=mediumpurple169133202,line width=0.004pt,postaction={pattern=north east lines}] (axis cs:-0.0833333333333333,0.3559028375) rectangle (axis cs:0.0833333333333333,0.443289068);
\draw[draw=black,fill=plum190163215,line width=0.004pt,postaction={pattern=north east lines}] (axis cs:-0.0833333333333333,0.443289068) rectangle (axis cs:0.0833333333333333,0.6186279805);
\draw[draw=black,fill=mediumpurple148103189,line width=0.004pt] (axis cs:-0.0833333333333333,0.6186279805) rectangle (axis cs:0.0833333333333333,0.6719930955);
\draw[draw=black,fill=mediumpurple169133202,line width=0.004pt] (axis cs:-0.0833333333333333,0.6719930955) rectangle (axis cs:0.0833333333333333,0.792448798);
\draw[draw=black,fill=plum190163215,line width=0.004pt] (axis cs:-0.0833333333333333,0.792448798) rectangle (axis cs:0.0833333333333333,1.0277880955);
\draw[draw=black,fill=lavender233224241,line width=0.004pt,postaction={pattern=crosshatch}] (axis cs:0.916666666666667,0) rectangle (axis cs:1.08333333333333,0.316878601);
\draw[draw=black,fill=mediumpurple148103189,line width=0.004pt,postaction={pattern=north east lines}] (axis cs:0.916666666666667,0.316878601) rectangle (axis cs:1.08333333333333,0.34870027);
\draw[draw=black,fill=mediumpurple169133202,line width=0.004pt,postaction={pattern=north east lines}] (axis cs:0.916666666666667,0.34870027) rectangle (axis cs:1.08333333333333,0.4246208985);
\draw[draw=black,fill=plum190163215,line width=0.004pt,postaction={pattern=north east lines}] (axis cs:0.916666666666667,0.4246208985) rectangle (axis cs:1.08333333333333,0.5665687855);
\draw[draw=black,fill=mediumpurple148103189,line width=0.004pt] (axis cs:0.916666666666667,0.5665687855) rectangle (axis cs:1.08333333333333,0.619709677);
\draw[draw=black,fill=mediumpurple169133202,line width=0.004pt] (axis cs:0.916666666666667,0.619709677) rectangle (axis cs:1.08333333333333,0.746839338);
\draw[draw=black,fill=plum190163215,line width=0.004pt] (axis cs:0.916666666666667,0.746839338) rectangle (axis cs:1.08333333333333,1.047342582);
\draw[draw=black,fill=lavender233224241,line width=0.004pt,postaction={pattern=crosshatch}] (axis cs:1.91666666666667,0) rectangle (axis cs:2.08333333333333,0.322396057);
\draw[draw=black,fill=mediumpurple148103189,line width=0.004pt,postaction={pattern=north east lines}] (axis cs:1.91666666666667,0.322396057) rectangle (axis cs:2.08333333333333,0.35207707);
\draw[draw=black,fill=mediumpurple169133202,line width=0.004pt,postaction={pattern=north east lines}] (axis cs:1.91666666666667,0.35207707) rectangle (axis cs:2.08333333333333,0.4273433365);
\draw[draw=black,fill=plum190163215,line width=0.004pt,postaction={pattern=north east lines}] (axis cs:1.91666666666667,0.4273433365) rectangle (axis cs:2.08333333333333,0.5662799185);
\draw[draw=black,fill=mediumpurple148103189,line width=0.004pt] (axis cs:1.91666666666667,0.5662799185) rectangle (axis cs:2.08333333333333,0.62292775);
\draw[draw=black,fill=mediumpurple169133202,line width=0.004pt] (axis cs:1.91666666666667,0.62292775) rectangle (axis cs:2.08333333333333,0.7411521745);
\draw[draw=black,fill=plum190163215,line width=0.004pt] (axis cs:1.91666666666667,0.7411521745) rectangle (axis cs:2.08333333333333,1.0756606445);
\draw[draw=black,fill=lavender233224241,line width=0.004pt,postaction={pattern=crosshatch}] (axis cs:2.91666666666667,0) rectangle (axis cs:3.08333333333333,0.3135838485);
\draw[draw=black,fill=mediumpurple148103189,line width=0.004pt,postaction={pattern=north east lines}] (axis cs:2.91666666666667,0.3135838485) rectangle (axis cs:3.08333333333333,0.343568668);
\draw[draw=black,fill=mediumpurple169133202,line width=0.004pt,postaction={pattern=north east lines}] (axis cs:2.91666666666667,0.343568668) rectangle (axis cs:3.08333333333333,0.417248557);
\draw[draw=black,fill=plum190163215,line width=0.004pt,postaction={pattern=north east lines}] (axis cs:2.91666666666667,0.417248557) rectangle (axis cs:3.08333333333333,0.551845153);
\draw[draw=black,fill=mediumpurple148103189,line width=0.004pt] (axis cs:2.91666666666667,0.551845153) rectangle (axis cs:3.08333333333333,0.605658658);
\draw[draw=black,fill=mediumpurple169133202,line width=0.004pt] (axis cs:2.91666666666667,0.605658658) rectangle (axis cs:3.08333333333333,0.728477756);
\draw[draw=black,fill=plum190163215,line width=0.004pt] (axis cs:2.91666666666667,0.728477756) rectangle (axis cs:3.08333333333333,1.116099435);
\draw[draw=black,fill=lavender233224241,line width=0.004pt,postaction={pattern=crosshatch}] (axis cs:3.91666666666667,0) rectangle (axis cs:4.08333333333333,0.3167856195);
\draw[draw=black,fill=mediumpurple148103189,line width=0.004pt,postaction={pattern=north east lines}] (axis cs:3.91666666666667,0.3167856195) rectangle (axis cs:4.08333333333333,0.347585438);
\draw[draw=black,fill=mediumpurple169133202,line width=0.004pt,postaction={pattern=north east lines}] (axis cs:3.91666666666667,0.347585438) rectangle (axis cs:4.08333333333333,0.4262232585);
\draw[draw=black,fill=plum190163215,line width=0.004pt,postaction={pattern=north east lines}] (axis cs:3.91666666666667,0.4262232585) rectangle (axis cs:4.08333333333333,0.566199851);
\draw[draw=black,fill=mediumpurple148103189,line width=0.004pt] (axis cs:3.91666666666667,0.566199851) rectangle (axis cs:4.08333333333333,0.6194036225);
\draw[draw=black,fill=mediumpurple169133202,line width=0.004pt] (axis cs:3.91666666666667,0.6194036225) rectangle (axis cs:4.08333333333333,0.7476323125);
\draw[draw=black,fill=plum190163215,line width=0.004pt] (axis cs:3.91666666666667,0.7476323125) rectangle (axis cs:4.08333333333333,1.129873575);
\draw[draw=black,fill=lavender233224241,line width=0.004pt,postaction={pattern=crosshatch}] (axis cs:4.91666666666667,0) rectangle (axis cs:5.08333333333333,0.317855948);
\draw[draw=black,fill=mediumpurple148103189,line width=0.004pt,postaction={pattern=north east lines}] (axis cs:4.91666666666667,0.317855948) rectangle (axis cs:5.08333333333333,0.3459035055);
\draw[draw=black,fill=mediumpurple169133202,line width=0.004pt,postaction={pattern=north east lines}] (axis cs:4.91666666666667,0.3459035055) rectangle (axis cs:5.08333333333333,0.414425989);
\draw[draw=black,fill=plum190163215,line width=0.004pt,postaction={pattern=north east lines}] (axis cs:4.91666666666667,0.414425989) rectangle (axis cs:5.08333333333333,0.5448022405);
\draw[draw=black,fill=mediumpurple148103189,line width=0.004pt] (axis cs:4.91666666666667,0.5448022405) rectangle (axis cs:5.08333333333333,0.606055196);
\draw[draw=black,fill=mediumpurple169133202,line width=0.004pt] (axis cs:4.91666666666667,0.606055196) rectangle (axis cs:5.08333333333333,0.778055971);
\draw[draw=black,fill=plum190163215,line width=0.004pt] (axis cs:4.91666666666667,0.778055971) rectangle (axis cs:5.08333333333333,1.1698969815);
\draw[draw=black,fill=lavender233224241,line width=0.004pt,postaction={pattern=crosshatch}] (axis cs:5.91666666666667,0) rectangle (axis cs:6.08333333333333,0.313023139);
\draw[draw=black,fill=mediumpurple148103189,line width=0.004pt,postaction={pattern=north east lines}] (axis cs:5.91666666666667,0.313023139) rectangle (axis cs:6.08333333333333,0.341868088);
\draw[draw=black,fill=mediumpurple169133202,line width=0.004pt,postaction={pattern=north east lines}] (axis cs:5.91666666666667,0.341868088) rectangle (axis cs:6.08333333333333,0.412765279);
\draw[draw=black,fill=plum190163215,line width=0.004pt,postaction={pattern=north east lines}] (axis cs:5.91666666666667,0.412765279) rectangle (axis cs:6.08333333333333,0.5216306885);
\draw[draw=black,fill=mediumpurple148103189,line width=0.004pt] (axis cs:5.91666666666667,0.5216306885) rectangle (axis cs:6.08333333333333,0.5801603815);
\draw[draw=black,fill=mediumpurple169133202,line width=0.004pt] (axis cs:5.91666666666667,0.5801603815) rectangle (axis cs:6.08333333333333,0.7535133);
\draw[draw=black,fill=plum190163215,line width=0.004pt] (axis cs:5.91666666666667,0.7535133) rectangle (axis cs:6.08333333333333,1.131439556);
\draw[draw=black,fill=lavender233224241,line width=0.004pt,postaction={pattern=crosshatch}] (axis cs:6.91666666666667,0) rectangle (axis cs:7.08333333333333,0.3179651735);
\draw[draw=black,fill=mediumpurple148103189,line width=0.004pt,postaction={pattern=north east lines}] (axis cs:6.91666666666667,0.3179651735) rectangle (axis cs:7.08333333333333,0.346761818);
\draw[draw=black,fill=mediumpurple169133202,line width=0.004pt,postaction={pattern=north east lines}] (axis cs:6.91666666666667,0.346761818) rectangle (axis cs:7.08333333333333,0.412317423);
\draw[draw=black,fill=plum190163215,line width=0.004pt,postaction={pattern=north east lines}] (axis cs:6.91666666666667,0.412317423) rectangle (axis cs:7.08333333333333,0.5245089645);
\draw[draw=black,fill=mediumpurple148103189,line width=0.004pt] (axis cs:6.91666666666667,0.5245089645) rectangle (axis cs:7.08333333333333,0.584591277);
\draw[draw=black,fill=mediumpurple169133202,line width=0.004pt] (axis cs:6.91666666666667,0.584591277) rectangle (axis cs:7.08333333333333,0.820368144);
\draw[draw=black,fill=plum190163215,line width=0.004pt] (axis cs:6.91666666666667,0.820368144) rectangle (axis cs:7.08333333333333,1.243305256);
\draw[draw=black,fill=gainsboro232221219,line width=0.004pt,postaction={pattern=crosshatch}] (axis cs:0.166666666666667,0) rectangle (axis cs:0.333333333333333,0.3008346165);
\draw[draw=black,fill=sienna1408675,line width=0.004pt,postaction={pattern=north east lines}] (axis cs:0.166666666666667,0.3008346165) rectangle (axis cs:0.333333333333333,0.333720918);
\draw[draw=black,fill=gray163119111,line width=0.004pt,postaction={pattern=north east lines}] (axis cs:0.166666666666667,0.333720918) rectangle (axis cs:0.333333333333333,0.4134623095);
\draw[draw=black,fill=rosybrown186153147,line width=0.004pt,postaction={pattern=north east lines}] (axis cs:0.166666666666667,0.4134623095) rectangle (axis cs:0.333333333333333,0.5843398925);
\draw[draw=black,fill=sienna1408675,line width=0.004pt] (axis cs:0.166666666666667,0.5843398925) rectangle (axis cs:0.333333333333333,0.631966772);
\draw[draw=black,fill=gray163119111,line width=0.004pt] (axis cs:0.166666666666667,0.631966772) rectangle (axis cs:0.333333333333333,0.7423838605);
\draw[draw=black,fill=rosybrown186153147,line width=0.004pt] (axis cs:0.166666666666667,0.7423838605) rectangle (axis cs:0.333333333333333,0.9594231315);
\draw[draw=black,fill=gainsboro232221219,line width=0.004pt,postaction={pattern=crosshatch}] (axis cs:1.16666666666667,0) rectangle (axis cs:1.33333333333333,0.279012645);
\draw[draw=black,fill=sienna1408675,line width=0.004pt,postaction={pattern=north east lines}] (axis cs:1.16666666666667,0.279012645) rectangle (axis cs:1.33333333333333,0.312680024);
\draw[draw=black,fill=gray163119111,line width=0.004pt,postaction={pattern=north east lines}] (axis cs:1.16666666666667,0.312680024) rectangle (axis cs:1.33333333333333,0.383669453);
\draw[draw=black,fill=rosybrown186153147,line width=0.004pt,postaction={pattern=north east lines}] (axis cs:1.16666666666667,0.383669453) rectangle (axis cs:1.33333333333333,0.5083172245);
\draw[draw=black,fill=sienna1408675,line width=0.004pt] (axis cs:1.16666666666667,0.5083172245) rectangle (axis cs:1.33333333333333,0.556078921);
\draw[draw=black,fill=gray163119111,line width=0.004pt] (axis cs:1.16666666666667,0.556078921) rectangle (axis cs:1.33333333333333,0.702537835);
\draw[draw=black,fill=rosybrown186153147,line width=0.004pt] (axis cs:1.16666666666667,0.702537835) rectangle (axis cs:1.33333333333333,0.9969640975);
\draw[draw=black,fill=gainsboro232221219,line width=0.004pt,postaction={pattern=crosshatch}] (axis cs:2.16666666666667,0) rectangle (axis cs:2.33333333333333,0.265189705);
\draw[draw=black,fill=sienna1408675,line width=0.004pt,postaction={pattern=north east lines}] (axis cs:2.16666666666667,0.265189705) rectangle (axis cs:2.33333333333333,0.299237396);
\draw[draw=black,fill=gray163119111,line width=0.004pt,postaction={pattern=north east lines}] (axis cs:2.16666666666667,0.299237396) rectangle (axis cs:2.33333333333333,0.369222693);
\draw[draw=black,fill=rosybrown186153147,line width=0.004pt,postaction={pattern=north east lines}] (axis cs:2.16666666666667,0.369222693) rectangle (axis cs:2.33333333333333,0.4709542405);
\draw[draw=black,fill=sienna1408675,line width=0.004pt] (axis cs:2.16666666666667,0.4709542405) rectangle (axis cs:2.33333333333333,0.51969648);
\draw[draw=black,fill=gray163119111,line width=0.004pt] (axis cs:2.16666666666667,0.51969648) rectangle (axis cs:2.33333333333333,0.6593605255);
\draw[draw=black,fill=rosybrown186153147,line width=0.004pt] (axis cs:2.16666666666667,0.6593605255) rectangle (axis cs:2.33333333333333,1.1145057935);
\draw[draw=black,fill=gainsboro232221219,line width=0.004pt,postaction={pattern=crosshatch}] (axis cs:3.16666666666667,0) rectangle (axis cs:3.33333333333333,0.2805603745);
\draw[draw=black,fill=sienna1408675,line width=0.004pt,postaction={pattern=north east lines}] (axis cs:3.16666666666667,0.2805603745) rectangle (axis cs:3.33333333333333,0.3146145715);
\draw[draw=black,fill=gray163119111,line width=0.004pt,postaction={pattern=north east lines}] (axis cs:3.16666666666667,0.3146145715) rectangle (axis cs:3.33333333333333,0.387955555);
\draw[draw=black,fill=rosybrown186153147,line width=0.004pt,postaction={pattern=north east lines}] (axis cs:3.16666666666667,0.387955555) rectangle (axis cs:3.33333333333333,0.486291108);
\draw[draw=black,fill=sienna1408675,line width=0.004pt] (axis cs:3.16666666666667,0.486291108) rectangle (axis cs:3.33333333333333,0.5349632645);
\draw[draw=black,fill=gray163119111,line width=0.004pt] (axis cs:3.16666666666667,0.5349632645) rectangle (axis cs:3.33333333333333,0.652916645);
\draw[draw=black,fill=rosybrown186153147,line width=0.004pt] (axis cs:3.16666666666667,0.652916645) rectangle (axis cs:3.33333333333333,1.162997352);
\draw[draw=black,fill=gainsboro232221219,line width=0.004pt,postaction={pattern=crosshatch}] (axis cs:4.16666666666667,0) rectangle (axis cs:4.33333333333333,0.275035363);
\draw[draw=black,fill=sienna1408675,line width=0.004pt,postaction={pattern=north east lines}] (axis cs:4.16666666666667,0.275035363) rectangle (axis cs:4.33333333333333,0.3088360685);
\draw[draw=black,fill=gray163119111,line width=0.004pt,postaction={pattern=north east lines}] (axis cs:4.16666666666667,0.3088360685) rectangle (axis cs:4.33333333333333,0.3796859725);
\draw[draw=black,fill=rosybrown186153147,line width=0.004pt,postaction={pattern=north east lines}] (axis cs:4.16666666666667,0.3796859725) rectangle (axis cs:4.33333333333333,0.4766431515);
\draw[draw=black,fill=sienna1408675,line width=0.004pt] (axis cs:4.16666666666667,0.4766431515) rectangle (axis cs:4.33333333333333,0.5251813295);
\draw[draw=black,fill=gray163119111,line width=0.004pt] (axis cs:4.16666666666667,0.5251813295) rectangle (axis cs:4.33333333333333,0.649456219);
\draw[draw=black,fill=rosybrown186153147,line width=0.004pt] (axis cs:4.16666666666667,0.649456219) rectangle (axis cs:4.33333333333333,1.191776575);
\draw[draw=black,fill=gainsboro232221219,line width=0.004pt,postaction={pattern=crosshatch}] (axis cs:5.16666666666667,0) rectangle (axis cs:5.33333333333333,0.267176508);
\draw[draw=black,fill=sienna1408675,line width=0.004pt,postaction={pattern=north east lines}] (axis cs:5.16666666666667,0.267176508) rectangle (axis cs:5.33333333333333,0.3015425135);
\draw[draw=black,fill=gray163119111,line width=0.004pt,postaction={pattern=north east lines}] (axis cs:5.16666666666667,0.3015425135) rectangle (axis cs:5.33333333333333,0.3633124255);
\draw[draw=black,fill=rosybrown186153147,line width=0.004pt,postaction={pattern=north east lines}] (axis cs:5.16666666666667,0.3633124255) rectangle (axis cs:5.33333333333333,0.445678422);
\draw[draw=black,fill=sienna1408675,line width=0.004pt] (axis cs:5.16666666666667,0.445678422) rectangle (axis cs:5.33333333333333,0.4953025095);
\draw[draw=black,fill=gray163119111,line width=0.004pt] (axis cs:5.16666666666667,0.4953025095) rectangle (axis cs:5.33333333333333,0.680652817);
\draw[draw=black,fill=rosybrown186153147,line width=0.004pt] (axis cs:5.16666666666667,0.680652817) rectangle (axis cs:5.33333333333333,1.2845670235);
\draw[draw=black,fill=gainsboro232221219,line width=0.004pt,postaction={pattern=crosshatch}] (axis cs:6.16666666666667,0) rectangle (axis cs:6.33333333333333,0.2825943775);
\draw[draw=black,fill=sienna1408675,line width=0.004pt,postaction={pattern=north east lines}] (axis cs:6.16666666666667,0.2825943775) rectangle (axis cs:6.33333333333333,0.3160187045);
\draw[draw=black,fill=gray163119111,line width=0.004pt,postaction={pattern=north east lines}] (axis cs:6.16666666666667,0.3160187045) rectangle (axis cs:6.33333333333333,0.3791597495);
\draw[draw=black,fill=rosybrown186153147,line width=0.004pt,postaction={pattern=north east lines}] (axis cs:6.16666666666667,0.3791597495) rectangle (axis cs:6.33333333333333,0.46276415);
\draw[draw=black,fill=sienna1408675,line width=0.004pt] (axis cs:6.16666666666667,0.46276415) rectangle (axis cs:6.33333333333333,0.5144929745);
\draw[draw=black,fill=gray163119111,line width=0.004pt] (axis cs:6.16666666666667,0.5144929745) rectangle (axis cs:6.33333333333333,0.7288409835);
\draw[draw=black,fill=rosybrown186153147,line width=0.004pt] (axis cs:6.16666666666667,0.7288409835) rectangle (axis cs:6.33333333333333,1.3679652955);
\draw[draw=black,fill=gainsboro232221219,line width=0.004pt,postaction={pattern=crosshatch}] (axis cs:7.16666666666667,0) rectangle (axis cs:7.33333333333333,0.2857713625);
\draw[draw=black,fill=sienna1408675,line width=0.004pt,postaction={pattern=north east lines}] (axis cs:7.16666666666667,0.2857713625) rectangle (axis cs:7.33333333333333,0.31927086);
\draw[draw=black,fill=gray163119111,line width=0.004pt,postaction={pattern=north east lines}] (axis cs:7.16666666666667,0.31927086) rectangle (axis cs:7.33333333333333,0.381814572);
\draw[draw=black,fill=rosybrown186153147,line width=0.004pt,postaction={pattern=north east lines}] (axis cs:7.16666666666667,0.381814572) rectangle (axis cs:7.33333333333333,0.4668706365);
\draw[draw=black,fill=sienna1408675,line width=0.004pt] (axis cs:7.16666666666667,0.4668706365) rectangle (axis cs:7.33333333333333,0.5198062055);
\draw[draw=black,fill=gray163119111,line width=0.004pt] (axis cs:7.16666666666667,0.5198062055) rectangle (axis cs:7.33333333333333,0.8471634205);
\draw[draw=black,fill=rosybrown186153147,line width=0.004pt] (axis cs:7.16666666666667,0.8471634205) rectangle (axis cs:7.33333333333333,1.56425404);
\end{axis}

\end{tikzpicture}

%% file: figures/python_fig/weak-cluster-eff-nb.tex
\begin{tikzpicture}

\definecolor{cornflowerblue107174214}{RGB}{107,174,214}
\definecolor{darkgray}{RGB}{169,169,169}
\definecolor{darkgray150}{RGB}{150,150,150}
\definecolor{darkgray176}{RGB}{176,176,176}
\definecolor{dimgray99}{RGB}{99,99,99}
\definecolor{mediumpurple148103189}{RGB}{148,103,189}
\definecolor{sienna1408675}{RGB}{140,86,75}
\definecolor{silver189}{RGB}{189,189,189}

\begin{axis}[
axis y line =left, axis x line =bottom, axis x line shift=5pt, axis line style ={-},
log basis x={10},
minor xtick={},
minor ytick={},
tick align=center,
tick pos=left,
x grid style={darkgray176},
x grid style={draw=none},
xlabel near ticks,
xlabel={\(\displaystyle r\) = Number of nodes},
xmajorgrids,
xmin=0.780854050171763, xmax=180.315386683374,
xmode=log,
xtick style={color=black},
xtick={1,2,4,8,16,32,64,128},
xticklabels={$1$,$2$,$4$,$8$,$16$,$32$,$64$,$128$},
y axis line style={draw=none},
y grid style={darkgray176},
y grid style={dotted},
ylabel near ticks,
ylabel={\(\displaystyle \eta_{P,w}\)},
ymajorgrids,
ymin=0.1, ymax=1.1,
ytick style={color=darkgray},
ytick={0,0.2,0.4,0.6,0.8,1,1.2}
]
\addplot [very thin, silver189]
table {%
1 1
1.10623917447252 0.99893873573367
1.22376511113765 0.997767344792031
1.35377690629318 0.996474702625059
1.49760104723774 0.995048627608495
1.65670494618546 0.993475795313024
1.83271191201275 0.991741649151127
2.02741771259094 0.989830308090089
2.24280889668757 0.98772447238247
2.48108306237128 0.985405328582658
2.74467127871536 0.982852455496799
3.03626288956452 0.980043733160157
3.35883295243341 0.976955257456583
3.71567259249104 0.973561263593482
4.11042258132746 0.969834062324067
4.54711048310091 0.965743993564361
5.0301917470609 0.961259402877388
5.56459516570714 0.956346647175619
6.15577316238565 0.950970136899437
6.80975742139761 0.945092422825811
7.53322042820502 0.938674336493865
8.33354354761706 0.931675193930743
9.2188923345467 0.92405307283003
10.19829984572 0.915765173462264
11.2817588023525 0.906768273248348
12.4803235441126 0.897019283949513
13.8062228145891 0.886475918658869
15.2729845289947 0.875097473056957
16.8955737970867 0.862845721572446
18.6905456095288 0.849685924065455
20.6762137455262 0.835587932390936
22.8728376250683 0.820527378771975
25.3028290121995 0.804486919522851
27.990980678275 0.787457498681303
30.9647193582112 0.769439587095771
34.2543855806009 0.750444344208946
37.8935432267474 0.730494643084532
41.9193219769959 0.709625895136823
46.3727961382798 0.687886610538049
51.2994037179932 0.665338634261185
56.7494100199254 0.64205700674697
62.7784204922449 0.618129412413156
69.4479480600299 0.593655198247727
76.826040730738 0.568743967528345
84.987975875964 0.543513778679839
94.017028273117 0.518089004346814
104.005319743213 0.492597928598586
115.054759053482 0.467170178519669
127.278081674459 0.44193409834483
140.8 0.417014178482069
};
\addplot [very thin, darkgray150]
table {%
1 1
1.10623917447252 0.999469086146885
1.22376511113765 0.998882424815988
1.35377690629318 0.998234238896037
1.49760104723774 0.997518169570914
1.65670494618546 0.996727221518157
1.83271191201275 0.995853703791155
2.02741771259094 0.994889166242687
2.24280889668757 0.993824331395983
2.48108306237128 0.992649021735143
2.74467127871536 0.991352082473073
3.03626288956452 0.989921299966444
3.35883295243341 0.988343316088466
3.71567259249104 0.98660353904678
4.11042258132746 0.984686051351786
4.54711048310091 0.982573515906553
5.0301917470609 0.980247081510087
5.56459516570714 0.977686289448036
6.15577316238565 0.974868983295416
6.80975742139761 0.971771224580465
7.53322042820502 0.968367217561128
8.33354354761706 0.964629247047365
9.2188923345467 0.960527633960605
10.19829984572 0.956030714147755
11.2817588023525 0.951104846844958
12.4803235441126 0.945714460089154
13.8062228145891 0.939822141264419
15.2729845289947 0.93338878178988
16.8955737970867 0.926373785633853
18.6905456095288 0.918735351781145
20.6762137455262 0.910430840872381
22.8728376250683 0.901417235840153
25.3028290121995 0.89165170533414
27.990980678275 0.881092276893017
30.9647193582112 0.869698624024428
34.2543855806009 0.857432967453854
37.8935432267474 0.844261085700279
41.9193219769959 0.830153423805073
46.3727961382798 0.815086281558654
51.2994037179932 0.799043054154998
56.7494100199254 0.78201548924167
62.7784204922449 0.764004915393419
69.4479480600299 0.745023388874171
76.826040730738 0.725094699072452
84.987975875964 0.704255169195446
94.017028273117 0.682554188671868
104.005319743213 0.660054418085774
115.054759053482 0.636831616889909
127.278081674459 0.612974058734193
140.8 0.588581518540314
};
\addplot [very thin, dimgray99]
table {%
1 1
1.10623917447252 0.999787566788513
1.22376511113765 0.999552669971432
1.35377690629318 0.999292946465838
1.49760104723774 0.999005787348038
1.65670494618546 0.998688312890455
1.83271191201275 0.998337345200888
2.02741771259094 0.997949378265003
2.24280889668757 0.997520545183634
2.48108306237128 0.997046582388857
2.74467127871536 0.99652279061725
3.03626288956452 0.995943992416316
3.35883295243341 0.995304485961618
3.71567259249104 0.994597994969032
4.11042258132746 0.993817614500275
4.54711048310091 0.99295575248223
5.0301917470609 0.992004066793912
5.56459516570714 0.990953397821724
6.15577316238565 0.989793696447119
6.80975742139761 0.988513947514544
7.53322042820502 0.987102088935683
8.33354354761706 0.98554492672348
9.2188923345467 0.983828046421508
10.19829984572 0.981935721606872
11.2817588023525 0.979850820404625
12.4803235441126 0.977554711265207
13.8062228145891 0.975027169630858
15.2729845289947 0.972246287558608
16.8955737970867 0.969188388882478
18.6905456095288 0.965827953089813
20.6762137455262 0.962137551758023
22.8728376250683 0.958087802146196
25.3028290121995 0.953647343353102
27.990980678275 0.948782841323897
30.9647193582112 0.943459029886935
34.2543855806009 0.937638795892143
37.8935432267474 0.93128331735
41.9193219769959 0.924352264164938
46.3727961382798 0.916804071527659
51.2994037179932 0.908596296165043
56.7494100199254 0.899686065311476
62.7784204922449 0.890030627310117
69.4479480600299 0.879588011015559
76.826040730738 0.868317798489083
84.987975875964 0.856182011710114
94.017028273117 0.843146109068765
104.005319743213 0.829180081218727
115.054759053482 0.814259628523541
127.278081674459 0.798367394022745
140.8 0.781494216942802
};
\addplot [very thin, black]
table {%
1 1
1.10623917447252 1
1.22376511113765 1
1.35377690629318 1
1.49760104723774 1
1.65670494618546 1
1.83271191201275 1
2.02741771259094 1
2.24280889668757 1
2.48108306237128 1
2.74467127871536 1
3.03626288956452 1
3.35883295243341 1
3.71567259249104 1
4.11042258132746 1
4.54711048310091 1
5.0301917470609 1
5.56459516570714 1
6.15577316238565 1
6.80975742139761 1
7.53322042820502 1
8.33354354761706 1
9.2188923345467 1
10.19829984572 1
11.2817588023525 1
12.4803235441126 1
13.8062228145891 1
15.2729845289947 1
16.8955737970867 1
18.6905456095288 1
20.6762137455262 1
22.8728376250683 1
25.3028290121995 1
27.990980678275 1
30.9647193582112 1
34.2543855806009 1
37.8935432267474 1
41.9193219769959 1
46.3727961382798 1
51.2994037179932 1
56.7494100199254 1
62.7784204922449 1
69.4479480600299 1
76.826040730738 1
84.987975875964 1
94.017028273117 1
104.005319743213 1
115.054759053482 1
127.278081674459 1
140.8 1
};
\addplot [very thin, silver189]
table {%
1 1
1.10623917447252 0.99893873573367
1.22376511113765 0.997767344792031
1.35377690629318 0.996474702625059
1.49760104723774 0.995048627608495
1.65670494618546 0.993475795313024
1.83271191201275 0.991741649151127
2.02741771259094 0.989830308090089
2.24280889668757 0.98772447238247
2.48108306237128 0.985405328582658
2.74467127871536 0.982852455496799
3.03626288956452 0.980043733160157
3.35883295243341 0.976955257456583
3.71567259249104 0.973561263593482
4.11042258132746 0.969834062324067
4.54711048310091 0.965743993564361
5.0301917470609 0.961259402877388
5.56459516570714 0.956346647175619
6.15577316238565 0.950970136899437
6.80975742139761 0.945092422825811
7.53322042820502 0.938674336493865
8.33354354761706 0.931675193930743
9.2188923345467 0.92405307283003
10.19829984572 0.915765173462264
11.2817588023525 0.906768273248348
12.4803235441126 0.897019283949513
13.8062228145891 0.886475918658869
15.2729845289947 0.875097473056957
16.8955737970867 0.862845721572446
18.6905456095288 0.849685924065455
20.6762137455262 0.835587932390936
22.8728376250683 0.820527378771975
25.3028290121995 0.804486919522851
27.990980678275 0.787457498681303
30.9647193582112 0.769439587095771
34.2543855806009 0.750444344208946
37.8935432267474 0.730494643084532
41.9193219769959 0.709625895136823
46.3727961382798 0.687886610538049
51.2994037179932 0.665338634261185
56.7494100199254 0.64205700674697
62.7784204922449 0.618129412413156
69.4479480600299 0.593655198247727
76.826040730738 0.568743967528345
84.987975875964 0.543513778679839
94.017028273117 0.518089004346814
104.005319743213 0.492597928598586
115.054759053482 0.467170178519669
127.278081674459 0.44193409834483
140.8 0.417014178482069
};
\addplot [very thin, darkgray150]
table {%
1 1
1.10623917447252 0.999469086146885
1.22376511113765 0.998882424815988
1.35377690629318 0.998234238896037
1.49760104723774 0.997518169570914
1.65670494618546 0.996727221518157
1.83271191201275 0.995853703791155
2.02741771259094 0.994889166242687
2.24280889668757 0.993824331395983
2.48108306237128 0.992649021735143
2.74467127871536 0.991352082473073
3.03626288956452 0.989921299966444
3.35883295243341 0.988343316088466
3.71567259249104 0.98660353904678
4.11042258132746 0.984686051351786
4.54711048310091 0.982573515906553
5.0301917470609 0.980247081510087
5.56459516570714 0.977686289448036
6.15577316238565 0.974868983295416
6.80975742139761 0.971771224580465
7.53322042820502 0.968367217561128
8.33354354761706 0.964629247047365
9.2188923345467 0.960527633960605
10.19829984572 0.956030714147755
11.2817588023525 0.951104846844958
12.4803235441126 0.945714460089154
13.8062228145891 0.939822141264419
15.2729845289947 0.93338878178988
16.8955737970867 0.926373785633853
18.6905456095288 0.918735351781145
20.6762137455262 0.910430840872381
22.8728376250683 0.901417235840153
25.3028290121995 0.89165170533414
27.990980678275 0.881092276893017
30.9647193582112 0.869698624024428
34.2543855806009 0.857432967453854
37.8935432267474 0.844261085700279
41.9193219769959 0.830153423805073
46.3727961382798 0.815086281558654
51.2994037179932 0.799043054154998
56.7494100199254 0.78201548924167
62.7784204922449 0.764004915393419
69.4479480600299 0.745023388874171
76.826040730738 0.725094699072452
84.987975875964 0.704255169195446
94.017028273117 0.682554188671868
104.005319743213 0.660054418085774
115.054759053482 0.636831616889909
127.278081674459 0.612974058734193
140.8 0.588581518540314
};
\addplot [very thin, dimgray99]
table {%
1 1
1.10623917447252 0.999787566788513
1.22376511113765 0.999552669971432
1.35377690629318 0.999292946465838
1.49760104723774 0.999005787348038
1.65670494618546 0.998688312890455
1.83271191201275 0.998337345200888
2.02741771259094 0.997949378265003
2.24280889668757 0.997520545183634
2.48108306237128 0.997046582388857
2.74467127871536 0.99652279061725
3.03626288956452 0.995943992416316
3.35883295243341 0.995304485961618
3.71567259249104 0.994597994969032
4.11042258132746 0.993817614500275
4.54711048310091 0.99295575248223
5.0301917470609 0.992004066793912
5.56459516570714 0.990953397821724
6.15577316238565 0.989793696447119
6.80975742139761 0.988513947514544
7.53322042820502 0.987102088935683
8.33354354761706 0.98554492672348
9.2188923345467 0.983828046421508
10.19829984572 0.981935721606872
11.2817588023525 0.979850820404625
12.4803235441126 0.977554711265207
13.8062228145891 0.975027169630858
15.2729845289947 0.972246287558608
16.8955737970867 0.969188388882478
18.6905456095288 0.965827953089813
20.6762137455262 0.962137551758023
22.8728376250683 0.958087802146196
25.3028290121995 0.953647343353102
27.990980678275 0.948782841323897
30.9647193582112 0.943459029886935
34.2543855806009 0.937638795892143
37.8935432267474 0.93128331735
41.9193219769959 0.924352264164938
46.3727961382798 0.916804071527659
51.2994037179932 0.908596296165043
56.7494100199254 0.899686065311476
62.7784204922449 0.890030627310117
69.4479480600299 0.879588011015559
76.826040730738 0.868317798489083
84.987975875964 0.856182011710114
94.017028273117 0.843146109068765
104.005319743213 0.829180081218727
115.054759053482 0.814259628523541
127.278081674459 0.798367394022745
140.8 0.781494216942802
};
\addplot [very thin, black]
table {%
1 1
1.10623917447252 1
1.22376511113765 1
1.35377690629318 1
1.49760104723774 1
1.65670494618546 1
1.83271191201275 1
2.02741771259094 1
2.24280889668757 1
2.48108306237128 1
2.74467127871536 1
3.03626288956452 1
3.35883295243341 1
3.71567259249104 1
4.11042258132746 1
4.54711048310091 1
5.0301917470609 1
5.56459516570714 1
6.15577316238565 1
6.80975742139761 1
7.53322042820502 1
8.33354354761706 1
9.2188923345467 1
10.19829984572 1
11.2817588023525 1
12.4803235441126 1
13.8062228145891 1
15.2729845289947 1
16.8955737970867 1
18.6905456095288 1
20.6762137455262 1
22.8728376250683 1
25.3028290121995 1
27.990980678275 1
30.9647193582112 1
34.2543855806009 1
37.8935432267474 1
41.9193219769959 1
46.3727961382798 1
51.2994037179932 1
56.7494100199254 1
62.7784204922449 1
69.4479480600299 1
76.826040730738 1
84.987975875964 1
94.017028273117 1
104.005319743213 1
115.054759053482 1
127.278081674459 1
140.8 1
};
\addplot [very thin, silver189]
table {%
1 1
1.10623917447252 0.99893873573367
1.22376511113765 0.997767344792031
1.35377690629318 0.996474702625059
1.49760104723774 0.995048627608495
1.65670494618546 0.993475795313024
1.83271191201275 0.991741649151127
2.02741771259094 0.989830308090089
2.24280889668757 0.98772447238247
2.48108306237128 0.985405328582658
2.74467127871536 0.982852455496799
3.03626288956452 0.980043733160157
3.35883295243341 0.976955257456583
3.71567259249104 0.973561263593482
4.11042258132746 0.969834062324067
4.54711048310091 0.965743993564361
5.0301917470609 0.961259402877388
5.56459516570714 0.956346647175619
6.15577316238565 0.950970136899437
6.80975742139761 0.945092422825811
7.53322042820502 0.938674336493865
8.33354354761706 0.931675193930743
9.2188923345467 0.92405307283003
10.19829984572 0.915765173462264
11.2817588023525 0.906768273248348
12.4803235441126 0.897019283949513
13.8062228145891 0.886475918658869
15.2729845289947 0.875097473056957
16.8955737970867 0.862845721572446
18.6905456095288 0.849685924065455
20.6762137455262 0.835587932390936
22.8728376250683 0.820527378771975
25.3028290121995 0.804486919522851
27.990980678275 0.787457498681303
30.9647193582112 0.769439587095771
34.2543855806009 0.750444344208946
37.8935432267474 0.730494643084532
41.9193219769959 0.709625895136823
46.3727961382798 0.687886610538049
51.2994037179932 0.665338634261185
56.7494100199254 0.64205700674697
62.7784204922449 0.618129412413156
69.4479480600299 0.593655198247727
76.826040730738 0.568743967528345
84.987975875964 0.543513778679839
94.017028273117 0.518089004346814
104.005319743213 0.492597928598586
115.054759053482 0.467170178519669
127.278081674459 0.44193409834483
140.8 0.417014178482069
};
\addplot [very thin, darkgray150]
table {%
1 1
1.10623917447252 0.999469086146885
1.22376511113765 0.998882424815988
1.35377690629318 0.998234238896037
1.49760104723774 0.997518169570914
1.65670494618546 0.996727221518157
1.83271191201275 0.995853703791155
2.02741771259094 0.994889166242687
2.24280889668757 0.993824331395983
2.48108306237128 0.992649021735143
2.74467127871536 0.991352082473073
3.03626288956452 0.989921299966444
3.35883295243341 0.988343316088466
3.71567259249104 0.98660353904678
4.11042258132746 0.984686051351786
4.54711048310091 0.982573515906553
5.0301917470609 0.980247081510087
5.56459516570714 0.977686289448036
6.15577316238565 0.974868983295416
6.80975742139761 0.971771224580465
7.53322042820502 0.968367217561128
8.33354354761706 0.964629247047365
9.2188923345467 0.960527633960605
10.19829984572 0.956030714147755
11.2817588023525 0.951104846844958
12.4803235441126 0.945714460089154
13.8062228145891 0.939822141264419
15.2729845289947 0.93338878178988
16.8955737970867 0.926373785633853
18.6905456095288 0.918735351781145
20.6762137455262 0.910430840872381
22.8728376250683 0.901417235840153
25.3028290121995 0.89165170533414
27.990980678275 0.881092276893017
30.9647193582112 0.869698624024428
34.2543855806009 0.857432967453854
37.8935432267474 0.844261085700279
41.9193219769959 0.830153423805073
46.3727961382798 0.815086281558654
51.2994037179932 0.799043054154998
56.7494100199254 0.78201548924167
62.7784204922449 0.764004915393419
69.4479480600299 0.745023388874171
76.826040730738 0.725094699072452
84.987975875964 0.704255169195446
94.017028273117 0.682554188671868
104.005319743213 0.660054418085774
115.054759053482 0.636831616889909
127.278081674459 0.612974058734193
140.8 0.588581518540314
};
\addplot [very thin, dimgray99]
table {%
1 1
1.10623917447252 0.999787566788513
1.22376511113765 0.999552669971432
1.35377690629318 0.999292946465838
1.49760104723774 0.999005787348038
1.65670494618546 0.998688312890455
1.83271191201275 0.998337345200888
2.02741771259094 0.997949378265003
2.24280889668757 0.997520545183634
2.48108306237128 0.997046582388857
2.74467127871536 0.99652279061725
3.03626288956452 0.995943992416316
3.35883295243341 0.995304485961618
3.71567259249104 0.994597994969032
4.11042258132746 0.993817614500275
4.54711048310091 0.99295575248223
5.0301917470609 0.992004066793912
5.56459516570714 0.990953397821724
6.15577316238565 0.989793696447119
6.80975742139761 0.988513947514544
7.53322042820502 0.987102088935683
8.33354354761706 0.98554492672348
9.2188923345467 0.983828046421508
10.19829984572 0.981935721606872
11.2817588023525 0.979850820404625
12.4803235441126 0.977554711265207
13.8062228145891 0.975027169630858
15.2729845289947 0.972246287558608
16.8955737970867 0.969188388882478
18.6905456095288 0.965827953089813
20.6762137455262 0.962137551758023
22.8728376250683 0.958087802146196
25.3028290121995 0.953647343353102
27.990980678275 0.948782841323897
30.9647193582112 0.943459029886935
34.2543855806009 0.937638795892143
37.8935432267474 0.93128331735
41.9193219769959 0.924352264164938
46.3727961382798 0.916804071527659
51.2994037179932 0.908596296165043
56.7494100199254 0.899686065311476
62.7784204922449 0.890030627310117
69.4479480600299 0.879588011015559
76.826040730738 0.868317798489083
84.987975875964 0.856182011710114
94.017028273117 0.843146109068765
104.005319743213 0.829180081218727
115.054759053482 0.814259628523541
127.278081674459 0.798367394022745
140.8 0.781494216942802
};
\addplot [very thin, black]
table {%
1 1
1.10623917447252 1
1.22376511113765 1
1.35377690629318 1
1.49760104723774 1
1.65670494618546 1
1.83271191201275 1
2.02741771259094 1
2.24280889668757 1
2.48108306237128 1
2.74467127871536 1
3.03626288956452 1
3.35883295243341 1
3.71567259249104 1
4.11042258132746 1
4.54711048310091 1
5.0301917470609 1
5.56459516570714 1
6.15577316238565 1
6.80975742139761 1
7.53322042820502 1
8.33354354761706 1
9.2188923345467 1
10.19829984572 1
11.2817588023525 1
12.4803235441126 1
13.8062228145891 1
15.2729845289947 1
16.8955737970867 1
18.6905456095288 1
20.6762137455262 1
22.8728376250683 1
25.3028290121995 1
27.990980678275 1
30.9647193582112 1
34.2543855806009 1
37.8935432267474 1
41.9193219769959 1
46.3727961382798 1
51.2994037179932 1
56.7494100199254 1
62.7784204922449 1
69.4479480600299 1
76.826040730738 1
84.987975875964 1
94.017028273117 1
104.005319743213 1
115.054759053482 1
127.278081674459 1
140.8 1
};
\addplot [semithick, cornflowerblue107174214, mark=*, mark size=1.5, mark options={solid}]
table {%
1 1
2 0.934798530982808
4 0.84707463951746
8 0.753963781926207
16 0.666080426233053
32 0.661381568004694
64 0.682571484284584
128 0.505815909176086
};
\addplot [semithick, mediumpurple148103189, mark=*, mark size=1.5, mark options={solid}]
table {%
1 1
2 0.981329426649818
4 0.95549474711678
8 0.920875025351124
16 0.909648759154315
32 0.878528718128845
64 0.908389750075169
128 0.826657886741854
};
\addplot [semithick, sienna1408675, mark=*, mark size=1.5, mark options={solid}]
table {%
1 1
2 0.962344716229864
4 0.86085073500338
8 0.824957279438414
16 0.805036071043769
32 0.746884447403845
64 0.701350490875812
128 0.613342274954265
};
\end{axis}

\end{tikzpicture}

%% file: flups_6_conclusion.tex
\section{Conclusions}
\label{sec_conclusion}

Massively distributed FFT transforms have numerous applications and in particular in the PDE resolution realm. However, the contributions usually proposed in the computer science field fail to address important requirements to provide the user with a flexible, yet performant and scalable library.
Relying on our expertise in computational fluid dynamics, we propose improvements to the \flups software \cite{Caprace:2021} to bridge this gap: the treatment of both node-centered and cell-centered data layouts as well as faster communication strategies exploring several possible implementations.
The resulting interface is built such that the user only provides the number of points in the 3D Cartesian grid, the desired boundary conditions (even, odd, periodic or unbounded) and the library automatically orders the sequence of FFTs, extends the domain to handle unbounded directions, and performs the forward and backward transforms.

At a methodological level, we first modify the numbering conventions to accommodate the different types of FFTs provided by \fftw and required for the cell- and node-centered data layouts.
Then, we present three implementation strategies for a distributed FFT.
First, the well-known \ata implementation relies on the commonly used \code{MPI\_Ialltoallv} function which implies a very strong synchronization and exposes very few parallelizations to the MPI library.
Then, the \nb approach, which relies on non-blocking persistent requests and manual packing/unpacking, exploits the possibility of a very fine-grained parallelization. %
This method makes the synchronization more explicit, through the parameters $\nsendbatch$ and $\nsendpending$ and hence reduces the overhead of the implementation.
Finally, we explore the use of \code{MPI\_Datatypes} to reduce the memory footprint of the solver, an approach named \isr.
This optimization removes the need for manual packing and is implemented through non-blocking send and receives. Both the \nb and the \isr implementations share a very similar structure which allows us to attribute the difference in \revthree{performance} to the use of \code{MPI\_Datatypes}, a currently active subject for the different MPI implementations.
To prove the flexibility of the proposed library, we demonstrate the use of \flups to solve the Biot-Savart equation in \Sect{sec2_biot_savart}. This requires a special operation in spectral space, as well as different real-to-real FFTs in the backward and the forward transform. To our knowledge, no other library offers this level of convenience for the user.

The non-blocking approaches (\isr and \nb) are first compared to the \accfft library\cite{Gholami:2015} in \Sect{sec_5_accfft}. The \nb strategy demonstrates $27\%$ \typo{faster time-to-solution} over a large range of core count, while the \isr implementation is as fast as \accfft.
We conclude that our implementation is as fast, if not faster than one of the fastest implementation of the distributed FFT on CPU.
Moreover, the flexibility introduced for the user does not reduce the achieved performance.
Then in \Sect{sec:weak-scaling} and in \Sect{sec:strong-scaling} we focus on the scalability of our implementation both from a weak and a strong perspective. The \ata is observed to be significantly slower yet to achieve a good scalability, which is expected due to the implicit barrier on the sub-communicator. Regarding the non-blocking implementation, the \isr is the fastest on a small count of nodes, but the advantage vanishes over a larger partition as the \nb approach has a better scalability. Both the \nb and the \ata implementation achieve an impressive weak and strong scalability with a sequential part of the implementation $ \beta$  below $0.5\%$. 
The strong scalability results follow the trends observed in the weak scalability, with slightly better estimates for $\beta$.
Finally we apply our test case to three different leading European clusters: Lumi, Vega, and MeluXina in \Sect{sec:eu_comparison}.
The time-to-solution are compared between the clusters and as expected the better bandwidth available on MeluXina provides a faster time-to-solution on large partitions.

With the proposed improvements and changes, \flups is now a highly flexible and performant distributed FFT framework, tailor-made for scientists and in particular computational fluid dynamics applications.
 We ambition here to bridge the gap between the numerous contributions in the computer science field, focusing mainly on performance, and the actual need of the user, which is a highly efficient and versatile framework to be used with lots of different configurations. 
Our contribution in this work also aims to provide a reference in terms of performance metrics with scalability tests on large partitions and on different architectures.

In the future we will further develop \flups to exploit heterogeneous architectures with a particular focus on the \code{MPI+X} approach as proposed in the latest \code{MPI} standard.
The newest additions indeed provide opportunities to reduce the rank count, one identified bottleneck with large partitions, as well as to exploit the concept of \textit{streams} with threads and GPUs.
This future direction aims at addressing the missing GPU implementation of this work.

%% file: flups_7_appendices.tex
\appendices

\renewcommand{\theequation}{\alph{equation}}
\setcounter{equation}{0} 

\renewcommand{\thetable}{\Alph{table}}
\setcounter{table}{0} 
\renewcommand{\thefigure}{\Alph{figure}}
\setcounter{figure}{0} 

\section{Implementation: performance strategies}
\label{sec:pref-strategies}

Throughout the development of the library we have established different strategies to improve the performance. 
The first one is to proceed first with the \textit{intra-node} communications with a special communicator before the \textit{inter-node} ones. As this optimization is usually also done in the MPI implementation, we have not measured a significant difference in terms of time-to-solution.
In this section we describe two other strategies we have used. First a new way of distributing the data through the {MPI} ranks to avoid imbalance between the nodes and a specific order to proceed to the inter-node communications.

\subsection{Load balancing - distribution of unknowns} To simplify the notation in this section we note the integer division by $/$.
Different approaches exist to distributed $N$ unknowns on $P$ ranks:
\begin{itemize}
\item ranks from $0$ to $N\%P$ have $N/P + 1$ data and ranks $> N\%P$ have $N/P$ data as originally implemented in \flups;
\item a rank $r$ gets its first data index from $\left( r \; N \right)/P$.
\end{itemize}
In \code{flups}, the computation of the communications requires such a formula to be invertible, \ie{we have to compute the corresponding rank for a data index and the data that is attributed to a specific rank index}.
While the first approach can be easily inverted (getting the rank index $r$ for a specific data), the unknowns are poorly distributed over the ranks as the first $N\%P$ ranks will get more data. When considering multiple ranks per node it results in a subset of nodes having more data than the others and therefore in a significant imbalance between nodes.

This issue is solved with the second approach which spreads the excess data over the whole rank range. 
However, getting the rank from a data index has been impossible for us, or at least in an efficient way. To combine the benefit of the two approaches, we propose another distribution that both distributes the excess data over the whole rank range and can be easily inverted.

As with the other methods, all ranks will get at least $B = N/P$ unknowns, also referred to as the baseline.
To distribute the remaining $R = N\%P$ ones, we create $R$ groups of ranks.
Each group has $S = P/R$ ranks (where $S$ stands for stride), except the last one which might be a special case.
The last rank of each group will get $1$ excess data, then each group has a total of $(S \; B + 1)$ data. Here again the last group might be special and not get any $+1$ as highlighted in the example below.
The first index for data attributed to a rank $r$ is then obtained as 
\be
\label{eq:idx}
r \; \plr{\dfrac{N}{P}} + \min \left\{ \dfrac{r}{S} \;;\; R \right\}
\eed

To inverse the relation and obtain the rank corresponding to a given data index $i$, we first identify the group index and then add the rank index within the group:
\begin{enumerate}
\item the group index where the data $i$ is located is given by 
\be
g_i = \min \left\{ i / \left( S\;B +1 \right) \;,\; R \right\}
\eec
where the $\min$ ensures that edge cases do not over-estimate the group id;
\item within the group, the local data index is now given by $i_L =  \left( i - g_i \left( S \; B +1 \right)  \right)$;
\item then the local rank attributed to the local index $i_L$ is $r_L = i_L / B$, where we have to bound $r_L$ to $S-1$ to if $g_i < R$ with $r_L = \min \left\{ i_L/B , S-1 \right\}$.
\end{enumerate}
The rank attributed to the data index $i$ is finally obtained as $r =  g_i \; S  + r_L$.

To illustrate our approach we consider an edge case with $N=32$, $P=7$. 
Then following the formula we obtain $B=4$, $R = 32\%4 = 4$, and $S=1$.
The distribution is then given by $\left[ 5 \;,\; 5\;,\; 5 \;,\; 5\;,\; 4\;,\; 4 \;,\; 4 \right]$.
If we want to get the rank for the data id of $i=14$, then we obtain $g_i = 14/5 = 2$, $i_L = 4$, and $r_L = \min \left\{1 , 0 \right\} = 0$. The final rank is then $r = 2$.
Similarly if we want to get the rank for the data id $i=27$, then $g_i = 27/5 = \min \left\{ 5 \;,\; 4 \right\} = 4$, $i_L = 7$, and $r_L = 1$. The final rank is then $r = 4+1 = 5 $.
We note here that with this specific example the distribution is not better than the one already used in \flups. However for more regular configurations we obtain a more homogeneous distribution (e.g. $N=32$, $P=6$).

\subsubsection{Order of the send requests based on the destination rank}  
As the communication is done in an all-to-all manner, each rank in a sub-communicator interacts with all the other ranks belonging to the same sub-communicator. 
An intuitive way to implement such a scheme is to have all the ranks start their request following the rank indexing of the sub-communicator. For example, all the send requests from all the ranks inside the sub-communicator will first be addressed to rank 0, then to rank 1, and so on. 
As stated in~\cite{Chunduri:2019}, this may lead to endpoint congestion and lower the code \revthree{performance}. 
To reduce the network congestion, the ranks communicate with others in ascending order, starting with their neighbor in the indexes list. In that case, the rank $r$ first sends its request to the rank having the index $r+1$, then to the rank indexed $r+2$, etc. 
The repartition of the active send requests across the different receivers is hence improved and it reduces the communication overheads.

\section{Analytical expressions used for the validation}
\label{app:validation:anal}
This section contains the details analytical expressions used for the validation of \flups in \Sect{sec_validation}.
The Poisson equation is solved on a cubic domain of spatial extent $[0, L]$ in all directions.

\subsection{Domain with symmetric and periodic BCs.}
\label{app:validation:spectral}
\be
\phi_{ref}(x,y,z) = \cos\plr{\pi \frac{x}{L}} \sin\plr{\frac{5\pi}{2} \frac{y}{L}} \sin\plr{8\pi \frac{z}{L}}
\eed

\subsection{Fully unbounded boundary conditions}
\label{app:validation:unbounded}
\begin{multline}
\phi_{ref}(x,y,z) = \\
\exp\plr{10 \plr{3 - \frac{1}{1-\plr{\frac{2x}{L} - 1}^{2}} - \frac{1}{1-\plr{\frac{2y}{L} - 1}^{2}} - \frac{1}{1-\plr{\frac{2z}{L} - 1}^{2}} }}.
\end{multline}

\subsection{Domain with two semi-infinite directions and one fully unbounded BC.}
\label{app:validation:mix}
\begin{multline*}
\phi_{ref}(x,y,z) = \\
\blr{\exp\plr{10 \plr{1 - \frac{1}{1 - \plr{ \frac{2x - 1.4L}{L} }^{2} } }}  + \exp\plr{10 \plr{1 - \frac{1}{1 - \plr{ \frac{2x - 2.6L}{L} }^{2} } }}}\\ 
\exp\plr{10 \plr{1 - \frac{1}{1-\plr{\frac{2y}{L} - 1}^{2}} }} \\ 
\blr{\exp\plr{10 \plr{1 - \frac{1}{1 - \plr{ \frac{2z - 0.6L}{L} }^{2} } }}  - \exp\plr{10 \plr{1 - \frac{1}{1 - \plr{ \frac{2z + 0.6L}{L} }^{2} } }}}.
\end{multline*}

\section{Convergence for the Bio-Savart solver}
\label{app:biosavart:conv}
This section provides the convergence results for the finite difference approximation of order $2$ and $6$ in the case of the Bio-Savart solver presented in \Sect{sec2_biot_savart}.

\begin{figure}[h!tp]
\centering
\input{figures/erratum_biot_savart/valid_tube_order_2}
\caption{Convergence of the Biot-Savart solver using 2nd order differentiation \code{CHAT2} ({\protect\ThickLineCircle{colchat2}{colchat2}}),  \code{HEJ2}({\protect\ThickLineCircle{colhej2}{colhej2}}), \code{HEJ4}({\protect\ThickLineCircle{colhej4}{colhej4}}), \code{HEJ6}({\protect\ThickLineCircle{colhej6}{colhej6}}), \code{HEJ8}({\protect\ThickLineCircle{colhej8}{colhej8}}), \code{HEJ10}({\protect\ThickLineCircle{colhej10}{colhej10}}),  \code{HEJ0}({\protect\ThickLineCircle{colhej0}{colhej0}})}
\label{fig:validation-bs-tube-FD2}
\end{figure}

\begin{figure}[h!tp]
\centering
\input{figures/erratum_biot_savart/valid_tube_order_4}
\caption{Convergence of the Biot-Savart solver using 4th order differentiation \code{CHAT2} ({\protect\ThickLineCircle{colchat2}{colchat2}}),  \code{HEJ2}({\protect\ThickLineCircle{colhej2}{colhej2}}), \code{HEJ4}({\protect\ThickLineCircle{colhej4}{colhej4}}), \code{HEJ6}({\protect\ThickLineCircle{colhej6}{colhej6}}), \code{HEJ8}({\protect\ThickLineCircle{colhej8}{colhej8}}), \code{HEJ10}({\protect\ThickLineCircle{colhej10}{colhej10}}),  \code{HEJ0}({\protect\ThickLineCircle{colhej0}{colhej0}})}
\label{fig:validation-bs-tube-FD4}
\end{figure}

\section{Performance metrics for weak and strong scalability}
\label{sec_app_perfs}
A software running for $T$ seconds on $P$ resources in parallel can be characterized by the percentage of the time spent in parallel regions, $\alpha$, and the percentage spent in serial regions, $\beta = 1 - \alpha$ such that $T = \alpha \; T + \left( 1 - \alpha \right) T$. 
When going from $P_0$ resources to $P_1$ with $r = P_1 / P_0$ the ratio between resources, only the parallel regions will benefit from the gain and the execution time becomes
\be
T_1 = \dfrac{\alpha \; T_0}{r} + \left(1 - \alpha \right) T_0
\eed

Amdahl's law \cite{Amdahl:1967} defines the \textit{speedup} as the ratio of both measured times:
\be
s_{P} = \dfrac{T_{0}}{T_{1}} = \dfrac{1}{\dfrac{\alpha}{r} + \left(1 - \alpha \right)}
\eec
and uses $s_p$ as a performance metric. This approach is usually referred to as strong scaling.

However, most practical applications scale the problem size with the available resources. In such a context, the time spent on $P_1$ resources scales with $r$ (as the problem size scales with $r$) while only the parallel region benefits from the additional resources:
\be
T_1 = r \left[ \dfrac{\alpha  \;T_0}{r} + \left(1 - \alpha \right) T_0 \right] =  \alpha  \;T_0 +r \; \left(1 - \alpha \right) \; T_0
\eed
Therefore serial regions will lead to a longer execution time, while time spent in parallel regions will remain constant.
This line of thought, usually called the weak scalability, leads to Gustafson's law \cite{Gustafson:1988} which defines the \textit{efficiency} as the ratio of both times
\be
\eta_{P,w} = \dfrac{T_0}{T_1} = \frac{1}{\alpha   + r \; \left(1 - \alpha \right)}
\eed

We note that an equivalent efficiency can also be obtained from Amdahl's law as the ratio between the speedup $s_P$ and theoretical gain that should have been obtained during the strong scalability, $r$:
\be
\eta_{P,s} = \dfrac{s_P}{r} = \dfrac{1}{ \alpha+r \; \left(1 - \alpha \right)}
\eed

Finally similar expressions can be obtained for $\beta = 1 - \alpha$, the serial percentage of the software.
We also highlight that both the strong and the weak scalability are driven by quality of implementation which can be measured by $\beta$.
A perfect scalability would lead to $\beta = 0$ and a perfect parallelization of the software.

\section{Details on the testcases}
\subsection{Comparison with \accfft}
\label{sec_detail_testcase_accfft}
\Cref{table_layout_accfft} details the layout used for the comparison with \accfft described in \Cref{sec_5_accfft}.

\begin{table}[ht!]
\centering
\begin{tabular}{ c | c c c}
nodes & $P_x$ & $P_y$ & $P_z$ \\
\hline
1 & 1 & 8 & 16 \\
2 & 1 & 16 & 16 \\
8 & 1 & 32 & 32 \\
32 & 1 & 64 & 64 \\
128 & 1 & 128 & 128 \\
\end{tabular}
\caption{Process distribution for the comparison with \accfft in \Cref{sec_5_accfft}}
\label{table_layout_accfft}
\end{table}

\subsection{Weak scalability}
\label{sec_detail_testcase_weak}
\Cref{table_layout_weak} details the layout used for the weak scaling analysis described in \Cref{sec:weak-scaling}.
The numbers in the table represents the process distribution in the three dimensions, the number of unknowns per process is kept constant to $96^3$.

\begin{table}[ht!]
\centering
\begin{tabular}{ c | c c c}
nodes & $P_x$ & $P_y$ & $P_z$ \\
\hline
1 & 4 & 4 & 8 \\
2 & 4 & 8 & 8 \\
4 & 8 & 8 & 8 \\
8 & 8 & 8 & 16 \\
16 & 8 & 16 & 16 \\
32 & 16 & 16 & 16 \\
64 & 16 & 16 & 32 \\
128 & 16 & 32 & 32 \\
256 & 32 & 32 & 32 \\
384 & 32 & 32 & 48 \\
\end{tabular}
\caption{Process distribution for the weak scaling analysis of \Cref{sec:weak-scaling}}
\label{table_layout_weak}
\end{table}

\subsection{Strong scalability}
\label{sec_detail_testcase_strong}
\Cref{table_layout_strong} details the layout used for the strong scaling analysis described in \Cref{sec:strong-scaling}.
The numbers in the table represents the process distribution in the three dimensions, the total number of unknowns is kept constant to $1280^3$.

\begin{table}[ht!]
\centering
\begin{tabular}{ c | c c c}
nodes & $P_x$ & $P_y$ & $P_z$ \\
\hline
1 & 4 & 4 & 8 \\
2 & 4 & 8 & 8 \\
4 & 8 & 8 & 8 \\
8 & 8 & 8 & 16 \\
16 & 8 & 16 & 16 \\
32 & 16 & 16 & 16 \\
64 & 16 & 16 & 32 \\
128 & 16 & 32 & 32 \\
256 & 32 & 32 & 32 \\
384 & 32 & 32 & 48 \\
\end{tabular}
\caption{Process distribution for the strong scaling analysis of \Cref{sec:strong-scaling}}
\label{table_layout_strong}
\end{table}

\section{Comparison of main European systems}
\label{app:eu-comp}

This section includes additional time-to-solution and weak efficiencies for the \isr and the \ata version of flups on different European systems, summarized in \Tref{tab:table-clusters}.
The results as presented in \Frefs{fig:benchmark-isr}{fig:benchmark-isr-weak-eff} for the \isr version and \Frefs{fig:benchmark-a2a}{fig:benchmark-a2a-weak-eff} for the \ata implementation.
For reference, we also provide the throughput per rank in \Tref{tab:eu-table-nb}, \Tref{tab:eu-table-isr} and \Tref{tab:eu-table-a2a}. The numbers have been multiplied by the factor $3/14$ for better comparison with the other results.

\begin{table*}[h!tp]
\centering
\begin{tabular}{ c c c c c c }
\hline
Name & Location & CPU & Interconnect & Transport Layer & OSU latency\\ 
\hline
\hline 

Lumi 	& Finland 		&  AMD EPYC 7763  & 200 Gb/s Slingshot-11       & \code{libfabric} 15.0.0 - \code{CXI} & 2.05 $\mu s$  \\
MeluXina  & Luxembourg  & AMD EPYC 7H12  & 200 Gb/s Infiniband HDR  & 	 \code{ucx} 1.13.1       & 1.45 $\mu s$ \\
Vega 	& Slovenia 	& AMD EPYC 7H12  & 100 Gb/s Infiniband HDR  & 	\code{ucx}  1.13.1       & 1.99 $\mu s$ \\
\hline 
\end{tabular}
  \caption{List of systems used for scalability testing}
  \label{tab:table-clusters}
\end{table*}

\begin{table}
\begin{center}
\input{figures/python_tab/comp_clusters_nb_time_unknowns_rank.tex}
\end{center}
\caption{Implementation \nb: throughput per rank [\MBps] for a solve on the three parallel architectures. To account for the domain-doubling technique for unbounded BCs a normalization factor of $14/3$ has been applied.}
\label{tab:eu-table-nb} 
\end{table}

\begin{figure}[h!tp]
	\centering
	\input{figures/python_fig/weak-cluster-comp-isr}
	\caption{\isr: Comparison of  three different architectures, Vega ({\protect\BarLegend{colvega}{colvega}}), MeluXina ({\protect\BarLegend{colmlx}{colmlx}}) and Lumi ({\protect\BarLegend{collumi}{collumi}}). Timings of weak scalability tests, performed with $96^{3}$ unknowns per rank on a fully unbounded domain.}
	\label{fig:benchmark-isr}
\end{figure}

\begin{figure}[h!tp]
	\centering		
	\begin{tikzpicture}
		\node[anchor=south west] (image1) at (0,0) {
			\input{figures/python_fig/weak-cluster-eff-isr}
                 };
                 \useasboundingbox;                     
		\node[anchor=south west,fill=white] at (7.7,6) (note) {$\beta=0\%$};
		\node[anchor=south west,fill=white] at (7.7,4.7) (note) {$\beta=0.2\%$};
		\node[anchor=south west,fill=white] at (7.7,3.5) (note) {$\beta=0.5\%$};
		\node[anchor=south west,fill=white] at (7.7,2.5) (note) {$\beta=1\%$};
	\end{tikzpicture}
	\caption{Implementation \isr: Weak efficiency $\eta_{P,w}$ on Vega ({\protect\ThickLineCircle{colvega}{colvega}}), MeluXina ({\protect\ThickLineCircle{colmlx}{colmlx}}), and Lumi ({\protect\ThickLineCircle{collumi}{collumi}}). Weak scalability tests performed with $96^{3}$ unknowns per rank on a fully unbounded test case.}
	\label{fig:benchmark-isr-weak-eff}
\end{figure}

\begin{table}[htp]
\begin{center}
\input{figures/python_tab/comp_clusters_isr_time_unknowns_rank.tex}
\end{center}
\caption{Implementation \isr: throughput per rank [\MBps] for a solve on the three parallel architectures. To account for the domain-doubling technique for unbounded BCs a normalization factor of $14/3$ has been applied.}
\label{tab:eu-table-isr} 
\end{table}

\begin{figure}[h!tp]
	\centering
	\input{figures/python_fig/weak-cluster-comp-a2a}
	\caption{\ata: Comparison of  three different architectures, Vega ({\protect\BarLegend{colvega}{colvega}}), MeluXina ({\protect\BarLegend{colmlx}{colmlx}}) and Lumi ({\protect\BarLegend{collumi}{collumi}}). Timings of weak scalability tests, performed with $96^{3}$ unknowns per rank on a fully unbounded domain.}
	\label{fig:benchmark-a2a}
\end{figure}

\begin{figure}[h!tp]
	\centering		
	\begin{tikzpicture}
		\node[anchor=south west] (image1) at (0,0) {
			\input{figures/python_fig/weak-cluster-eff-a2a}
                 };
                 \useasboundingbox;                     
		\node[anchor=south west,fill=white] at (7.6,6) (note) {$\beta=0\%$};
		\node[anchor=south west,fill=white] at (7.6,4.7) (note) {$\beta=0.2\%$};
		\node[anchor=south west,fill=white] at (7.6,3.5) (note) {$\beta=0.5\%$};
		\node[anchor=south west,fill=white] at (7.6,2.5) (note) {$\beta=1\%$};
	\end{tikzpicture}
	\caption{Implementation \ata: Weak efficiency $\eta_{P,w}$ on Vega ({\protect\ThickLineCircle{colvega}{colvega}}), MeluXina ({\protect\ThickLineCircle{colmlx}{colmlx}}), and Lumi ({\protect\ThickLineCircle{collumi}{collumi}}). Weak scalability tests performed with $96^{3}$ unknowns per rank on a fully unbounded test case.}
	\label{fig:benchmark-a2a-weak-eff}
\end{figure}

\begin{table}[htp]
\begin{center}
\input{figures/python_tab/comp_clusters_a2a_time_unknowns_rank.tex}
\end{center}
\caption{Implementation \ata: throughput per rank [\MBps] for a solve on the three parallel architectures. To account for the domain-doubling technique for unbounded BCs a normalization factor of $14/3$ has been applied.}
\label{tab:eu-table-a2a} 
\end{table}

%% file: figures/erratum_biot_savart/valid_tube_order_2.tex
\begin{tikzpicture}

\definecolor{darkgray}{RGB}{169,169,169}
\definecolor{darkgray176}{RGB}{176,176,176}
\definecolor{lightgreen161217155}{RGB}{161,217,155}
\definecolor{lightsteelblue188189220}{RGB}{188,189,220}
\definecolor{orangered2308513}{RGB}{230,85,13}
\definecolor{sandybrown253174107}{RGB}{253,174,107}
\definecolor{seagreen4916384}{RGB}{49,163,84}
\definecolor{slateblue117107177}{RGB}{117,107,177}
\definecolor{steelblue49130189}{RGB}{49,130,189}

\begin{axis}[
axis y line =left, axis x line =bottom, axis line style ={-},
log basis x={10},
log basis y={10},
minor xtick={},
minor ytick={},
tick align=center,
tick pos=left,
x grid style={darkgray176},
x grid style={draw=none},
xlabel near ticks,
xlabel={N\(\displaystyle _{points}\)},
xmajorgrids,
xmin=26.8449462631643, xmax=2518.79811332614,
xmode=log,
xtick style={color=black},
xtick={32,64,128,256,512,1024,2048},
xticklabels={$32^3$,$64^3$,$128^3$,$256^3$,$512^3$,$1024^3$,$2048^3$},
y axis line style={draw=none},
y grid style={darkgray176},
y grid style={dotted},
ylabel near ticks,
ylabel={\(\displaystyle E_{\infty}\)},
ymajorgrids,
ymin=5e-07, ymax=5,
ymode=log,
ytick style={color=darkgray},
ytick={1e-08,1e-07,1e-06,1e-05,0.0001,0.001,0.01,0.1,1,10,100}
]
\addplot [semithick, steelblue49130189, mark=*, mark size=1.5, mark options={solid}]
table {%
33 0.1002936301105
65 0.02892184592254
129 0.007364824473642
257 0.001849769831879
513 0.0004629803142637
1025 0.0001157868079251
2049 2.895094036504e-05
};
\addplot [semithick, orangered2308513, mark=*, mark size=1.5, mark options={solid}]
table {%
33 0.7320265641503
65 0.3005434681957
129 0.09220517657953
257 0.02447130885305
513 0.006204910790951
1025 0.001557351859239
2049 0.0003896902849596
};
\addplot [semithick, black, dashed]
table {%
128 0.0605670352413785
128 0.0605670352413785
256 0.0151417588103446
512 0.00378543970258616
1024 0.000946359925646539
2048 0.000236589981411635
2048 0.000236589981411635
};
\addplot [semithick, sandybrown253174107, mark=*, mark size=1.5, mark options={solid}]
table {%
33 0.4036129729964
65 0.08682197980525
129 0.01165743963267
257 0.001817063655263
513 0.000374318907808
1025 8.841640874757e-05
2049 2.177834808492e-05
};
\addplot [semithick, seagreen4916384, mark=*, mark size=1.5, mark options={solid}]
table {%
33 0.2384353879877
65 0.03688477297371
129 0.005956318985347
257 0.001393680204539
513 0.0003467768784418
1025 8.667605979151e-05
2049 2.166927723946e-05
};
\addplot [semithick, lightgreen161217155, mark=*, mark size=1.5, mark options={solid}]
table {%
33 0.1584544232673
65 0.02533445793591
129 0.005560044851691
257 0.001385982332419
513 0.0003466496999533
1025 8.667404455864e-05
2049 2.166924564073e-05
};
\addplot [semithick, slateblue117107177, mark=*, mark size=1.5, mark options={solid}]
table {%
33 0.121591003546
65 0.02269534955622
129 0.005533226504797
257 0.001385846584646
513 0.000346649133655
1025 8.667404230978e-05
2049 2.166924563185e-05
};
\addplot [semithick, black, dashed]
table {%
128 0.00340871497817954
128 0.00340871497817954
256 0.000852178744544886
512 0.000213044686136222
1024 5.32611715340554e-05
2048 1.33152928835138e-05
2048 1.33152928835138e-05
};
\addplot [semithick, lightsteelblue188189220, mark=*, mark size=1.5, mark options={solid}]
table {%
33 0.0770372084154
65 0.02203548751726
129 0.005509869954314
257 0.001338839812597
513 0.0003303740227614
1025 8.228167758584e-05
2049 2.055120318212e-05
};
\draw (axis cs:1024,0.000946359925646539) ++(2pt,-2pt) node[
  scale=1.04166666666667,
  anchor=north west,
  text=black,
  rotate=337.2
]{$p = 2$};
\draw (axis cs:1024,5.32611715340554e-05) ++(2pt,-2pt) node[
  scale=1.04166666666667,
  anchor=north west,
  text=black,
  rotate=337.2
]{$p = 2$};
\end{axis}

\end{tikzpicture}

%% file: figures/erratum_biot_savart/valid_tube_order_4.tex
\begin{tikzpicture}

\definecolor{darkgray}{RGB}{169,169,169}
\definecolor{darkgray176}{RGB}{176,176,176}
\definecolor{lightgreen161217155}{RGB}{161,217,155}
\definecolor{lightsteelblue188189220}{RGB}{188,189,220}
\definecolor{orangered2308513}{RGB}{230,85,13}
\definecolor{sandybrown253174107}{RGB}{253,174,107}
\definecolor{seagreen4916384}{RGB}{49,163,84}
\definecolor{slateblue117107177}{RGB}{117,107,177}
\definecolor{steelblue49130189}{RGB}{49,130,189}

\begin{axis}[
axis y line =left, axis x line =bottom, axis line style ={-},
log basis x={10},
log basis y={10},
minor xtick={},
minor ytick={},
tick align=center,
tick pos=left,
x grid style={darkgray176},
x grid style={draw=none},
xlabel near ticks,
xlabel={N\(\displaystyle _{points}\)},
xmajorgrids,
xmin=26.8449462631643, xmax=2518.79811332614,
xmode=log,
xtick style={color=black},
xtick={32,64,128,256,512,1024,2048},
xticklabels={$32^3$,$64^3$,$128^3$,$256^3$,$512^3$,$1024^3$,$2048^3$},
y axis line style={draw=none},
y grid style={darkgray176},
y grid style={dotted},
ylabel near ticks,
ylabel={\(\displaystyle E_{\infty}\)},
ymajorgrids,
ymin=5e-11, ymax=10,
ymode=log,
ytick style={color=darkgray},
ytick={1e-13,1e-11,1e-09,1e-07,1e-05,0.001,0.1,10,1000}
]
\addplot [semithick, steelblue49130189, mark=*, mark size=1.5, mark options={solid}]
table {%
33 0.03543002715661
65 0.008202335182184
129 0.001912651956991
257 0.0004703958862897
513 0.0001170642972688
1025 2.923947913813e-05
2049 7.307715006677e-06
};
\addplot [semithick, orangered2308513, mark=*, mark size=1.5, mark options={solid}]
table {%
33 0.7130260129902
65 0.2879301701702
129 0.0875044241303
257 0.02314566992518
513 0.005864098666384
1025 0.001471270854556
2049 0.0003681313996038
};
\addplot [semithick, black, dashed]
table {%
128 0.0572162774229101
128 0.0572162774229101
256 0.0143040693557275
512 0.00357601733893188
1024 0.00089400433473297
2048 0.000223501083683242
2048 0.000223501083683242
};
\addplot [semithick, sandybrown253174107, mark=*, mark size=1.5, mark options={solid}]
table {%
33 0.3663157584688
65 0.06851429815619
129 0.006393523464014
257 0.0004488439138552
513 2.890922118204e-05
1025 1.820687723697e-06
2049 1.140236267627e-07
};
\addplot [semithick, black, dashed]
table {%
128 0.00454125160753318
128 0.00454125160753318
256 0.000283828225470824
512 1.77392640919265e-05
1024 1.1087040057454e-06
2048 6.92940003590878e-08
2048 6.92940003590878e-08
};
\addplot [semithick, seagreen4916384, mark=*, mark size=1.5, mark options={solid}]
table {%
33 0.1880862831172
65 0.01700467400794
129 0.0005491868183181
257 1.352411361666e-05
513 4.255161835109e-07
1025 1.964065399207e-08
2049 1.117561376773e-09
};
\addplot [semithick, lightgreen161217155, mark=*, mark size=1.5, mark options={solid}]
table {%
33 0.1141755475825
65 0.005188520161972
129 0.0001071484305406
257 4.598291415747e-06
513 2.770523529483e-07
1025 1.729183440879e-08
2049 1.080717071389e-09
};
\addplot [semithick, slateblue117107177, mark=*, mark size=1.5, mark options={solid}]
table {%
33 0.0718237136608
65 0.002122878848376
129 7.212742272955e-05
257 4.413761364952e-06
513 2.762734679918e-07
1025 1.728877285778e-08
2049 1.08070508098e-09
};
\addplot [semithick, black, dashed]
table {%
128 2.73731574124392e-05
128 2.73731574124392e-05
256 1.71082233827745e-06
512 1.0692639614234e-07
1024 6.68289975889628e-09
2048 4.17681234931018e-10
2048 4.17681234931018e-10
};
\addplot [semithick, lightsteelblue188189220, mark=*, mark size=1.5, mark options={solid}]
table {%
33 0.01449187529344
65 0.001116996247952
129 0.0001083768893706
257 4.660575194904e-05
513 1.62059938118e-05
1025 4.400428341267e-06
2049 1.122865710856e-06
};
\draw (axis cs:1024,0.00089400433473297) ++(2pt,-2pt) node[
  scale=1.04166666666667,
  anchor=north west,
  text=black,
  rotate=345.4
]{$p = 2$};
\draw (axis cs:1024,1.1087040057454e-06) ++(2pt,-2pt) node[
  scale=1.04166666666667,
  anchor=north west,
  text=black,
  rotate=332.5
]{$p = 4$};
\draw (axis cs:1024,6.68289975889628e-09) ++(2pt,-2pt) node[
  scale=1.04166666666667,
  anchor=north west,
  text=black,
  rotate=332.5
]{$p = 4$};
\end{axis}

\end{tikzpicture}

%% file: figures/python_tab/comp_clusters_nb_time_unknowns_rank.tex
\begin{tabular}{lrrrrr}
\hline
  N nodes   &   \centercell{$1$ } &   \centercell{$2$ } &   \centercell{$8$ } &   \centercell{$64$ } &   \centercell{$128$ } \\
\hline
 Vega       &               33.82 &               31.62 &               25.50 &                23.09 &                 17.11 \\
 MeluXina   &               32.14 &               31.54 &               29.59 &                29.19 &                 26.57 \\
 Lumi       &               34.43 &               33.13 &               28.40 &                24.15 &                 21.12 \\
\hline
\end{tabular}

%% file: figures/python_fig/weak-cluster-comp-isr.tex
\begin{tikzpicture}

\definecolor{cornflowerblue107174214}{RGB}{107,174,214}
\definecolor{darkgray}{RGB}{169,169,169}
\definecolor{darkgray176}{RGB}{176,176,176}
\definecolor{gainsboro232221219}{RGB}{232,221,219}
\definecolor{gray163119111}{RGB}{163,119,111}
\definecolor{lavender225238246}{RGB}{225,238,246}
\definecolor{lavender233224241}{RGB}{233,224,241}
\definecolor{lightblue166206230}{RGB}{166,206,230}
\definecolor{mediumpurple148103189}{RGB}{148,103,189}
\definecolor{mediumpurple169133202}{RGB}{169,133,202}
\definecolor{plum190163215}{RGB}{190,163,215}
\definecolor{rosybrown186153147}{RGB}{186,153,147}
\definecolor{sienna1408675}{RGB}{140,86,75}
\definecolor{skyblue136190222}{RGB}{136,190,222}

\begin{axis}[
axis line style={draw=none},
tick pos=left,
x grid style={darkgray176},
x grid style={draw=none},
xlabel near ticks,
xlabel={N nodes},
xmajorgrids,
xmin=-0.716666666666667, xmax=7.71666666666667,
xtick style={color=darkgray},
xtick={0,1,2,3,4,5,6,7},
xticklabels={$1$,$2$,$4$,$8$,$16$,$32$,$64$,$128$},
xticklabels={1,2,4,8,16,32,64,128},
y grid style={darkgray176},
ylabel near ticks,
ylabel={time/solve - [sec]},
ymajorgrids,
ymin=0, ymax=2.5,
ytick style={color=darkgray}
]
\draw[draw=black,fill=lavender225238246,line width=0.004pt,postaction={pattern=crosshatch}] (axis cs:-0.333333333333333,0) rectangle (axis cs:-0.166666666666667,0.3054909535);
\draw[draw=black,fill=cornflowerblue107174214,line width=0.004pt,postaction={pattern=north east lines}] (axis cs:-0.333333333333333,0.3054909535) rectangle (axis cs:-0.166666666666667,0.317927586);
\draw[draw=black,fill=skyblue136190222,line width=0.004pt,postaction={pattern=north east lines}] (axis cs:-0.333333333333333,0.317927586) rectangle (axis cs:-0.166666666666667,0.3591915535);
\draw[draw=black,fill=lightblue166206230,line width=0.004pt,postaction={pattern=north east lines}] (axis cs:-0.333333333333333,0.3591915535) rectangle (axis cs:-0.166666666666667,0.441097513);
\draw[draw=black,fill=cornflowerblue107174214,line width=0.004pt] (axis cs:-0.333333333333333,0.441097513) rectangle (axis cs:-0.166666666666667,0.4804012485);
\draw[draw=black,fill=skyblue136190222,line width=0.004pt] (axis cs:-0.333333333333333,0.4804012485) rectangle (axis cs:-0.166666666666667,0.5843578185);
\draw[draw=black,fill=lightblue166206230,line width=0.004pt] (axis cs:-0.333333333333333,0.5843578185) rectangle (axis cs:-0.166666666666667,0.783511238);
\draw[draw=black,fill=lavender225238246,line width=0.004pt,postaction={pattern=crosshatch}] (axis cs:0.666666666666667,0) rectangle (axis cs:0.833333333333333,0.308992876);
\draw[draw=black,fill=cornflowerblue107174214,line width=0.004pt,postaction={pattern=north east lines}] (axis cs:0.666666666666667,0.308992876) rectangle (axis cs:0.833333333333333,0.3221837505);
\draw[draw=black,fill=skyblue136190222,line width=0.004pt,postaction={pattern=north east lines}] (axis cs:0.666666666666667,0.3221837505) rectangle (axis cs:0.833333333333333,0.3640010135);
\draw[draw=black,fill=lightblue166206230,line width=0.004pt,postaction={pattern=north east lines}] (axis cs:0.666666666666667,0.3640010135) rectangle (axis cs:0.833333333333333,0.438496926);
\draw[draw=black,fill=cornflowerblue107174214,line width=0.004pt] (axis cs:0.666666666666667,0.438496926) rectangle (axis cs:0.833333333333333,0.4771811955);
\draw[draw=black,fill=skyblue136190222,line width=0.004pt] (axis cs:0.666666666666667,0.4771811955) rectangle (axis cs:0.833333333333333,0.59510205);
\draw[draw=black,fill=lightblue166206230,line width=0.004pt] (axis cs:0.666666666666667,0.59510205) rectangle (axis cs:0.833333333333333,0.8494521995);
\draw[draw=black,fill=lavender225238246,line width=0.004pt,postaction={pattern=crosshatch}] (axis cs:1.66666666666667,0) rectangle (axis cs:1.83333333333333,0.3029592865);
\draw[draw=black,fill=cornflowerblue107174214,line width=0.004pt,postaction={pattern=north east lines}] (axis cs:1.66666666666667,0.3029592865) rectangle (axis cs:1.83333333333333,0.3162317455);
\draw[draw=black,fill=skyblue136190222,line width=0.004pt,postaction={pattern=north east lines}] (axis cs:1.66666666666667,0.3162317455) rectangle (axis cs:1.83333333333333,0.3553845145);
\draw[draw=black,fill=lightblue166206230,line width=0.004pt,postaction={pattern=north east lines}] (axis cs:1.66666666666667,0.3553845145) rectangle (axis cs:1.83333333333333,0.409343639);
\draw[draw=black,fill=cornflowerblue107174214,line width=0.004pt] (axis cs:1.66666666666667,0.409343639) rectangle (axis cs:1.83333333333333,0.4515973985);
\draw[draw=black,fill=skyblue136190222,line width=0.004pt] (axis cs:1.66666666666667,0.4515973985) rectangle (axis cs:1.83333333333333,0.5863563335);
\draw[draw=black,fill=lightblue166206230,line width=0.004pt] (axis cs:1.66666666666667,0.5863563335) rectangle (axis cs:1.83333333333333,1.0203035745);
\draw[draw=black,fill=lavender225238246,line width=0.004pt,postaction={pattern=crosshatch}] (axis cs:2.66666666666667,0) rectangle (axis cs:2.83333333333333,0.2978016875);
\draw[draw=black,fill=cornflowerblue107174214,line width=0.004pt,postaction={pattern=north east lines}] (axis cs:2.66666666666667,0.2978016875) rectangle (axis cs:2.83333333333333,0.3109882185);
\draw[draw=black,fill=skyblue136190222,line width=0.004pt,postaction={pattern=north east lines}] (axis cs:2.66666666666667,0.3109882185) rectangle (axis cs:2.83333333333333,0.348909097);
\draw[draw=black,fill=lightblue166206230,line width=0.004pt,postaction={pattern=north east lines}] (axis cs:2.66666666666667,0.348909097) rectangle (axis cs:2.83333333333333,0.3903250335);
\draw[draw=black,fill=cornflowerblue107174214,line width=0.004pt] (axis cs:2.66666666666667,0.3903250335) rectangle (axis cs:2.83333333333333,0.4333398325);
\draw[draw=black,fill=skyblue136190222,line width=0.004pt] (axis cs:2.66666666666667,0.4333398325) rectangle (axis cs:2.83333333333333,0.559195752);
\draw[draw=black,fill=lightblue166206230,line width=0.004pt] (axis cs:2.66666666666667,0.559195752) rectangle (axis cs:2.83333333333333,1.1569656925);
\draw[draw=black,fill=lavender225238246,line width=0.004pt,postaction={pattern=crosshatch}] (axis cs:3.66666666666667,0) rectangle (axis cs:3.83333333333333,0.2802891905);
\draw[draw=black,fill=cornflowerblue107174214,line width=0.004pt,postaction={pattern=north east lines}] (axis cs:3.66666666666667,0.2802891905) rectangle (axis cs:3.83333333333333,0.2939512415);
\draw[draw=black,fill=skyblue136190222,line width=0.004pt,postaction={pattern=north east lines}] (axis cs:3.66666666666667,0.2939512415) rectangle (axis cs:3.83333333333333,0.333328735);
\draw[draw=black,fill=lightblue166206230,line width=0.004pt,postaction={pattern=north east lines}] (axis cs:3.66666666666667,0.333328735) rectangle (axis cs:3.83333333333333,0.372134517);
\draw[draw=black,fill=cornflowerblue107174214,line width=0.004pt] (axis cs:3.66666666666667,0.372134517) rectangle (axis cs:3.83333333333333,0.4149351065);
\draw[draw=black,fill=skyblue136190222,line width=0.004pt] (axis cs:3.66666666666667,0.4149351065) rectangle (axis cs:3.83333333333333,0.609927278);
\draw[draw=black,fill=lightblue166206230,line width=0.004pt] (axis cs:3.66666666666667,0.609927278) rectangle (axis cs:3.83333333333333,1.4312978755);
\draw[draw=black,fill=lavender225238246,line width=0.004pt,postaction={pattern=crosshatch}] (axis cs:4.66666666666667,0) rectangle (axis cs:4.83333333333333,0.2976058585);
\draw[draw=black,fill=cornflowerblue107174214,line width=0.004pt,postaction={pattern=north east lines}] (axis cs:4.66666666666667,0.2976058585) rectangle (axis cs:4.83333333333333,0.311461009);
\draw[draw=black,fill=skyblue136190222,line width=0.004pt,postaction={pattern=north east lines}] (axis cs:4.66666666666667,0.311461009) rectangle (axis cs:4.83333333333333,0.3443480455);
\draw[draw=black,fill=lightblue166206230,line width=0.004pt,postaction={pattern=north east lines}] (axis cs:4.66666666666667,0.3443480455) rectangle (axis cs:4.83333333333333,0.3788212665);
\draw[draw=black,fill=cornflowerblue107174214,line width=0.004pt] (axis cs:4.66666666666667,0.3788212665) rectangle (axis cs:4.83333333333333,0.4274812725);
\draw[draw=black,fill=skyblue136190222,line width=0.004pt] (axis cs:4.66666666666667,0.4274812725) rectangle (axis cs:4.83333333333333,0.672227424);
\draw[draw=black,fill=lightblue166206230,line width=0.004pt] (axis cs:4.66666666666667,0.672227424) rectangle (axis cs:4.83333333333333,1.522666636);
\draw[draw=black,fill=lavender225238246,line width=0.004pt,postaction={pattern=crosshatch}] (axis cs:5.66666666666667,0) rectangle (axis cs:5.83333333333333,0.286468077);
\draw[draw=black,fill=cornflowerblue107174214,line width=0.004pt,postaction={pattern=north east lines}] (axis cs:5.66666666666667,0.286468077) rectangle (axis cs:5.83333333333333,0.3002376785);
\draw[draw=black,fill=skyblue136190222,line width=0.004pt,postaction={pattern=north east lines}] (axis cs:5.66666666666667,0.3002376785) rectangle (axis cs:5.83333333333333,0.3312000995);
\draw[draw=black,fill=lightblue166206230,line width=0.004pt,postaction={pattern=north east lines}] (axis cs:5.66666666666667,0.3312000995) rectangle (axis cs:5.83333333333333,0.3626690305);
\draw[draw=black,fill=cornflowerblue107174214,line width=0.004pt] (axis cs:5.66666666666667,0.3626690305) rectangle (axis cs:5.83333333333333,0.41166095);
\draw[draw=black,fill=skyblue136190222,line width=0.004pt] (axis cs:5.66666666666667,0.41166095) rectangle (axis cs:5.83333333333333,0.626281469);
\draw[draw=black,fill=lightblue166206230,line width=0.004pt] (axis cs:5.66666666666667,0.626281469) rectangle (axis cs:5.83333333333333,1.563023535);
\draw[draw=black,fill=lavender225238246,line width=0.004pt,postaction={pattern=crosshatch}] (axis cs:6.66666666666667,0) rectangle (axis cs:6.83333333333333,0.2969293005);
\draw[draw=black,fill=cornflowerblue107174214,line width=0.004pt,postaction={pattern=north east lines}] (axis cs:6.66666666666667,0.2969293005) rectangle (axis cs:6.83333333333333,0.311601827);
\draw[draw=black,fill=skyblue136190222,line width=0.004pt,postaction={pattern=north east lines}] (axis cs:6.66666666666667,0.311601827) rectangle (axis cs:6.83333333333333,0.3460175575);
\draw[draw=black,fill=lightblue166206230,line width=0.004pt,postaction={pattern=north east lines}] (axis cs:6.66666666666667,0.3460175575) rectangle (axis cs:6.83333333333333,0.379734239);
\draw[draw=black,fill=cornflowerblue107174214,line width=0.004pt] (axis cs:6.66666666666667,0.379734239) rectangle (axis cs:6.83333333333333,0.4288361205);
\draw[draw=black,fill=skyblue136190222,line width=0.004pt] (axis cs:6.66666666666667,0.4288361205) rectangle (axis cs:6.83333333333333,0.877867615);
\draw[draw=black,fill=lightblue166206230,line width=0.004pt] (axis cs:6.66666666666667,0.877867615) rectangle (axis cs:6.83333333333333,2.0271372555);
\draw[draw=black,fill=lavender233224241,line width=0.004pt,postaction={pattern=crosshatch}] (axis cs:-0.0833333333333333,0) rectangle (axis cs:0.0833333333333333,0.3131803425);
\draw[draw=black,fill=mediumpurple148103189,line width=0.004pt,postaction={pattern=north east lines}] (axis cs:-0.0833333333333333,0.3131803425) rectangle (axis cs:0.0833333333333333,0.3262096605);
\draw[draw=black,fill=mediumpurple169133202,line width=0.004pt,postaction={pattern=north east lines}] (axis cs:-0.0833333333333333,0.3262096605) rectangle (axis cs:0.0833333333333333,0.3695229205);
\draw[draw=black,fill=plum190163215,line width=0.004pt,postaction={pattern=north east lines}] (axis cs:-0.0833333333333333,0.3695229205) rectangle (axis cs:0.0833333333333333,0.454220782);
\draw[draw=black,fill=mediumpurple148103189,line width=0.004pt] (axis cs:-0.0833333333333333,0.454220782) rectangle (axis cs:0.0833333333333333,0.494498673);
\draw[draw=black,fill=mediumpurple169133202,line width=0.004pt] (axis cs:-0.0833333333333333,0.494498673) rectangle (axis cs:0.0833333333333333,0.602061583);
\draw[draw=black,fill=plum190163215,line width=0.004pt] (axis cs:-0.0833333333333333,0.602061583) rectangle (axis cs:0.0833333333333333,0.8068698685);
\draw[draw=black,fill=lavender233224241,line width=0.004pt,postaction={pattern=crosshatch}] (axis cs:0.916666666666667,0) rectangle (axis cs:1.08333333333333,0.3219412555);
\draw[draw=black,fill=mediumpurple148103189,line width=0.004pt,postaction={pattern=north east lines}] (axis cs:0.916666666666667,0.3219412555) rectangle (axis cs:1.08333333333333,0.335544663);
\draw[draw=black,fill=mediumpurple169133202,line width=0.004pt,postaction={pattern=north east lines}] (axis cs:0.916666666666667,0.335544663) rectangle (axis cs:1.08333333333333,0.3803152365);
\draw[draw=black,fill=plum190163215,line width=0.004pt,postaction={pattern=north east lines}] (axis cs:0.916666666666667,0.3803152365) rectangle (axis cs:1.08333333333333,0.4717677835);
\draw[draw=black,fill=mediumpurple148103189,line width=0.004pt] (axis cs:0.916666666666667,0.4717677835) rectangle (axis cs:1.08333333333333,0.511654312);
\draw[draw=black,fill=mediumpurple169133202,line width=0.004pt] (axis cs:0.916666666666667,0.511654312) rectangle (axis cs:1.08333333333333,0.6273247245);
\draw[draw=black,fill=plum190163215,line width=0.004pt] (axis cs:0.916666666666667,0.6273247245) rectangle (axis cs:1.08333333333333,0.8604125095);
\draw[draw=black,fill=lavender233224241,line width=0.004pt,postaction={pattern=crosshatch}] (axis cs:1.91666666666667,0) rectangle (axis cs:2.08333333333333,0.3167791635);
\draw[draw=black,fill=mediumpurple148103189,line width=0.004pt,postaction={pattern=north east lines}] (axis cs:1.91666666666667,0.3167791635) rectangle (axis cs:2.08333333333333,0.3307854995);
\draw[draw=black,fill=mediumpurple169133202,line width=0.004pt,postaction={pattern=north east lines}] (axis cs:1.91666666666667,0.3307854995) rectangle (axis cs:2.08333333333333,0.373768301);
\draw[draw=black,fill=plum190163215,line width=0.004pt,postaction={pattern=north east lines}] (axis cs:1.91666666666667,0.373768301) rectangle (axis cs:2.08333333333333,0.4373708245);
\draw[draw=black,fill=mediumpurple148103189,line width=0.004pt] (axis cs:1.91666666666667,0.4373708245) rectangle (axis cs:2.08333333333333,0.481189448);
\draw[draw=black,fill=mediumpurple169133202,line width=0.004pt] (axis cs:1.91666666666667,0.481189448) rectangle (axis cs:2.08333333333333,0.609244789);
\draw[draw=black,fill=plum190163215,line width=0.004pt] (axis cs:1.91666666666667,0.609244789) rectangle (axis cs:2.08333333333333,0.9621874315);
\draw[draw=black,fill=lavender233224241,line width=0.004pt,postaction={pattern=crosshatch}] (axis cs:2.91666666666667,0) rectangle (axis cs:3.08333333333333,0.3169454135);
\draw[draw=black,fill=mediumpurple148103189,line width=0.004pt,postaction={pattern=north east lines}] (axis cs:2.91666666666667,0.3169454135) rectangle (axis cs:3.08333333333333,0.330893929);
\draw[draw=black,fill=mediumpurple169133202,line width=0.004pt,postaction={pattern=north east lines}] (axis cs:2.91666666666667,0.330893929) rectangle (axis cs:3.08333333333333,0.3734772125);
\draw[draw=black,fill=plum190163215,line width=0.004pt,postaction={pattern=north east lines}] (axis cs:2.91666666666667,0.3734772125) rectangle (axis cs:3.08333333333333,0.4306532825);
\draw[draw=black,fill=mediumpurple148103189,line width=0.004pt] (axis cs:2.91666666666667,0.4306532825) rectangle (axis cs:3.08333333333333,0.4750181235);
\draw[draw=black,fill=mediumpurple169133202,line width=0.004pt] (axis cs:2.91666666666667,0.4750181235) rectangle (axis cs:3.08333333333333,0.5952132435);
\draw[draw=black,fill=plum190163215,line width=0.004pt] (axis cs:2.91666666666667,0.5952132435) rectangle (axis cs:3.08333333333333,1.016622728);
\draw[draw=black,fill=lavender233224241,line width=0.004pt,postaction={pattern=crosshatch}] (axis cs:3.91666666666667,0) rectangle (axis cs:4.08333333333333,0.3156866215);
\draw[draw=black,fill=mediumpurple148103189,line width=0.004pt,postaction={pattern=north east lines}] (axis cs:3.91666666666667,0.3156866215) rectangle (axis cs:4.08333333333333,0.329673557);
\draw[draw=black,fill=mediumpurple169133202,line width=0.004pt,postaction={pattern=north east lines}] (axis cs:3.91666666666667,0.329673557) rectangle (axis cs:4.08333333333333,0.37084059);
\draw[draw=black,fill=plum190163215,line width=0.004pt,postaction={pattern=north east lines}] (axis cs:3.91666666666667,0.37084059) rectangle (axis cs:4.08333333333333,0.427272454);
\draw[draw=black,fill=mediumpurple148103189,line width=0.004pt] (axis cs:3.91666666666667,0.427272454) rectangle (axis cs:4.08333333333333,0.4721085735);
\draw[draw=black,fill=mediumpurple169133202,line width=0.004pt] (axis cs:3.91666666666667,0.4721085735) rectangle (axis cs:4.08333333333333,0.6209276545);
\draw[draw=black,fill=plum190163215,line width=0.004pt] (axis cs:3.91666666666667,0.6209276545) rectangle (axis cs:4.08333333333333,1.0887241035);
\draw[draw=black,fill=lavender233224241,line width=0.004pt,postaction={pattern=crosshatch}] (axis cs:4.91666666666667,0) rectangle (axis cs:5.08333333333333,0.3068474565);
\draw[draw=black,fill=mediumpurple148103189,line width=0.004pt,postaction={pattern=north east lines}] (axis cs:4.91666666666667,0.3068474565) rectangle (axis cs:5.08333333333333,0.321241185);
\draw[draw=black,fill=mediumpurple169133202,line width=0.004pt,postaction={pattern=north east lines}] (axis cs:4.91666666666667,0.321241185) rectangle (axis cs:5.08333333333333,0.357739982);
\draw[draw=black,fill=plum190163215,line width=0.004pt,postaction={pattern=north east lines}] (axis cs:4.91666666666667,0.357739982) rectangle (axis cs:5.08333333333333,0.4013377585);
\draw[draw=black,fill=mediumpurple148103189,line width=0.004pt] (axis cs:4.91666666666667,0.4013377585) rectangle (axis cs:5.08333333333333,0.449570246);
\draw[draw=black,fill=mediumpurple169133202,line width=0.004pt] (axis cs:4.91666666666667,0.449570246) rectangle (axis cs:5.08333333333333,0.6390834765);
\draw[draw=black,fill=plum190163215,line width=0.004pt] (axis cs:4.91666666666667,0.6390834765) rectangle (axis cs:5.08333333333333,1.210944598);
\draw[draw=black,fill=lavender233224241,line width=0.004pt,postaction={pattern=crosshatch}] (axis cs:5.91666666666667,0) rectangle (axis cs:6.08333333333333,0.306652815);
\draw[draw=black,fill=mediumpurple148103189,line width=0.004pt,postaction={pattern=north east lines}] (axis cs:5.91666666666667,0.306652815) rectangle (axis cs:6.08333333333333,0.321189027);
\draw[draw=black,fill=mediumpurple169133202,line width=0.004pt,postaction={pattern=north east lines}] (axis cs:5.91666666666667,0.321189027) rectangle (axis cs:6.08333333333333,0.35803039);
\draw[draw=black,fill=plum190163215,line width=0.004pt,postaction={pattern=north east lines}] (axis cs:5.91666666666667,0.35803039) rectangle (axis cs:6.08333333333333,0.395563924);
\draw[draw=black,fill=mediumpurple148103189,line width=0.004pt] (axis cs:5.91666666666667,0.395563924) rectangle (axis cs:6.08333333333333,0.443450597);
\draw[draw=black,fill=mediumpurple169133202,line width=0.004pt] (axis cs:5.91666666666667,0.443450597) rectangle (axis cs:6.08333333333333,0.622660325);
\draw[draw=black,fill=plum190163215,line width=0.004pt] (axis cs:5.91666666666667,0.622660325) rectangle (axis cs:6.08333333333333,1.2268986195);
\draw[draw=black,fill=lavender233224241,line width=0.004pt,postaction={pattern=crosshatch}] (axis cs:6.91666666666667,0) rectangle (axis cs:7.08333333333333,0.3021572655);
\draw[draw=black,fill=mediumpurple148103189,line width=0.004pt,postaction={pattern=north east lines}] (axis cs:6.91666666666667,0.3021572655) rectangle (axis cs:7.08333333333333,0.3165073005);
\draw[draw=black,fill=mediumpurple169133202,line width=0.004pt,postaction={pattern=north east lines}] (axis cs:6.91666666666667,0.3165073005) rectangle (axis cs:7.08333333333333,0.349636095);
\draw[draw=black,fill=plum190163215,line width=0.004pt,postaction={pattern=north east lines}] (axis cs:6.91666666666667,0.349636095) rectangle (axis cs:7.08333333333333,0.3892873675);
\draw[draw=black,fill=mediumpurple148103189,line width=0.004pt] (axis cs:6.91666666666667,0.3892873675) rectangle (axis cs:7.08333333333333,0.4403530925);
\draw[draw=black,fill=mediumpurple169133202,line width=0.004pt] (axis cs:6.91666666666667,0.4403530925) rectangle (axis cs:7.08333333333333,0.698434321);
\draw[draw=black,fill=plum190163215,line width=0.004pt] (axis cs:6.91666666666667,0.698434321) rectangle (axis cs:7.08333333333333,1.3868527435);
\draw[draw=black,fill=gainsboro232221219,line width=0.004pt,postaction={pattern=crosshatch}] (axis cs:0.166666666666667,0) rectangle (axis cs:0.333333333333333,0.2936796525);
\draw[draw=black,fill=sienna1408675,line width=0.004pt,postaction={pattern=north east lines}] (axis cs:0.166666666666667,0.2936796525) rectangle (axis cs:0.333333333333333,0.3074714195);
\draw[draw=black,fill=gray163119111,line width=0.004pt,postaction={pattern=north east lines}] (axis cs:0.166666666666667,0.3074714195) rectangle (axis cs:0.333333333333333,0.352489211);
\draw[draw=black,fill=rosybrown186153147,line width=0.004pt,postaction={pattern=north east lines}] (axis cs:0.166666666666667,0.352489211) rectangle (axis cs:0.333333333333333,0.4458582915);
\draw[draw=black,fill=sienna1408675,line width=0.004pt] (axis cs:0.166666666666667,0.4458582915) rectangle (axis cs:0.333333333333333,0.4974541675);
\draw[draw=black,fill=gray163119111,line width=0.004pt] (axis cs:0.166666666666667,0.4974541675) rectangle (axis cs:0.333333333333333,0.628247295);
\draw[draw=black,fill=rosybrown186153147,line width=0.004pt] (axis cs:0.166666666666667,0.628247295) rectangle (axis cs:0.333333333333333,0.856325876);
\draw[draw=black,fill=gainsboro232221219,line width=0.004pt,postaction={pattern=crosshatch}] (axis cs:1.16666666666667,0) rectangle (axis cs:1.33333333333333,0.277358019);
\draw[draw=black,fill=sienna1408675,line width=0.004pt,postaction={pattern=north east lines}] (axis cs:1.16666666666667,0.277358019) rectangle (axis cs:1.33333333333333,0.291879441);
\draw[draw=black,fill=gray163119111,line width=0.004pt,postaction={pattern=north east lines}] (axis cs:1.16666666666667,0.291879441) rectangle (axis cs:1.33333333333333,0.335585042);
\draw[draw=black,fill=rosybrown186153147,line width=0.004pt,postaction={pattern=north east lines}] (axis cs:1.16666666666667,0.335585042) rectangle (axis cs:1.33333333333333,0.396042137);
\draw[draw=black,fill=sienna1408675,line width=0.004pt] (axis cs:1.16666666666667,0.396042137) rectangle (axis cs:1.33333333333333,0.446106274);
\draw[draw=black,fill=gray163119111,line width=0.004pt] (axis cs:1.16666666666667,0.446106274) rectangle (axis cs:1.33333333333333,0.596688962);
\draw[draw=black,fill=rosybrown186153147,line width=0.004pt] (axis cs:1.16666666666667,0.596688962) rectangle (axis cs:1.33333333333333,0.949350335);
\draw[draw=black,fill=gainsboro232221219,line width=0.004pt,postaction={pattern=crosshatch}] (axis cs:2.16666666666667,0) rectangle (axis cs:2.33333333333333,0.270165271);
\draw[draw=black,fill=sienna1408675,line width=0.004pt,postaction={pattern=north east lines}] (axis cs:2.16666666666667,0.270165271) rectangle (axis cs:2.33333333333333,0.2856113575);
\draw[draw=black,fill=gray163119111,line width=0.004pt,postaction={pattern=north east lines}] (axis cs:2.16666666666667,0.2856113575) rectangle (axis cs:2.33333333333333,0.3293631675);
\draw[draw=black,fill=rosybrown186153147,line width=0.004pt,postaction={pattern=north east lines}] (axis cs:2.16666666666667,0.3293631675) rectangle (axis cs:2.33333333333333,0.3797460765);
\draw[draw=black,fill=sienna1408675,line width=0.004pt] (axis cs:2.16666666666667,0.3797460765) rectangle (axis cs:2.33333333333333,0.4306017635);
\draw[draw=black,fill=gray163119111,line width=0.004pt] (axis cs:2.16666666666667,0.4306017635) rectangle (axis cs:2.33333333333333,0.5844520385);
\draw[draw=black,fill=rosybrown186153147,line width=0.004pt] (axis cs:2.16666666666667,0.5844520385) rectangle (axis cs:2.33333333333333,1.099169487);
\draw[draw=black,fill=gainsboro232221219,line width=0.004pt,postaction={pattern=crosshatch}] (axis cs:3.16666666666667,0) rectangle (axis cs:3.33333333333333,0.274551853);
\draw[draw=black,fill=sienna1408675,line width=0.004pt,postaction={pattern=north east lines}] (axis cs:3.16666666666667,0.274551853) rectangle (axis cs:3.33333333333333,0.2900887665);
\draw[draw=black,fill=gray163119111,line width=0.004pt,postaction={pattern=north east lines}] (axis cs:3.16666666666667,0.2900887665) rectangle (axis cs:3.33333333333333,0.3338131835);
\draw[draw=black,fill=rosybrown186153147,line width=0.004pt,postaction={pattern=north east lines}] (axis cs:3.16666666666667,0.3338131835) rectangle (axis cs:3.33333333333333,0.375564728);
\draw[draw=black,fill=sienna1408675,line width=0.004pt] (axis cs:3.16666666666667,0.375564728) rectangle (axis cs:3.33333333333333,0.4266844675);
\draw[draw=black,fill=gray163119111,line width=0.004pt] (axis cs:3.16666666666667,0.4266844675) rectangle (axis cs:3.33333333333333,0.5625342685);
\draw[draw=black,fill=rosybrown186153147,line width=0.004pt] (axis cs:3.16666666666667,0.5625342685) rectangle (axis cs:3.33333333333333,1.142740948);
\draw[draw=black,fill=gainsboro232221219,line width=0.004pt,postaction={pattern=crosshatch}] (axis cs:4.16666666666667,0) rectangle (axis cs:4.33333333333333,0.261596793);
\draw[draw=black,fill=sienna1408675,line width=0.004pt,postaction={pattern=north east lines}] (axis cs:4.16666666666667,0.261596793) rectangle (axis cs:4.33333333333333,0.276953877);
\draw[draw=black,fill=gray163119111,line width=0.004pt,postaction={pattern=north east lines}] (axis cs:4.16666666666667,0.276953877) rectangle (axis cs:4.33333333333333,0.3171933935);
\draw[draw=black,fill=rosybrown186153147,line width=0.004pt,postaction={pattern=north east lines}] (axis cs:4.16666666666667,0.3171933935) rectangle (axis cs:4.33333333333333,0.3539265035);
\draw[draw=black,fill=sienna1408675,line width=0.004pt] (axis cs:4.16666666666667,0.3539265035) rectangle (axis cs:4.33333333333333,0.405027958);
\draw[draw=black,fill=gray163119111,line width=0.004pt] (axis cs:4.16666666666667,0.405027958) rectangle (axis cs:4.33333333333333,0.5563508895);
\draw[draw=black,fill=rosybrown186153147,line width=0.004pt] (axis cs:4.16666666666667,0.5563508895) rectangle (axis cs:4.33333333333333,1.193947391);
\draw[draw=black,fill=gainsboro232221219,line width=0.004pt,postaction={pattern=crosshatch}] (axis cs:5.16666666666667,0) rectangle (axis cs:5.33333333333333,0.2590661625);
\draw[draw=black,fill=sienna1408675,line width=0.004pt,postaction={pattern=north east lines}] (axis cs:5.16666666666667,0.2590661625) rectangle (axis cs:5.33333333333333,0.274724489);
\draw[draw=black,fill=gray163119111,line width=0.004pt,postaction={pattern=north east lines}] (axis cs:5.16666666666667,0.274724489) rectangle (axis cs:5.33333333333333,0.314519332);
\draw[draw=black,fill=rosybrown186153147,line width=0.004pt,postaction={pattern=north east lines}] (axis cs:5.16666666666667,0.314519332) rectangle (axis cs:5.33333333333333,0.3555323545);
\draw[draw=black,fill=sienna1408675,line width=0.004pt] (axis cs:5.16666666666667,0.3555323545) rectangle (axis cs:5.33333333333333,0.4090520065);
\draw[draw=black,fill=gray163119111,line width=0.004pt] (axis cs:5.16666666666667,0.4090520065) rectangle (axis cs:5.33333333333333,0.645366313);
\draw[draw=black,fill=rosybrown186153147,line width=0.004pt] (axis cs:5.16666666666667,0.645366313) rectangle (axis cs:5.33333333333333,1.276642822);
\draw[draw=black,fill=gainsboro232221219,line width=0.004pt,postaction={pattern=crosshatch}] (axis cs:6.16666666666667,0) rectangle (axis cs:6.33333333333333,0.280053726);
\draw[draw=black,fill=sienna1408675,line width=0.004pt,postaction={pattern=north east lines}] (axis cs:6.16666666666667,0.280053726) rectangle (axis cs:6.33333333333333,0.2953583645);
\draw[draw=black,fill=gray163119111,line width=0.004pt,postaction={pattern=north east lines}] (axis cs:6.16666666666667,0.2953583645) rectangle (axis cs:6.33333333333333,0.3344572875);
\draw[draw=black,fill=rosybrown186153147,line width=0.004pt,postaction={pattern=north east lines}] (axis cs:6.16666666666667,0.3344572875) rectangle (axis cs:6.33333333333333,0.3748665285);
\draw[draw=black,fill=sienna1408675,line width=0.004pt] (axis cs:6.16666666666667,0.3748665285) rectangle (axis cs:6.33333333333333,0.42931044);
\draw[draw=black,fill=gray163119111,line width=0.004pt] (axis cs:6.16666666666667,0.42931044) rectangle (axis cs:6.33333333333333,0.688312019);
\draw[draw=black,fill=rosybrown186153147,line width=0.004pt] (axis cs:6.16666666666667,0.688312019) rectangle (axis cs:6.33333333333333,1.3712857285);
\draw[draw=black,fill=gainsboro232221219,line width=0.004pt,postaction={pattern=crosshatch}] (axis cs:7.16666666666667,0) rectangle (axis cs:7.33333333333333,0.276174431);
\draw[draw=black,fill=sienna1408675,line width=0.004pt,postaction={pattern=north east lines}] (axis cs:7.16666666666667,0.276174431) rectangle (axis cs:7.33333333333333,0.2915974685);
\draw[draw=black,fill=gray163119111,line width=0.004pt,postaction={pattern=north east lines}] (axis cs:7.16666666666667,0.2915974685) rectangle (axis cs:7.33333333333333,0.333230151);
\draw[draw=black,fill=rosybrown186153147,line width=0.004pt,postaction={pattern=north east lines}] (axis cs:7.16666666666667,0.333230151) rectangle (axis cs:7.33333333333333,0.374450117);
\draw[draw=black,fill=sienna1408675,line width=0.004pt] (axis cs:7.16666666666667,0.374450117) rectangle (axis cs:7.33333333333333,0.429812215);
\draw[draw=black,fill=gray163119111,line width=0.004pt] (axis cs:7.16666666666667,0.429812215) rectangle (axis cs:7.33333333333333,0.855143396);
\draw[draw=black,fill=rosybrown186153147,line width=0.004pt] (axis cs:7.16666666666667,0.855143396) rectangle (axis cs:7.33333333333333,1.593361227);
\end{axis}

\end{tikzpicture}

%% file: figures/python_fig/weak-cluster-eff-isr.tex
\begin{tikzpicture}

\definecolor{cornflowerblue107174214}{RGB}{107,174,214}
\definecolor{darkgray}{RGB}{169,169,169}
\definecolor{darkgray150}{RGB}{150,150,150}
\definecolor{darkgray176}{RGB}{176,176,176}
\definecolor{dimgray99}{RGB}{99,99,99}
\definecolor{mediumpurple148103189}{RGB}{148,103,189}
\definecolor{sienna1408675}{RGB}{140,86,75}
\definecolor{silver189}{RGB}{189,189,189}

\begin{axis}[
axis y line =left, axis x line =bottom, axis x line shift=5pt, axis line style ={-},
log basis x={10},
minor xtick={},
minor ytick={},
tick align=center,
tick pos=left,
x grid style={darkgray176},
x grid style={draw=none},
xlabel near ticks,
xlabel={\(\displaystyle r\) = Number of nodes},
xmajorgrids,
xmin=0.780854050171763, xmax=180.315386683374,
xmode=log,
xtick style={color=black},
xtick={1,2,4,8,16,32,64,128},
xticklabels={$1$,$2$,$4$,$8$,$16$,$32$,$64$,$128$},
y axis line style={draw=none},
y grid style={darkgray176},
y grid style={dotted},
ylabel near ticks,
ylabel={\(\displaystyle \eta_{P,w}\)},
ymajorgrids,
ymin=0.1, ymax=1.1,
ytick style={color=darkgray},
ytick={0,0.2,0.4,0.6,0.8,1,1.2}
]
\addplot [very thin, silver189]
table {%
1 1
1.10623917447252 0.99893873573367
1.22376511113765 0.997767344792031
1.35377690629318 0.996474702625059
1.49760104723774 0.995048627608495
1.65670494618546 0.993475795313024
1.83271191201275 0.991741649151127
2.02741771259094 0.989830308090089
2.24280889668757 0.98772447238247
2.48108306237128 0.985405328582658
2.74467127871536 0.982852455496799
3.03626288956452 0.980043733160157
3.35883295243341 0.976955257456583
3.71567259249104 0.973561263593482
4.11042258132746 0.969834062324067
4.54711048310091 0.965743993564361
5.0301917470609 0.961259402877388
5.56459516570714 0.956346647175619
6.15577316238565 0.950970136899437
6.80975742139761 0.945092422825811
7.53322042820502 0.938674336493865
8.33354354761706 0.931675193930743
9.2188923345467 0.92405307283003
10.19829984572 0.915765173462264
11.2817588023525 0.906768273248348
12.4803235441126 0.897019283949513
13.8062228145891 0.886475918658869
15.2729845289947 0.875097473056957
16.8955737970867 0.862845721572446
18.6905456095288 0.849685924065455
20.6762137455262 0.835587932390936
22.8728376250683 0.820527378771975
25.3028290121995 0.804486919522851
27.990980678275 0.787457498681303
30.9647193582112 0.769439587095771
34.2543855806009 0.750444344208946
37.8935432267474 0.730494643084532
41.9193219769959 0.709625895136823
46.3727961382798 0.687886610538049
51.2994037179932 0.665338634261185
56.7494100199254 0.64205700674697
62.7784204922449 0.618129412413156
69.4479480600299 0.593655198247727
76.826040730738 0.568743967528345
84.987975875964 0.543513778679839
94.017028273117 0.518089004346814
104.005319743213 0.492597928598586
115.054759053482 0.467170178519669
127.278081674459 0.44193409834483
140.8 0.417014178482069
};
\addplot [very thin, darkgray150]
table {%
1 1
1.10623917447252 0.999469086146885
1.22376511113765 0.998882424815988
1.35377690629318 0.998234238896037
1.49760104723774 0.997518169570914
1.65670494618546 0.996727221518157
1.83271191201275 0.995853703791155
2.02741771259094 0.994889166242687
2.24280889668757 0.993824331395983
2.48108306237128 0.992649021735143
2.74467127871536 0.991352082473073
3.03626288956452 0.989921299966444
3.35883295243341 0.988343316088466
3.71567259249104 0.98660353904678
4.11042258132746 0.984686051351786
4.54711048310091 0.982573515906553
5.0301917470609 0.980247081510087
5.56459516570714 0.977686289448036
6.15577316238565 0.974868983295416
6.80975742139761 0.971771224580465
7.53322042820502 0.968367217561128
8.33354354761706 0.964629247047365
9.2188923345467 0.960527633960605
10.19829984572 0.956030714147755
11.2817588023525 0.951104846844958
12.4803235441126 0.945714460089154
13.8062228145891 0.939822141264419
15.2729845289947 0.93338878178988
16.8955737970867 0.926373785633853
18.6905456095288 0.918735351781145
20.6762137455262 0.910430840872381
22.8728376250683 0.901417235840153
25.3028290121995 0.89165170533414
27.990980678275 0.881092276893017
30.9647193582112 0.869698624024428
34.2543855806009 0.857432967453854
37.8935432267474 0.844261085700279
41.9193219769959 0.830153423805073
46.3727961382798 0.815086281558654
51.2994037179932 0.799043054154998
56.7494100199254 0.78201548924167
62.7784204922449 0.764004915393419
69.4479480600299 0.745023388874171
76.826040730738 0.725094699072452
84.987975875964 0.704255169195446
94.017028273117 0.682554188671868
104.005319743213 0.660054418085774
115.054759053482 0.636831616889909
127.278081674459 0.612974058734193
140.8 0.588581518540314
};
\addplot [very thin, dimgray99]
table {%
1 1
1.10623917447252 0.999787566788513
1.22376511113765 0.999552669971432
1.35377690629318 0.999292946465838
1.49760104723774 0.999005787348038
1.65670494618546 0.998688312890455
1.83271191201275 0.998337345200888
2.02741771259094 0.997949378265003
2.24280889668757 0.997520545183634
2.48108306237128 0.997046582388857
2.74467127871536 0.99652279061725
3.03626288956452 0.995943992416316
3.35883295243341 0.995304485961618
3.71567259249104 0.994597994969032
4.11042258132746 0.993817614500275
4.54711048310091 0.99295575248223
5.0301917470609 0.992004066793912
5.56459516570714 0.990953397821724
6.15577316238565 0.989793696447119
6.80975742139761 0.988513947514544
7.53322042820502 0.987102088935683
8.33354354761706 0.98554492672348
9.2188923345467 0.983828046421508
10.19829984572 0.981935721606872
11.2817588023525 0.979850820404625
12.4803235441126 0.977554711265207
13.8062228145891 0.975027169630858
15.2729845289947 0.972246287558608
16.8955737970867 0.969188388882478
18.6905456095288 0.965827953089813
20.6762137455262 0.962137551758023
22.8728376250683 0.958087802146196
25.3028290121995 0.953647343353102
27.990980678275 0.948782841323897
30.9647193582112 0.943459029886935
34.2543855806009 0.937638795892143
37.8935432267474 0.93128331735
41.9193219769959 0.924352264164938
46.3727961382798 0.916804071527659
51.2994037179932 0.908596296165043
56.7494100199254 0.899686065311476
62.7784204922449 0.890030627310117
69.4479480600299 0.879588011015559
76.826040730738 0.868317798489083
84.987975875964 0.856182011710114
94.017028273117 0.843146109068765
104.005319743213 0.829180081218727
115.054759053482 0.814259628523541
127.278081674459 0.798367394022745
140.8 0.781494216942802
};
\addplot [very thin, black]
table {%
1 1
1.10623917447252 1
1.22376511113765 1
1.35377690629318 1
1.49760104723774 1
1.65670494618546 1
1.83271191201275 1
2.02741771259094 1
2.24280889668757 1
2.48108306237128 1
2.74467127871536 1
3.03626288956452 1
3.35883295243341 1
3.71567259249104 1
4.11042258132746 1
4.54711048310091 1
5.0301917470609 1
5.56459516570714 1
6.15577316238565 1
6.80975742139761 1
7.53322042820502 1
8.33354354761706 1
9.2188923345467 1
10.19829984572 1
11.2817588023525 1
12.4803235441126 1
13.8062228145891 1
15.2729845289947 1
16.8955737970867 1
18.6905456095288 1
20.6762137455262 1
22.8728376250683 1
25.3028290121995 1
27.990980678275 1
30.9647193582112 1
34.2543855806009 1
37.8935432267474 1
41.9193219769959 1
46.3727961382798 1
51.2994037179932 1
56.7494100199254 1
62.7784204922449 1
69.4479480600299 1
76.826040730738 1
84.987975875964 1
94.017028273117 1
104.005319743213 1
115.054759053482 1
127.278081674459 1
140.8 1
};
\addplot [very thin, silver189]
table {%
1 1
1.10623917447252 0.99893873573367
1.22376511113765 0.997767344792031
1.35377690629318 0.996474702625059
1.49760104723774 0.995048627608495
1.65670494618546 0.993475795313024
1.83271191201275 0.991741649151127
2.02741771259094 0.989830308090089
2.24280889668757 0.98772447238247
2.48108306237128 0.985405328582658
2.74467127871536 0.982852455496799
3.03626288956452 0.980043733160157
3.35883295243341 0.976955257456583
3.71567259249104 0.973561263593482
4.11042258132746 0.969834062324067
4.54711048310091 0.965743993564361
5.0301917470609 0.961259402877388
5.56459516570714 0.956346647175619
6.15577316238565 0.950970136899437
6.80975742139761 0.945092422825811
7.53322042820502 0.938674336493865
8.33354354761706 0.931675193930743
9.2188923345467 0.92405307283003
10.19829984572 0.915765173462264
11.2817588023525 0.906768273248348
12.4803235441126 0.897019283949513
13.8062228145891 0.886475918658869
15.2729845289947 0.875097473056957
16.8955737970867 0.862845721572446
18.6905456095288 0.849685924065455
20.6762137455262 0.835587932390936
22.8728376250683 0.820527378771975
25.3028290121995 0.804486919522851
27.990980678275 0.787457498681303
30.9647193582112 0.769439587095771
34.2543855806009 0.750444344208946
37.8935432267474 0.730494643084532
41.9193219769959 0.709625895136823
46.3727961382798 0.687886610538049
51.2994037179932 0.665338634261185
56.7494100199254 0.64205700674697
62.7784204922449 0.618129412413156
69.4479480600299 0.593655198247727
76.826040730738 0.568743967528345
84.987975875964 0.543513778679839
94.017028273117 0.518089004346814
104.005319743213 0.492597928598586
115.054759053482 0.467170178519669
127.278081674459 0.44193409834483
140.8 0.417014178482069
};
\addplot [very thin, darkgray150]
table {%
1 1
1.10623917447252 0.999469086146885
1.22376511113765 0.998882424815988
1.35377690629318 0.998234238896037
1.49760104723774 0.997518169570914
1.65670494618546 0.996727221518157
1.83271191201275 0.995853703791155
2.02741771259094 0.994889166242687
2.24280889668757 0.993824331395983
2.48108306237128 0.992649021735143
2.74467127871536 0.991352082473073
3.03626288956452 0.989921299966444
3.35883295243341 0.988343316088466
3.71567259249104 0.98660353904678
4.11042258132746 0.984686051351786
4.54711048310091 0.982573515906553
5.0301917470609 0.980247081510087
5.56459516570714 0.977686289448036
6.15577316238565 0.974868983295416
6.80975742139761 0.971771224580465
7.53322042820502 0.968367217561128
8.33354354761706 0.964629247047365
9.2188923345467 0.960527633960605
10.19829984572 0.956030714147755
11.2817588023525 0.951104846844958
12.4803235441126 0.945714460089154
13.8062228145891 0.939822141264419
15.2729845289947 0.93338878178988
16.8955737970867 0.926373785633853
18.6905456095288 0.918735351781145
20.6762137455262 0.910430840872381
22.8728376250683 0.901417235840153
25.3028290121995 0.89165170533414
27.990980678275 0.881092276893017
30.9647193582112 0.869698624024428
34.2543855806009 0.857432967453854
37.8935432267474 0.844261085700279
41.9193219769959 0.830153423805073
46.3727961382798 0.815086281558654
51.2994037179932 0.799043054154998
56.7494100199254 0.78201548924167
62.7784204922449 0.764004915393419
69.4479480600299 0.745023388874171
76.826040730738 0.725094699072452
84.987975875964 0.704255169195446
94.017028273117 0.682554188671868
104.005319743213 0.660054418085774
115.054759053482 0.636831616889909
127.278081674459 0.612974058734193
140.8 0.588581518540314
};
\addplot [very thin, dimgray99]
table {%
1 1
1.10623917447252 0.999787566788513
1.22376511113765 0.999552669971432
1.35377690629318 0.999292946465838
1.49760104723774 0.999005787348038
1.65670494618546 0.998688312890455
1.83271191201275 0.998337345200888
2.02741771259094 0.997949378265003
2.24280889668757 0.997520545183634
2.48108306237128 0.997046582388857
2.74467127871536 0.99652279061725
3.03626288956452 0.995943992416316
3.35883295243341 0.995304485961618
3.71567259249104 0.994597994969032
4.11042258132746 0.993817614500275
4.54711048310091 0.99295575248223
5.0301917470609 0.992004066793912
5.56459516570714 0.990953397821724
6.15577316238565 0.989793696447119
6.80975742139761 0.988513947514544
7.53322042820502 0.987102088935683
8.33354354761706 0.98554492672348
9.2188923345467 0.983828046421508
10.19829984572 0.981935721606872
11.2817588023525 0.979850820404625
12.4803235441126 0.977554711265207
13.8062228145891 0.975027169630858
15.2729845289947 0.972246287558608
16.8955737970867 0.969188388882478
18.6905456095288 0.965827953089813
20.6762137455262 0.962137551758023
22.8728376250683 0.958087802146196
25.3028290121995 0.953647343353102
27.990980678275 0.948782841323897
30.9647193582112 0.943459029886935
34.2543855806009 0.937638795892143
37.8935432267474 0.93128331735
41.9193219769959 0.924352264164938
46.3727961382798 0.916804071527659
51.2994037179932 0.908596296165043
56.7494100199254 0.899686065311476
62.7784204922449 0.890030627310117
69.4479480600299 0.879588011015559
76.826040730738 0.868317798489083
84.987975875964 0.856182011710114
94.017028273117 0.843146109068765
104.005319743213 0.829180081218727
115.054759053482 0.814259628523541
127.278081674459 0.798367394022745
140.8 0.781494216942802
};
\addplot [very thin, black]
table {%
1 1
1.10623917447252 1
1.22376511113765 1
1.35377690629318 1
1.49760104723774 1
1.65670494618546 1
1.83271191201275 1
2.02741771259094 1
2.24280889668757 1
2.48108306237128 1
2.74467127871536 1
3.03626288956452 1
3.35883295243341 1
3.71567259249104 1
4.11042258132746 1
4.54711048310091 1
5.0301917470609 1
5.56459516570714 1
6.15577316238565 1
6.80975742139761 1
7.53322042820502 1
8.33354354761706 1
9.2188923345467 1
10.19829984572 1
11.2817588023525 1
12.4803235441126 1
13.8062228145891 1
15.2729845289947 1
16.8955737970867 1
18.6905456095288 1
20.6762137455262 1
22.8728376250683 1
25.3028290121995 1
27.990980678275 1
30.9647193582112 1
34.2543855806009 1
37.8935432267474 1
41.9193219769959 1
46.3727961382798 1
51.2994037179932 1
56.7494100199254 1
62.7784204922449 1
69.4479480600299 1
76.826040730738 1
84.987975875964 1
94.017028273117 1
104.005319743213 1
115.054759053482 1
127.278081674459 1
140.8 1
};
\addplot [very thin, silver189]
table {%
1 1
1.10623917447252 0.99893873573367
1.22376511113765 0.997767344792031
1.35377690629318 0.996474702625059
1.49760104723774 0.995048627608495
1.65670494618546 0.993475795313024
1.83271191201275 0.991741649151127
2.02741771259094 0.989830308090089
2.24280889668757 0.98772447238247
2.48108306237128 0.985405328582658
2.74467127871536 0.982852455496799
3.03626288956452 0.980043733160157
3.35883295243341 0.976955257456583
3.71567259249104 0.973561263593482
4.11042258132746 0.969834062324067
4.54711048310091 0.965743993564361
5.0301917470609 0.961259402877388
5.56459516570714 0.956346647175619
6.15577316238565 0.950970136899437
6.80975742139761 0.945092422825811
7.53322042820502 0.938674336493865
8.33354354761706 0.931675193930743
9.2188923345467 0.92405307283003
10.19829984572 0.915765173462264
11.2817588023525 0.906768273248348
12.4803235441126 0.897019283949513
13.8062228145891 0.886475918658869
15.2729845289947 0.875097473056957
16.8955737970867 0.862845721572446
18.6905456095288 0.849685924065455
20.6762137455262 0.835587932390936
22.8728376250683 0.820527378771975
25.3028290121995 0.804486919522851
27.990980678275 0.787457498681303
30.9647193582112 0.769439587095771
34.2543855806009 0.750444344208946
37.8935432267474 0.730494643084532
41.9193219769959 0.709625895136823
46.3727961382798 0.687886610538049
51.2994037179932 0.665338634261185
56.7494100199254 0.64205700674697
62.7784204922449 0.618129412413156
69.4479480600299 0.593655198247727
76.826040730738 0.568743967528345
84.987975875964 0.543513778679839
94.017028273117 0.518089004346814
104.005319743213 0.492597928598586
115.054759053482 0.467170178519669
127.278081674459 0.44193409834483
140.8 0.417014178482069
};
\addplot [very thin, darkgray150]
table {%
1 1
1.10623917447252 0.999469086146885
1.22376511113765 0.998882424815988
1.35377690629318 0.998234238896037
1.49760104723774 0.997518169570914
1.65670494618546 0.996727221518157
1.83271191201275 0.995853703791155
2.02741771259094 0.994889166242687
2.24280889668757 0.993824331395983
2.48108306237128 0.992649021735143
2.74467127871536 0.991352082473073
3.03626288956452 0.989921299966444
3.35883295243341 0.988343316088466
3.71567259249104 0.98660353904678
4.11042258132746 0.984686051351786
4.54711048310091 0.982573515906553
5.0301917470609 0.980247081510087
5.56459516570714 0.977686289448036
6.15577316238565 0.974868983295416
6.80975742139761 0.971771224580465
7.53322042820502 0.968367217561128
8.33354354761706 0.964629247047365
9.2188923345467 0.960527633960605
10.19829984572 0.956030714147755
11.2817588023525 0.951104846844958
12.4803235441126 0.945714460089154
13.8062228145891 0.939822141264419
15.2729845289947 0.93338878178988
16.8955737970867 0.926373785633853
18.6905456095288 0.918735351781145
20.6762137455262 0.910430840872381
22.8728376250683 0.901417235840153
25.3028290121995 0.89165170533414
27.990980678275 0.881092276893017
30.9647193582112 0.869698624024428
34.2543855806009 0.857432967453854
37.8935432267474 0.844261085700279
41.9193219769959 0.830153423805073
46.3727961382798 0.815086281558654
51.2994037179932 0.799043054154998
56.7494100199254 0.78201548924167
62.7784204922449 0.764004915393419
69.4479480600299 0.745023388874171
76.826040730738 0.725094699072452
84.987975875964 0.704255169195446
94.017028273117 0.682554188671868
104.005319743213 0.660054418085774
115.054759053482 0.636831616889909
127.278081674459 0.612974058734193
140.8 0.588581518540314
};
\addplot [very thin, dimgray99]
table {%
1 1
1.10623917447252 0.999787566788513
1.22376511113765 0.999552669971432
1.35377690629318 0.999292946465838
1.49760104723774 0.999005787348038
1.65670494618546 0.998688312890455
1.83271191201275 0.998337345200888
2.02741771259094 0.997949378265003
2.24280889668757 0.997520545183634
2.48108306237128 0.997046582388857
2.74467127871536 0.99652279061725
3.03626288956452 0.995943992416316
3.35883295243341 0.995304485961618
3.71567259249104 0.994597994969032
4.11042258132746 0.993817614500275
4.54711048310091 0.99295575248223
5.0301917470609 0.992004066793912
5.56459516570714 0.990953397821724
6.15577316238565 0.989793696447119
6.80975742139761 0.988513947514544
7.53322042820502 0.987102088935683
8.33354354761706 0.98554492672348
9.2188923345467 0.983828046421508
10.19829984572 0.981935721606872
11.2817588023525 0.979850820404625
12.4803235441126 0.977554711265207
13.8062228145891 0.975027169630858
15.2729845289947 0.972246287558608
16.8955737970867 0.969188388882478
18.6905456095288 0.965827953089813
20.6762137455262 0.962137551758023
22.8728376250683 0.958087802146196
25.3028290121995 0.953647343353102
27.990980678275 0.948782841323897
30.9647193582112 0.943459029886935
34.2543855806009 0.937638795892143
37.8935432267474 0.93128331735
41.9193219769959 0.924352264164938
46.3727961382798 0.916804071527659
51.2994037179932 0.908596296165043
56.7494100199254 0.899686065311476
62.7784204922449 0.890030627310117
69.4479480600299 0.879588011015559
76.826040730738 0.868317798489083
84.987975875964 0.856182011710114
94.017028273117 0.843146109068765
104.005319743213 0.829180081218727
115.054759053482 0.814259628523541
127.278081674459 0.798367394022745
140.8 0.781494216942802
};
\addplot [very thin, black]
table {%
1 1
1.10623917447252 1
1.22376511113765 1
1.35377690629318 1
1.49760104723774 1
1.65670494618546 1
1.83271191201275 1
2.02741771259094 1
2.24280889668757 1
2.48108306237128 1
2.74467127871536 1
3.03626288956452 1
3.35883295243341 1
3.71567259249104 1
4.11042258132746 1
4.54711048310091 1
5.0301917470609 1
5.56459516570714 1
6.15577316238565 1
6.80975742139761 1
7.53322042820502 1
8.33354354761706 1
9.2188923345467 1
10.19829984572 1
11.2817588023525 1
12.4803235441126 1
13.8062228145891 1
15.2729845289947 1
16.8955737970867 1
18.6905456095288 1
20.6762137455262 1
22.8728376250683 1
25.3028290121995 1
27.990980678275 1
30.9647193582112 1
34.2543855806009 1
37.8935432267474 1
41.9193219769959 1
46.3727961382798 1
51.2994037179932 1
56.7494100199254 1
62.7784204922449 1
69.4479480600299 1
76.826040730738 1
84.987975875964 1
94.017028273117 1
104.005319743213 1
115.054759053482 1
127.278081674459 1
140.8 1
};
\addplot [semithick, cornflowerblue107174214, mark=*, mark size=1.5, mark options={solid}]
table {%
1 1
2 0.922372369464917
4 0.767919722700138
8 0.677212162019229
16 0.547413121623124
32 0.51456518418126
64 0.501279232497289
128 0.386511192507655
};
\addplot [semithick, mediumpurple148103189, mark=*, mark size=1.5, mark options={solid}]
table {%
1 1
2 0.937770964033154
4 0.838578682369746
8 0.793676795016528
16 0.741115096015691
32 0.666314437367844
64 0.657649992978902
128 0.581799237360776
};
\addplot [semithick, sienna1408675, mark=*, mark size=1.5, mark options={solid}]
table {%
1 1
2 0.902012507321652
4 0.779066273334423
8 0.749361329441045
16 0.71722245255947
32 0.670763866951034
64 0.624469326999198
128 0.537433609836422
};
\end{axis}

\end{tikzpicture}

%% file: figures/python_tab/comp_clusters_isr_time_unknowns_rank.tex
\begin{tabular}{lrrrrr}
\hline
  N nodes   &   \centercell{$1$ } &   \centercell{$2$ } &   \centercell{$8$ } &   \centercell{$64$ } &   \centercell{$128$ } \\
\hline
 Vega       &               42.16 &               38.88 &               28.55 &                21.13 &                 16.29 \\
 MeluXina   &               40.94 &               38.39 &               32.49 &                26.92 &                 23.82 \\
 Lumi       &               38.57 &               34.79 &               28.90 &                24.09 &                 20.73 \\
\hline
\end{tabular}

%% file: figures/python_fig/weak-cluster-comp-a2a.tex
\begin{tikzpicture}

\definecolor{cornflowerblue107174214}{RGB}{107,174,214}
\definecolor{darkgray}{RGB}{169,169,169}
\definecolor{darkgray176}{RGB}{176,176,176}
\definecolor{gainsboro232221219}{RGB}{232,221,219}
\definecolor{gray163119111}{RGB}{163,119,111}
\definecolor{lavender225238246}{RGB}{225,238,246}
\definecolor{lavender233224241}{RGB}{233,224,241}
\definecolor{lightblue166206230}{RGB}{166,206,230}
\definecolor{mediumpurple148103189}{RGB}{148,103,189}
\definecolor{mediumpurple169133202}{RGB}{169,133,202}
\definecolor{plum190163215}{RGB}{190,163,215}
\definecolor{rosybrown186153147}{RGB}{186,153,147}
\definecolor{sienna1408675}{RGB}{140,86,75}
\definecolor{skyblue136190222}{RGB}{136,190,222}

\begin{axis}[
axis line style={draw=none},
tick pos=left,
x grid style={darkgray176},
x grid style={draw=none},
xlabel near ticks,
xlabel={N nodes},
xmajorgrids,
xmin=-0.716666666666667, xmax=7.71666666666667,
xtick style={color=darkgray},
xtick={0,1,2,3,4,5,6,7},
xticklabels={$1$,$2$,$4$,$8$,$16$,$32$,$64$,$128$},
xticklabels={1,2,4,8,16,32,64,128},
y grid style={darkgray176},
ylabel near ticks,
ylabel={time/solve - [sec]},
ymajorgrids,
ymin=0, ymax=2.5,
ytick style={color=darkgray}
]
\draw[draw=black,fill=lavender225238246,line width=0.004pt,postaction={pattern=crosshatch}] (axis cs:-0.333333333333333,0) rectangle (axis cs:-0.166666666666667,0.3077952285);
\draw[draw=black,fill=cornflowerblue107174214,line width=0.004pt,postaction={pattern=north east lines}] (axis cs:-0.333333333333333,0.3077952285) rectangle (axis cs:-0.166666666666667,0.359887644);
\draw[draw=black,fill=skyblue136190222,line width=0.004pt,postaction={pattern=north east lines}] (axis cs:-0.333333333333333,0.359887644) rectangle (axis cs:-0.166666666666667,0.4844812945);
\draw[draw=black,fill=lightblue166206230,line width=0.004pt,postaction={pattern=north east lines}] (axis cs:-0.333333333333333,0.4844812945) rectangle (axis cs:-0.166666666666667,0.7499377115);
\draw[draw=black,fill=cornflowerblue107174214,line width=0.004pt] (axis cs:-0.333333333333333,0.7499377115) rectangle (axis cs:-0.166666666666667,0.788794285);
\draw[draw=black,fill=skyblue136190222,line width=0.004pt] (axis cs:-0.333333333333333,0.788794285) rectangle (axis cs:-0.166666666666667,0.874493165);
\draw[draw=black,fill=lightblue166206230,line width=0.004pt] (axis cs:-0.333333333333333,0.874493165) rectangle (axis cs:-0.166666666666667,1.04582161);
\draw[draw=black,fill=lavender225238246,line width=0.004pt,postaction={pattern=crosshatch}] (axis cs:0.666666666666667,0) rectangle (axis cs:0.833333333333333,0.3101553315);
\draw[draw=black,fill=cornflowerblue107174214,line width=0.004pt,postaction={pattern=north east lines}] (axis cs:0.666666666666667,0.3101553315) rectangle (axis cs:0.833333333333333,0.3578436155);
\draw[draw=black,fill=skyblue136190222,line width=0.004pt,postaction={pattern=north east lines}] (axis cs:0.666666666666667,0.3578436155) rectangle (axis cs:0.833333333333333,0.460091329);
\draw[draw=black,fill=lightblue166206230,line width=0.004pt,postaction={pattern=north east lines}] (axis cs:0.666666666666667,0.460091329) rectangle (axis cs:0.833333333333333,0.698084027);
\draw[draw=black,fill=cornflowerblue107174214,line width=0.004pt] (axis cs:0.666666666666667,0.698084027) rectangle (axis cs:0.833333333333333,0.728508956);
\draw[draw=black,fill=skyblue136190222,line width=0.004pt] (axis cs:0.666666666666667,0.728508956) rectangle (axis cs:0.833333333333333,0.8197191175);
\draw[draw=black,fill=lightblue166206230,line width=0.004pt] (axis cs:0.666666666666667,0.8197191175) rectangle (axis cs:0.833333333333333,1.128114344);
\draw[draw=black,fill=lavender225238246,line width=0.004pt,postaction={pattern=crosshatch}] (axis cs:1.66666666666667,0) rectangle (axis cs:1.83333333333333,0.311395638);
\draw[draw=black,fill=cornflowerblue107174214,line width=0.004pt,postaction={pattern=north east lines}] (axis cs:1.66666666666667,0.311395638) rectangle (axis cs:1.83333333333333,0.3614037325);
\draw[draw=black,fill=skyblue136190222,line width=0.004pt,postaction={pattern=north east lines}] (axis cs:1.66666666666667,0.3614037325) rectangle (axis cs:1.83333333333333,0.472342964);
\draw[draw=black,fill=lightblue166206230,line width=0.004pt,postaction={pattern=north east lines}] (axis cs:1.66666666666667,0.472342964) rectangle (axis cs:1.83333333333333,0.721404638);
\draw[draw=black,fill=cornflowerblue107174214,line width=0.004pt] (axis cs:1.66666666666667,0.721404638) rectangle (axis cs:1.83333333333333,0.762989715);
\draw[draw=black,fill=skyblue136190222,line width=0.004pt] (axis cs:1.66666666666667,0.762989715) rectangle (axis cs:1.83333333333333,0.862570725);
\draw[draw=black,fill=lightblue166206230,line width=0.004pt] (axis cs:1.66666666666667,0.862570725) rectangle (axis cs:1.83333333333333,1.3620190835);
\draw[draw=black,fill=lavender225238246,line width=0.004pt,postaction={pattern=crosshatch}] (axis cs:2.66666666666667,0) rectangle (axis cs:2.83333333333333,0.3108601215);
\draw[draw=black,fill=cornflowerblue107174214,line width=0.004pt,postaction={pattern=north east lines}] (axis cs:2.66666666666667,0.3108601215) rectangle (axis cs:2.83333333333333,0.36090916);
\draw[draw=black,fill=skyblue136190222,line width=0.004pt,postaction={pattern=north east lines}] (axis cs:2.66666666666667,0.36090916) rectangle (axis cs:2.83333333333333,0.473282722);
\draw[draw=black,fill=lightblue166206230,line width=0.004pt,postaction={pattern=north east lines}] (axis cs:2.66666666666667,0.473282722) rectangle (axis cs:2.83333333333333,0.703307836);
\draw[draw=black,fill=cornflowerblue107174214,line width=0.004pt] (axis cs:2.66666666666667,0.703307836) rectangle (axis cs:2.83333333333333,0.7448852895);
\draw[draw=black,fill=skyblue136190222,line width=0.004pt] (axis cs:2.66666666666667,0.7448852895) rectangle (axis cs:2.83333333333333,0.842990109);
\draw[draw=black,fill=lightblue166206230,line width=0.004pt] (axis cs:2.66666666666667,0.842990109) rectangle (axis cs:2.83333333333333,1.470904753);
\draw[draw=black,fill=lavender225238246,line width=0.004pt,postaction={pattern=crosshatch}] (axis cs:3.66666666666667,0) rectangle (axis cs:3.83333333333333,0.3204927665);
\draw[draw=black,fill=cornflowerblue107174214,line width=0.004pt,postaction={pattern=north east lines}] (axis cs:3.66666666666667,0.3204927665) rectangle (axis cs:3.83333333333333,0.3718069055);
\draw[draw=black,fill=skyblue136190222,line width=0.004pt,postaction={pattern=north east lines}] (axis cs:3.66666666666667,0.3718069055) rectangle (axis cs:3.83333333333333,0.4815648945);
\draw[draw=black,fill=lightblue166206230,line width=0.004pt,postaction={pattern=north east lines}] (axis cs:3.66666666666667,0.4815648945) rectangle (axis cs:3.83333333333333,0.694980218);
\draw[draw=black,fill=cornflowerblue107174214,line width=0.004pt] (axis cs:3.66666666666667,0.694980218) rectangle (axis cs:3.83333333333333,0.7371530105);
\draw[draw=black,fill=skyblue136190222,line width=0.004pt] (axis cs:3.66666666666667,0.7371530105) rectangle (axis cs:3.83333333333333,0.8695991185);
\draw[draw=black,fill=lightblue166206230,line width=0.004pt] (axis cs:3.66666666666667,0.8695991185) rectangle (axis cs:3.83333333333333,1.6025072055);
\draw[draw=black,fill=lavender225238246,line width=0.004pt,postaction={pattern=crosshatch}] (axis cs:4.66666666666667,0) rectangle (axis cs:4.83333333333333,0.322026676);
\draw[draw=black,fill=cornflowerblue107174214,line width=0.004pt,postaction={pattern=north east lines}] (axis cs:4.66666666666667,0.322026676) rectangle (axis cs:4.83333333333333,0.375220741);
\draw[draw=black,fill=skyblue136190222,line width=0.004pt,postaction={pattern=north east lines}] (axis cs:4.66666666666667,0.375220741) rectangle (axis cs:4.83333333333333,0.4900413385);
\draw[draw=black,fill=lightblue166206230,line width=0.004pt,postaction={pattern=north east lines}] (axis cs:4.66666666666667,0.4900413385) rectangle (axis cs:4.83333333333333,0.7112089765);
\draw[draw=black,fill=cornflowerblue107174214,line width=0.004pt] (axis cs:4.66666666666667,0.7112089765) rectangle (axis cs:4.83333333333333,0.7572450605);
\draw[draw=black,fill=skyblue136190222,line width=0.004pt] (axis cs:4.66666666666667,0.7572450605) rectangle (axis cs:4.83333333333333,0.9871859385);
\draw[draw=black,fill=lightblue166206230,line width=0.004pt] (axis cs:4.66666666666667,0.9871859385) rectangle (axis cs:4.83333333333333,1.7464380505);
\draw[draw=black,fill=lavender225238246,line width=0.004pt,postaction={pattern=crosshatch}] (axis cs:5.66666666666667,0) rectangle (axis cs:5.83333333333333,0.319324329);
\draw[draw=black,fill=cornflowerblue107174214,line width=0.004pt,postaction={pattern=north east lines}] (axis cs:5.66666666666667,0.319324329) rectangle (axis cs:5.83333333333333,0.3719426895);
\draw[draw=black,fill=skyblue136190222,line width=0.004pt,postaction={pattern=north east lines}] (axis cs:5.66666666666667,0.3719426895) rectangle (axis cs:5.83333333333333,0.4796900945);
\draw[draw=black,fill=lightblue166206230,line width=0.004pt,postaction={pattern=north east lines}] (axis cs:5.66666666666667,0.4796900945) rectangle (axis cs:5.83333333333333,0.669910524);
\draw[draw=black,fill=cornflowerblue107174214,line width=0.004pt] (axis cs:5.66666666666667,0.669910524) rectangle (axis cs:5.83333333333333,0.7159827605);
\draw[draw=black,fill=skyblue136190222,line width=0.004pt] (axis cs:5.66666666666667,0.7159827605) rectangle (axis cs:5.83333333333333,0.964000475);
\draw[draw=black,fill=lightblue166206230,line width=0.004pt] (axis cs:5.66666666666667,0.964000475) rectangle (axis cs:5.83333333333333,1.824304797);
\draw[draw=black,fill=lavender225238246,line width=0.004pt,postaction={pattern=crosshatch}] (axis cs:6.66666666666667,0) rectangle (axis cs:6.83333333333333,0.334655829);
\draw[draw=black,fill=cornflowerblue107174214,line width=0.004pt,postaction={pattern=north east lines}] (axis cs:6.66666666666667,0.334655829) rectangle (axis cs:6.83333333333333,0.3883637425);
\draw[draw=black,fill=skyblue136190222,line width=0.004pt,postaction={pattern=north east lines}] (axis cs:6.66666666666667,0.3883637425) rectangle (axis cs:6.83333333333333,0.4980077005);
\draw[draw=black,fill=lightblue166206230,line width=0.004pt,postaction={pattern=north east lines}] (axis cs:6.66666666666667,0.4980077005) rectangle (axis cs:6.83333333333333,0.7012923255);
\draw[draw=black,fill=cornflowerblue107174214,line width=0.004pt] (axis cs:6.66666666666667,0.7012923255) rectangle (axis cs:6.83333333333333,0.7479599575);
\draw[draw=black,fill=skyblue136190222,line width=0.004pt] (axis cs:6.66666666666667,0.7479599575) rectangle (axis cs:6.83333333333333,1.1591522245);
\draw[draw=black,fill=lightblue166206230,line width=0.004pt] (axis cs:6.66666666666667,1.1591522245) rectangle (axis cs:6.83333333333333,2.3010647505);
\draw[draw=black,fill=lavender233224241,line width=0.004pt,postaction={pattern=crosshatch}] (axis cs:-0.0833333333333333,0) rectangle (axis cs:0.0833333333333333,0.318895546);
\draw[draw=black,fill=mediumpurple148103189,line width=0.004pt,postaction={pattern=north east lines}] (axis cs:-0.0833333333333333,0.318895546) rectangle (axis cs:0.0833333333333333,0.3745067295);
\draw[draw=black,fill=mediumpurple169133202,line width=0.004pt,postaction={pattern=north east lines}] (axis cs:-0.0833333333333333,0.3745067295) rectangle (axis cs:0.0833333333333333,0.5043525595);
\draw[draw=black,fill=plum190163215,line width=0.004pt,postaction={pattern=north east lines}] (axis cs:-0.0833333333333333,0.5043525595) rectangle (axis cs:0.0833333333333333,0.778857539);
\draw[draw=black,fill=mediumpurple148103189,line width=0.004pt] (axis cs:-0.0833333333333333,0.778857539) rectangle (axis cs:0.0833333333333333,0.820389006);
\draw[draw=black,fill=mediumpurple169133202,line width=0.004pt] (axis cs:-0.0833333333333333,0.820389006) rectangle (axis cs:0.0833333333333333,0.91084508);
\draw[draw=black,fill=plum190163215,line width=0.004pt] (axis cs:-0.0833333333333333,0.91084508) rectangle (axis cs:0.0833333333333333,1.088675068);
\draw[draw=black,fill=lavender233224241,line width=0.004pt,postaction={pattern=crosshatch}] (axis cs:0.916666666666667,0) rectangle (axis cs:1.08333333333333,0.3186631645);
\draw[draw=black,fill=mediumpurple148103189,line width=0.004pt,postaction={pattern=north east lines}] (axis cs:0.916666666666667,0.3186631645) rectangle (axis cs:1.08333333333333,0.368269101);
\draw[draw=black,fill=mediumpurple169133202,line width=0.004pt,postaction={pattern=north east lines}] (axis cs:0.916666666666667,0.368269101) rectangle (axis cs:1.08333333333333,0.481165614);
\draw[draw=black,fill=plum190163215,line width=0.004pt,postaction={pattern=north east lines}] (axis cs:0.916666666666667,0.481165614) rectangle (axis cs:1.08333333333333,0.744513847);
\draw[draw=black,fill=mediumpurple148103189,line width=0.004pt] (axis cs:0.916666666666667,0.744513847) rectangle (axis cs:1.08333333333333,0.7758217555);
\draw[draw=black,fill=mediumpurple169133202,line width=0.004pt] (axis cs:0.916666666666667,0.7758217555) rectangle (axis cs:1.08333333333333,0.867516589);
\draw[draw=black,fill=plum190163215,line width=0.004pt] (axis cs:0.916666666666667,0.867516589) rectangle (axis cs:1.08333333333333,1.1063788225);
\draw[draw=black,fill=lavender233224241,line width=0.004pt,postaction={pattern=crosshatch}] (axis cs:1.91666666666667,0) rectangle (axis cs:2.08333333333333,0.3203934215);
\draw[draw=black,fill=mediumpurple148103189,line width=0.004pt,postaction={pattern=north east lines}] (axis cs:1.91666666666667,0.3203934215) rectangle (axis cs:2.08333333333333,0.374975716);
\draw[draw=black,fill=mediumpurple169133202,line width=0.004pt,postaction={pattern=north east lines}] (axis cs:1.91666666666667,0.374975716) rectangle (axis cs:2.08333333333333,0.496181119);
\draw[draw=black,fill=plum190163215,line width=0.004pt,postaction={pattern=north east lines}] (axis cs:1.91666666666667,0.496181119) rectangle (axis cs:2.08333333333333,0.766342383);
\draw[draw=black,fill=mediumpurple148103189,line width=0.004pt] (axis cs:1.91666666666667,0.766342383) rectangle (axis cs:2.08333333333333,0.809961216);
\draw[draw=black,fill=mediumpurple169133202,line width=0.004pt] (axis cs:1.91666666666667,0.809961216) rectangle (axis cs:2.08333333333333,0.9051660255);
\draw[draw=black,fill=plum190163215,line width=0.004pt] (axis cs:1.91666666666667,0.9051660255) rectangle (axis cs:2.08333333333333,1.188427556);
\draw[draw=black,fill=lavender233224241,line width=0.004pt,postaction={pattern=crosshatch}] (axis cs:2.91666666666667,0) rectangle (axis cs:3.08333333333333,0.3216185865);
\draw[draw=black,fill=mediumpurple148103189,line width=0.004pt,postaction={pattern=north east lines}] (axis cs:2.91666666666667,0.3216185865) rectangle (axis cs:3.08333333333333,0.376055395);
\draw[draw=black,fill=mediumpurple169133202,line width=0.004pt,postaction={pattern=north east lines}] (axis cs:2.91666666666667,0.376055395) rectangle (axis cs:3.08333333333333,0.4966798365);
\draw[draw=black,fill=plum190163215,line width=0.004pt,postaction={pattern=north east lines}] (axis cs:2.91666666666667,0.4966798365) rectangle (axis cs:3.08333333333333,0.743724191);
\draw[draw=black,fill=mediumpurple148103189,line width=0.004pt] (axis cs:2.91666666666667,0.743724191) rectangle (axis cs:3.08333333333333,0.786988862);
\draw[draw=black,fill=mediumpurple169133202,line width=0.004pt] (axis cs:2.91666666666667,0.786988862) rectangle (axis cs:3.08333333333333,0.882010453);
\draw[draw=black,fill=plum190163215,line width=0.004pt] (axis cs:2.91666666666667,0.882010453) rectangle (axis cs:3.08333333333333,1.214015122);
\draw[draw=black,fill=lavender233224241,line width=0.004pt,postaction={pattern=crosshatch}] (axis cs:3.91666666666667,0) rectangle (axis cs:4.08333333333333,0.3223934345);
\draw[draw=black,fill=mediumpurple148103189,line width=0.004pt,postaction={pattern=north east lines}] (axis cs:3.91666666666667,0.3223934345) rectangle (axis cs:4.08333333333333,0.377215677);
\draw[draw=black,fill=mediumpurple169133202,line width=0.004pt,postaction={pattern=north east lines}] (axis cs:3.91666666666667,0.377215677) rectangle (axis cs:4.08333333333333,0.4969262105);
\draw[draw=black,fill=plum190163215,line width=0.004pt,postaction={pattern=north east lines}] (axis cs:3.91666666666667,0.4969262105) rectangle (axis cs:4.08333333333333,0.7407181345);
\draw[draw=black,fill=mediumpurple148103189,line width=0.004pt] (axis cs:3.91666666666667,0.7407181345) rectangle (axis cs:4.08333333333333,0.7843835285);
\draw[draw=black,fill=mediumpurple169133202,line width=0.004pt] (axis cs:3.91666666666667,0.7843835285) rectangle (axis cs:4.08333333333333,0.887952079);
\draw[draw=black,fill=plum190163215,line width=0.004pt] (axis cs:3.91666666666667,0.887952079) rectangle (axis cs:4.08333333333333,1.224180271);
\draw[draw=black,fill=lavender233224241,line width=0.004pt,postaction={pattern=crosshatch}] (axis cs:4.91666666666667,0) rectangle (axis cs:5.08333333333333,0.3226837415);
\draw[draw=black,fill=mediumpurple148103189,line width=0.004pt,postaction={pattern=north east lines}] (axis cs:4.91666666666667,0.3226837415) rectangle (axis cs:5.08333333333333,0.376428799);
\draw[draw=black,fill=mediumpurple169133202,line width=0.004pt,postaction={pattern=north east lines}] (axis cs:4.91666666666667,0.376428799) rectangle (axis cs:5.08333333333333,0.4947181175);
\draw[draw=black,fill=plum190163215,line width=0.004pt,postaction={pattern=north east lines}] (axis cs:4.91666666666667,0.4947181175) rectangle (axis cs:5.08333333333333,0.7367447625);
\draw[draw=black,fill=mediumpurple148103189,line width=0.004pt] (axis cs:4.91666666666667,0.7367447625) rectangle (axis cs:5.08333333333333,0.7826634395);
\draw[draw=black,fill=mediumpurple169133202,line width=0.004pt] (axis cs:4.91666666666667,0.7826634395) rectangle (axis cs:5.08333333333333,0.9322976675);
\draw[draw=black,fill=plum190163215,line width=0.004pt] (axis cs:4.91666666666667,0.9322976675) rectangle (axis cs:5.08333333333333,1.267331512);
\draw[draw=black,fill=lavender233224241,line width=0.004pt,postaction={pattern=crosshatch}] (axis cs:5.91666666666667,0) rectangle (axis cs:6.08333333333333,0.325898754);
\draw[draw=black,fill=mediumpurple148103189,line width=0.004pt,postaction={pattern=north east lines}] (axis cs:5.91666666666667,0.325898754) rectangle (axis cs:6.08333333333333,0.379624845);
\draw[draw=black,fill=mediumpurple169133202,line width=0.004pt,postaction={pattern=north east lines}] (axis cs:5.91666666666667,0.379624845) rectangle (axis cs:6.08333333333333,0.495069975);
\draw[draw=black,fill=plum190163215,line width=0.004pt,postaction={pattern=north east lines}] (axis cs:5.91666666666667,0.495069975) rectangle (axis cs:6.08333333333333,0.7177937175);
\draw[draw=black,fill=mediumpurple148103189,line width=0.004pt] (axis cs:5.91666666666667,0.7177937175) rectangle (axis cs:6.08333333333333,0.763620278);
\draw[draw=black,fill=mediumpurple169133202,line width=0.004pt] (axis cs:5.91666666666667,0.763620278) rectangle (axis cs:6.08333333333333,0.920273322);
\draw[draw=black,fill=plum190163215,line width=0.004pt] (axis cs:5.91666666666667,0.920273322) rectangle (axis cs:6.08333333333333,1.273096441);
\draw[draw=black,fill=lavender233224241,line width=0.004pt,postaction={pattern=crosshatch}] (axis cs:6.91666666666667,0) rectangle (axis cs:7.08333333333333,0.3272236035);
\draw[draw=black,fill=mediumpurple148103189,line width=0.004pt,postaction={pattern=north east lines}] (axis cs:6.91666666666667,0.3272236035) rectangle (axis cs:7.08333333333333,0.3811352675);
\draw[draw=black,fill=mediumpurple169133202,line width=0.004pt,postaction={pattern=north east lines}] (axis cs:6.91666666666667,0.3811352675) rectangle (axis cs:7.08333333333333,0.4948544945);
\draw[draw=black,fill=plum190163215,line width=0.004pt,postaction={pattern=north east lines}] (axis cs:6.91666666666667,0.4948544945) rectangle (axis cs:7.08333333333333,0.711501018);
\draw[draw=black,fill=mediumpurple148103189,line width=0.004pt] (axis cs:6.91666666666667,0.711501018) rectangle (axis cs:7.08333333333333,0.7575044165);
\draw[draw=black,fill=mediumpurple169133202,line width=0.004pt] (axis cs:6.91666666666667,0.7575044165) rectangle (axis cs:7.08333333333333,0.9607041545);
\draw[draw=black,fill=plum190163215,line width=0.004pt] (axis cs:6.91666666666667,0.9607041545) rectangle (axis cs:7.08333333333333,1.357799428);
\draw[draw=black,fill=gainsboro232221219,line width=0.004pt,postaction={pattern=crosshatch}] (axis cs:0.166666666666667,0) rectangle (axis cs:0.333333333333333,0.3081722705);
\draw[draw=black,fill=sienna1408675,line width=0.004pt,postaction={pattern=north east lines}] (axis cs:0.166666666666667,0.3081722705) rectangle (axis cs:0.333333333333333,0.354787387);
\draw[draw=black,fill=gray163119111,line width=0.004pt,postaction={pattern=north east lines}] (axis cs:0.166666666666667,0.354787387) rectangle (axis cs:0.333333333333333,0.463477397);
\draw[draw=black,fill=rosybrown186153147,line width=0.004pt,postaction={pattern=north east lines}] (axis cs:0.166666666666667,0.463477397) rectangle (axis cs:0.333333333333333,0.708565712);
\draw[draw=black,fill=sienna1408675,line width=0.004pt] (axis cs:0.166666666666667,0.708565712) rectangle (axis cs:0.333333333333333,0.7430286895);
\draw[draw=black,fill=gray163119111,line width=0.004pt] (axis cs:0.166666666666667,0.7430286895) rectangle (axis cs:0.333333333333333,0.8192861155);
\draw[draw=black,fill=rosybrown186153147,line width=0.004pt] (axis cs:0.166666666666667,0.8192861155) rectangle (axis cs:0.333333333333333,0.96610149);
\draw[draw=black,fill=gainsboro232221219,line width=0.004pt,postaction={pattern=crosshatch}] (axis cs:1.16666666666667,0) rectangle (axis cs:1.33333333333333,0.2597993035);
\draw[draw=black,fill=sienna1408675,line width=0.004pt,postaction={pattern=north east lines}] (axis cs:1.16666666666667,0.2597993035) rectangle (axis cs:1.33333333333333,0.307315054);
\draw[draw=black,fill=gray163119111,line width=0.004pt,postaction={pattern=north east lines}] (axis cs:1.16666666666667,0.307315054) rectangle (axis cs:1.33333333333333,0.399253897);
\draw[draw=black,fill=rosybrown186153147,line width=0.004pt,postaction={pattern=north east lines}] (axis cs:1.16666666666667,0.399253897) rectangle (axis cs:1.33333333333333,0.559011626);
\draw[draw=black,fill=sienna1408675,line width=0.004pt] (axis cs:1.16666666666667,0.559011626) rectangle (axis cs:1.33333333333333,0.5931711075);
\draw[draw=black,fill=gray163119111,line width=0.004pt] (axis cs:1.16666666666667,0.5931711075) rectangle (axis cs:1.33333333333333,0.7250643105);
\draw[draw=black,fill=rosybrown186153147,line width=0.004pt] (axis cs:1.16666666666667,0.7250643105) rectangle (axis cs:1.33333333333333,0.9724803775);
\draw[draw=black,fill=gainsboro232221219,line width=0.004pt,postaction={pattern=crosshatch}] (axis cs:2.16666666666667,0) rectangle (axis cs:2.33333333333333,0.2735967935);
\draw[draw=black,fill=sienna1408675,line width=0.004pt,postaction={pattern=north east lines}] (axis cs:2.16666666666667,0.2735967935) rectangle (axis cs:2.33333333333333,0.3217979645);
\draw[draw=black,fill=gray163119111,line width=0.004pt,postaction={pattern=north east lines}] (axis cs:2.16666666666667,0.3217979645) rectangle (axis cs:2.33333333333333,0.41542512);
\draw[draw=black,fill=rosybrown186153147,line width=0.004pt,postaction={pattern=north east lines}] (axis cs:2.16666666666667,0.41542512) rectangle (axis cs:2.33333333333333,0.576107525);
\draw[draw=black,fill=sienna1408675,line width=0.004pt] (axis cs:2.16666666666667,0.576107525) rectangle (axis cs:2.33333333333333,0.609305808);
\draw[draw=black,fill=gray163119111,line width=0.004pt] (axis cs:2.16666666666667,0.609305808) rectangle (axis cs:2.33333333333333,0.7474442495);
\draw[draw=black,fill=rosybrown186153147,line width=0.004pt] (axis cs:2.16666666666667,0.7474442495) rectangle (axis cs:2.33333333333333,1.095791506);
\draw[draw=black,fill=gainsboro232221219,line width=0.004pt,postaction={pattern=crosshatch}] (axis cs:3.16666666666667,0) rectangle (axis cs:3.33333333333333,0.2737337985);
\draw[draw=black,fill=sienna1408675,line width=0.004pt,postaction={pattern=north east lines}] (axis cs:3.16666666666667,0.2737337985) rectangle (axis cs:3.33333333333333,0.3218548345);
\draw[draw=black,fill=gray163119111,line width=0.004pt,postaction={pattern=north east lines}] (axis cs:3.16666666666667,0.3218548345) rectangle (axis cs:3.33333333333333,0.420169464);
\draw[draw=black,fill=rosybrown186153147,line width=0.004pt,postaction={pattern=north east lines}] (axis cs:3.16666666666667,0.420169464) rectangle (axis cs:3.33333333333333,0.589021759);
\draw[draw=black,fill=sienna1408675,line width=0.004pt] (axis cs:3.16666666666667,0.589021759) rectangle (axis cs:3.33333333333333,0.6222254975);
\draw[draw=black,fill=gray163119111,line width=0.004pt] (axis cs:3.16666666666667,0.6222254975) rectangle (axis cs:3.33333333333333,0.7367322805);
\draw[draw=black,fill=rosybrown186153147,line width=0.004pt] (axis cs:3.16666666666667,0.7367322805) rectangle (axis cs:3.33333333333333,1.1624897815);
\draw[draw=black,fill=gainsboro232221219,line width=0.004pt,postaction={pattern=crosshatch}] (axis cs:4.16666666666667,0) rectangle (axis cs:4.33333333333333,0.262145163);
\draw[draw=black,fill=sienna1408675,line width=0.004pt,postaction={pattern=north east lines}] (axis cs:4.16666666666667,0.262145163) rectangle (axis cs:4.33333333333333,0.31072758);
\draw[draw=black,fill=gray163119111,line width=0.004pt,postaction={pattern=north east lines}] (axis cs:4.16666666666667,0.31072758) rectangle (axis cs:4.33333333333333,0.4088862315);
\draw[draw=black,fill=rosybrown186153147,line width=0.004pt,postaction={pattern=north east lines}] (axis cs:4.16666666666667,0.4088862315) rectangle (axis cs:4.33333333333333,0.571293866);
\draw[draw=black,fill=sienna1408675,line width=0.004pt] (axis cs:4.16666666666667,0.571293866) rectangle (axis cs:4.33333333333333,0.604486933);
\draw[draw=black,fill=gray163119111,line width=0.004pt] (axis cs:4.16666666666667,0.604486933) rectangle (axis cs:4.33333333333333,0.7460714605);
\draw[draw=black,fill=rosybrown186153147,line width=0.004pt] (axis cs:4.16666666666667,0.7460714605) rectangle (axis cs:4.33333333333333,1.2455962685);
\draw[draw=black,fill=gainsboro232221219,line width=0.004pt,postaction={pattern=crosshatch}] (axis cs:5.16666666666667,0) rectangle (axis cs:5.33333333333333,0.259451015);
\draw[draw=black,fill=sienna1408675,line width=0.004pt,postaction={pattern=north east lines}] (axis cs:5.16666666666667,0.259451015) rectangle (axis cs:5.33333333333333,0.3078960355);
\draw[draw=black,fill=gray163119111,line width=0.004pt,postaction={pattern=north east lines}] (axis cs:5.16666666666667,0.3078960355) rectangle (axis cs:5.33333333333333,0.406106659);
\draw[draw=black,fill=rosybrown186153147,line width=0.004pt,postaction={pattern=north east lines}] (axis cs:5.16666666666667,0.406106659) rectangle (axis cs:5.33333333333333,0.569086052);
\draw[draw=black,fill=sienna1408675,line width=0.004pt] (axis cs:5.16666666666667,0.569086052) rectangle (axis cs:5.33333333333333,0.604398871);
\draw[draw=black,fill=gray163119111,line width=0.004pt] (axis cs:5.16666666666667,0.604398871) rectangle (axis cs:5.33333333333333,0.86902287);
\draw[draw=black,fill=rosybrown186153147,line width=0.004pt] (axis cs:5.16666666666667,0.86902287) rectangle (axis cs:5.33333333333333,1.441037401);
\draw[draw=black,fill=gainsboro232221219,line width=0.004pt,postaction={pattern=crosshatch}] (axis cs:6.16666666666667,0) rectangle (axis cs:6.33333333333333,0.273278921);
\draw[draw=black,fill=sienna1408675,line width=0.004pt,postaction={pattern=north east lines}] (axis cs:6.16666666666667,0.273278921) rectangle (axis cs:6.33333333333333,0.321670656);
\draw[draw=black,fill=gray163119111,line width=0.004pt,postaction={pattern=north east lines}] (axis cs:6.16666666666667,0.321670656) rectangle (axis cs:6.33333333333333,0.4238464555);
\draw[draw=black,fill=rosybrown186153147,line width=0.004pt,postaction={pattern=north east lines}] (axis cs:6.16666666666667,0.4238464555) rectangle (axis cs:6.33333333333333,0.5831252805);
\draw[draw=black,fill=sienna1408675,line width=0.004pt] (axis cs:6.16666666666667,0.5831252805) rectangle (axis cs:6.33333333333333,0.6183598045);
\draw[draw=black,fill=gray163119111,line width=0.004pt] (axis cs:6.16666666666667,0.6183598045) rectangle (axis cs:6.33333333333333,0.818809669);
\draw[draw=black,fill=rosybrown186153147,line width=0.004pt] (axis cs:6.16666666666667,0.818809669) rectangle (axis cs:6.33333333333333,1.440619368);
\draw[draw=black,fill=gainsboro232221219,line width=0.004pt,postaction={pattern=crosshatch}] (axis cs:7.16666666666667,0) rectangle (axis cs:7.33333333333333,0.2775361325);
\draw[draw=black,fill=sienna1408675,line width=0.004pt,postaction={pattern=north east lines}] (axis cs:7.16666666666667,0.2775361325) rectangle (axis cs:7.33333333333333,0.3262236745);
\draw[draw=black,fill=gray163119111,line width=0.004pt,postaction={pattern=north east lines}] (axis cs:7.16666666666667,0.3262236745) rectangle (axis cs:7.33333333333333,0.425504794);
\draw[draw=black,fill=rosybrown186153147,line width=0.004pt,postaction={pattern=north east lines}] (axis cs:7.16666666666667,0.425504794) rectangle (axis cs:7.33333333333333,0.580805584);
\draw[draw=black,fill=sienna1408675,line width=0.004pt] (axis cs:7.16666666666667,0.580805584) rectangle (axis cs:7.33333333333333,0.6161121925);
\draw[draw=black,fill=gray163119111,line width=0.004pt] (axis cs:7.16666666666667,0.6161121925) rectangle (axis cs:7.33333333333333,0.985071037);
\draw[draw=black,fill=rosybrown186153147,line width=0.004pt] (axis cs:7.16666666666667,0.985071037) rectangle (axis cs:7.33333333333333,2.0578114015);
\end{axis}

\end{tikzpicture}

%% file: figures/python_fig/weak-cluster-eff-a2a.tex
\begin{tikzpicture}

\definecolor{cornflowerblue107174214}{RGB}{107,174,214}
\definecolor{darkgray}{RGB}{169,169,169}
\definecolor{darkgray150}{RGB}{150,150,150}
\definecolor{darkgray176}{RGB}{176,176,176}
\definecolor{dimgray99}{RGB}{99,99,99}
\definecolor{mediumpurple148103189}{RGB}{148,103,189}
\definecolor{sienna1408675}{RGB}{140,86,75}
\definecolor{silver189}{RGB}{189,189,189}

\begin{axis}[
axis y line =left, axis x line =bottom, axis x line shift=5pt, axis line style ={-},
log basis x={10},
minor xtick={},
minor ytick={},
tick align=center,
tick pos=left,
x grid style={darkgray176},
x grid style={draw=none},
xlabel near ticks,
xlabel={\(\displaystyle r\) = Number of nodes},
xmajorgrids,
xmin=0.780854050171763, xmax=180.315386683374,
xmode=log,
xtick style={color=black},
xtick={1,2,4,8,16,32,64,128},
xticklabels={$1$,$2$,$4$,$8$,$16$,$32$,$64$,$128$},
y axis line style={draw=none},
y grid style={darkgray176},
y grid style={dotted},
ylabel near ticks,
ylabel={\(\displaystyle \eta_{P,w}\)},
ymajorgrids,
ymin=0.1, ymax=1.1,
ytick style={color=darkgray},
ytick={0,0.2,0.4,0.6,0.8,1,1.2}
]
\addplot [very thin, silver189]
table {%
1 1
1.10623917447252 0.99893873573367
1.22376511113765 0.997767344792031
1.35377690629318 0.996474702625059
1.49760104723774 0.995048627608495
1.65670494618546 0.993475795313024
1.83271191201275 0.991741649151127
2.02741771259094 0.989830308090089
2.24280889668757 0.98772447238247
2.48108306237128 0.985405328582658
2.74467127871536 0.982852455496799
3.03626288956452 0.980043733160157
3.35883295243341 0.976955257456583
3.71567259249104 0.973561263593482
4.11042258132746 0.969834062324067
4.54711048310091 0.965743993564361
5.0301917470609 0.961259402877388
5.56459516570714 0.956346647175619
6.15577316238565 0.950970136899437
6.80975742139761 0.945092422825811
7.53322042820502 0.938674336493865
8.33354354761706 0.931675193930743
9.2188923345467 0.92405307283003
10.19829984572 0.915765173462264
11.2817588023525 0.906768273248348
12.4803235441126 0.897019283949513
13.8062228145891 0.886475918658869
15.2729845289947 0.875097473056957
16.8955737970867 0.862845721572446
18.6905456095288 0.849685924065455
20.6762137455262 0.835587932390936
22.8728376250683 0.820527378771975
25.3028290121995 0.804486919522851
27.990980678275 0.787457498681303
30.9647193582112 0.769439587095771
34.2543855806009 0.750444344208946
37.8935432267474 0.730494643084532
41.9193219769959 0.709625895136823
46.3727961382798 0.687886610538049
51.2994037179932 0.665338634261185
56.7494100199254 0.64205700674697
62.7784204922449 0.618129412413156
69.4479480600299 0.593655198247727
76.826040730738 0.568743967528345
84.987975875964 0.543513778679839
94.017028273117 0.518089004346814
104.005319743213 0.492597928598586
115.054759053482 0.467170178519669
127.278081674459 0.44193409834483
140.8 0.417014178482069
};
\addplot [very thin, darkgray150]
table {%
1 1
1.10623917447252 0.999469086146885
1.22376511113765 0.998882424815988
1.35377690629318 0.998234238896037
1.49760104723774 0.997518169570914
1.65670494618546 0.996727221518157
1.83271191201275 0.995853703791155
2.02741771259094 0.994889166242687
2.24280889668757 0.993824331395983
2.48108306237128 0.992649021735143
2.74467127871536 0.991352082473073
3.03626288956452 0.989921299966444
3.35883295243341 0.988343316088466
3.71567259249104 0.98660353904678
4.11042258132746 0.984686051351786
4.54711048310091 0.982573515906553
5.0301917470609 0.980247081510087
5.56459516570714 0.977686289448036
6.15577316238565 0.974868983295416
6.80975742139761 0.971771224580465
7.53322042820502 0.968367217561128
8.33354354761706 0.964629247047365
9.2188923345467 0.960527633960605
10.19829984572 0.956030714147755
11.2817588023525 0.951104846844958
12.4803235441126 0.945714460089154
13.8062228145891 0.939822141264419
15.2729845289947 0.93338878178988
16.8955737970867 0.926373785633853
18.6905456095288 0.918735351781145
20.6762137455262 0.910430840872381
22.8728376250683 0.901417235840153
25.3028290121995 0.89165170533414
27.990980678275 0.881092276893017
30.9647193582112 0.869698624024428
34.2543855806009 0.857432967453854
37.8935432267474 0.844261085700279
41.9193219769959 0.830153423805073
46.3727961382798 0.815086281558654
51.2994037179932 0.799043054154998
56.7494100199254 0.78201548924167
62.7784204922449 0.764004915393419
69.4479480600299 0.745023388874171
76.826040730738 0.725094699072452
84.987975875964 0.704255169195446
94.017028273117 0.682554188671868
104.005319743213 0.660054418085774
115.054759053482 0.636831616889909
127.278081674459 0.612974058734193
140.8 0.588581518540314
};
\addplot [very thin, dimgray99]
table {%
1 1
1.10623917447252 0.999787566788513
1.22376511113765 0.999552669971432
1.35377690629318 0.999292946465838
1.49760104723774 0.999005787348038
1.65670494618546 0.998688312890455
1.83271191201275 0.998337345200888
2.02741771259094 0.997949378265003
2.24280889668757 0.997520545183634
2.48108306237128 0.997046582388857
2.74467127871536 0.99652279061725
3.03626288956452 0.995943992416316
3.35883295243341 0.995304485961618
3.71567259249104 0.994597994969032
4.11042258132746 0.993817614500275
4.54711048310091 0.99295575248223
5.0301917470609 0.992004066793912
5.56459516570714 0.990953397821724
6.15577316238565 0.989793696447119
6.80975742139761 0.988513947514544
7.53322042820502 0.987102088935683
8.33354354761706 0.98554492672348
9.2188923345467 0.983828046421508
10.19829984572 0.981935721606872
11.2817588023525 0.979850820404625
12.4803235441126 0.977554711265207
13.8062228145891 0.975027169630858
15.2729845289947 0.972246287558608
16.8955737970867 0.969188388882478
18.6905456095288 0.965827953089813
20.6762137455262 0.962137551758023
22.8728376250683 0.958087802146196
25.3028290121995 0.953647343353102
27.990980678275 0.948782841323897
30.9647193582112 0.943459029886935
34.2543855806009 0.937638795892143
37.8935432267474 0.93128331735
41.9193219769959 0.924352264164938
46.3727961382798 0.916804071527659
51.2994037179932 0.908596296165043
56.7494100199254 0.899686065311476
62.7784204922449 0.890030627310117
69.4479480600299 0.879588011015559
76.826040730738 0.868317798489083
84.987975875964 0.856182011710114
94.017028273117 0.843146109068765
104.005319743213 0.829180081218727
115.054759053482 0.814259628523541
127.278081674459 0.798367394022745
140.8 0.781494216942802
};
\addplot [very thin, black]
table {%
1 1
1.10623917447252 1
1.22376511113765 1
1.35377690629318 1
1.49760104723774 1
1.65670494618546 1
1.83271191201275 1
2.02741771259094 1
2.24280889668757 1
2.48108306237128 1
2.74467127871536 1
3.03626288956452 1
3.35883295243341 1
3.71567259249104 1
4.11042258132746 1
4.54711048310091 1
5.0301917470609 1
5.56459516570714 1
6.15577316238565 1
6.80975742139761 1
7.53322042820502 1
8.33354354761706 1
9.2188923345467 1
10.19829984572 1
11.2817588023525 1
12.4803235441126 1
13.8062228145891 1
15.2729845289947 1
16.8955737970867 1
18.6905456095288 1
20.6762137455262 1
22.8728376250683 1
25.3028290121995 1
27.990980678275 1
30.9647193582112 1
34.2543855806009 1
37.8935432267474 1
41.9193219769959 1
46.3727961382798 1
51.2994037179932 1
56.7494100199254 1
62.7784204922449 1
69.4479480600299 1
76.826040730738 1
84.987975875964 1
94.017028273117 1
104.005319743213 1
115.054759053482 1
127.278081674459 1
140.8 1
};
\addplot [very thin, silver189]
table {%
1 1
1.10623917447252 0.99893873573367
1.22376511113765 0.997767344792031
1.35377690629318 0.996474702625059
1.49760104723774 0.995048627608495
1.65670494618546 0.993475795313024
1.83271191201275 0.991741649151127
2.02741771259094 0.989830308090089
2.24280889668757 0.98772447238247
2.48108306237128 0.985405328582658
2.74467127871536 0.982852455496799
3.03626288956452 0.980043733160157
3.35883295243341 0.976955257456583
3.71567259249104 0.973561263593482
4.11042258132746 0.969834062324067
4.54711048310091 0.965743993564361
5.0301917470609 0.961259402877388
5.56459516570714 0.956346647175619
6.15577316238565 0.950970136899437
6.80975742139761 0.945092422825811
7.53322042820502 0.938674336493865
8.33354354761706 0.931675193930743
9.2188923345467 0.92405307283003
10.19829984572 0.915765173462264
11.2817588023525 0.906768273248348
12.4803235441126 0.897019283949513
13.8062228145891 0.886475918658869
15.2729845289947 0.875097473056957
16.8955737970867 0.862845721572446
18.6905456095288 0.849685924065455
20.6762137455262 0.835587932390936
22.8728376250683 0.820527378771975
25.3028290121995 0.804486919522851
27.990980678275 0.787457498681303
30.9647193582112 0.769439587095771
34.2543855806009 0.750444344208946
37.8935432267474 0.730494643084532
41.9193219769959 0.709625895136823
46.3727961382798 0.687886610538049
51.2994037179932 0.665338634261185
56.7494100199254 0.64205700674697
62.7784204922449 0.618129412413156
69.4479480600299 0.593655198247727
76.826040730738 0.568743967528345
84.987975875964 0.543513778679839
94.017028273117 0.518089004346814
104.005319743213 0.492597928598586
115.054759053482 0.467170178519669
127.278081674459 0.44193409834483
140.8 0.417014178482069
};
\addplot [very thin, darkgray150]
table {%
1 1
1.10623917447252 0.999469086146885
1.22376511113765 0.998882424815988
1.35377690629318 0.998234238896037
1.49760104723774 0.997518169570914
1.65670494618546 0.996727221518157
1.83271191201275 0.995853703791155
2.02741771259094 0.994889166242687
2.24280889668757 0.993824331395983
2.48108306237128 0.992649021735143
2.74467127871536 0.991352082473073
3.03626288956452 0.989921299966444
3.35883295243341 0.988343316088466
3.71567259249104 0.98660353904678
4.11042258132746 0.984686051351786
4.54711048310091 0.982573515906553
5.0301917470609 0.980247081510087
5.56459516570714 0.977686289448036
6.15577316238565 0.974868983295416
6.80975742139761 0.971771224580465
7.53322042820502 0.968367217561128
8.33354354761706 0.964629247047365
9.2188923345467 0.960527633960605
10.19829984572 0.956030714147755
11.2817588023525 0.951104846844958
12.4803235441126 0.945714460089154
13.8062228145891 0.939822141264419
15.2729845289947 0.93338878178988
16.8955737970867 0.926373785633853
18.6905456095288 0.918735351781145
20.6762137455262 0.910430840872381
22.8728376250683 0.901417235840153
25.3028290121995 0.89165170533414
27.990980678275 0.881092276893017
30.9647193582112 0.869698624024428
34.2543855806009 0.857432967453854
37.8935432267474 0.844261085700279
41.9193219769959 0.830153423805073
46.3727961382798 0.815086281558654
51.2994037179932 0.799043054154998
56.7494100199254 0.78201548924167
62.7784204922449 0.764004915393419
69.4479480600299 0.745023388874171
76.826040730738 0.725094699072452
84.987975875964 0.704255169195446
94.017028273117 0.682554188671868
104.005319743213 0.660054418085774
115.054759053482 0.636831616889909
127.278081674459 0.612974058734193
140.8 0.588581518540314
};
\addplot [very thin, dimgray99]
table {%
1 1
1.10623917447252 0.999787566788513
1.22376511113765 0.999552669971432
1.35377690629318 0.999292946465838
1.49760104723774 0.999005787348038
1.65670494618546 0.998688312890455
1.83271191201275 0.998337345200888
2.02741771259094 0.997949378265003
2.24280889668757 0.997520545183634
2.48108306237128 0.997046582388857
2.74467127871536 0.99652279061725
3.03626288956452 0.995943992416316
3.35883295243341 0.995304485961618
3.71567259249104 0.994597994969032
4.11042258132746 0.993817614500275
4.54711048310091 0.99295575248223
5.0301917470609 0.992004066793912
5.56459516570714 0.990953397821724
6.15577316238565 0.989793696447119
6.80975742139761 0.988513947514544
7.53322042820502 0.987102088935683
8.33354354761706 0.98554492672348
9.2188923345467 0.983828046421508
10.19829984572 0.981935721606872
11.2817588023525 0.979850820404625
12.4803235441126 0.977554711265207
13.8062228145891 0.975027169630858
15.2729845289947 0.972246287558608
16.8955737970867 0.969188388882478
18.6905456095288 0.965827953089813
20.6762137455262 0.962137551758023
22.8728376250683 0.958087802146196
25.3028290121995 0.953647343353102
27.990980678275 0.948782841323897
30.9647193582112 0.943459029886935
34.2543855806009 0.937638795892143
37.8935432267474 0.93128331735
41.9193219769959 0.924352264164938
46.3727961382798 0.916804071527659
51.2994037179932 0.908596296165043
56.7494100199254 0.899686065311476
62.7784204922449 0.890030627310117
69.4479480600299 0.879588011015559
76.826040730738 0.868317798489083
84.987975875964 0.856182011710114
94.017028273117 0.843146109068765
104.005319743213 0.829180081218727
115.054759053482 0.814259628523541
127.278081674459 0.798367394022745
140.8 0.781494216942802
};
\addplot [very thin, black]
table {%
1 1
1.10623917447252 1
1.22376511113765 1
1.35377690629318 1
1.49760104723774 1
1.65670494618546 1
1.83271191201275 1
2.02741771259094 1
2.24280889668757 1
2.48108306237128 1
2.74467127871536 1
3.03626288956452 1
3.35883295243341 1
3.71567259249104 1
4.11042258132746 1
4.54711048310091 1
5.0301917470609 1
5.56459516570714 1
6.15577316238565 1
6.80975742139761 1
7.53322042820502 1
8.33354354761706 1
9.2188923345467 1
10.19829984572 1
11.2817588023525 1
12.4803235441126 1
13.8062228145891 1
15.2729845289947 1
16.8955737970867 1
18.6905456095288 1
20.6762137455262 1
22.8728376250683 1
25.3028290121995 1
27.990980678275 1
30.9647193582112 1
34.2543855806009 1
37.8935432267474 1
41.9193219769959 1
46.3727961382798 1
51.2994037179932 1
56.7494100199254 1
62.7784204922449 1
69.4479480600299 1
76.826040730738 1
84.987975875964 1
94.017028273117 1
104.005319743213 1
115.054759053482 1
127.278081674459 1
140.8 1
};
\addplot [very thin, silver189]
table {%
1 1
1.10623917447252 0.99893873573367
1.22376511113765 0.997767344792031
1.35377690629318 0.996474702625059
1.49760104723774 0.995048627608495
1.65670494618546 0.993475795313024
1.83271191201275 0.991741649151127
2.02741771259094 0.989830308090089
2.24280889668757 0.98772447238247
2.48108306237128 0.985405328582658
2.74467127871536 0.982852455496799
3.03626288956452 0.980043733160157
3.35883295243341 0.976955257456583
3.71567259249104 0.973561263593482
4.11042258132746 0.969834062324067
4.54711048310091 0.965743993564361
5.0301917470609 0.961259402877388
5.56459516570714 0.956346647175619
6.15577316238565 0.950970136899437
6.80975742139761 0.945092422825811
7.53322042820502 0.938674336493865
8.33354354761706 0.931675193930743
9.2188923345467 0.92405307283003
10.19829984572 0.915765173462264
11.2817588023525 0.906768273248348
12.4803235441126 0.897019283949513
13.8062228145891 0.886475918658869
15.2729845289947 0.875097473056957
16.8955737970867 0.862845721572446
18.6905456095288 0.849685924065455
20.6762137455262 0.835587932390936
22.8728376250683 0.820527378771975
25.3028290121995 0.804486919522851
27.990980678275 0.787457498681303
30.9647193582112 0.769439587095771
34.2543855806009 0.750444344208946
37.8935432267474 0.730494643084532
41.9193219769959 0.709625895136823
46.3727961382798 0.687886610538049
51.2994037179932 0.665338634261185
56.7494100199254 0.64205700674697
62.7784204922449 0.618129412413156
69.4479480600299 0.593655198247727
76.826040730738 0.568743967528345
84.987975875964 0.543513778679839
94.017028273117 0.518089004346814
104.005319743213 0.492597928598586
115.054759053482 0.467170178519669
127.278081674459 0.44193409834483
140.8 0.417014178482069
};
\addplot [very thin, darkgray150]
table {%
1 1
1.10623917447252 0.999469086146885
1.22376511113765 0.998882424815988
1.35377690629318 0.998234238896037
1.49760104723774 0.997518169570914
1.65670494618546 0.996727221518157
1.83271191201275 0.995853703791155
2.02741771259094 0.994889166242687
2.24280889668757 0.993824331395983
2.48108306237128 0.992649021735143
2.74467127871536 0.991352082473073
3.03626288956452 0.989921299966444
3.35883295243341 0.988343316088466
3.71567259249104 0.98660353904678
4.11042258132746 0.984686051351786
4.54711048310091 0.982573515906553
5.0301917470609 0.980247081510087
5.56459516570714 0.977686289448036
6.15577316238565 0.974868983295416
6.80975742139761 0.971771224580465
7.53322042820502 0.968367217561128
8.33354354761706 0.964629247047365
9.2188923345467 0.960527633960605
10.19829984572 0.956030714147755
11.2817588023525 0.951104846844958
12.4803235441126 0.945714460089154
13.8062228145891 0.939822141264419
15.2729845289947 0.93338878178988
16.8955737970867 0.926373785633853
18.6905456095288 0.918735351781145
20.6762137455262 0.910430840872381
22.8728376250683 0.901417235840153
25.3028290121995 0.89165170533414
27.990980678275 0.881092276893017
30.9647193582112 0.869698624024428
34.2543855806009 0.857432967453854
37.8935432267474 0.844261085700279
41.9193219769959 0.830153423805073
46.3727961382798 0.815086281558654
51.2994037179932 0.799043054154998
56.7494100199254 0.78201548924167
62.7784204922449 0.764004915393419
69.4479480600299 0.745023388874171
76.826040730738 0.725094699072452
84.987975875964 0.704255169195446
94.017028273117 0.682554188671868
104.005319743213 0.660054418085774
115.054759053482 0.636831616889909
127.278081674459 0.612974058734193
140.8 0.588581518540314
};
\addplot [very thin, dimgray99]
table {%
1 1
1.10623917447252 0.999787566788513
1.22376511113765 0.999552669971432
1.35377690629318 0.999292946465838
1.49760104723774 0.999005787348038
1.65670494618546 0.998688312890455
1.83271191201275 0.998337345200888
2.02741771259094 0.997949378265003
2.24280889668757 0.997520545183634
2.48108306237128 0.997046582388857
2.74467127871536 0.99652279061725
3.03626288956452 0.995943992416316
3.35883295243341 0.995304485961618
3.71567259249104 0.994597994969032
4.11042258132746 0.993817614500275
4.54711048310091 0.99295575248223
5.0301917470609 0.992004066793912
5.56459516570714 0.990953397821724
6.15577316238565 0.989793696447119
6.80975742139761 0.988513947514544
7.53322042820502 0.987102088935683
8.33354354761706 0.98554492672348
9.2188923345467 0.983828046421508
10.19829984572 0.981935721606872
11.2817588023525 0.979850820404625
12.4803235441126 0.977554711265207
13.8062228145891 0.975027169630858
15.2729845289947 0.972246287558608
16.8955737970867 0.969188388882478
18.6905456095288 0.965827953089813
20.6762137455262 0.962137551758023
22.8728376250683 0.958087802146196
25.3028290121995 0.953647343353102
27.990980678275 0.948782841323897
30.9647193582112 0.943459029886935
34.2543855806009 0.937638795892143
37.8935432267474 0.93128331735
41.9193219769959 0.924352264164938
46.3727961382798 0.916804071527659
51.2994037179932 0.908596296165043
56.7494100199254 0.899686065311476
62.7784204922449 0.890030627310117
69.4479480600299 0.879588011015559
76.826040730738 0.868317798489083
84.987975875964 0.856182011710114
94.017028273117 0.843146109068765
104.005319743213 0.829180081218727
115.054759053482 0.814259628523541
127.278081674459 0.798367394022745
140.8 0.781494216942802
};
\addplot [very thin, black]
table {%
1 1
1.10623917447252 1
1.22376511113765 1
1.35377690629318 1
1.49760104723774 1
1.65670494618546 1
1.83271191201275 1
2.02741771259094 1
2.24280889668757 1
2.48108306237128 1
2.74467127871536 1
3.03626288956452 1
3.35883295243341 1
3.71567259249104 1
4.11042258132746 1
4.54711048310091 1
5.0301917470609 1
5.56459516570714 1
6.15577316238565 1
6.80975742139761 1
7.53322042820502 1
8.33354354761706 1
9.2188923345467 1
10.19829984572 1
11.2817588023525 1
12.4803235441126 1
13.8062228145891 1
15.2729845289947 1
16.8955737970867 1
18.6905456095288 1
20.6762137455262 1
22.8728376250683 1
25.3028290121995 1
27.990980678275 1
30.9647193582112 1
34.2543855806009 1
37.8935432267474 1
41.9193219769959 1
46.3727961382798 1
51.2994037179932 1
56.7494100199254 1
62.7784204922449 1
69.4479480600299 1
76.826040730738 1
84.987975875964 1
94.017028273117 1
104.005319743213 1
115.054759053482 1
127.278081674459 1
140.8 1
};
\addplot [semithick, cornflowerblue107174214, mark=*, mark size=1.5, mark options={solid}]
table {%
1 1
2 0.9270528431469
4 0.767846517475025
8 0.711005663600572
16 0.652615854961908
32 0.598831209444036
64 0.573271315034535
128 0.45449464634698
};
\addplot [semithick, mediumpurple148103189, mark=*, mark size=1.5, mark options={solid}]
table {%
1 1
2 0.983998469475404
4 0.916063467649853
8 0.896755772042187
16 0.889309437335312
32 0.859029431282697
64 0.855139510990118
128 0.801793730023578
};
\addplot [semithick, sienna1408675, mark=*, mark size=1.5, mark options={solid}]
table {%
1 1
2 0.993440600296328
4 0.881647178966178
8 0.831062350288711
16 0.77561366747142
32 0.670420829695037
64 0.67061536965259
128 0.469480093897711
};
\end{axis}

\end{tikzpicture}

%% file: figures/python_tab/comp_clusters_a2a_time_unknowns_rank.tex
\begin{tabular}{lrrrrr}
\hline
  N nodes   &   \centercell{$1$ } &   \centercell{$2$ } &   \centercell{$8$ } &   \centercell{$64$ } &   \centercell{$128$ } \\
\hline
 Vega       &               31.58 &               29.28 &               22.46 &                18.11 &                 14.35 \\
 MeluXina   &               30.34 &               29.85 &               27.21 &                25.94 &                 24.33 \\
 Lumi       &               34.19 &               33.96 &               28.41 &                22.93 &                 16.05 \\
\hline
\end{tabular}